\documentclass[linenumbers]{aastex63}
\usepackage{longtable}
\usepackage{booktabs}
\usepackage{CJK}

\pdfoutput=1

\received{**}
\revised{**}
\accepted{**}
\submitjournal{ApJ}

\shorttitle{Chemical clock}
\shortauthors{Wang et al.}

\begin{document}
\begin{CJK*}{UTF8}{gbsn}

\title{A possible chemical clock in high-mass star-forming regions: $N\rm (HC_{3}N)$/$N\rm (N_{2}H^{+})$?}

\correspondingauthor{J. S. Zhang}
\email{jszhang@gzhu.edu.cn}

\author[0000-0001-9155-0777]{Y. X. Wang(汪友鑫)}
\affiliation{Center for Astrophysics, Guangzhou University, Guangzhou, 510006, PR China}

\author[0000-0002-5161-8180]{J. S. Zhang(张江水)}
\affiliation{Center for Astrophysics, Guangzhou University, Guangzhou, 510006, PR China}

\author[0000-0002-5634-131X]{H. Z. Yu(余鸿智)}
\affiliation{Ural Federal University, 19 Mira Street, 620002 Ekaterinburg, Russia}
\affiliation{Center for Astrophysics, Guangzhou University, Guangzhou, 510006, PR China}

\author[0000-0002-5412-352X]{Y. Wang(王尧)}
\affiliation{Purple Mountain Observatory and Key Laboratory of Radio Astronomy, Chinese Academy of Sciences, 10 Yuanhua Road, Nanjing 210023, PR China}

\author[0000-0001-5574-0549]{Y. T. Yan(闫耀庭)}
\affiliation{Max-Planck-Institut f{\"u}r Radioastronomie, Auf dem H{\"u}gel 69, D-53121 Bonn, Germany}

\author[0000-0001-8980-9663]{J. L. Chen(陈家梁)}
\affiliation{Center for Astrophysics, Guangzhou University, Guangzhou, 510006, PR China}

\author{J. Y. Zhao(赵洁瑜)}
\affiliation{Center for Astrophysics, Guangzhou University, Guangzhou, 510006, PR China}

\author[0000-0002-5230-8010]{Y. P. Zou(邹益鹏)}
\affiliation{Center for Astrophysics, Guangzhou University, Guangzhou, 510006, PR China}

\begin{abstract}
We conducted observations of multiple HC$_{3}$N (J = 10$-$9, 12$-$11 and 16$-$15) lines and N$_{2}$H$^{+}$ (J = 1$-$0) line toward a large sample of 61 ultracompact (UC) H{\sc ii} regions, through the Institut de Radioastronomie Millm$\acute{e}$trique 30 m and the Arizona Radio Observatory 12 m telescopes. The N$_{2}$H$^{+}$ J = 1$-$0 line is detected in 60 sources and HC$_{3}$N is detected in 59 sources, including 40 sources with three lines, 9 sources with two lines, and 10 sources with one line. Using the rotational diagram, the rotational temperature and column density of HC$_{3}$N were estimated toward sources with at least two HC$_{3}$N lines. For 10 sources with only one HC$_{3}$N line, their parameters were estimated, taking one average value of $T_{\rm rot}$. For N$_{2}$H$^{+}$, we estimated the optical depth of the N$_{2}$H$^{+}$ J = 1$-$0 line, based on the line intensity ratio of its hyperfine structure lines. Then the excitation temperature and column density were calculated. When combining our results in UC H{\sc ii} regions and previous observation results on high-mass starless cores (HMSCs) and high-mass protostellar cores (HMPOs), the $N\rm (HC_{3}N)$/$N\rm (N_{2}H^{+})$ ratio clearly increases with the evolution of high mass star-forming regions (HMSFRs). Moreover, positive correlations between the ratio and other evolutionary indicators  (dust temperature, bolometric luminosity, and luminosity-to-mass ratio) are found. Thus we propose the ratio of $N\rm (HC_{3}N)$/$N\rm (N_{2}H^{+})$ as a reliable chemical clock of HMSFRs.
\end{abstract}


\section{Introduction} 

High-mass stars (\textgreater 8 M$_{\odot}$ and \textgreater 10$^{3}$ L$_{\odot}$) play an important role in the dynamics and evolution of the interstellar medium. They play a major role in the energy budget of any galaxy, driving their evolution \citep{1996GReGr..28.1309M,2012ARA&A..50..531K}. They also produce all heavy elements in the universe, which are fed back into the interstellar medium via feedback processes of high-mass stars, such as shocks, ultraviolet (UV) radiation, and stellar winds. The heavy elements will be used as the raw material to form new generations of stars and planets \citep[see reviews by][and references therein]{2007ARA&A..45..481Z,2014prpl.conf..243K}. Therefore, it is extremely important to understand the high-mass star-formation process. However, high-mass stars are rare in galaxies, accounting for about 3$\%$ of the total number of stars \citep{2001MNRAS.322..231K,2003ApJ...586L.133C}. The majority of high-mass stars are at large distances away \citep[$\textgreater$ 2 kpc; e.g.][]{2018MNRAS.473.1059U}. They also evolve very quickly, reaching the main sequence while still deeply embedded \citep[e.g.][]{2018ARA&A..56...41M,2020ApJ...900...82P}. All of these complications hinder the understanding of high-mass star formation processes \citep{2022MNRAS.510.3389U}. 

So far, it is generally accepted that high-mass star-forming regions (HMSFRs) evolve from high-mass starless cores (HMSCs) to high-mass protostellar cores (HMPOs), then to ultracompact (UC) H{\sc ii} regions \citep[e.g.][]{2007prpl.conf..165B,2014A&A...563A..97G,2015A&A...579A..80G,2015ApJS..219....2J,2015A&A...576A.131H}. HMSCs, as the first stage of HMSFR, are formed by cold, dense gas and dust with a temperature of $\sim$ 10 K. They primarily emit radiation at submillimeter wavelengths, while emitting almost nothing at mid-infrared wavelengths \citep[e.g.][]{2005ApJ...634L..57S,2006ApJ...641..389R,2010A&A...518L..78B,2017ApJS..231...11Y,2020A&A...637A..40Z}. HMPOs are the second phase stage of HMSFRs, when an active protostar(s) is formed. Emission at mid-infrared wavelengths has been demonstrated, but no radio continuum emission. At this stage, there is a formed disk that transfers the infalling material from the envelope to the HMPO \citep[e.g.][]{2002ApJ...566..931S,2002ApJ...566..945B,2006ApJ...638..241E,2007prpl.conf..197C,2009A&A...498..147G,2021ApJ...912..108H}. Finally, an UC H{\sc ii} region (size $\leq$ 0.1 pc, electron density \textgreater 10$^{4}$ cm$^{-3}$ and emission measure $\geq$ 10$^{7}$ pc cm$^{-6}$) is formed when a central high-mass star begins to ionize the surrounding gas \citep[e.g.][]{1989ApJ...340..265W,2005AJ....130..711K,2007prpl.conf..181H,2018A&A...617A..60S}. A Lyman continuum flux would be derived from the free-free emission (bremsstrahlung emission) emitted by the hydrogen, which makes UC H{\sc ii} regions directly observable in radio continuum bands \citep[e.g.][]{2013MNRAS.435..400U,2016ApJ...833...18H,2018AA...615A.103K,2019MNRAS.487.1057D}. However, this classification method, according to the physical conditions mentioned above, is extremely crude and may lead to potential overlap among these phases.

For detailed understanding of HMSFRs, it is necessary to determine the chemical composition of high-mass stars at different stages of evolution, owing to the fact that the chemical composition carries a large amount of information about the star-forming region \citep[e.g.][]{2011ARA&A..49..471M,2012A&ARv..20...56C,2020ARA&A..58..727J,2021PhR...893....1O}. Combining chemical features with physical conditions should be more helpful in distinguishing the different phases of HMSFRs. 

To date, many attempts have been made to study the chemical evolution of HMSFRs, known as the chemical clock. The chemical clock is a parameter expressing the chemical evolution by ratios of the column density for two molecules \citep[e.g.][]{2004A&A...422..159W,2017ApJS..228...12T,2019ApJ...872..154T,2021SCPMA..6479511X}. The Millimeter Astronomy Legacy Team 90 GHz Survey (MALT90)\footnote{\url{http://malt90.bu.edu/}} provides the largest sample of molecular lines in HMSFRs \citep{2011ApJS..197...25F,2013PASA...30...57J}. Based on the MALT90 survey data, a number of studies were conducted to investigate the chemical clock of HMSFRs. For example, \cite{2013ApJ...777..157H} calculated the column density ratio between N$_{2}$H$^{+}$ and HCO$^{+}$, showing a slight increase from HMSCs to HMPOs, then to UC H{\sc ii} regions. \cite{2015MNRAS.451.2507Y} also carried out an investigation of the $N\rm (N_{2}H^{+})$/$N\rm (H^{13}CO^{+})$ ratio and the  $N\rm (CCH)$/$N\rm (H^{13}CO^{+})$ ratio, which showed a marginal decrease from HMPOs to UC H{\sc ii} regions. Based on H$^{13}$CN and HN$^{13}$C line observations, \cite{2015ApJS..219....2J} found that a slight statistically increasing tendency in the HCN/HNC abundance ratio with the evolution of HMSFRs and suggested that the HCN/HNC abundance ratio can be used as a chemical clock. Moreover, \cite{2019ApJ...872..154T} carried out a survey observations of HC$_{3}$N (J = 9$-$8 and 10$-$9), N$_{2}$H$^{+}$ (J = 1$-$0) and CCS (J$_{N}$ = 6$_{7}-$5$_{6}$) toward 17 HMSCs and 28 HMPOs using the Nobeyama 45 m telescope. They obtained that the $N\rm (HC_{3}N)$/$N\rm (N_{2}H^{+})$ ratio clearly increases from HMSCs to HMPOs and therefore proposed that the $N\rm (HC_{3}N)$/$N\rm (N_{2}H^{+})$ ratio may be used as a reliable chemical clock in HMSFRs. However, this survey was not sufficient to consider the $N\rm (HC_{3}N)$/$N\rm (N_{2}H^{+})$ ratio as a good chemical clock due to the lack of observational evidence for the UC H{\sc ii} regions. Therefore, we have conducted survey observations of HC$_{3}$N and N$_{2}$H$^{+}$ molecular line transitions toward UC H{\sc ii} regions.

In this paper, we presented results of the survey of HC$_{3}$N (J = 10$-$9, 12$-$11, 16$-$15) and N$_{2}$H$^{+}$ (J = 1$-$0) toward 61 UC H{\sc ii} regions using the Institut de Radioastronomie Millm$\acute{e}$trique (IRAM) 30 m telescope\footnote{\url{https://www.iram-institute.org/EN/content-page-55-7-55-0-0-0.html}} and the Arizona Radio Observatory (ARO) 12 m telescope\footnote{\url{https://aro.as.arizona.edu/?q=facilities/uarizona-aro-12-meter-telescope}}. Our sample and observation information are summarized in Section 2. In Section 3, we present the fitting results of detected spectral lines and made the analyze on the spectral features (line wing, line width). The physical parameters (column density, temperature) are also estimated for our detections. In Section 4, we investigate the $N\rm (HC_{3}N)$/$N\rm (N_{2}H^{+})$ ratio at different stages of HMSFRs and make a comparative analysis on the ratio and other evolutionary indicators, in order to check the $N\rm (HC_{3}N)$/$N\rm (N_{2}H^{+})$ ratio as a reliable chemical clock of HMSFRs. A short summary is listed in Section 5.

\section{Source selection and observations}
\subsection{Source selection}
UC H{\sc ii} regions are the most luminous objects in the Milky Way at the far-infrared wavelengths, and have radio continuum emission.
One large UC {\sc ii} region sample of about 600 sources has been established \citep[][]{2005AJ....129..348G,2005AJ....130..156G,2007A&A...461...11U,2009AA...501..539U,2013MNRAS.435..400U,2015A&A...579A..71C,2016ApJ...833...18H,2018AA...615A.103K,2019MNRAS.487.1057D}, based on infrared survey \citep[][]{2001AJ....121.2819P}, radio surveys \citep[][]{2012PASP..124..939H,2013ApJS..205....1P}, and submillimeter surveys \citep[][]{2006A&A...453.1003T,2009A&A...504..415S,2014A&A...565A..75C}.

Toward a large sample of HMSFRs with accurate distances, thanks to the results of the Bar and Spiral Structure Legacy (BeSSeL\footnote{\url{http://bessel.vlbi-astrometry.org/}}) project \citep{2014ApJ...783..130R,2019ApJ...885..131R},  we have performed one systematic survey on CS and its isotopic molecules, to make direct measurements of Galactic carbon and sulfur isotope ratios \citep[][]{2022Yan}. Crossmatching this sample with the catalog of UC H{\sc ii} regions, we obtain 61 UC H{\sc ii} regions with accurate distances. These 61 UC H{\sc ii} regions serve as our targets to test the chemical clock $N\rm (HC_{3}N)$/$N\rm (N_{2}H^{+})$ in HMSFRs. The detailed information of these UC H{\sc ii} regions are summarized in Table \ref{tab1}.

\subsection{Observations}
We carried out observations of multiple lines of HC$_{3}$N (J = 10$-$9, 12$-$11, 16$-$15) and N$_{2}$H$^{+}$ (J = 1$-$0) lines with the IRAM 30 m telescope and the ARO 12 m telescope. Table \ref{tab2} lists the parameters of targeted lines, including the rest frequency $\nu$, the upper energy level $E_{\rm u}$, the product of the line strength, and the square of the electric dipole moment $S\mu^{2}$, which are taken from the Cologne Database for Molecular Spectroscopy (CDMS)\footnote{\url{https://cdms.astro.uni-koeln.de/cdms/portal/}} catalog, and the half-power beamwidth (HPBW) for each transition. 

\subsubsection{IRAM 30 m observations}
The observations of the HC$_{3}$N (J = 10$-$9 and 16$-$15) and N$_{2}$H$^{+}$ (J = 1-0) lines were carried out toward our sample with IRAM 30 m telescope during August 18$-$24, 2020 within project 022-20 (PI Hongzhi Yu). The Eight Mixer Receiver (EMIR) heterodyne receivers (E090 and E150) were employed covering the frequency ranges of 90.5$-$98.2 GHz and 138.3$-$146 GHz. We applied a fast Fourier transform spectrometer (FTS) backend in the wideband mode (16 GHz, containing four IF subbands with 4 GHz each in dual polarization), with a velocity resolution of $\sim$ 0.64 km s$^{-1}$ at 90 GHz, $\sim$ 0.62 km s$^{-1}$ at 93 GHz, and $\sim$ 0.40 km s$^{-1}$ at 145 GHz. The standard position-switching mode and the chopper-wheel calibration method were adopted in our observations, with the off$-$source position set at 30$^\prime$ in azimuth. The typical system temperature was 92$-$150 K for HC$_{3}$N (J = 10$-$9) and 120$-$300 K for HC$_{3}$N (J = 16$-$15). The main-beam brightness temperature $T_{\rm mb}$ can be obtained from the antenna temperature $T_{\rm A}^{\star}$ divided by the ratio of the main-beam efficiency $B_{\rm eff}$ and the forward efficiency $F_{\rm eff}$. The ratio of $B_{\rm eff}$/$F_{\rm eff}$ is about 85$\%$ at 90 and 93 GHz, and 78$\%$ at 145 GHz\footnote{\url{https://publicwiki.iram.es/Iram30mEfficiencies}}. 

\subsubsection{ARO 12 m observations}
Toward those 49 sources with detections of both HC$_{3}$N lines (J = 10$-$9 and 16$-$15) lines by the IRAM 30 m telescope, we performed complementary observations on their HC$_{3}$N J = 12$-$11 line in 2021 May, using the ARO 12 m telescope (Yu$\_$20A$\_$1, PI, Hongzhi Yu). The HC$_{3}$N J = 10$-$9 line of 33 sources among them was also observed, to check how much different beam sizes affect our results. A 3 mm sideband separating (SBS) and the dual polarization receiver and the two backends of the ARO Wideband Spectrometer (AROWS) mode were adopted, with a velocity resolution of $\sim$ 0.52 km s$^{-1}$ at 90 GHz and $\sim$ 0.43 km s$^{-1}$ at 109 GHz. The center frequencies were set at 97.75 GHz and 108.80 GHz, respectively. A standard position-switching mode was adopted in our observations by using 30$^\prime$ offset in azimuth. The typical system temperature was 85$-$175 K for HC$_{3}$N (J = 10$-$9) and 116$-$243 K for HC$_{3}$N (J = 12$-$11). Calibrated by the equation $T_{\rm mb}$ = $T_{\rm A}^{\star}$/$\eta_{b}$, $T_{\rm A}^{\star}$ can be converted to the $T_{\rm mb}$. Determined from continuum observations of the Jupiter, the $T_{\rm mb}$ is $\sim$ 94$\%$ at 90 GHz and 92$\%$ at 109 GHz.

\section{Results and analyses}
\subsection{Spectra fitting results}
The CLASS/GILDAS\footnote{\url{https://www.iram.fr/IRAMFR/GILDAS/}} software was used to process the data. We first checked all spectra and then averaged the spectra for two polarizations of each transition for each source. A first-order linear fit was used to subtract the spectral baseline. 

We set the criteria for line detection as a signal-to-noise ratio (S/N) above 5 ($T_{\rm mb}$/rms $\textgreater$ 5). Out of 61 UC H{\sc ii} regions observed by the IRAM 30 m telescope, N$_{2}$H$^{+}$ (J = 1$-$0) is detected in all sources, except G060.57$-$00.18. Forty-nine sources among them were detected in both HC$_{3}$N J = 10$-$9 and J = 16$-$15 lines and 10 sources among them were detected in HC$_{3}$N J = 10$-$9. Our complementary observations by the ARO 12 m telescope detected the HC$_{3}$N (J = 12$-$11) line in 40 sources out of those 49 sources.

For our sources, we fitted the spectra of HC$_{3}$N with a single Gaussian profile to obtain their spectral parameters, which are summarized in Table \ref{tab3}. The HC$_{3}$N J = 16$-$15 line in G033.39+00.00 and the HC$_{3}$N J = 10$-$9 line in G097.53+03.18 show a large velocity distribution that deviates from the Gaussian profile, so we use the "Print area" command in CLASS to obtain the total integrated intensity. The spectra of HC$_{3}$N lines of all 61 sources are shown in Figure \ref{fig1}. Our ARO 12 m observations also detected HC$_{3}$N (J = 10$-$9) in 33 sources. For comparison, their HC$_{3}$N J = 10$-$9 spectra from the IRAM 30 m telescope and the ARO 12 m telescope are displayed in Figure \ref{fig2}.

Although N$_{2}$H$^{+}$ (J = 1$-$0) theoretically has seven hyperfine transitions at different frequencies, the relatively broad line width ($\geq$ 2 km s$^{-1}$) observed toward HMSFRs results in the seven components blending into three groups with roughly Gaussian shapes \citep[e.g.][]{2009MNRAS.394..323P}. Thus we tried to fit the spectra of N$_{2}$H$^{+}$ (J = 1$-$0) with three Gaussian profiles. For seven sources (G001.00$-$00.23, G028.39+00.08, G030.70$-$00.06, G030.81$-$00.05, G032.79+00.19, G045.49+00.12, and G097.53+03.18) with blending velocity components, the total integrated intensity of them was obtained by the "Print area" method. The N$_{2}$H$^{+}$ J = 1$-$0 spectra of our sample are presented in Figure \ref{fig3} and the fitting results are summarized in Table \ref{tab4}.

\subsection{Features of spectral lines}
\subsubsection{Line wing}
Line wing features can be found clearly in the HC$_{3}$N spectra of 23 sources and most of them show both blue and red wings (for more details see Table \ref{tab5}). We crossmatched them with the outflow catalogs identified by the CO molecule \citep{2004AA...426..503W,2008AJ....136.2391C,2015MNRAS.453..645M,2016AJ....152...92L,2018ApJS..235....3Y,2018ApJ...867..167L,2022AA...658A.160Y} and found that 20 out of 23 sources show CO molecular outflows. The detailed outflow information for these sources is also shown in Table \ref{tab5}. 

Molecular outflows are one type of feedback of high-mass star formation to the surrounding molecular cloud, and often lead to the production of shock activities \citep[e.g.][]{2007prpl.conf..245A}. To verify the presence of shock activities in these sources, we investigated observations of other shock tracers toward these sources. The H$_{2}$O maser, produced in the circumstellar disk, is caused by shocks from molecular outflows and stellar winds from young stars in HMSFRs \citep{2005Ap&SS.295...53T}. The Class \uppercase\expandafter{\romannumeral1} CH$_{3}$OH maser, associated with shocks, is known to be collision pumped \citep[e.g.][]{2016A&A...592A..31L}. We find that these sources mostly (19 of 23) show Class \uppercase\expandafter{\romannumeral1} CH$_{3}$OH (methanol) maser emission \citep[e.g.][]{2017ApJS..231...20Y,2019ApJS..244....2K} and all of them are associated with H$_{2}$O maser emission except G034.79$-$01.38 \citep[e.g.][]{2010MNRAS.406.1487B,2014MNRAS.443.2923T,2022AJ....163..124L,2011MNRAS.416..178B,2015MNRAS.453.4203X,2005AA...434..613S,2019ApJS..244....2K,2013ApJ...764...61C}. Therefore, we proposed the existence of shock activities in these sources (Table \ref{tab5}), which needs further mapping observations with high resolution.  

\subsubsection{Line width}

The line width from the Gaussian fit of observed lines gives the information about the state of turbulence and the region where the gas is mainly being emitting from \citep{2002ApJ...566..945B}, and it consists of thermal and nonthermal components.
The velocity dispersion of molecules attributed by the thermal motion can be estimated by the following formula: 
\begin{equation}\label{6}
\Delta V_{thermal} = (\frac{8kTln2}{m})^{1/2},  
\end{equation}
where $m$ is the molecular mass of the gas with a value of 51 amu and 29 amu for HC$_{3}$N and N$_{2}$H$^{+}$, respectively, $k$ is the boltzmann constant, and $T$ is the gas temperature. Then the nonthermal line width can be obtained by
\begin{equation}\label{7}
\Delta V_{non-thermal} = (\Delta V_{FWHM}^{2} - \Delta V_{thermal}^{2})^{1/2},
\end{equation}
where $\Delta V_{FWHM}$ is the observed line width.

Our calculated results show that the thermal line width can be negligible, with a maximum value of 0.17 km s$^{-1}$ for HC$_{3}$N and 0.19 km s$^{-1}$ for N$_{2}$H$^{+}$, which is less than 10$\%$ of the total line width of our sources. Figure \ref{fig4} shows plots of the comparison of the line width between HC$_{3}$N lines (J = 10$-$9, 12$-$11 and 16$-$15) and N$_{2}$H$^{+}$ (J = 1$-$0) line. We used the FWHM of N$_{2}$H$^{+}$ J = 1$-$0, F$_{1}$ = 0$-$1 as the intrinsic FWHM of N$_{2}$H$^{+}$, since this group spectra consists of only one component, without blending. It shows clearly that the line width of HC$_{3}$N tend to be larger than that of N$_{2}$H$^{+}$ toward our UC H{\sc ii} region sample. \cite{2021PASJ...73..467F} argued that the inner dense warm regions have more turbulence than the outer regions, by comparing the linewidths of the different transitions of HC$_{3}$N. Therefore, our results reflect that HC$_{3}$N is more likely to exist in inner and more active star-forming regions compared to N$_{2}$H$^{+}$, considering the line width mainly from the nonthermal component caused by turbulence.

\subsection{Physical parameters of HC\texorpdfstring{$_{3}$}NN and N\texorpdfstring{$_{2}$}HH$^{+}$}
\subsubsection{Rotational temperature and column density of HC\texorpdfstring{$_{3}$}NN}
Most of the lines of HC$_{3}$N are usually optically thin \citep{1976ApJ...205...82M}. Therefore, the optical depth effect is insignificant and is not used for further analysis. Under local thermodynamic equilibrium (LTE) assumption, the rotational temperature and the column density can be usually determined by the rotational diagram method with the following formula \citep{1999ApJ...517..209G}:
\begin{equation}\label{1}
\centering
ln\frac{3kW}{8\pi ^{3}\nu S\mu ^{2}} = ln\frac{N}{Q({T_{rot}})}- \frac{E_{u}}{kT_{rot}},
\end{equation}
where $k$ and $W$ are the Boltzmann constant and the integrated line intensity, respectively, $Q(T_{\rm rot})$ = 4.58 $T_{\rm rot}$ + 0.28 is the partition function for HC$_{3}$N \citep{2022A&A...663A.177W}, $T_{\rm rot}$ and $N$ are the rotational temperature and the column density, respectively, and the $S\mu^{2}$, $\nu$ and $E_{\rm u}$ values are taken from the CDMS catalog (see Section 2.2).

Since our data come from two different telescopes, the IRAM 30 m and ARO 12 m, with beam sizes of $\sim$ 27$^{\prime\prime}$ and $\sim$69$^{\prime\prime}$, the beam dilution effect should be investigated. On the IRAM 30 m and ARO 12 m HC$_{3}$N (J = 10$-$9) spectra of 33 sources is shown in Figure \ref{fig2}. It shows a stronger signal from the IRAM 30 m than that from the ARO 12 m, indicating that the source size should not be larger than the HPBW of ARO 12 m and there is nonnegligible beam dilution effect. Thus the brightness temperature ($T_{\rm B}$) can be derived from the main-beam brightness temperature ($T_{\rm mb}$) dilution:
\begin{equation}\label{2}
T_{\rm B} = T_{\rm mb}/\eta_{\rm BD} = T_{\rm mb}\frac{(\theta_{\rm s}^{2} + \theta_{\rm beam}^{2})}{\theta_{\rm s}^{2}},
\end{equation}
where $\eta_{BD}$ is the beam-filling factor and $\theta_{s}$ and $\theta_{beam}$ are the source size and the HPBW, respectively \citep{2017A&A...606A..74Z,2020ApJ...899..145Y}. According to $T_{\rm mb}$ values measured by the IRAM 30 m and ARO 12 m, the source size (HC$_{3}$N J = 10$-$9) is estimated for those 33 sources except G010.32$-$00.15 and G109.87+02.11 (unreasonable results of size, larger than the HPBW of ARO 12 m, may be nonuniform with a clumpy structure). The results show that the sizes of the sources are larger than the HPBW of the IRAM 30 m telescope and smaller than that of the ARO 12 m telescope. Thus our ARO 12 m HC$_{3}$N (J = 12$-$11) data needs to be corrected for the beam dilution effect. According to Equation (4), $T_{\rm B}$ for our sample is derived from $T_{\rm mb}$ divided by the beam-filling factor, which can be determined from the source size. 

For 49 sources that have at least two HC$_{3}$N line detections, the rotational temperature and column density of HC$_{3}$N of them are derived by Equation (3), using the Levenberg-Marquardt algorithm provided in the python package {\it lmfit} \citep{2016ascl.soft06014N}. The rotational diagrams are shown in Figure \ref{fig5} and the derived rotational temperature and column density of HC$_{3}$N are summarized in Table \ref{tab6}. For 10 sources with only HC$_{3}$N (J = 10$-$9) detection, the average $T_{\rm rot}$ value is used as their $T_{\rm rot}$ values. Then the column density of HC$_{3}$N is estimated by Equation (3) (see Table \ref{tab6}).

\subsubsection{Excitation temperature and column density of N\texorpdfstring{$_{2}$}HH$^{+}$}

We used the line intensity ratio method to estimate the optical depth of N$_{2}$H$^{+}$, following the procedure described by \cite{2009MNRAS.394..323P}. The theoretical line-integrated intensity of Group 1/Group 2 (see details in Table \ref{tab2}) should be 1:5 under an optically thin condition, assuming equal line width for all individual hyperfine components. Assuming the beam-filling factor is 1, the optical depth of N$_{2}$H$^{+}$ can be determined with the following formula:
\begin{equation}\label{3}
\frac{\int T_{\rm B,group1}d\nu }{\int T_{\rm B,group2}d\nu} = \frac{T_{\rm mb,group1}}{T_{\rm mb,group2}} = \frac{1-e^{-0.2\tau_{2}}}{1-e^{\tau_{2}}},
\end{equation}
where $\tau_{2}$ is the optical depth of N$_{2}$H$^{+}$ group 2. We then calculated the excitation temperature of N$_{2}$H$^{+}$ with the following equation:
\begin{equation}\label{4}
T_{\rm ex} = 4.47/ln(1+[\frac{T_{\rm mb,group2}}{4.47(1-e^{-\tau_{2})}}+0.236]^{-1}).
\end{equation}
Finally, the column density of N$_{2}$H$^{+}$ was derived using the following formula \citep{1986ApJS...60..819C}:
\begin{equation}\label{5}
N = \frac{3kW}{8\pi^{3}\nu S\mu^{2}}\left(\frac{T_{ex}}{T_{ex}-T_{bg}}\right)\left(\frac{\tau_{2}}{1-exp(- \tau_{2})}\right)Q(T_{ex})exp(E_{u}/kT_{ex}),
\end{equation}
where $T_{\rm bg}$ (= 2.73 K) and $W$ are the background brightness temperature and the integrated intensity of N$_{2}$H$^{+}$ group 2, respectively. The partition function $Q(T_{\rm ex})$ for N$_{2}$H$^{+}$ used in our calculations is $Q(T_{\rm ex})$ = 4.198$T_{\rm ex}$ \citep{2009MNRAS.394..323P}. The derived parameters of N$_{2}$H$^{+}$, including the optical depth, the excitation temperature, and the column density, are also summarized in Table \ref{tab6}. We find that the excitation temperature of N$_{2}$H$^{+}$ is much lower than the rotational temperature of HC$_{3}$N. This indicates that N$_{2}$H$^{+}$ exists in outer and colder regions, while HC$_{3}$N traces inner and hotter ones, which is consistent with results in Section 3.2. However, N$_{2}$H$^{+}$ emission with a clumpy structure ($B_{\rm eff}$ $\textless$ 1) cannot be ruled out. In this case, the derived $T_{\rm ex}$ is just a lower limit to the real one. Higher angular resolution or mapping observations would be needed to check it.

\section{Discussions}
\subsection{N(HC\texorpdfstring{$_{3}$}NN)/N(N\texorpdfstring{$_{2}$}HH$^{+}$), a possible chemical clock?}

N$_{2}$H$^{+}$ is generally considered to be formed in cold and dense environments by the gas-phase reaction H$_{3}^{+}$ + N$_{2}$ $\rightarrow$ N$_{2}$H$^{+}$ + H$_{2}$. When the dust temperature exceeds $\sim$ 20 K, CO molecules will evaporate from the dust grains and react with N$_{2}$H$^{+}$, resulting in the destruction of N$_{2}$H$^{+}$ in the gas phase through the reaction N$_{2}$H$^{+}$ + CO $\rightarrow$ HCO$^{+}$ + N$_{2}$ \citep[e.g.][]{2013ApJ...765...18T}. The central protostar will release UV photons in the UC H{\sc ii} regions, which also contribute to the destruction of N$_{2}$H$^{+}$ by the dissociative recombination with electrons N$_{2}$H$^{+}$ + e$^{-}$ $\rightarrow$ N$_{2}$ + H or NH + N \citep[e.g.][]{2015MNRAS.446.2566Y}. The column density of N$_{2}$H$^{+}$ should decrease through the reaction with CO molecules or electrons induced by UV photons in the UC H{\sc ii} region phase. Using the N$_{2}$H$^{+}$ (J = 1$-$0) data of the Nobeyama 45 m telescope with a beam size of 17$^{\prime\prime}$ in HMSCs and HMPOs from \cite{2019ApJ...872..154T}, and following the procedure and equations in Section 3.3.2, we recalculated $N\rm (N_{2}H^{+})$ of HMSCs and HMPOs (see details in Table \ref{tab7}). Then we compared the N$_{2}$H$^{+}$ column density of our UC H{\sc ii} region sources with that of sources in HMSCs and HMPOs. The cumulative distribution function of $N\rm (N_{2}H^{+})$ for three samples was presented in Figure \ref{fig6} (top-left panel). Significant differences on the distribution of $N\rm (N_{2}H^{+})$ between UC H{\sc ii} regions and other evolutionary stages (HMSCs and HMPOs) can be found, which is supported by Kolmogorov-Smirnov (K-S) test statistical results (see details in Table \ref{tab8}). This is also supported by the values of the mean $N\rm (N_{2}H^{+})$ within uncertainty ranges and corresponding t-test statistical results (Table \ref{tab8}). It indicates that the N$_{2}$H$^{+}$ column density remains basically stable in HMSCs and HMPOs but begins to increase in UC H{\sc ii} regions, inconsistent with previous theoretical expectations. More theoretical and observational works are needed to further investigate the chemical properties and evolution of N$_{2}$H$^{+}$.

HC$_{3}$N is believed to be a typical hot-core tracer \citep[e.g.][]{2019MNRAS.484.4444U}. Its precursor molecule, CH$_{4}$, sublimates from dust grains and reacts with C$^{+}$ to produce C$_{2}$H$_{2}$ when the kinetic temperature reaches 25 K. It then reacts with CN in the gas phase through the following neutral-neutral reaction C$_{2}$H$_{2}$ + CN $\rightarrow$ HC$_{3}$N + H, resulting in the formation of HC$_{3}$N \citep[e.g.][]{2008ApJ...681.1385H}. As the temperature reaches $T$$\sim$ 55 K, C$_{2}$H$_{2}$ in the dust-grain surface can sublimate and further react with CN, leading to an increase in the abundance of HC$_{3}$N in the gas phase. When the temperature is greater than 90 K (above the sublimation temperature of HC$_{3}$N), HC$_{3}$N evaporates directly from the dust grains and its abundance reaches a peak \citep{2019ApJ...881...57T}. Thus the HC$_{3}$N abundance theoretically increases with the evolution of HMSFRs. \cite{2019ApJ...872..154T} derived $N\rm (HC_{3}N)$ in HMSCs and HMPOs with the rotational diagram method, using HC$_{3}$N (J = 5$-$4, 9$-$8 and 10$-$9) data from the Nobeyama 45 m telescope with beam sizes of 37$^{\prime\prime}$, 20$^{\prime\prime}$, and 18$^{\prime\prime}$ (Table \ref{tab7}). Comparisons on observed HC$_{3}$N abundance of those three samples (HMSCs and HMPOs from \cite{2019ApJ...872..154T} and UC H{\sc ii} regions from ours) show significant differences on their cumulative distributions (Figure \ref{fig6}, top-right panel), which is supported by K-S test statistical results (see details in Table \ref{tab8}). The difference in the mean value of HC$_{3}$N between three samples in different evolutionary stages is also significant, with a chance probability of less than 0.05 from the t-tests (Table \ref{tab8}). These results reflect that the column density of HC$_{3}$N gradually increases with evolution, i.e., from HMSCs to HMPOs to UC H{\sc ii} regions.

\cite{2019MNRAS.484.4444U} found that the integrated line intensity ratio of HC$_{3}$N and N$_{2}$H$^{+}$ increases with evolution based on a 3$-$mm molecular line survey toward 570 HMSFRs using the Mopra telescope, and thus suggested this ratio as a chemical clock. However, the line intensity ratio may be influenced by some factors, such as the optical depth. Compared with the line intensity ratio of HC$_{3}$N and N$_{2}$H$^{+}$, the ratio of their column densities should be more appropriate for expressing their relative abundance. In Figure \ref{fig6} (bottom-left panel), we plotted the $N\rm (HC_{3}N)$ against the ratio. It clearly shows that the $N\rm (HC_{3}N)$/$N\rm (N_{2}H^{+})$ ratio increases from HMSCs to HMPOs and then to UC H{\sc ii} regions, with an increasing value of the $N\rm (HC_{3}N)$. The difference in the ratio of the three samples can also be supported by the cumulative distribution results (Figure \ref{fig6}, bottom-right panel), and different mean ratio values (Table \ref{tab8}). All chance probabilities of tests (K-S test for distributions and t-test for the mean values) are less than 0.05 (Table \ref{tab8}), supporting a statistically significant difference in the HC$_{3}$N/N$_{2}$H$^{+}$ ratio among these three samples. It indicates that the ratio of $N\rm (HC_{3}N)$/$N\rm (N_{2}H^{+})$ increases in those three evolutionary stages and thus it could be adopted as a chemical clock in HMSFRs.

The linear beam size is different for sources at different distances. A larger beam size of sources at larger distances may include more relatively diffused low-density gas, which could affect the $N\rm (HC_{3}N)$/$N\rm (N_{2}H^{+})$ ratio results. To check this potential bias in our analysis, we plotted the $N\rm (HC_{3}N)$/$N\rm (N_{2}H^{+})$ ratio against the heliocentric distance in Figure \ref{fig7}, where the heliocentric distance of UC H{\sc ii} regions is determined from trigonometric parallax measurements \citep{2014ApJ...783..130R,2019ApJ...885..131R}. HMSCs as well as HMPOs are determined from the latest parallax-based distance calculator V2\footnote{\url{http://bessel.vlbi-astrometry.org/node/378}}. It shows those three samples have a similar distance distribution (mainly from 2-11 kpc) and there is no systematic dependence between the ratio and the distance, which means that any observational bias related to the beam dilution is not significant for our ratio results.

\subsection{The correlations between N(HC\texorpdfstring{$_{3}$}NN)/N(N\texorpdfstring{$_{2}$}HH$^{+}$) and other evolutionary indicators}

In the latest version of APEX Telescope Large Area Survey of the Galaxy (ATLASGAL) compact source catalog, \cite{2022MNRAS.510.3389U} classified 5007 HMSFRs into different evolutionary stages, based on multiwavelength (8$-$870 $\mu$m) images. And they found clear systematic increases in the dust temperature ($T_{\rm dust}$), bolometric luminosity ($L_{\rm bol}$), and luminosity-to-mass ($L_{\rm bol}$/$M_{\rm clump}$) ratio, which are consistent with the evolutionary stages identified. As discussed in Section 4.1, we find that the $N\rm (HC_{3}N)$/$N\rm (N_{2}H^{+})$ ratio also increases with the evolution of HMSFRs. It is therefore reasonable to believe that positive correlations exist between the $N\rm (HC_{3}N)$/$N\rm (N_{2}H^{+})$ ratio and other evolutionary indicators.

Toward those three evolutionary samples, we collected their physical parameters from \cite{2018MNRAS.473.1059U}, including the dust temperature, bolometric luminosity, and luminosity-to-mass ratio. Using these data, we plotted the $N\rm (HC_{3}N)$/$N\rm (N_{2}H^{+})$ ratio against the dust temperature, bolometric luminosity, and luminosity-to-mass ratio for those samples (Figure \ref{fig8}). Positive correlations could be found in all plots, which supports our previous proposition (section 4.1), i.e., that $N\rm (HC_{3}N)$/$N\rm (N_{2}H^{+})$ can be used as a reliable chemical clock in HMSFRs. 

\section{Conclusions}
Using the IRAM 30 m and ARO 12 m telescopes, we have conducted systematic measurements on N$_{2}$H$^{+}$ (J = 1$-$0) line and multiple HC$_{3}$N (J = 10$-$9, 12$-$11 and 16$-$15) lines toward 61 UC H{\sc ii} regions. The main results in this work can be summarized as follows:

\begin{enumerate}
\item Out of 61 UC H{\sc ii} regions observed by the IRAM 30 m telescope, N$_{2}$H$^{+}$ (J = 1$-$0) is detected in all sources, except G060.57$-$00.18. Forty-nine sources among them were detected in both HC$_{3}$N J = 10$-$9 and J = 16$-$15 lines and 10 sources among them were detected in HC$_{3}$N J = 10$-$9. Our complementary observations by the ARO 12 m telescope detected the HC$_{3}$N (J = 12$-$11) line in 40 sources among those 49 sources. 

\item The HC$_{3}$N spectra of 23 sources among those 61 sources show clear line wings, indicating the presence of molecular outflows. And the majority of these 23 sources show maser emission associated with shock, which reflects the existence of shock activities in these sources. Our calculations on the line width show that the thermal line width can be negligible in both HC$_{3}$N and N$_{2}$H$^{+}$ lines. Comparisons show that the line width of HC$_{3}$N is larger than that of N$_{2}$H$^{+}$, suggesting that HC$_{3}$N is more likely from inner and more active star-forming regions compared to N$_{2}$H$^{+}$.

\item Toward 49 sources with at least two HC$_{3}$N line detections, we determined the column density and the rotational temperature of HC$_{3}$N, using the rotational diagram method. For 10 sources with only one line detection, their $N\rm (HC_{3}N)$ was derived with the average $T_{\rm rot}$. And we estimated the optical depth of N$_{2}$H$^{+}$ (J = 1$-$0) of this sample using the line intensity ratio method and then obtained their excitation temperatures and column densities of N$_{2}$H$^{+}$.
 
\item Through comparative analysis on our data in UC H{\sc ii} regions and those in HMSCs and HMPOs, we found that the column density of HC$_{3}$N increases from HMSCs to HMPOs, and then to UC H{\sc ii} regions, while that of N$_{2}$H$^{+}$ stays basically stable. And the column density ratio of HC$_{3}$N and N$_{2}$H$^{+}$ was confirmed to increase with HMSFR evolution. Moreover, positive correlations were found between the ratio and other evolutionary indicators (the dust temperature, bolometric luminosity, and luminosity-to-mass ratio). This supports the proposal that the ratio of $N\rm (HC_{3}N)$ and $N\rm (N_{2}H^{+})$ can be a reliable chemical clock for HMSFRs.

\end{enumerate}

\acknowledgments

This work is supported by the Natural Science Foundation of China (Nos. 12041302, 11590782). We thank the operators and staff at the IRAM 30 m and ARO 12 m telescopes for their assistance during our observations. We also thank Dr. J.Z. Wang, and Dr. X. Chen for their nice comments and suggestions. Y.T.Y. is a member of the International Max Planck Research School (IMPRS) for Astronomy and Astrophysics at the Universities of Bonn and Cologne. Y.T.Y. would like to thank the China Scholarship Council (CSC) for support.

\bibliography{cef}{}
\bibliographystyle{aasjournal}

\begin{startlongtable}
\begin{deluxetable*}{cccccccccc}
\tablecaption{Detailed information of our UC H{\sc ii} region sample. \label{tab1}}
\tablewidth{0pt}
\tablehead{
\colhead{Source} & \colhead{R.A.} & \colhead{DEC.} & \colhead{D} & \colhead{CH$_{3}$OH} & \colhead{CH$_{3}$OH} & \colhead{H$_{2}$O} & \colhead{OH} & \colhead{Evolutionary stage} & \colhead{Ref.}\\
\colhead{} & \colhead{J2000} & \colhead{J2000} & \colhead{(kpc)} & \colhead{Class \uppercase\expandafter{\romannumeral2}} & \colhead{Class \uppercase\expandafter{\romannumeral1}} & \colhead{} & \colhead{} & \colhead{} & \colhead{}
}
\decimalcolnumbers
\startdata
G359.13$+$00.03	&	17:43:25.61	&	$-$29:39:17.55	&	2.92	&	Y	&	N	&	Y	&	Y	&	UC H{\sc ii}	&	[1]	\\
G000.31$-$00.20	&	17:47:09.11	&	$-$28:46:16.28	&	11.11	&	Y	&	N	&	Y	&	Y	&	UC H{\sc ii}	&	[1]	\\
G001.00$-$00.23	&	17:48:55.28	&	$-$28:11:48.24	&	3.47	&	Y	&	…	&	Y	&	…	&	UC H{\sc ii}	&	[1]	\\
G006.79$-$00.25	&	18:01:57.75	&	$-$23:12:34.25	&	20.41	&	Y	&	Y	&	Y	&	Y	&	UC H{\sc ii}	&	[1]	\\
G007.47$+$00.05	&	18:02:13.18	&	$-$22:27:58.98	&	3.30	&	N	&	N	&	Y	&	…	&	UC H{\sc ii}	&	[2]	\\
G009.21$-$00.20	&	18:06:52.84	&	$-$21:04:27.88	&	2.92	&	Y	&	N	&	Y	&	…	&	UC H{\sc ii}	&	[1]	\\
G010.32$-$00.15	&	18:09:01.45	&	$-$20:05:07.85	&	2.76	&	Y	&	…	&	Y	&	…	&	UC H{\sc ii}	&	[1]	\\
G010.62$-$00.33	&	18:10:17.98	&	$-$19:54:04.65	&	4.07	&	Y	&	N	&	Y	&	…	&	UC H{\sc ii}	&	[1]	\\
\enddata
\tablecomments{(1): Source name; (2$-$3): Right ascension and Declination; (4): The heliocentric distance, which is taken from \cite{2014ApJ...783..130R,2019ApJ...885..131R}; (5$-$8): Class \uppercase\expandafter{\romannumeral2} CH$_{3}$OH maser, Class \uppercase\expandafter{\romannumeral1} CH$_{3}$OH maser, H$_{2}$O maser and OH maser; (8): Evolutionary stage; (6): References. [1] \cite{2016ApJ...833...18H}, [2] \cite{2005AJ....129..348G}, [3] \cite{2018AA...615A.103K}, [4] \cite{2009AA...501..539U}.}
\end{deluxetable*}
\end{startlongtable}

\begin{deluxetable*}{cccccc}
\tablecaption{Information of our targeted lines. \label{tab2}}
\tablewidth{0pt}
\tablehead{
\colhead{Telescope} & \colhead{Transition} & \colhead{Rest frequency} & \colhead{$E_{\rm u}/k$} & \colhead{$S\mu^{2}$} & \colhead{HPBW}\\
\colhead{} & \colhead{} & \colhead{(GHz)} & \colhead{(K)} &
\colhead{(D)} & \colhead{($^{\prime\prime}$)}
}
\decimalcolnumbers
\startdata
IRAM 30 m	&	HC$_{3}$N (J = 10-9)	&	90.97902	&	24.01	&	139.25	&	27.6	\\
	&	HC$_{3}$N (J = 16-15)	&	145.56096	&	59.38	&	222.79	&	17.3	\\
	&	N$_{2}$H$^{+}$ (1$_{0}$-0$_{1}$, group 1)$^{a}$	&	93.17188	&	4.47	&	37.25	&	27.0	\\
	&	N$_{2}$H$^{+}$ (1$_{2}$-0$_{1}$, group 2)$^{a}$	&	93.17370	&	4.47	&	62.09	&	27.0	\\
	&	N$_{2}$H$^{+}$ (1$_{1}$-0$_{1}$, group 3)$^{a}$	&	93.17613	&	4.47	&	12.42	&	27.0	\\
ARO 12 m	&	HC$_{3}$N (J = 10-9)	&	90.97902	&	24.01	&	139.25	&	69.1	\\
	&	HC$_{3}$N (J = 12-11)	&	109.17363	&	34.06	&	167.09	&	57.6	\\
\enddata
\tablecomments{(1): Telescope; (2): Transition; (3): Rest frequency; (4$-$5): The upper level energy and the product of the line strength and the square of the electric dipole moment, which are taken from CDMS; (6): Half power beam width. \\
$^{a}$The seven hyperfine lines of the N$_{2}$H$^{+}$ (J = 1$-$0) transition are blended into three groups: group 1 includes only the F$_{1}$F = 01$-$12 transition; group 2 includes the F$_{1}$F = 21$-$11, 23$-$12 and 22$-$11 transitions; and group 3 includes the F$_{1}$F = 11$-$10, 12$-$12 and 10$-$11 transitions \citep{2013ApJS..206...22C}.}
\end{deluxetable*}

\newpage
\begin{startlongtable}
\begin{deluxetable*}{ccccccccc}
\tablecaption{Spectral line parameters of HC$_{3}$N. \label{tab3}}
\tablewidth{0pt}
\tablehead{
\colhead{Source} & \colhead{Telescope} & \colhead{Transition} & \colhead{Integrated time} & \colhead{rms} & \colhead{$V_{\rm LSR}$} & \colhead{$\Delta \nu$} & \colhead{$\int T_{mb}d_{\nu }$} & \colhead{$T_{\rm mb}$}\\
\colhead{} & \colhead{} & \colhead{HC$_{3}$N} & \colhead{(min)} & \colhead{(K)} & \colhead{(km s$^{-1}$)} & \colhead{(km s$^{-1}$)} & \colhead{(K km s$^{-1}$)} & \colhead{(K)}
}
\decimalcolnumbers
\startdata
G000.31$-$00.20	&	IRAM 30 m	&	J = 10$-$9	&	3.9	&	0.06	&	18.43$\pm$0.02	&	1.91$\pm$0.05	&	4.99$\pm$0.11	&	2.46	\\
	&	ARO 12 m	&	J = 10$-$9	&	10.0	&	0.02	&	18.43$\pm$0.02	&	2.18$\pm$0.06	&	1.58$\pm$0.04	&	0.68	\\
	&	ARO 12 m	&	J = 12$-$11	&	6.0	&	0.05	&	18.34$\pm$0.04	&	1.85$\pm$0.09	&	1.58$\pm$0.06	&	0.80	\\
	&	IRAM 30 m	&	J = 16$-$15	&	3.9	&	0.11	&	18.20$\pm$0.04	&	2.17$\pm$0.11	&	3.77$\pm$0.15	&	1.63	\\
G006.79$-$00.25	&	IRAM 30 m	&	J = 10$-$9	&	61.8	&	0.01	&	21.10$\pm$0.01	&	2.62$\pm$0.02	&	7.43$\pm$0.04	&	2.66	\\
	&	ARO 12 m	&	J = 10$-$9	&	10.0	&	0.03	&	21.11$\pm$0.03	&	2.81$\pm$0.08	&	2.07$\pm$0.05	&	0.69	\\
	&	ARO 12 m	&	J = 12$-$11	&	6.0	&	0.06	&	20.95$\pm$0.07	&	2.57$\pm$0.17	&	1.97$\pm$0.11	&	0.72	\\
	&	IRAM 30 m	&	J = 16$-$15	&	61.8	&	0.02	&	20.86$\pm$0.01	&	2.62$\pm$0.03	&	4.05$\pm$0.04	&	1.45	\\
\enddata
\tablecomments{(1): Source name; (2): Telescope; (3): HC$_{3}$N transitions; (4$-$5): Integrated time and the corresponding root mean square (rms) value; (6): Local standard of rest (LSR) velocity; (7): Full width at half maximum (FWHM); (8): Integrated line intensity; (9): Peak $T_{\rm mb}$ value.\\}
\end{deluxetable*}
\end{startlongtable}

\newpage
\begin{startlongtable}
\begin{deluxetable*}{cccccccc}
\tablecaption{Spectral line parameters of N$_{2}$H$^{+}$.\label{tab4}}
\tablewidth{10pt}
\tablehead{
\colhead{Source} & \colhead{Transition} &\colhead{Integrated time} & \colhead{rms} & \colhead{$V_{\rm LSR}$} & \colhead{$\Delta \nu$} & \colhead{$\int T_{mb}d_{\nu }$} & \colhead{$T_{\rm mb}$}\\
\colhead{} & \colhead{} & \colhead{(min)} & \colhead{(K)} & \colhead{(km s$^{-1}$)} & \colhead{(km s$^{-1}$)} & \colhead{(K km s$^{-1}$)} & \colhead{(K)}
}
\decimalcolnumbers
\startdata
G000.31$-$00.20	&	N$_{2}$H$^{+}$ J = 1$-$0 F1 = 0$-$1	&	0.07	&	9.46$\pm$0.07	&	2.16$\pm$0.17	&	3.89$\pm$0.25	&	1.70	\\
	&	N$_{2}$H$^{+}$ J = 1$-$0 F1 = 2$-$1	&		&	17.66$\pm$0.02	&	2.82$\pm$0.05	&	17.36$\pm$0.28	&	5.78	\\
	&	N$_{2}$H$^{+}$ J = 1$-$0 F1 = 1$-$1	&		&	23.38$\pm$0.04	&	2.48$\pm$0.08	&	8.82$\pm$0.25	&	3.34	\\
G006.79$-$00.25	&	N$_{2}$H$^{+}$ J = 1$-$0 F1 = 0$-$1	&	0.01	&	12.20$\pm$0.01	&	2.80$\pm$0.02	&	5.33$\pm$0.03	&	1.79	\\
	&	N$_{2}$H$^{+}$ J = 1$-$0 F1 = 2$-$1	&		&	20.22$\pm$0.01	&	3.38$\pm$0.01	&	21.28$\pm$0.03	&	5.91	\\
	&	N$_{2}$H$^{+}$ J = 1$-$0 F1 = 1$-$1	&		&	26.01$\pm$0.01	&	3.29$\pm$0.01	&	12.12$\pm$0.03	&	3.46	\\
G010.32$-$00.15	&	N$_{2}$H$^{+}$ J = 1$-$0 F1 = 0$-$1	&	0.01	&	3.11$\pm$0.01	&	2.17$\pm$0.02	&	2.64$\pm$0.02	&	1.15	\\
	&	N$_{2}$H$^{+}$ J = 1$-$0 F1 = 2$-$1	&		&	11.40$\pm$0.01	&	2.86$\pm$0.01	&	12.03$\pm$0.03	&	3.95	\\
\enddata
\tablecomments{(1): Source name; (2): N$_{2}$H$^{+}$ (J = 1$-$0) transitions; (3$-$4): Integrated time and the corresponding rms value; (5): The LSR velocity; (6): FWHM; (7): Integrated line intensity; (8): Peak $T_{\rm mb}$ value.\\}
\end{deluxetable*}
\end{startlongtable}

\begin{deluxetable*}{ccccccccc}
\tablecaption{Outflow and maser information of sources with line wings. \label{tab5}}
\tablewidth{0pt}
\tablehead{
\colhead{Source} & \colhead{Blue wing} & \colhead{Red wing} & \colhead{Outflow} & \colhead{Ref.} & \colhead{H$_{2}$O maser} & \colhead{Ref.} & \colhead{Class \uppercase\expandafter{\romannumeral1} CH$_{3}$OH maser} & \colhead{Ref.} 
}
\decimalcolnumbers
\startdata
G006.79$-$00.25	&	Y	&	Y	&	Y	&	[1]	&	Y	&	[a]	&	Y	&	[$\alpha$]	\\
G010.32$-$00.15	&	Y	&	Y	&	Y	&	[1]	&	Y	&	[a]	&
…	&	…	\\
G015.66$-$00.49	&	Y	&	Y	&	…	&	…	&	Y	&	[b]	&	Y	&	[$\alpha$]	\\
G016.86$-$02.15	&	Y	&	Y	&	Y	&	[2]	&	Y	&	[b]	&	Y	&	[$\alpha$]	\\
G017.02$-$02.40	&	Y	&	Y	&	Y	&	[3]	&	Y	&	[b]	&	Y	&	[$\alpha$]	\\
G019.36$-$00.03	&	Y	&	Y	&	Y	&	[2]	&	Y	&	[c]	&	Y	&	[$\alpha$]	\\
G023.43$-$00.18	&	Y	&	Y	&	Y	&	[2]	&	Y	&	[d]	&	Y	&	[$\alpha$]	\\
G024.78+00.08	&	Y	&	Y	&	Y	&	[2]	&	Y	&	[d]	&	Y	&	[$\alpha$]	\\
\enddata
\tablecomments{(1): Source name; (2$-$3): Blue and red wings; (4$-$5): Dipolar CO outflows, with corresponding references. [1] \cite{2022AA...658A.160Y}, [2] \cite{2016AJ....152...92L}, [3] \cite{2004AA...426..503W}, [4] \cite{2018ApJS..235....3Y}, [5] \cite{2018ApJ...867..167L}, [6] \cite{2008AJ....136.2391C} and [7] \cite{2015MNRAS.453..645M}; (6$-$9): Water maser and Class \uppercase\expandafter{\romannumeral1} CH$_{3}$OH maser information, with corresponding references. [a] \cite{2010MNRAS.406.1487B}, [b] \cite{2014MNRAS.443.2923T}, [c] \cite{2022AJ....163..124L}, [d] \cite{2011MNRAS.416..178B}, [e] \cite{2015MNRAS.453.4203X}, [f] \cite{2005AA...434..613S}, [g] \cite{2019ApJS..244....2K} and [h] \cite{2013ApJ...764...61C}; 
[$\alpha$] \cite{2017ApJS..231...20Y} and [$\beta$] \cite{2019ApJS..244....2K}.
}
\end{deluxetable*}

\clearpage
\begin{startlongtable}
\begin{deluxetable*}{ccccccc}
\tablecaption{Derived parameters of HC$_{3}$N and N$_{2}$H$^{+}$ in UC H{\sc ii} regions.\label{tab6}}
\tablewidth{6pt}
\tablehead{
\colhead{Source} & \multicolumn{3}{c}{HC$_{3}$N} & \multicolumn{3}{c}{N$_{2}$H$^{+}$}  \\
\cmidrule(r){2-4} \cmidrule(r){5-7}
\noalign{\smallskip}
\colhead{} & \colhead{source size} & \colhead{$T_{\rm rot}$} & \colhead{$N$} & \colhead{$\tau$} & \colhead{$T_{\rm ex}$} & \colhead{$N$} \\
\colhead{} & \colhead{(")} & \colhead{(K)} & \colhead{($\times$ 10$^{13}$ cm$^{-2}$)} & \colhead{} & \colhead{(K)} & \colhead{($\times$ 10$^{13}$ cm$^{-2}$)} 
} 
\decimalcolnumbers
\startdata
G000.31$-$00.20	&	42.71	&	28.27$\pm$2.78	&	2.10$\pm$0.14	&	1.08$\pm$0.14	&	11.90$\pm$0.02	&	7.73$\pm$0.13	\\
G001.00$-$00.23	&	…	&	33.47$^{**}$	&	1.03$\pm$0.05$^{**}$	&	…	&	…	&	…	\\
G006.79$-$00.25	&	40.85	&	22.49$\pm$1.47	&	3.07$\pm$0.20	&	1.17$\pm$0.02	&	11.72$\pm$0.01	&	9.78$\pm$0.01	\\
G007.47+00.05	&	…	&	27.28$^{*}$	&	0.20$^{*}$	&	0.89$\pm$1.05	&	3.14$\pm$0.68	&	1.31$\pm$0.32	\\
G009.21$-$00.20	&	…	&	33.47$^{**}$	&	0.78$\pm$0.04$^{**}$	&	2.99$\pm$0.17	&	6.83$\pm$0.01	&	12.35$\pm$0.10	\\
G010.32$-$00.15	&	…	&	35.09$^{*}$	&	1.67$^{*}$	&	1.05$\pm$0.03	&	9.19$\pm$0.01	&	5.00$\pm$0.01	\\
G010.62$-$00.33	&	…	&	33.47$^{**}$	&	1.15$\pm$0.03$^{**}$	&	2.05$\pm$0.09	&	8.73$\pm$0.01	&	10.34$\pm$0.06	\\
G011.10$-$00.11	&	…	&	24.70$^{*}$	&	0.89$^{*}$	&	3.05$\pm$0.21	&	5.98$\pm$0.01	&	9.07$\pm$0.10	\\
\enddata
\tablecomments{(1): Source name; (2): Source size (HC$_{3}$N J = 10$-$9). (3$-$4): The rotational temperature and the column density of HC$_{3}$N. $T_{\rm rot}$ and $N$ are obtained by the rotational diagram method with three HC$_{3}$N lines data. $T_{\rm rot}$ and $N$ with $^{*}$ are calculated by the rotational diagram with two HC$_{3}$N (10$-$9 $\&$ 16$-$15) lines data. $T_{\rm rot}$ and $N$ with $^{**}$ represent the average vaule of $T_{\rm rot}$ of HC$_{3}$N and the corresponding column density of HC$_{3}$N; (5$-$7): The optical depth, the excitation temperature and the column density of N$_{2}$H$^{+}$.}
\end{deluxetable*}
\end{startlongtable}

\clearpage
\begin{startlongtable}
\begin{deluxetable*}{ccccccc}
\tablecaption{Physical parameters of HC$_{3}$N and N$_{2}$H$^{+}$ in HMSCs and HMPOs.\label{tab7}}
\tablewidth{6pt}
\tablehead{
\colhead{Evolutionary stage} & \colhead{Source} & \colhead{D} & \colhead{HC$_{3}$N} & \multicolumn{3}{c}{N$_{2}$H$^{+}$} \\
\cmidrule(r){5-7}
\noalign{\smallskip}
\colhead{} & \colhead{} & \colhead{} & \colhead{N} & \colhead{$\tau$} & \colhead{$T_{\rm ex}$} & \colhead{N} \\
\colhead{} & \colhead{} & \colhead{(kpc)} & \colhead{($\times$ 10$^{13}$ cm$^{-2}$)} & \colhead{} & \colhead{(K)} & \colhead{($\times$ 10$^{13}$ cm$^{-2}$)} 
} 
\decimalcolnumbers
\startdata
HMSC	&	18385$-$0512$-$3	&	10.96	&	0.49$\pm$0.15	&	2.20$\pm$0.16	&	3.66$\pm$0.02	&	3.57$\pm$0.11	\\
	&	18445$-$0222$-$4	&	4.81	&	0.05$\pm$0.01	&	2.59$\pm$0.42	&	3.07$\pm$0.05	&	3.07$\pm$0.26	\\
	&	18447$-$0229$-$4	&	6.88	&	0.19$\pm$0.06	&	1.45$\pm$0.33	&	3.23$\pm$0.10	&	1.92$\pm$0.12	\\
	&	18447$-$0229$-$5	&	6.68	&	0.09$\pm$0.03	&	2.14$\pm$0.34	&	3.16$\pm$0.05	&	2.32$\pm$0.10	\\
	&	18454$-$0158$-$1	&	7.27	&	0.18$\pm$0.06	&	2.78$\pm$0.14	&	3.77$\pm$0.01	&	5.65$\pm$0.16	\\
	&	18454$-$0158$-$5	&	7.25	&	0.90$\pm$0.20	&	1.23$\pm$0.07	&	5.16$\pm$0.02	&	4.08$\pm$0.07	\\
	&	18454$-$0158$-$8	&	7.26	&	0.27$\pm$0.08	&	2.76$\pm$0.11	&	4.07$\pm$0.01	&	7.37$\pm$0.16	\\
	&	18454$-$0158$-$10	&	7.27	&	…	&	1.03$\pm$0.30	&	3.39$\pm$0.15	&	1.74$\pm$0.10	\\
\enddata
\tablecomments{(1): Evolutionary stage; (2): Source name; (3): The heliocentric distance, which is determined from the latest parallax-based distance calculator V2; (4): The column density of HC$_{3}$N, taken from \cite{2019ApJ...872..154T}. (5$-$7): The optical depth, the excitation temperature and the column density of N$_{2}$H$^{+}$, which are derived with the N$_{2}$H$^{+}$ (J = 1$-$0) data of \cite{2019ApJ...872..154T} and the method of \cite{2009MNRAS.394..323P}.}
\end{deluxetable*}
\end{startlongtable}

\begin{deluxetable}{cccc}[h]
\tablecaption{Average $N\rm (N_{2}H^{+})$, $N\rm (HC_{3}N)$ and $N\rm (HC_{3}N)$/$N\rm (N_{2}H^{+})$ values of HMSC, HMPO and UC H{\sc ii} region samples.\label{tab8}}
\tablewidth{10pt}
\tablehead{
\colhead{Evolutionary stage} & \colhead{$N\rm (N_{2}H^{+})$} &  \colhead{$N\rm (HC_{3}N)$}  & \colhead{$N\rm (HC_{3}N)$/$N\rm (N_{2}H^{+})$} \\
\colhead{}  & \colhead{($\times$ 10$^{13}$ cm$^{-2}$)} & \colhead{($\times$ 10$^{13}$ cm$^{-2}$)} & \colhead{} 
} 
\decimalcolnumbers
\startdata
HMSC	&	3.39$\pm$1.90	&	0.30$\pm$0.28	&	0.09$\pm$0.06	\\
HMPO	&	3.18$\pm$2.61	&	0.64$\pm$0.54	&	0.24$\pm$0.23	\\
UC H{\sc ii} region	&	5.51$\pm$4.55	&	1.65$\pm$1.54	&	0.42$\pm$0.41	\\
K-S test p-value$^{a}$	&	0.77	&	0.10	&	0.05	\\
K-S test p-value$^{b}$	&	0.01	&	0.02	&	0.04	\\
K-S test p-value$^{c}$	&	0.38	&	\textless 0.01	&	\textless 0.01	\\
t Test p-value$^{a}$	&	0.79	&	0.03	&	0.01	\\
t Test p-value$^{b}$	&	0.01	&	\textless 0.01	&	0.02	\\
t Test p-value$^{c}$	&	0.02	&	\textless 0.01	&	\textless 0.01	\\
\enddata
\tablecomments{K-S test and t-est chance probabilities for comparison between: $^{a}$ HMSC and HMPO; $^{b}$ HMPO and UC H{\sc ii} region; $^{c}$ HMSC and UC H{\sc ii} region.}
\end{deluxetable}

\begin{figure*}
    \centering
    \includegraphics[width=0.29\textwidth]{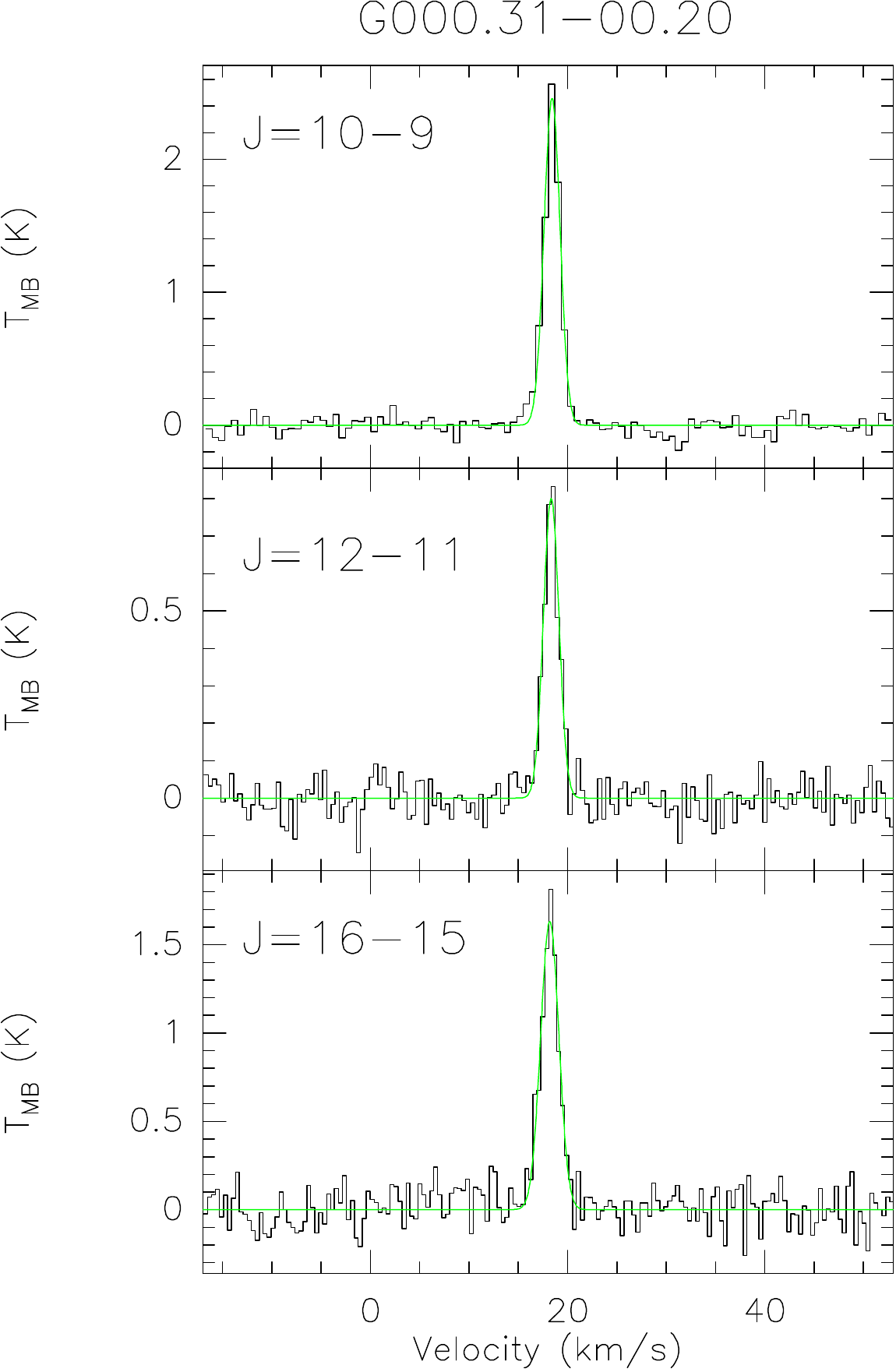}
    \includegraphics[width=0.29\textwidth]{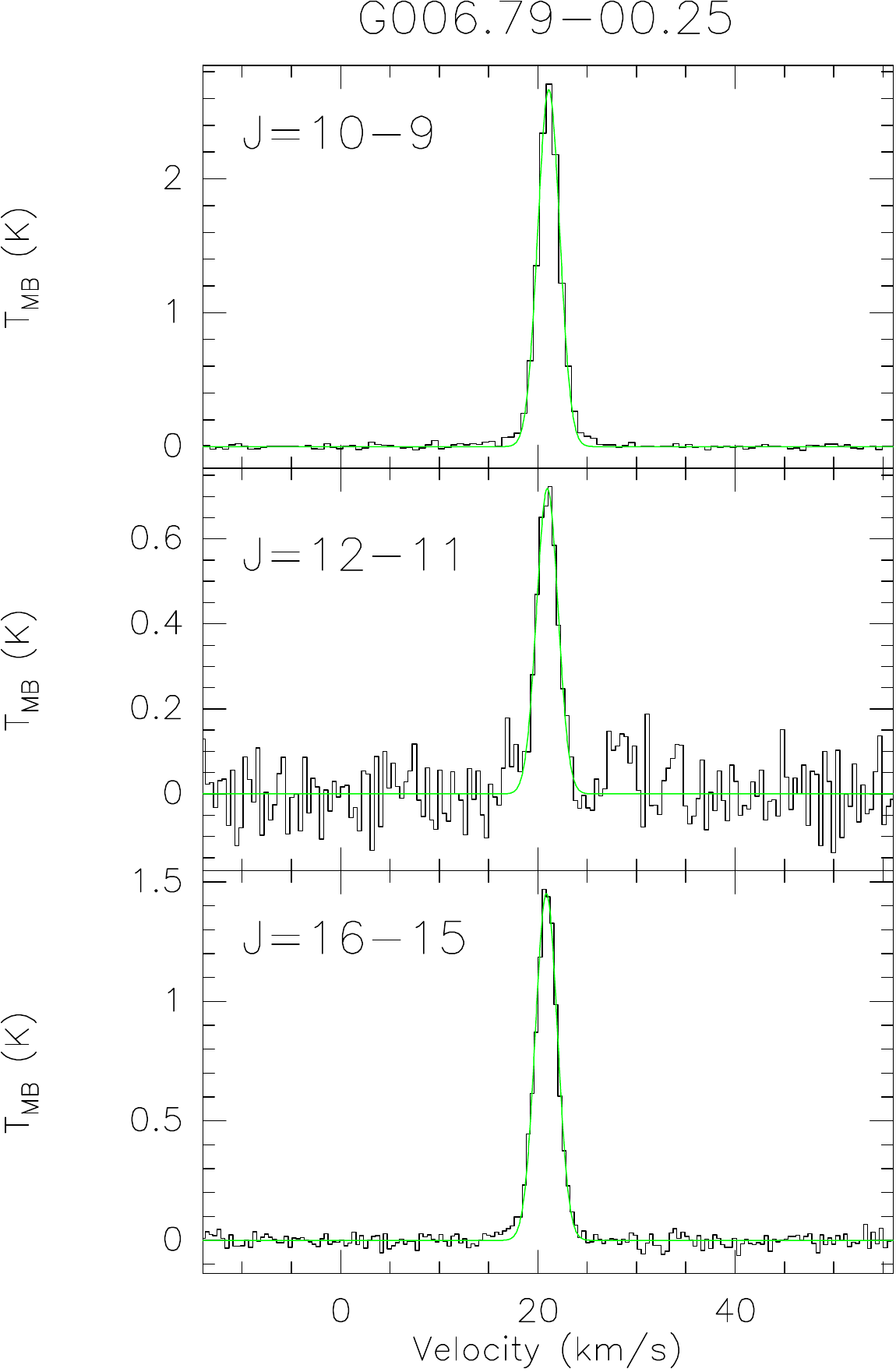}
    \includegraphics[width=0.29\textwidth]{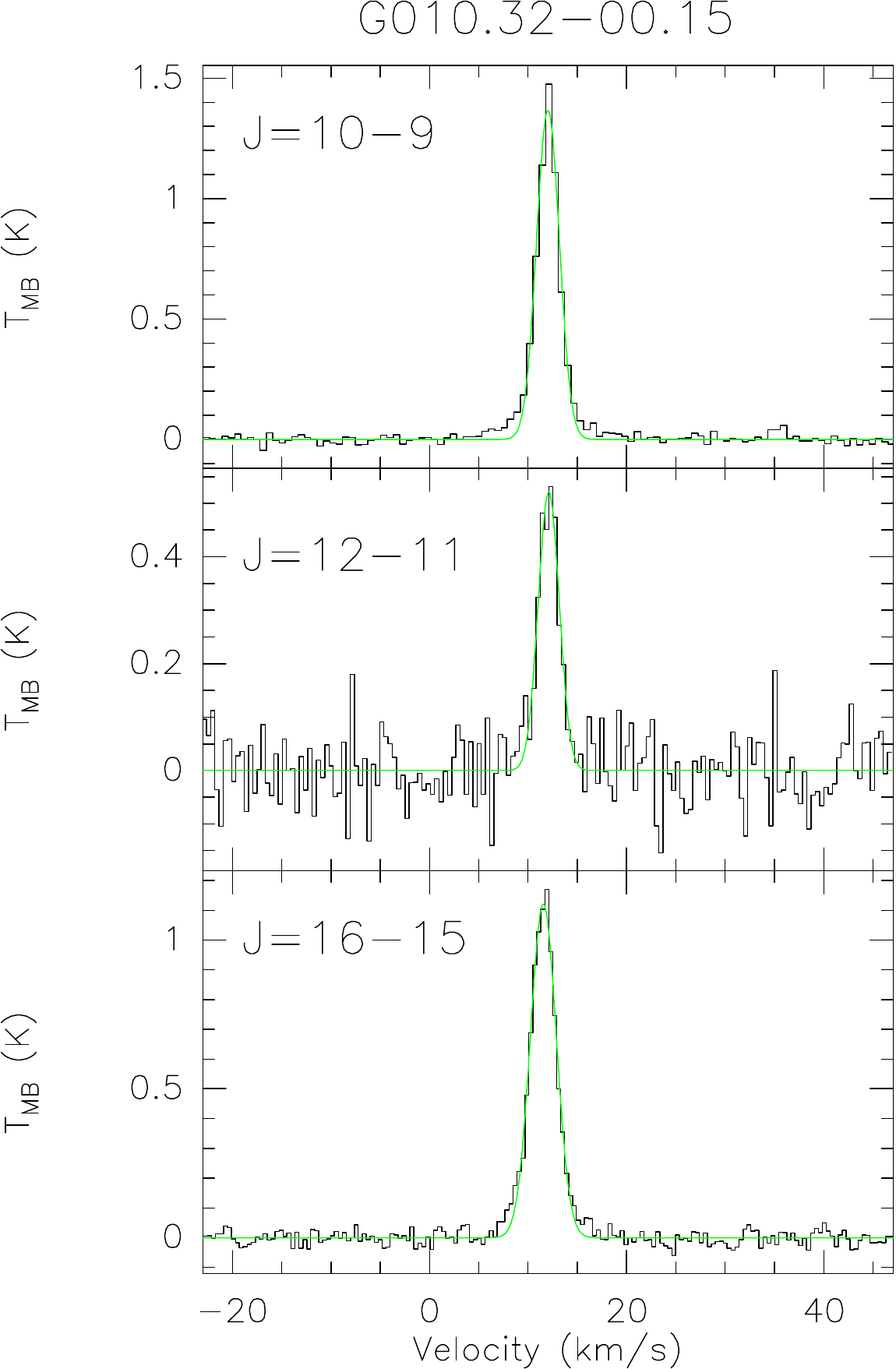} 
    \includegraphics[width=0.29\textwidth]{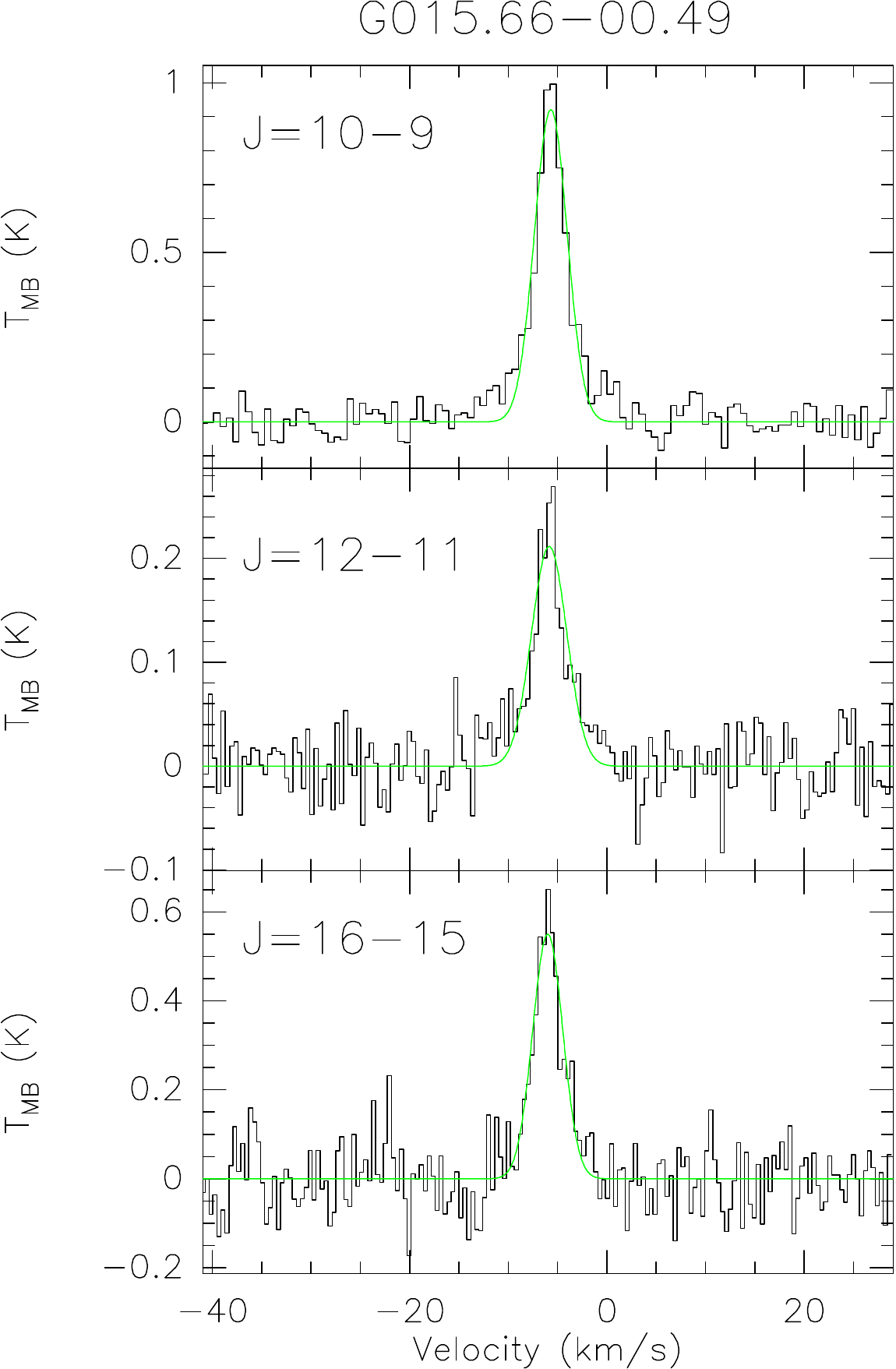}
    \includegraphics[width=0.29\textwidth]{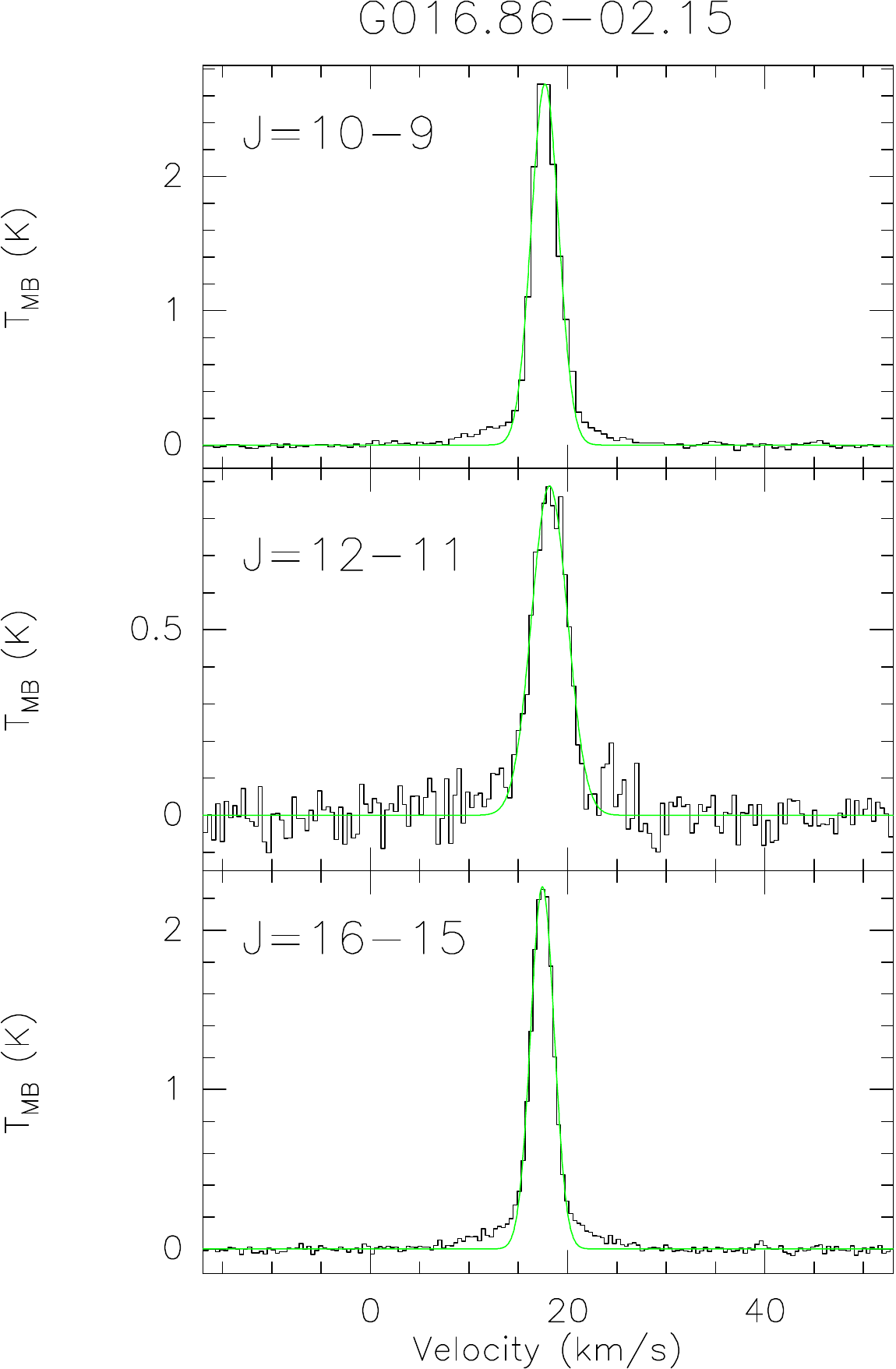}
    \includegraphics[width=0.29\textwidth]{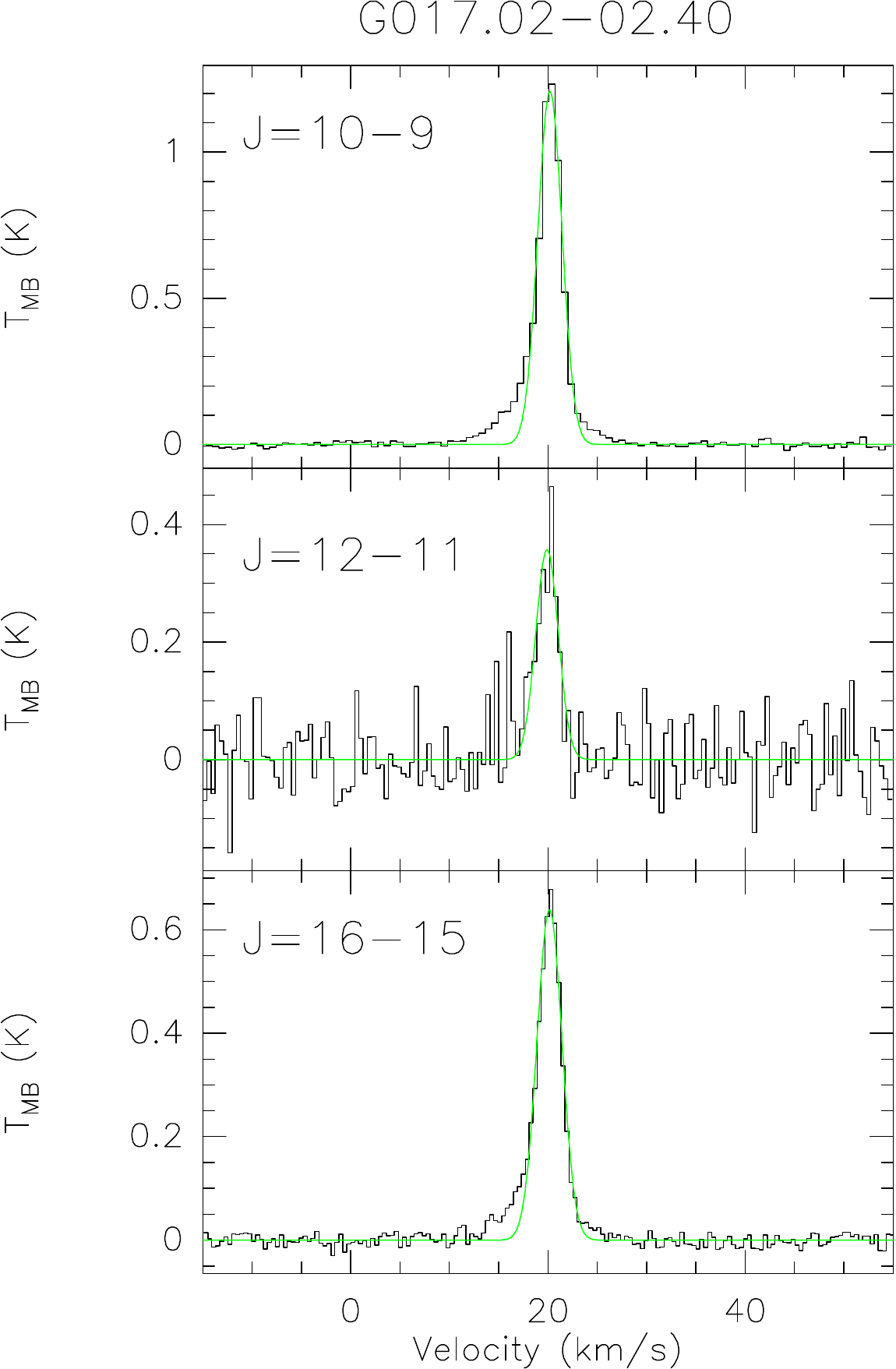}   
    \includegraphics[width=0.29\textwidth]{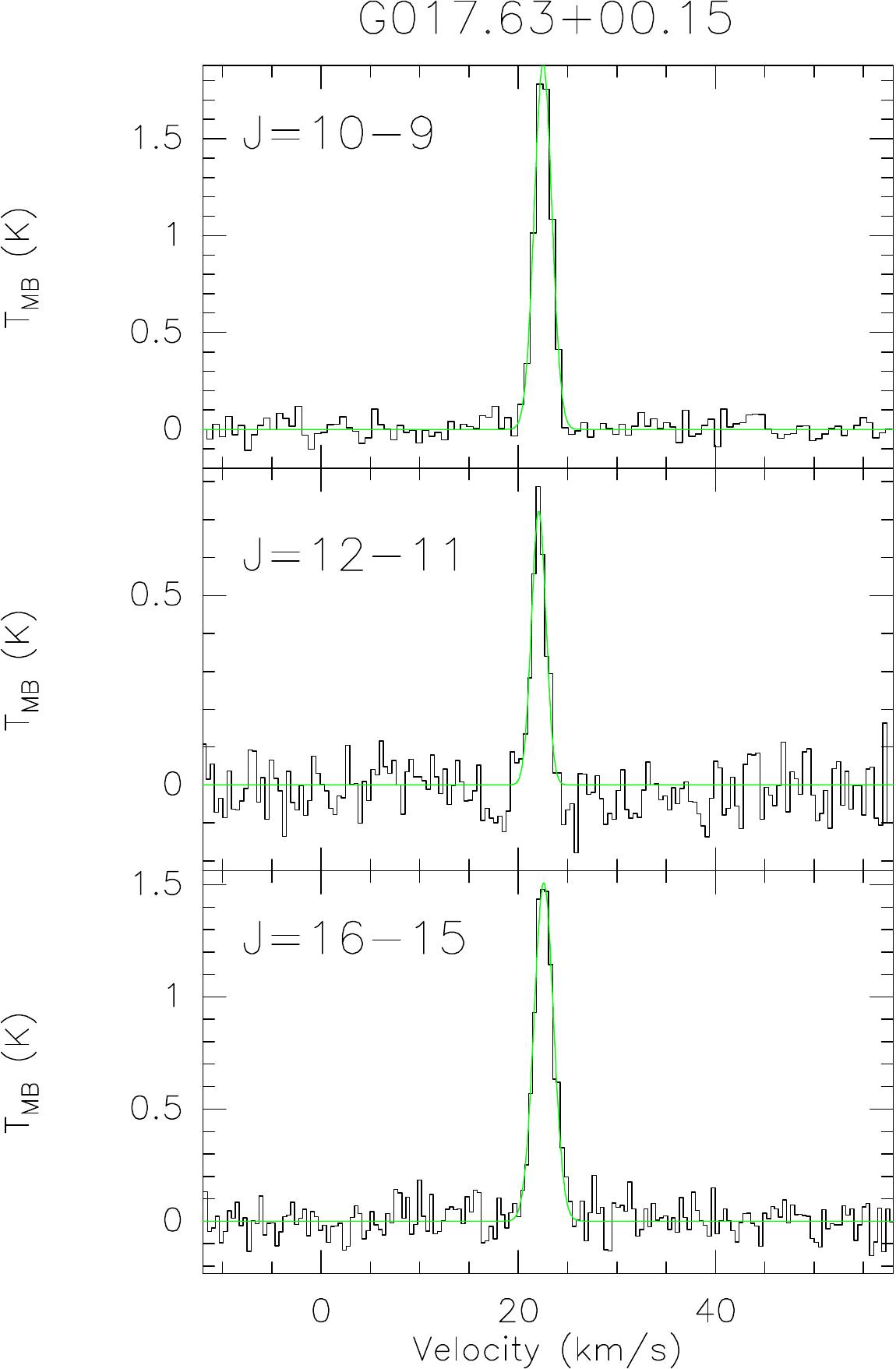}
    \includegraphics[width=0.29\textwidth]{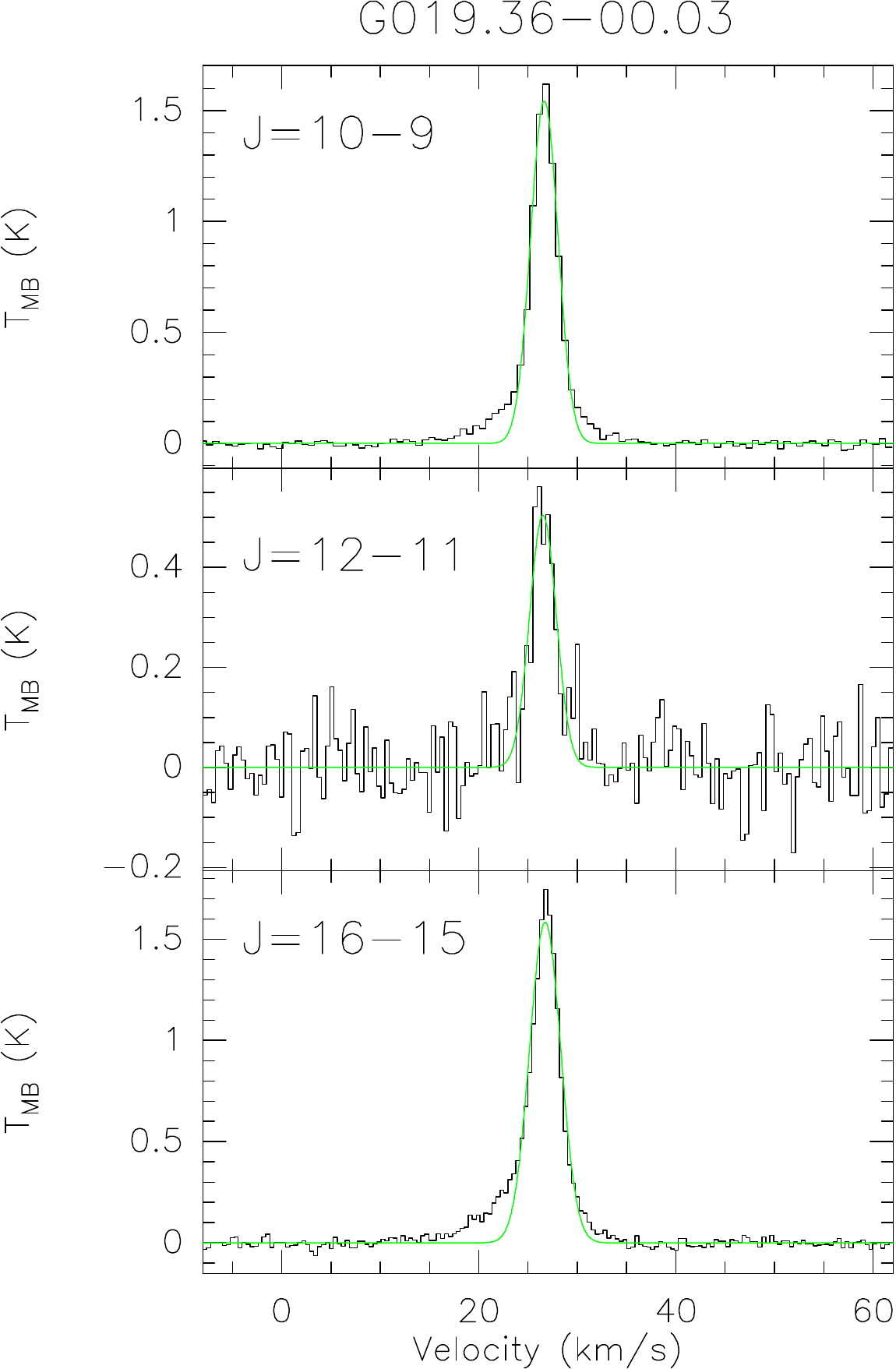}
    \includegraphics[width=0.29\textwidth]{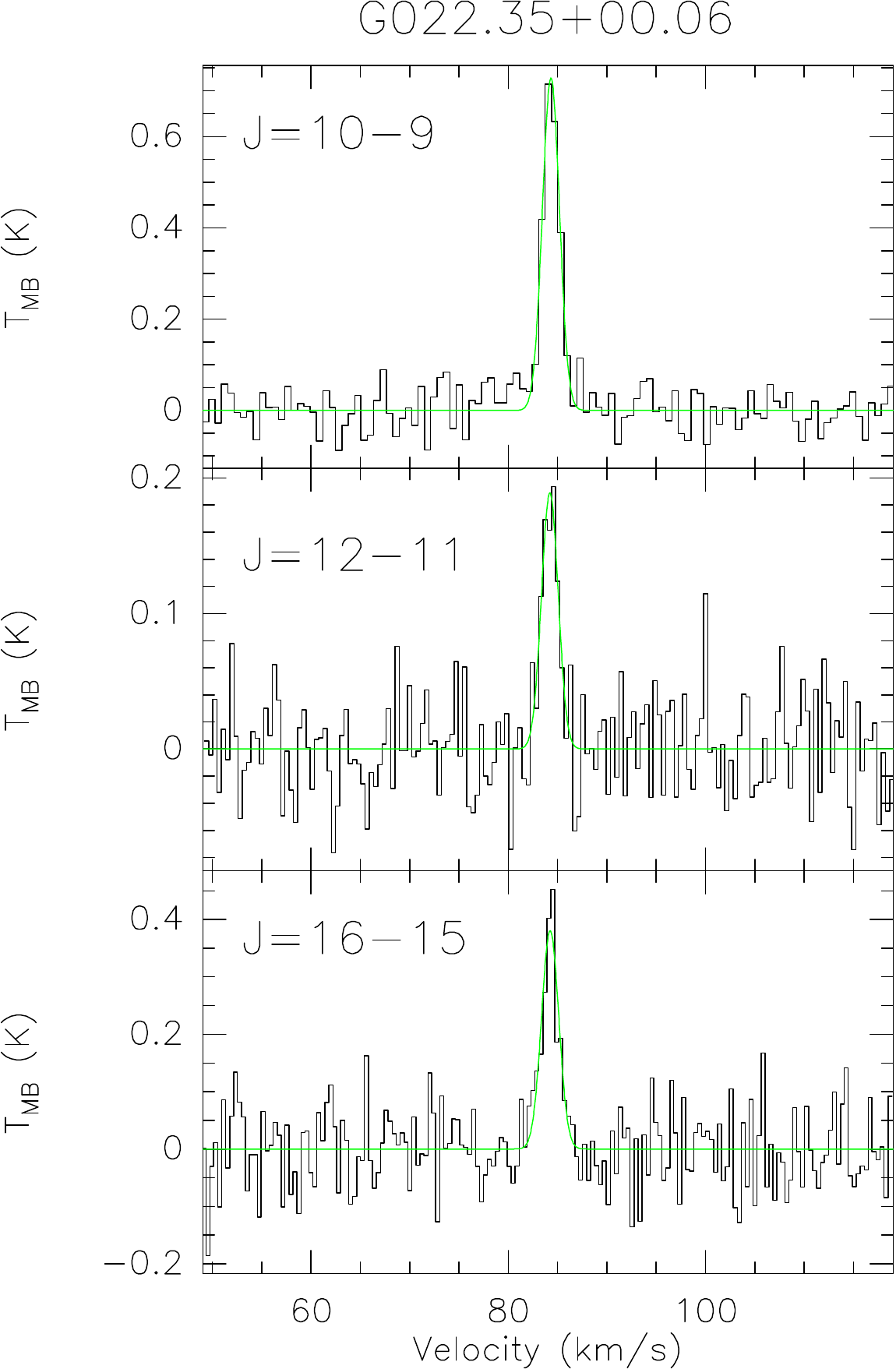} 
    \caption{HC$_{3}$N spectra with fitting lines of those 40 sources with detections of all targeted lines, J=10$-$9 and J=16$-$15 IRAM 30 m spectra (top and bottom panels) and J=12$-$11 ARO 12 m spectra (middle panels).}
\end{figure*}
    
\addtocounter{figure}{-1}
\begin{figure*} 
    \centering
    \includegraphics[width=0.29\textwidth]{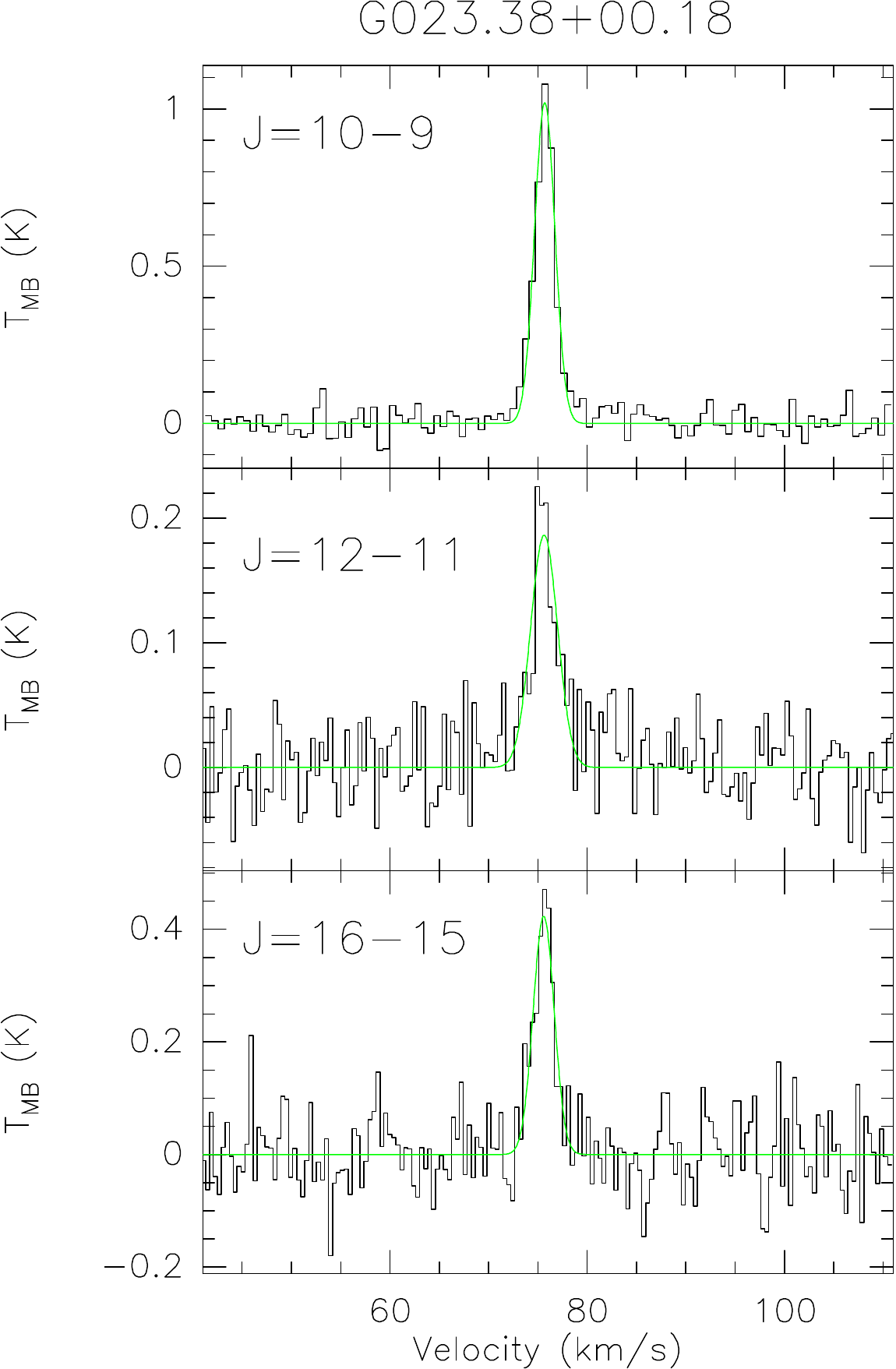}
    \includegraphics[width=0.29\textwidth]{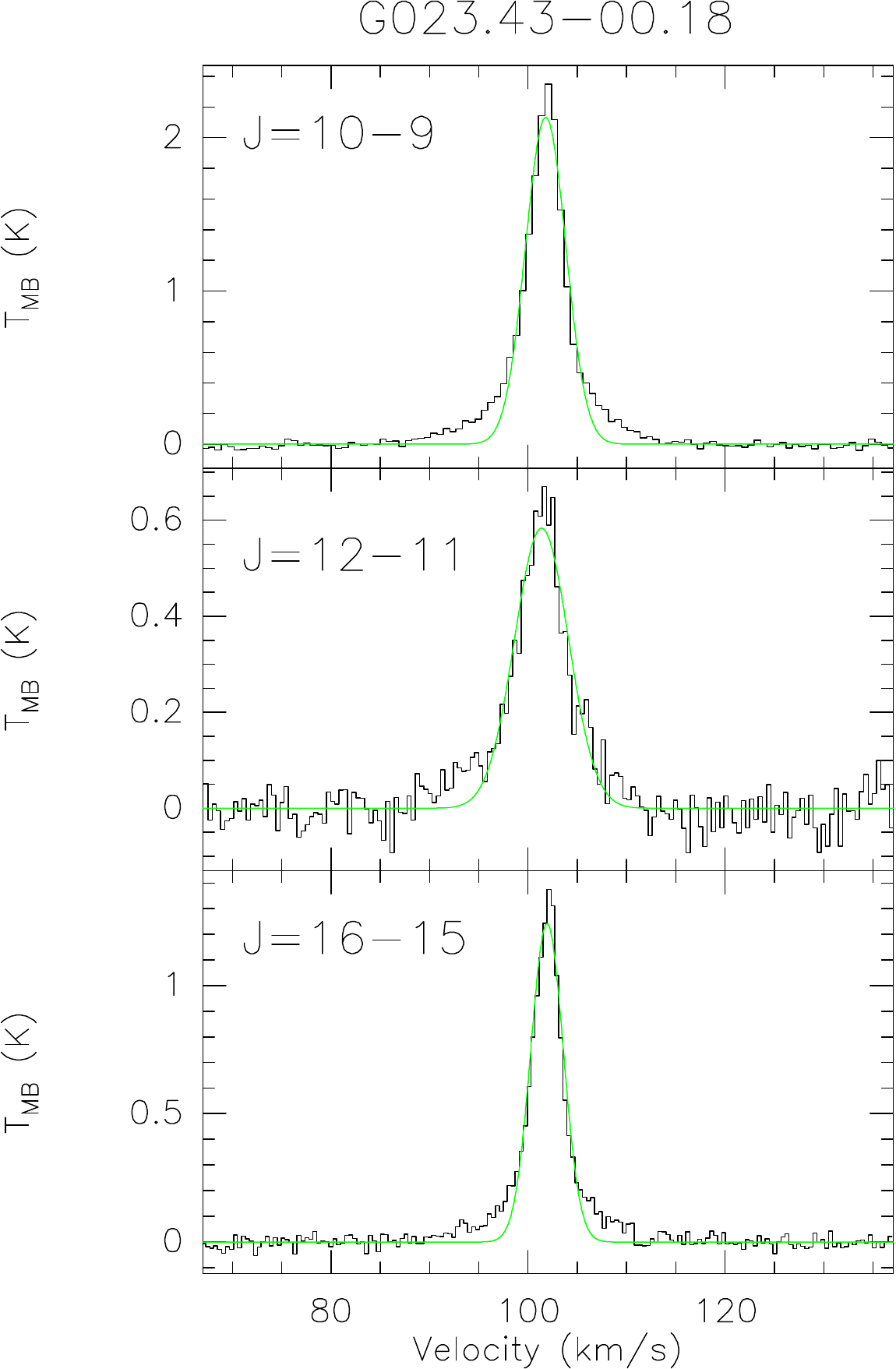}
    \includegraphics[width=0.29\textwidth]{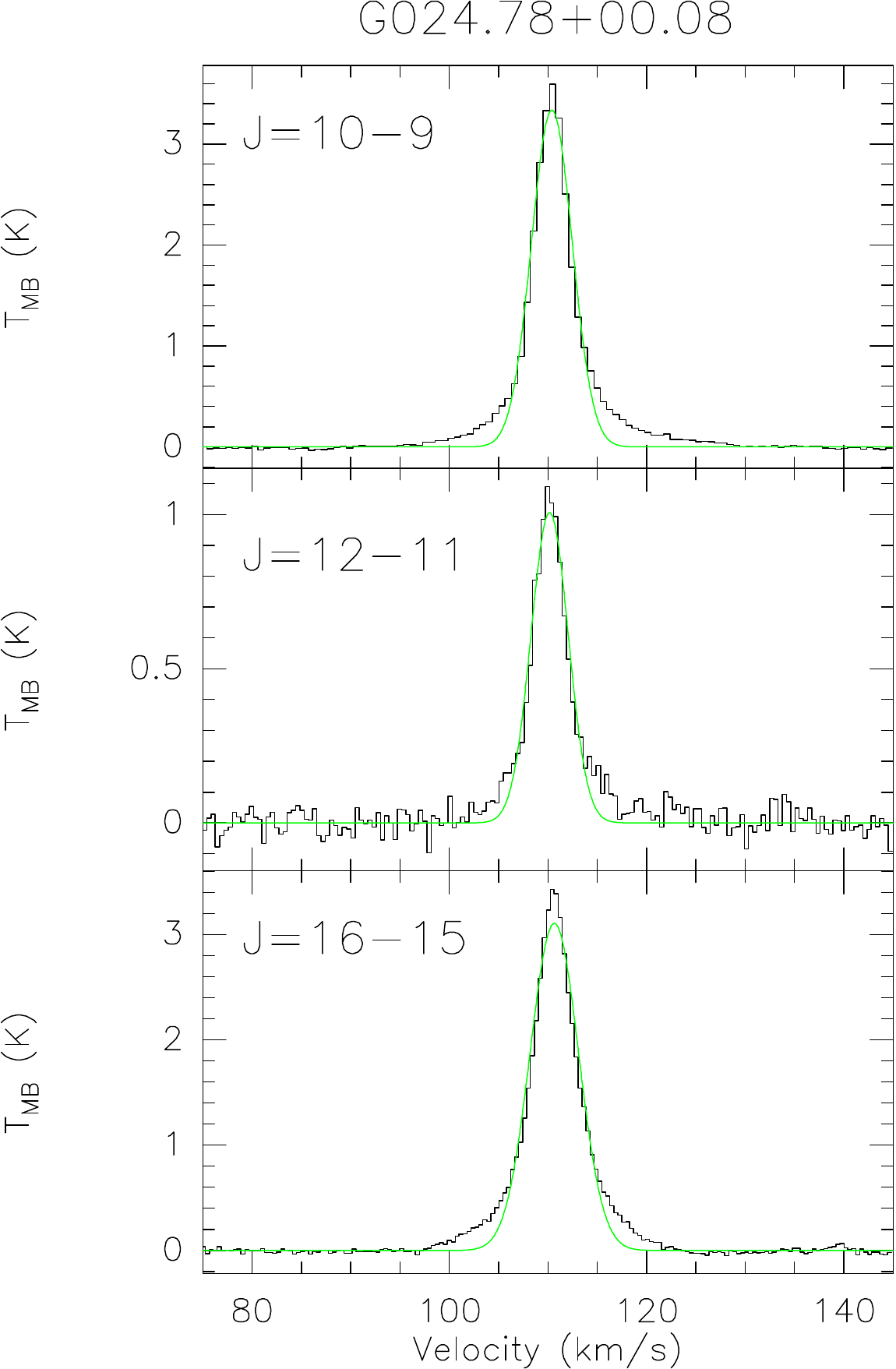}    
    \includegraphics[width=0.29\textwidth]{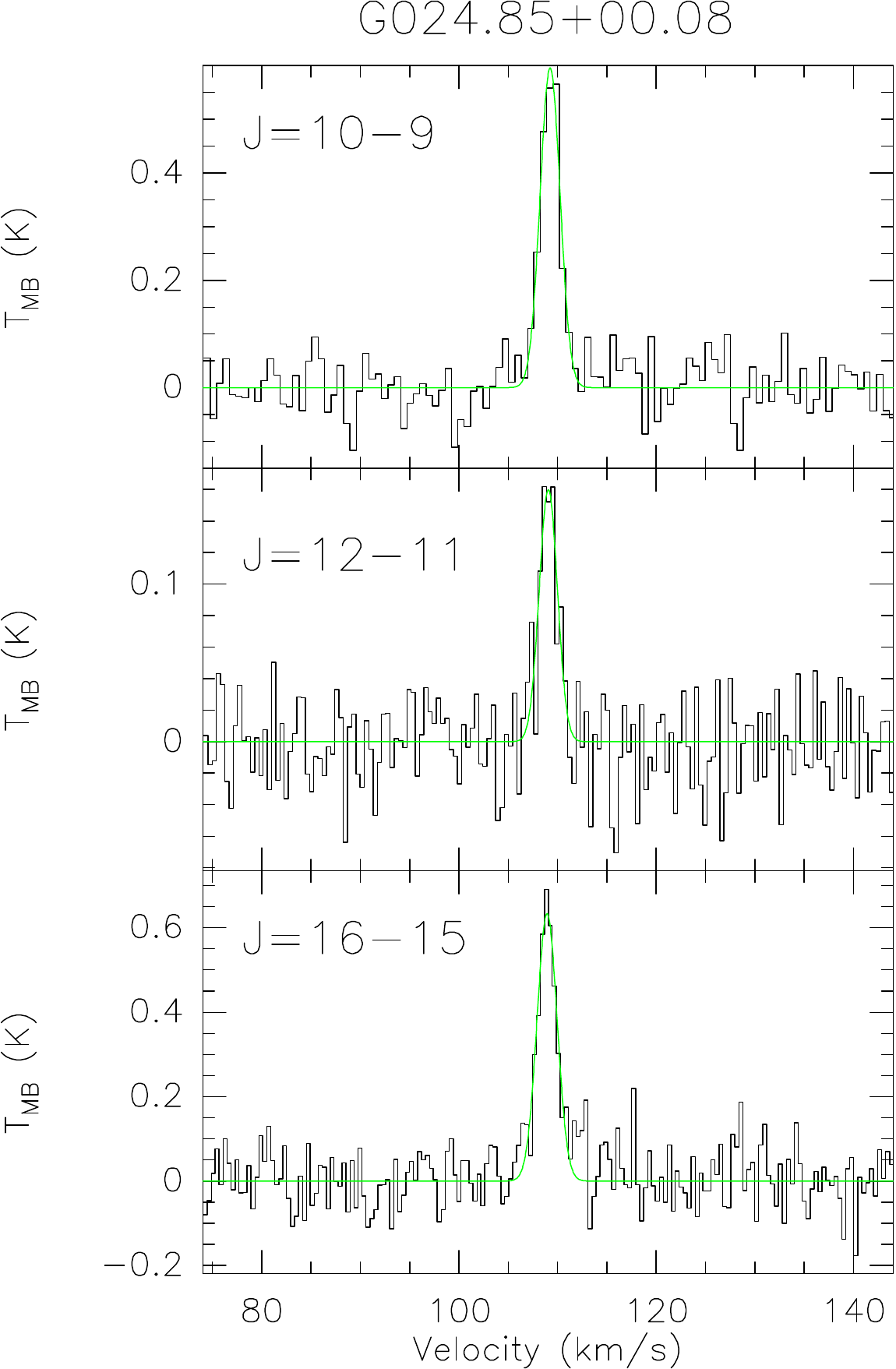}
    \includegraphics[width=0.29\textwidth]{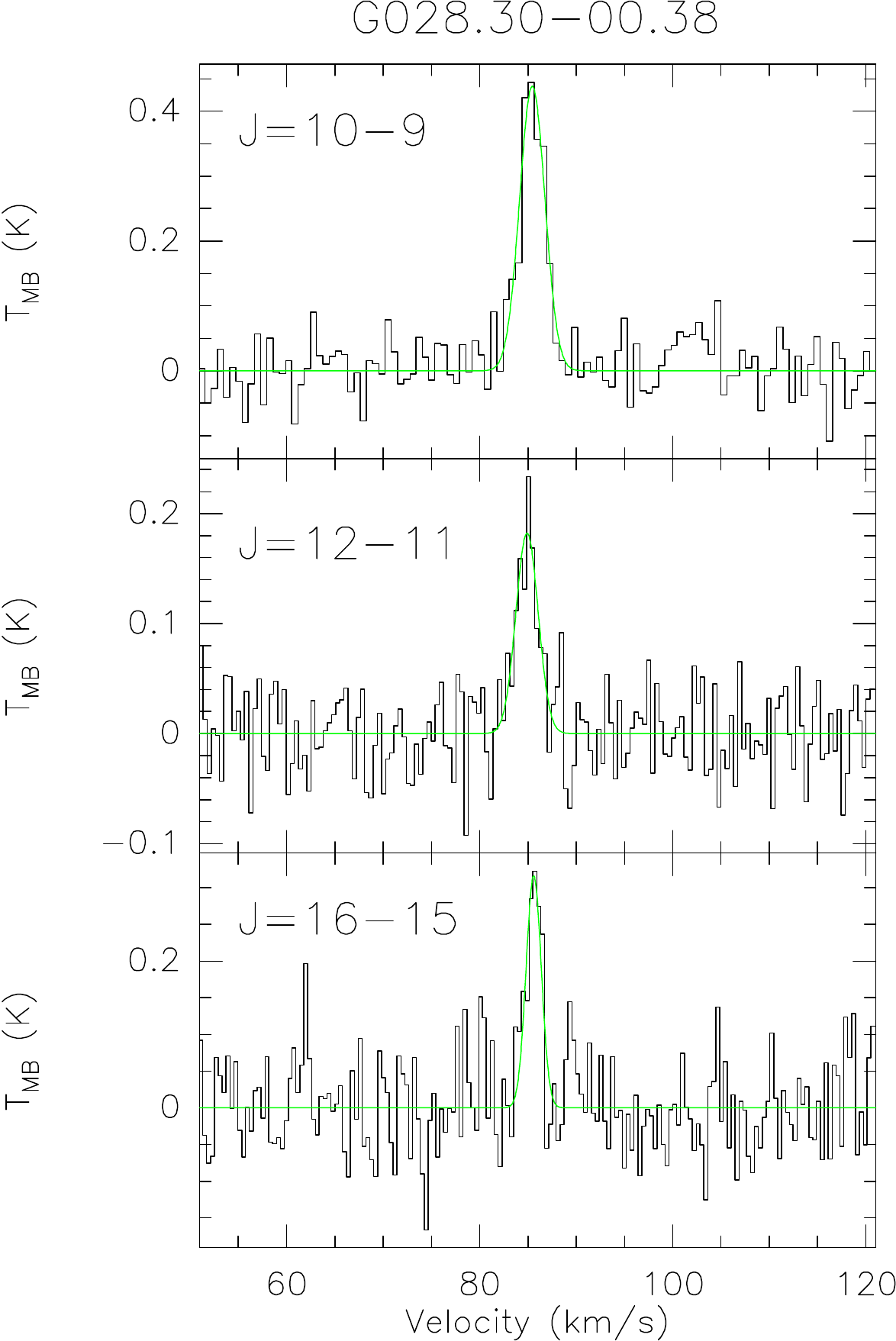}
    \includegraphics[width=0.29\textwidth]{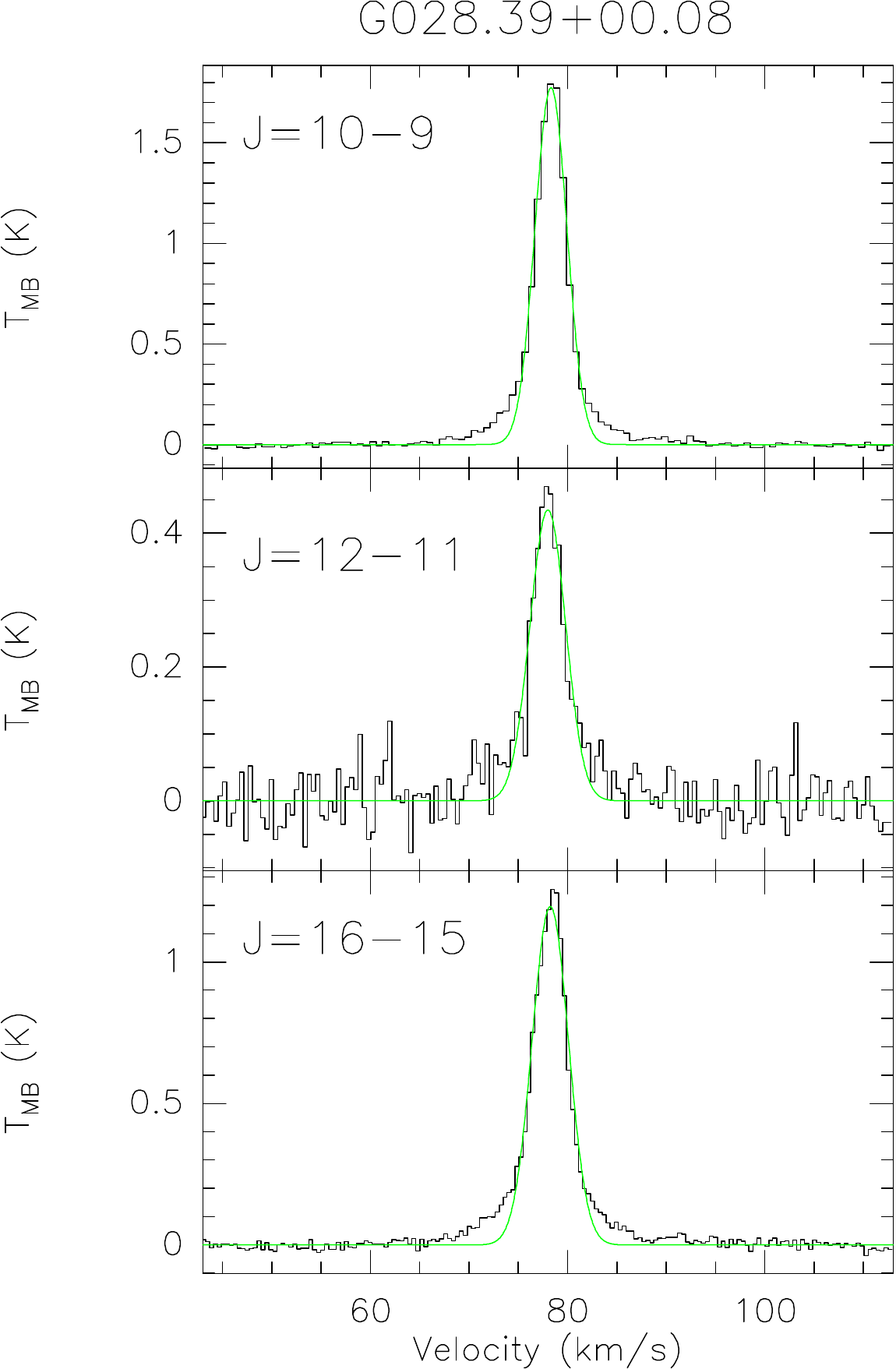} 
    \includegraphics[width=0.29\textwidth]{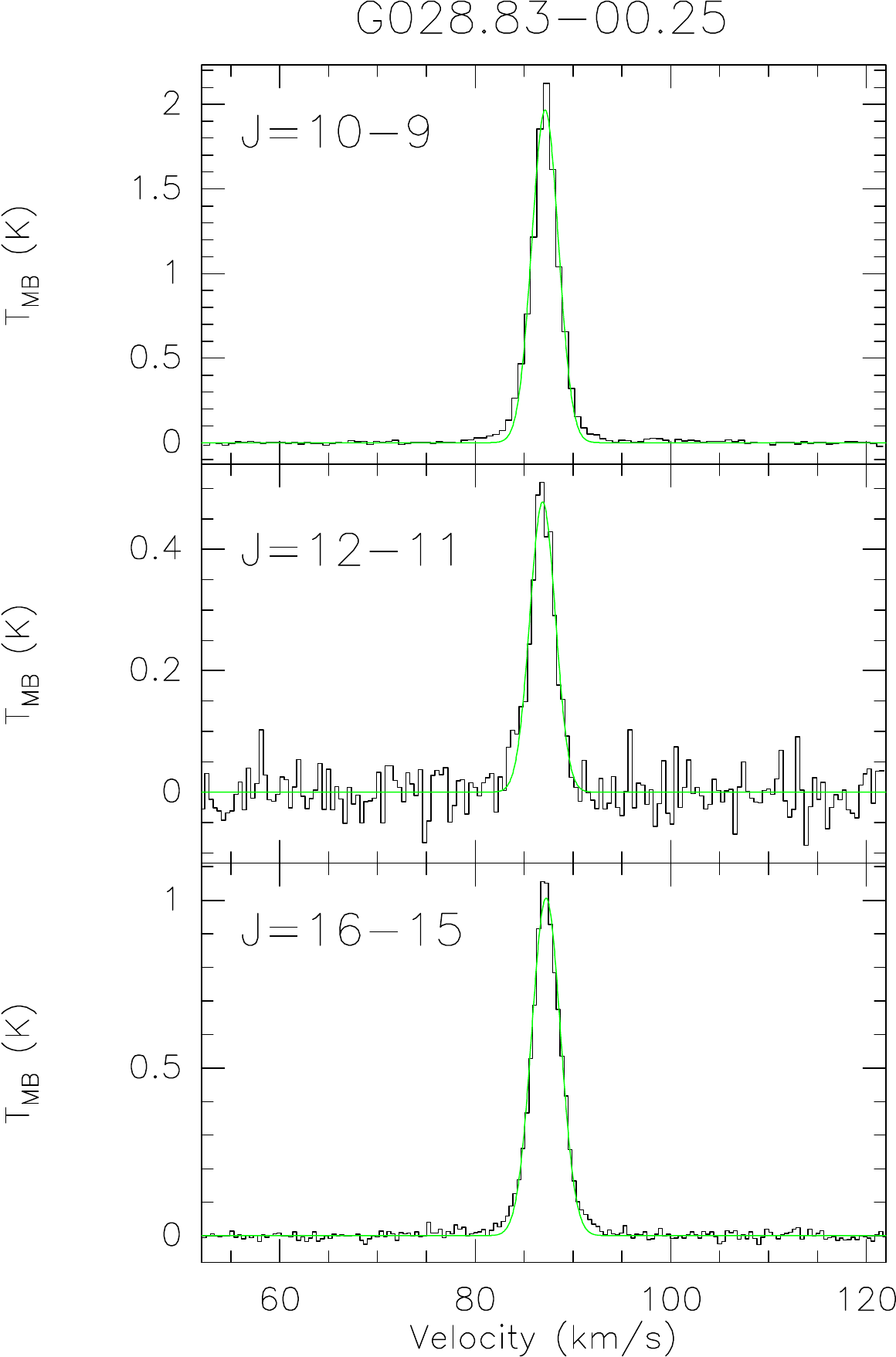}
    \includegraphics[width=0.29\textwidth]{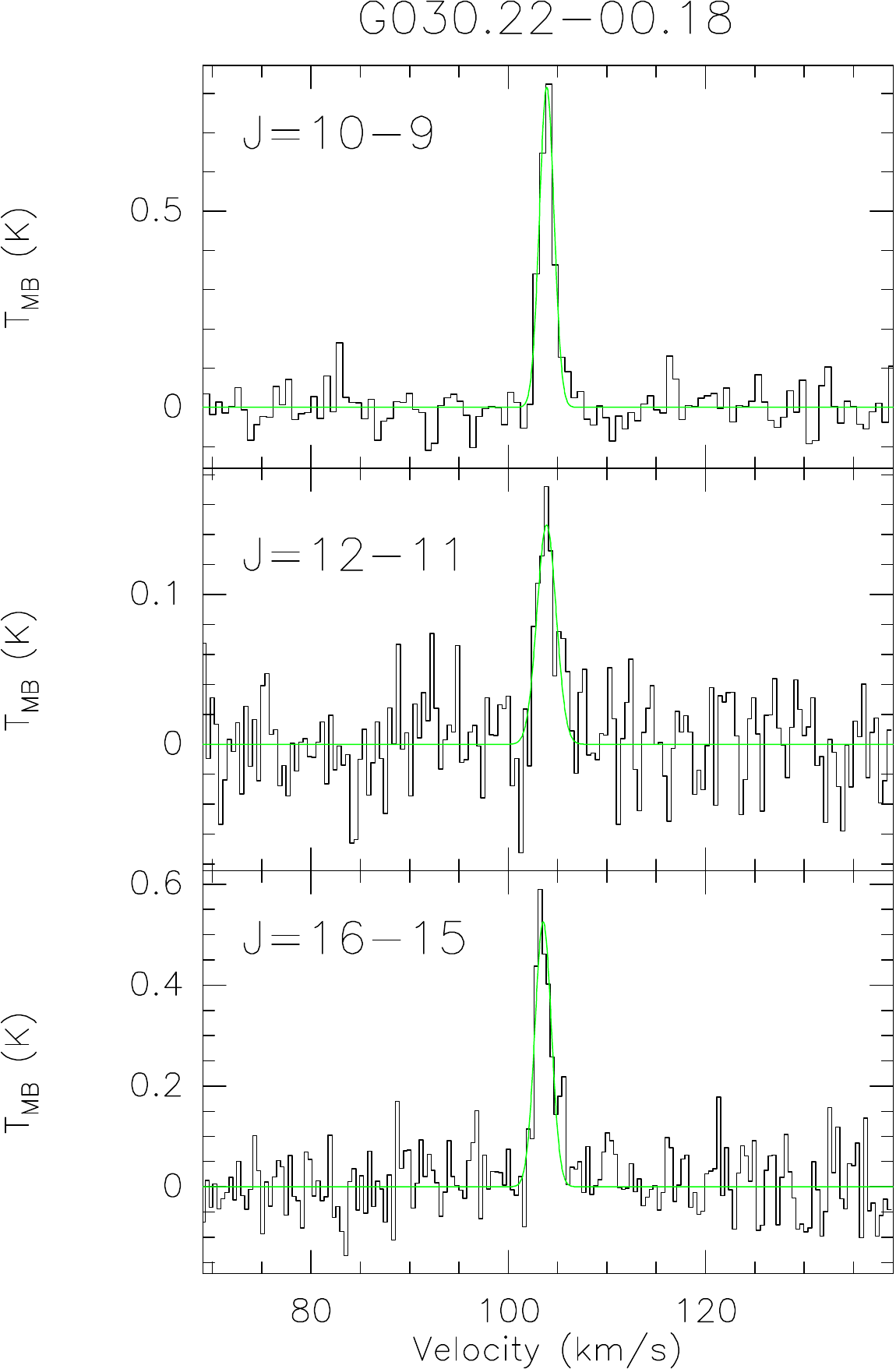}
    \includegraphics[width=0.29\textwidth]{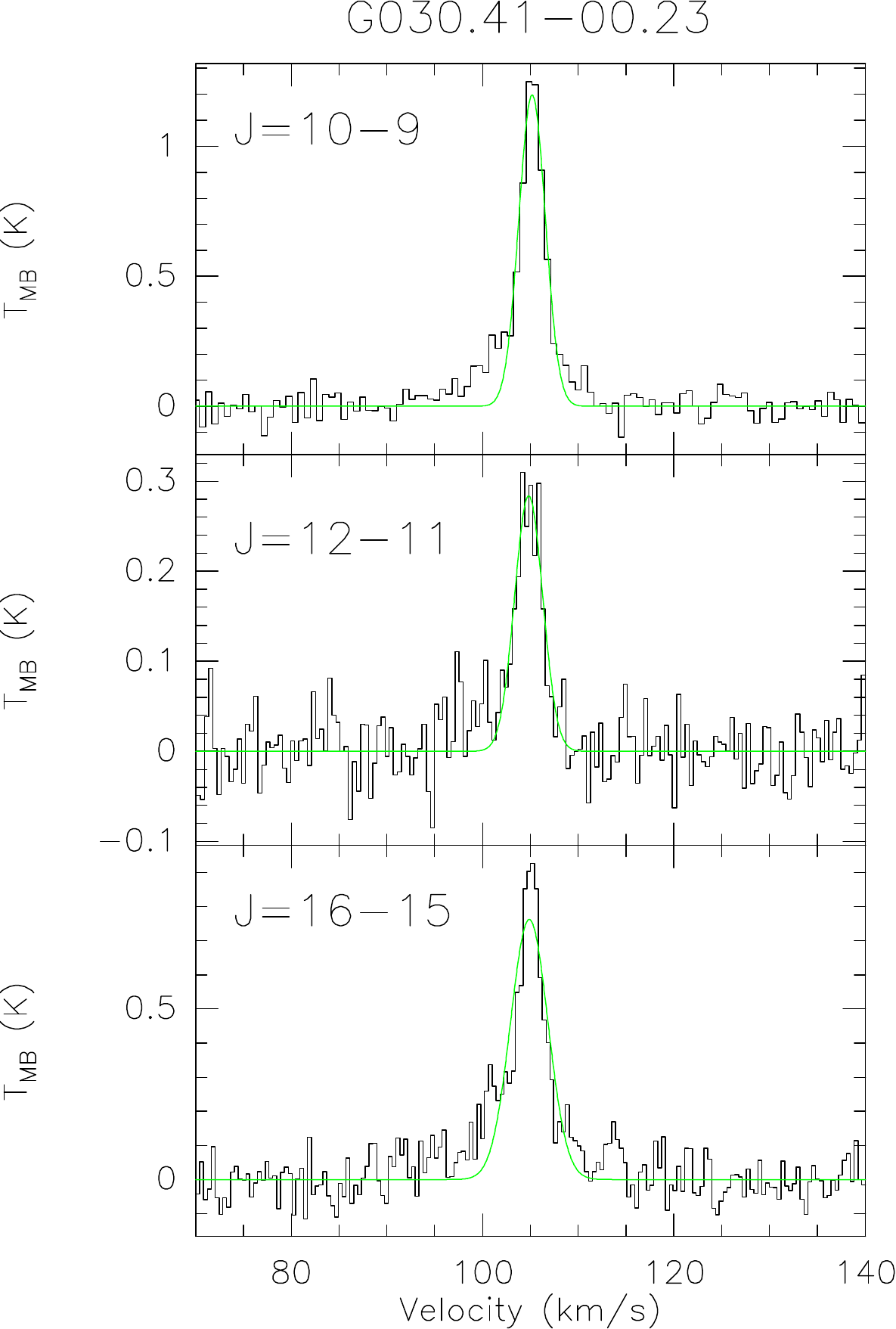}    
    \caption{Continued.}
\end{figure*}
    
\addtocounter{figure}{-1}
\begin{figure*}
    \centering
    \includegraphics[width=0.29\textwidth]{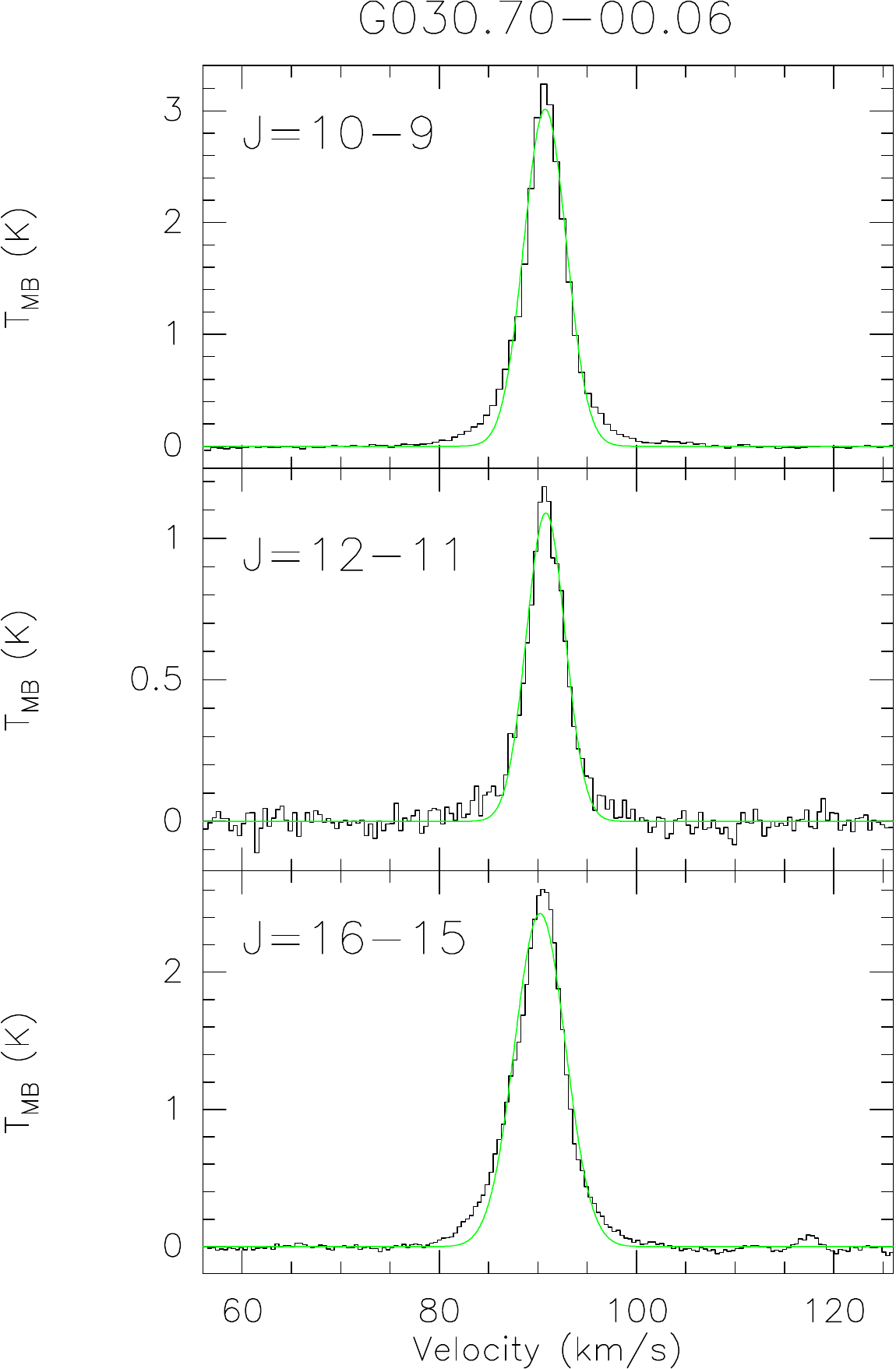}
    \includegraphics[width=0.29\textwidth]{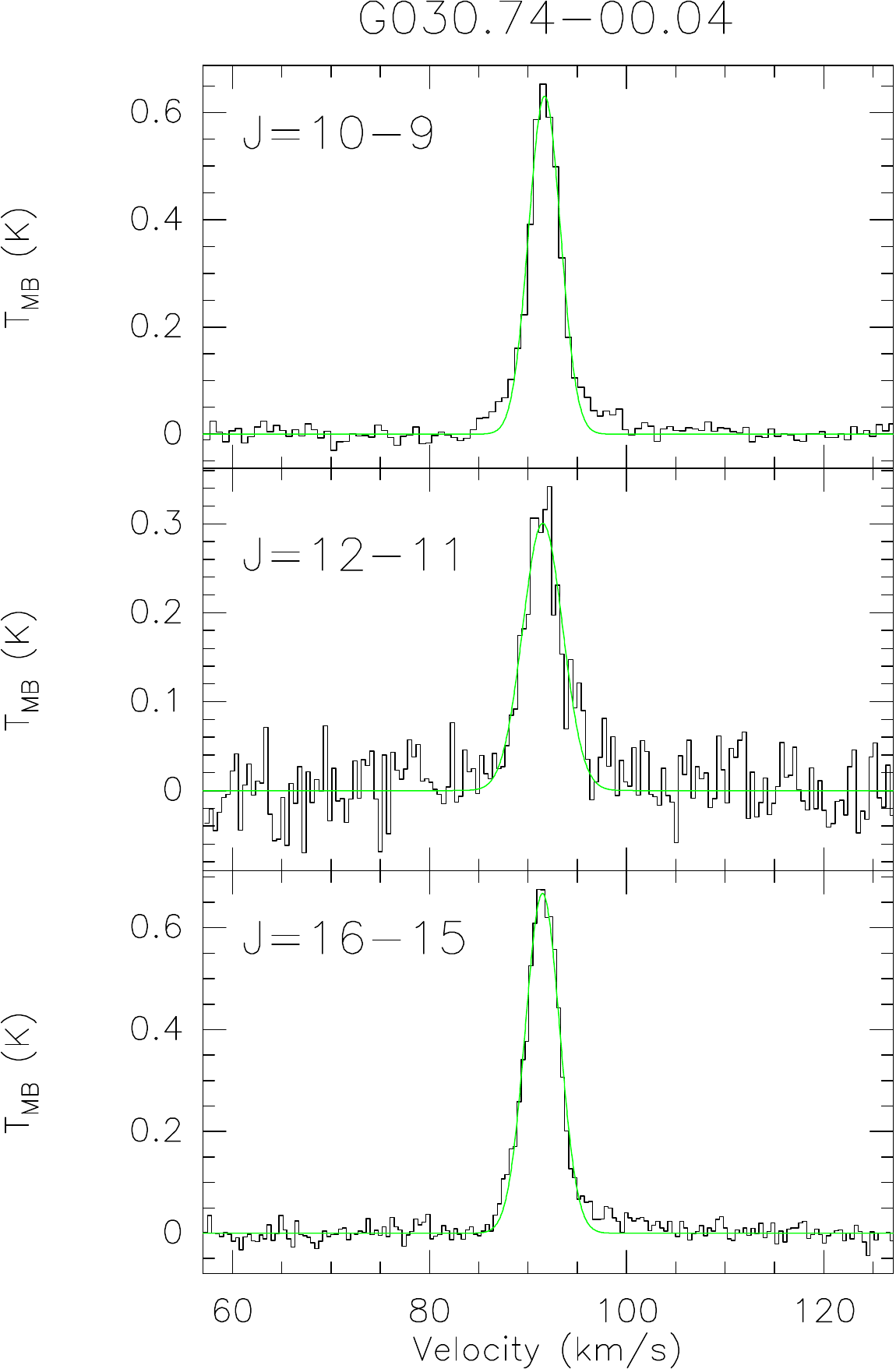}
    \includegraphics[width=0.29\textwidth]{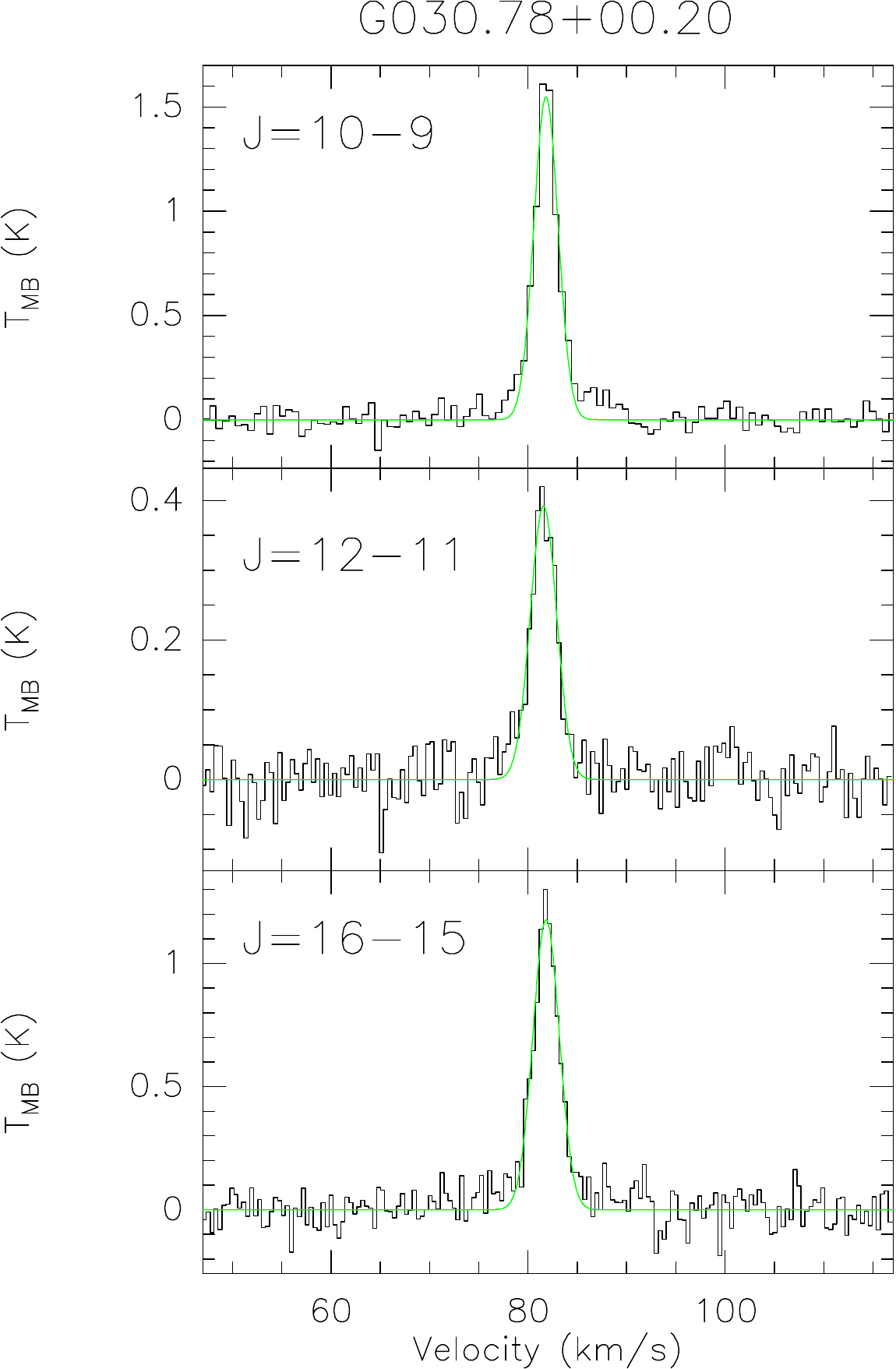} 
    \includegraphics[width=0.29\textwidth]{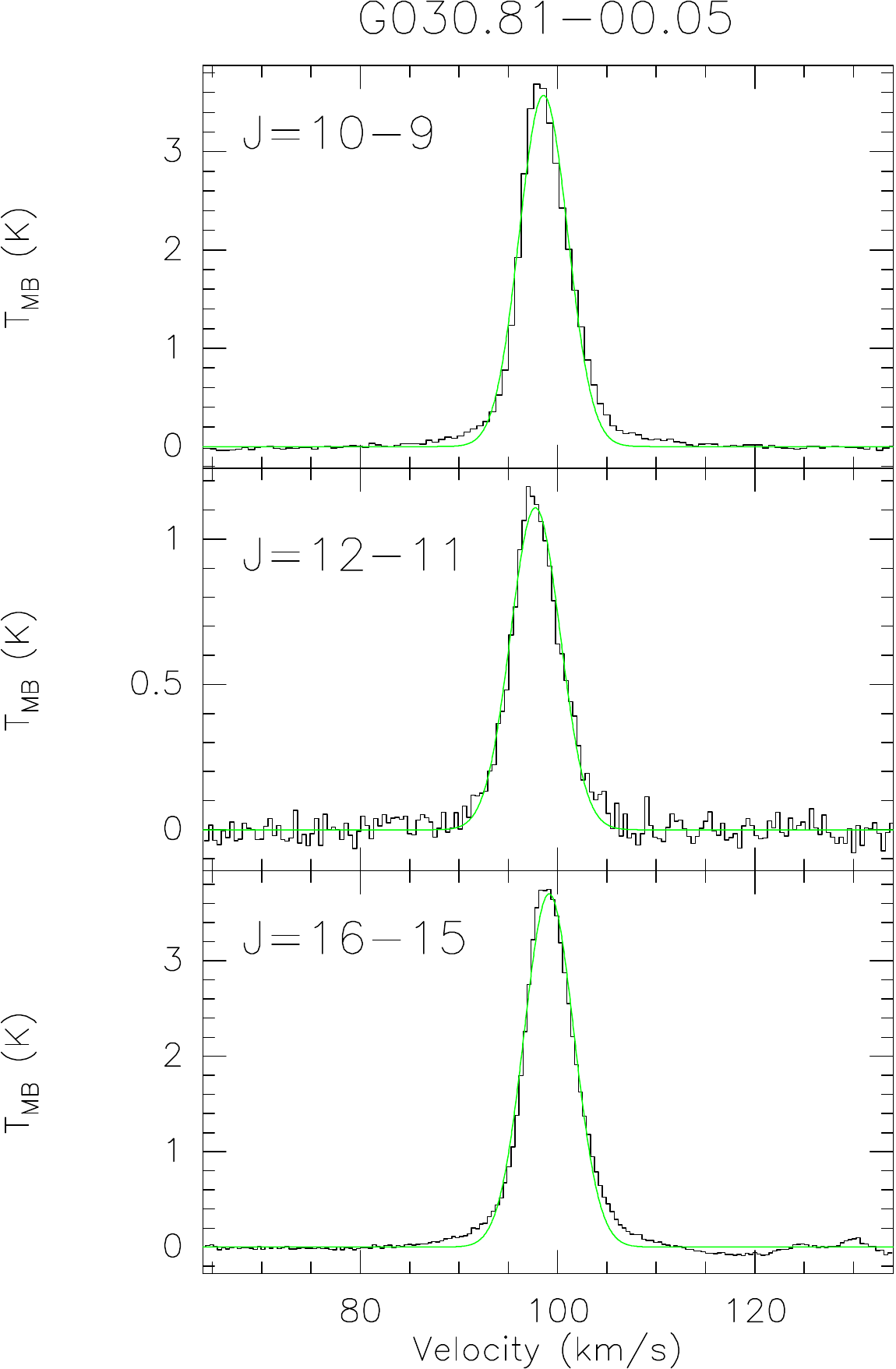}
    \includegraphics[width=0.29\textwidth]{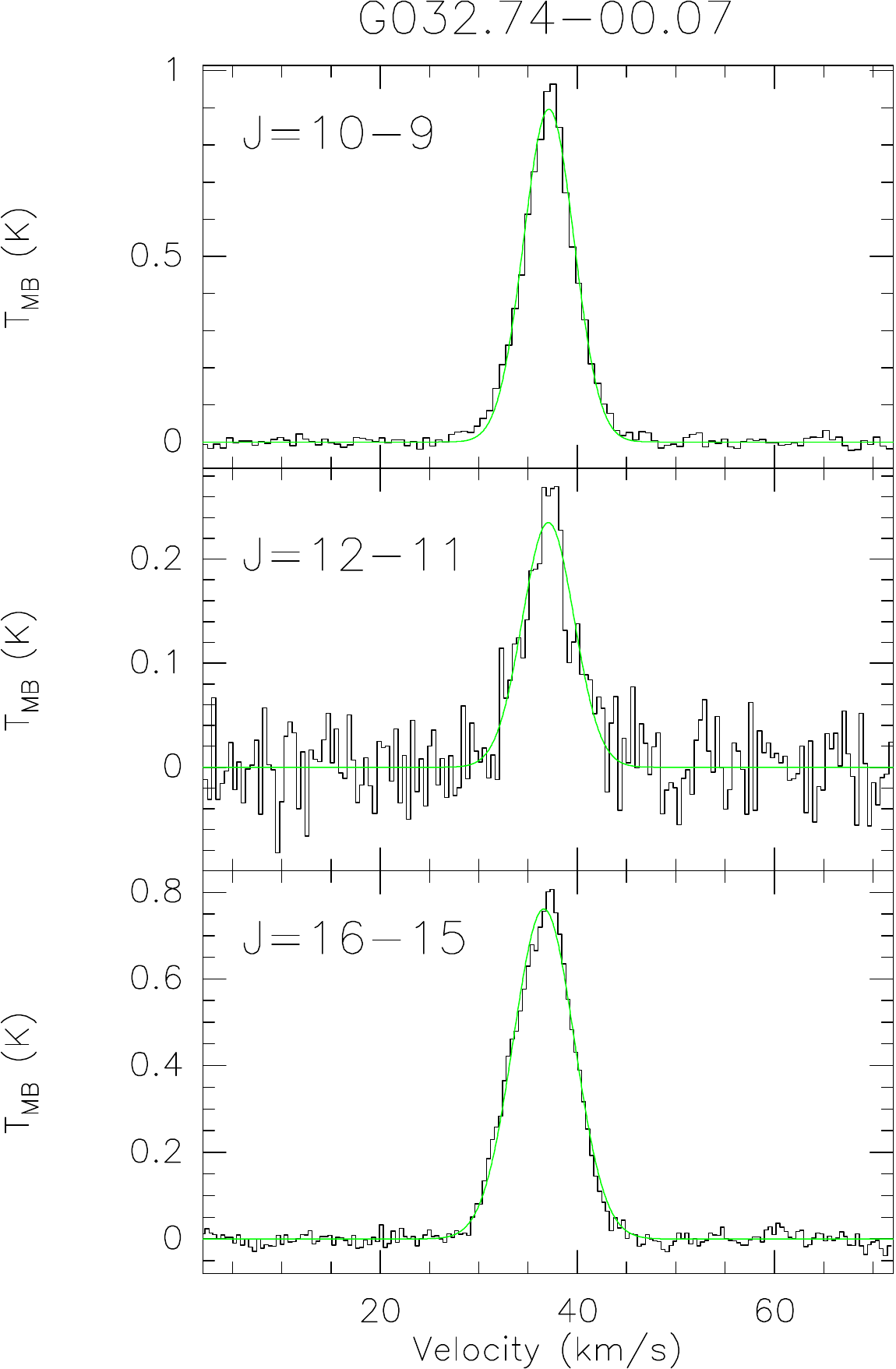}  
    \includegraphics[width=0.29\textwidth]{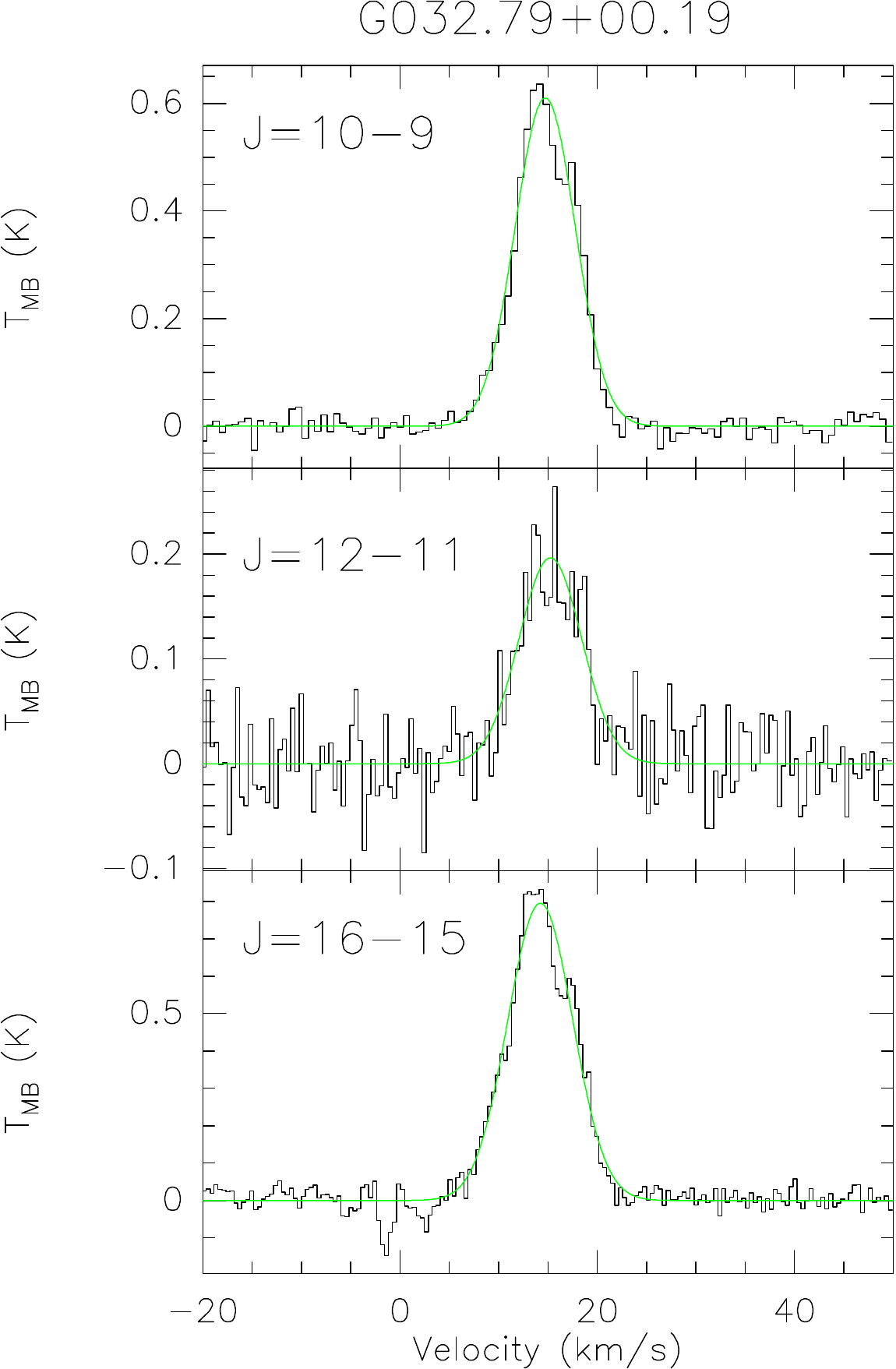} 
    \includegraphics[width=0.29\textwidth]{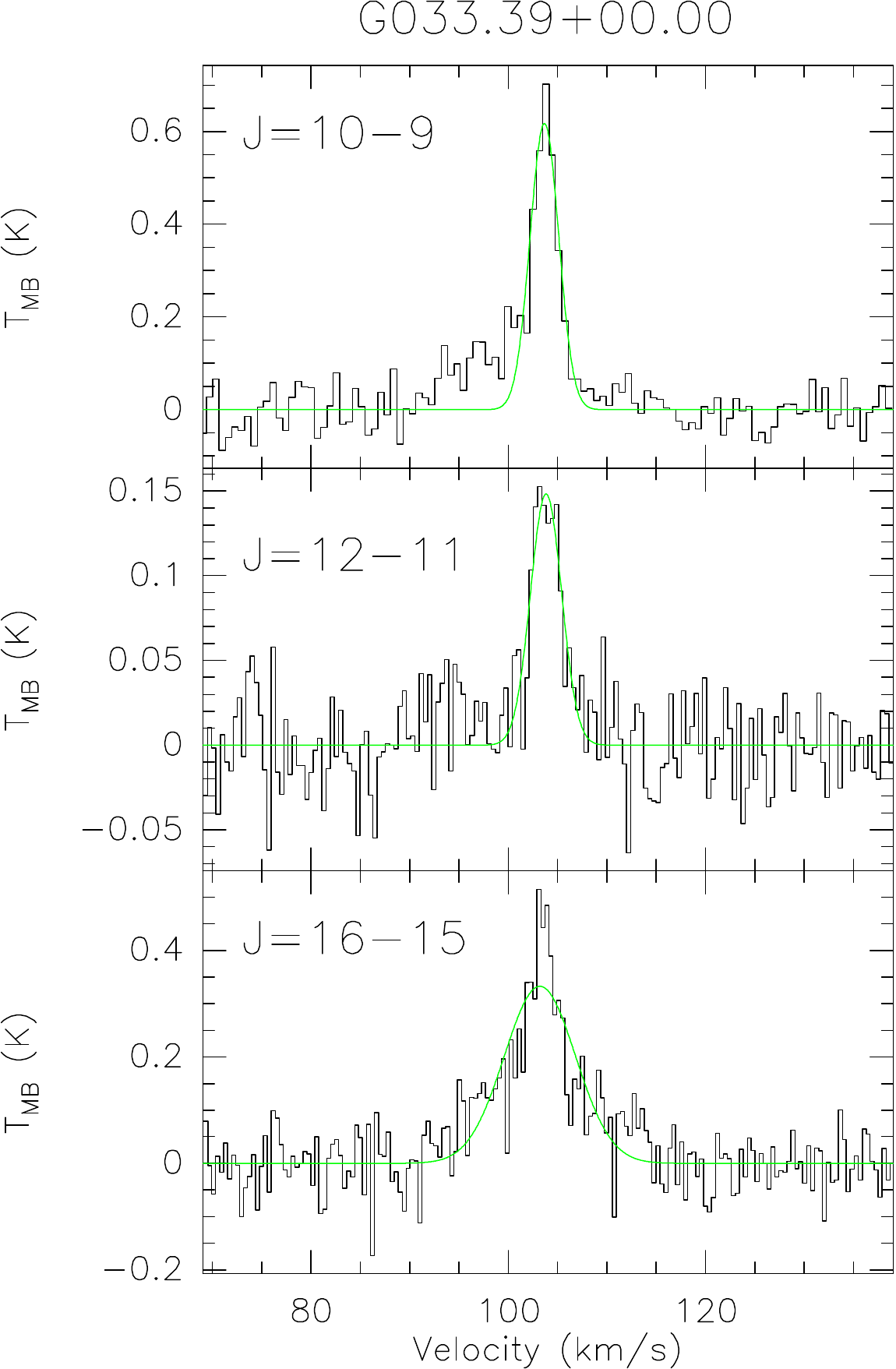}
    \includegraphics[width=0.29\textwidth]{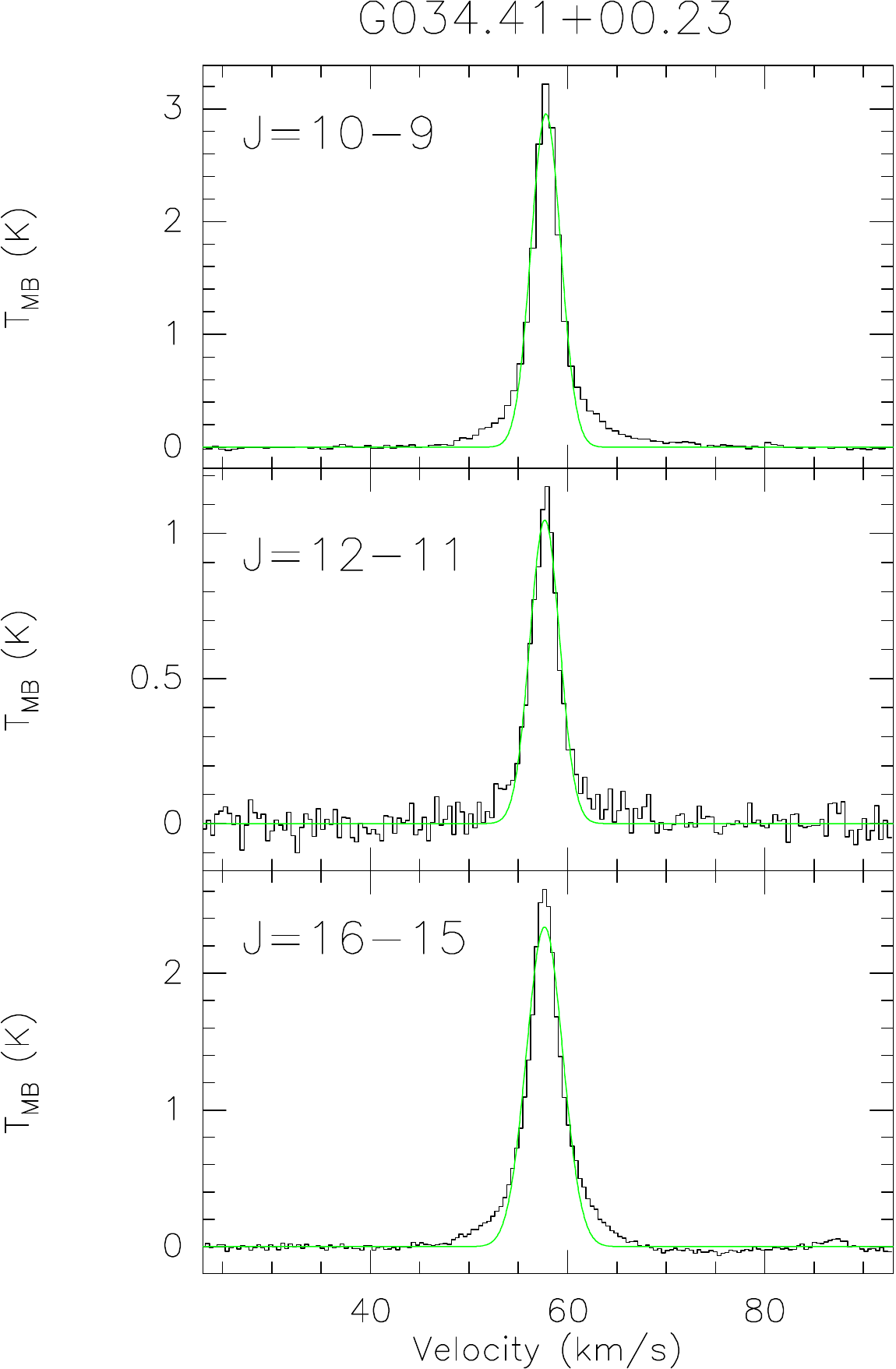}
    \includegraphics[width=0.29\textwidth]{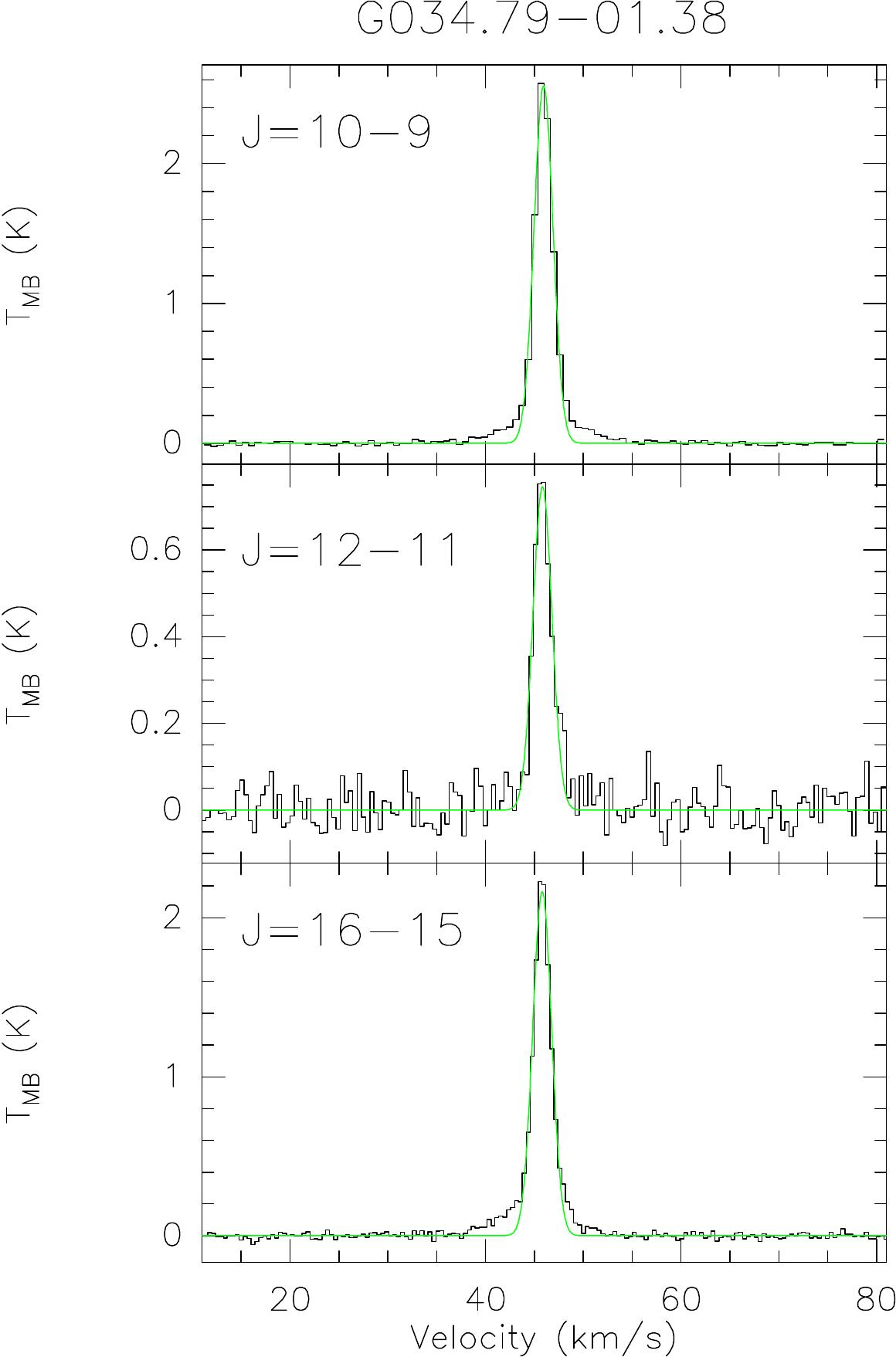}
    \caption{Continued.}
\end{figure*}   
    
\addtocounter{figure}{-1}
\begin{figure*}
    \centering
    \includegraphics[width=0.29\textwidth]{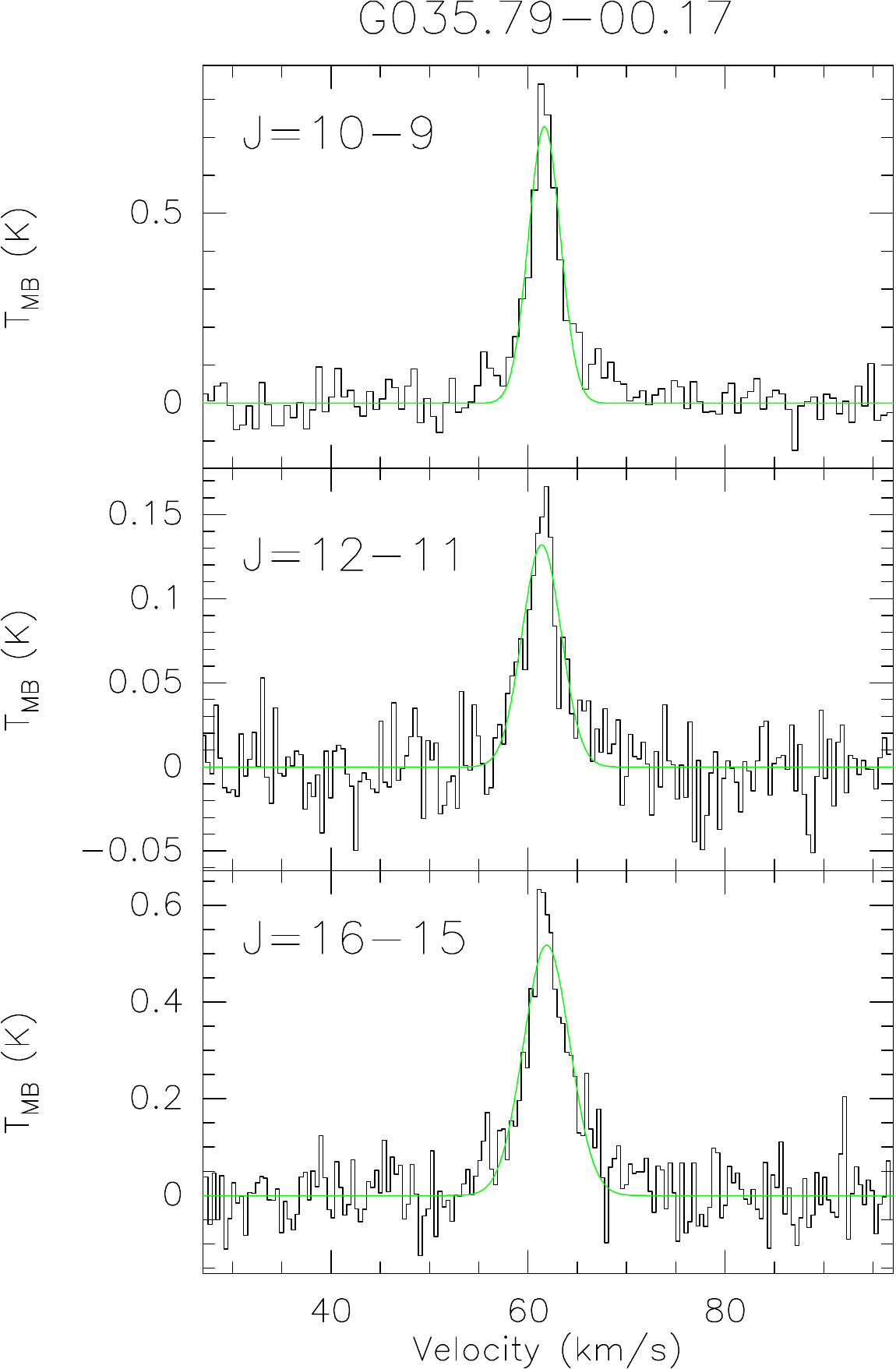}
    \includegraphics[width=0.29\textwidth]{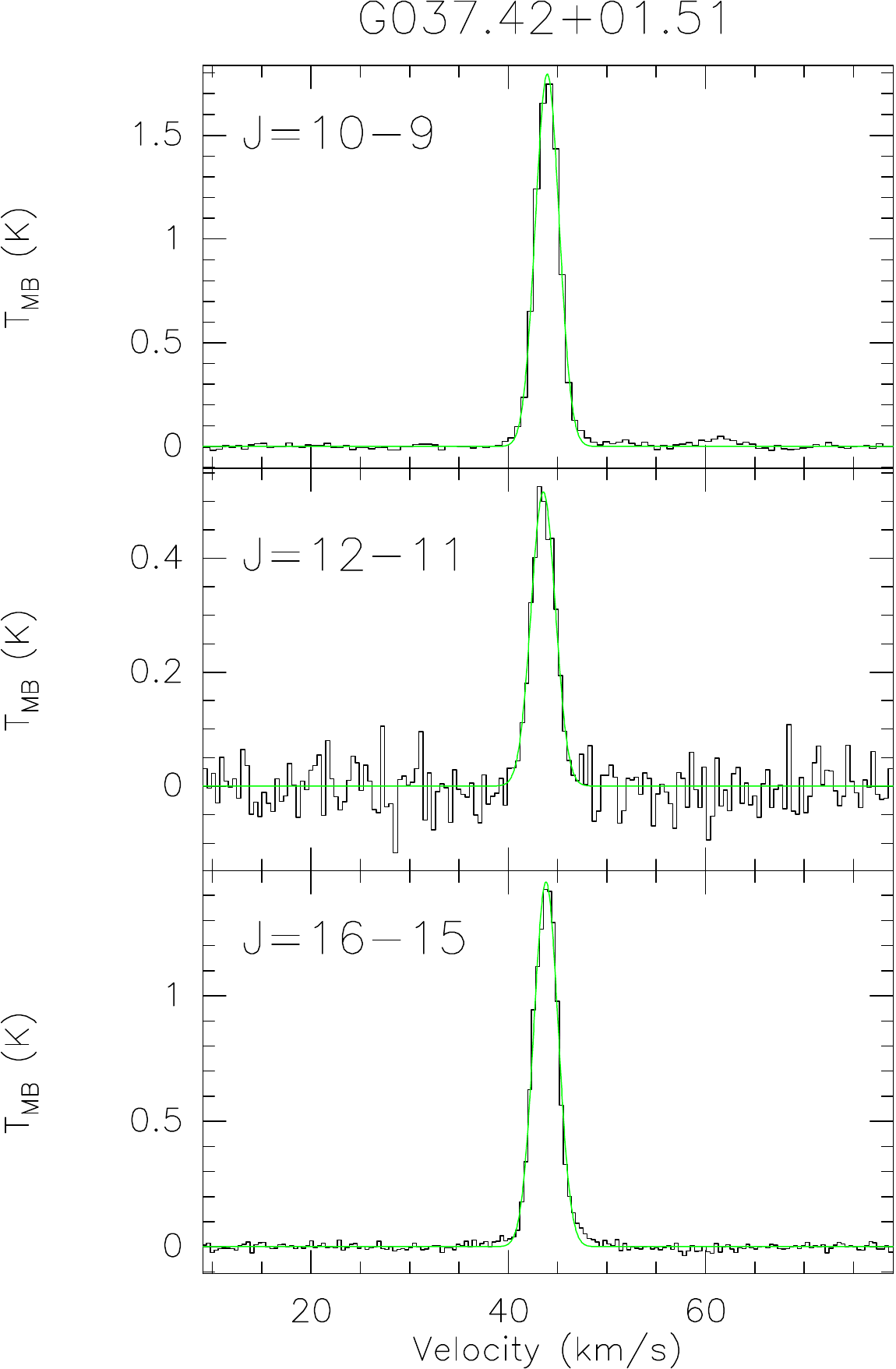}
    \includegraphics[width=0.29\textwidth]{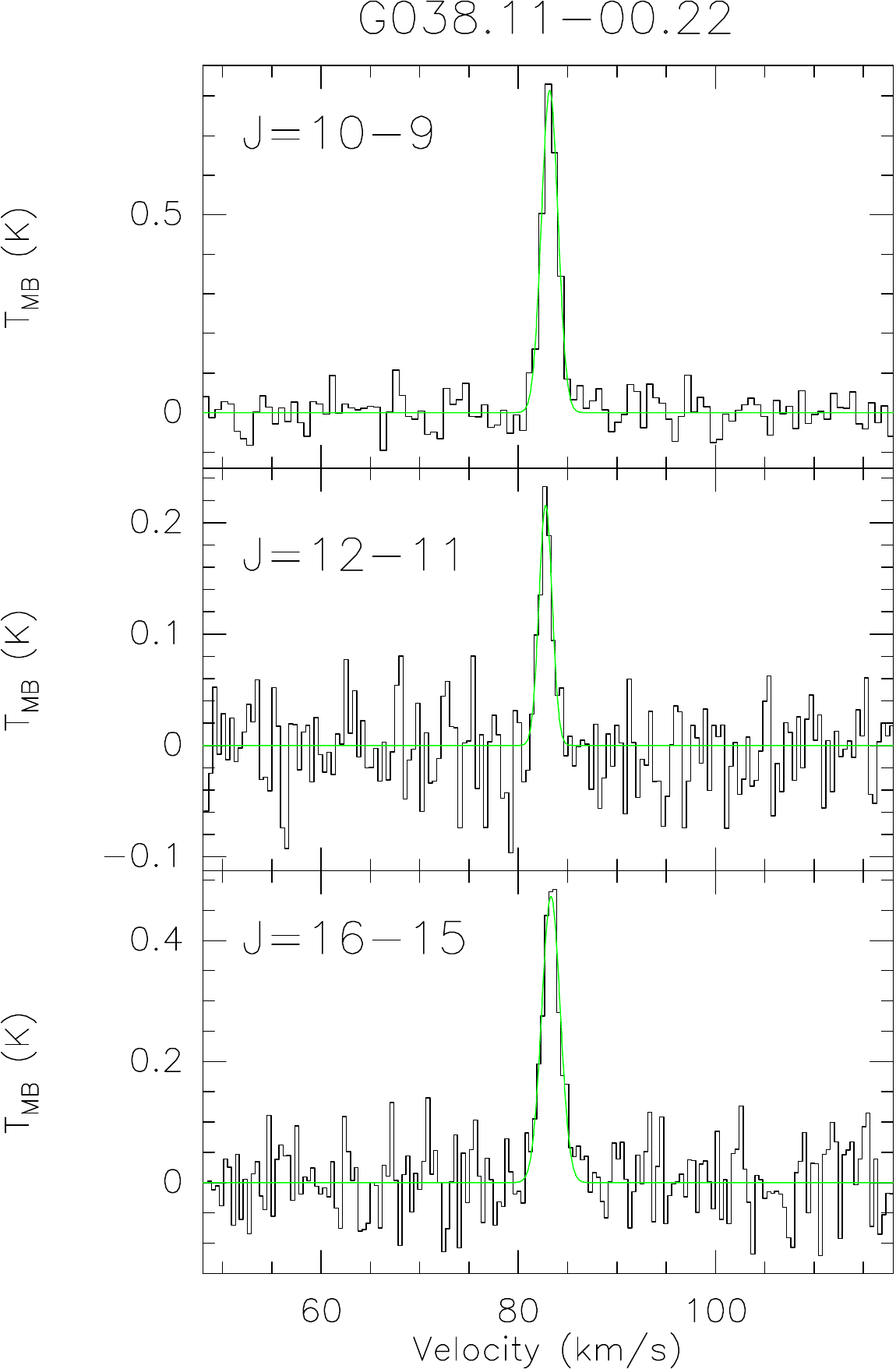} 
    \includegraphics[width=0.29\textwidth]{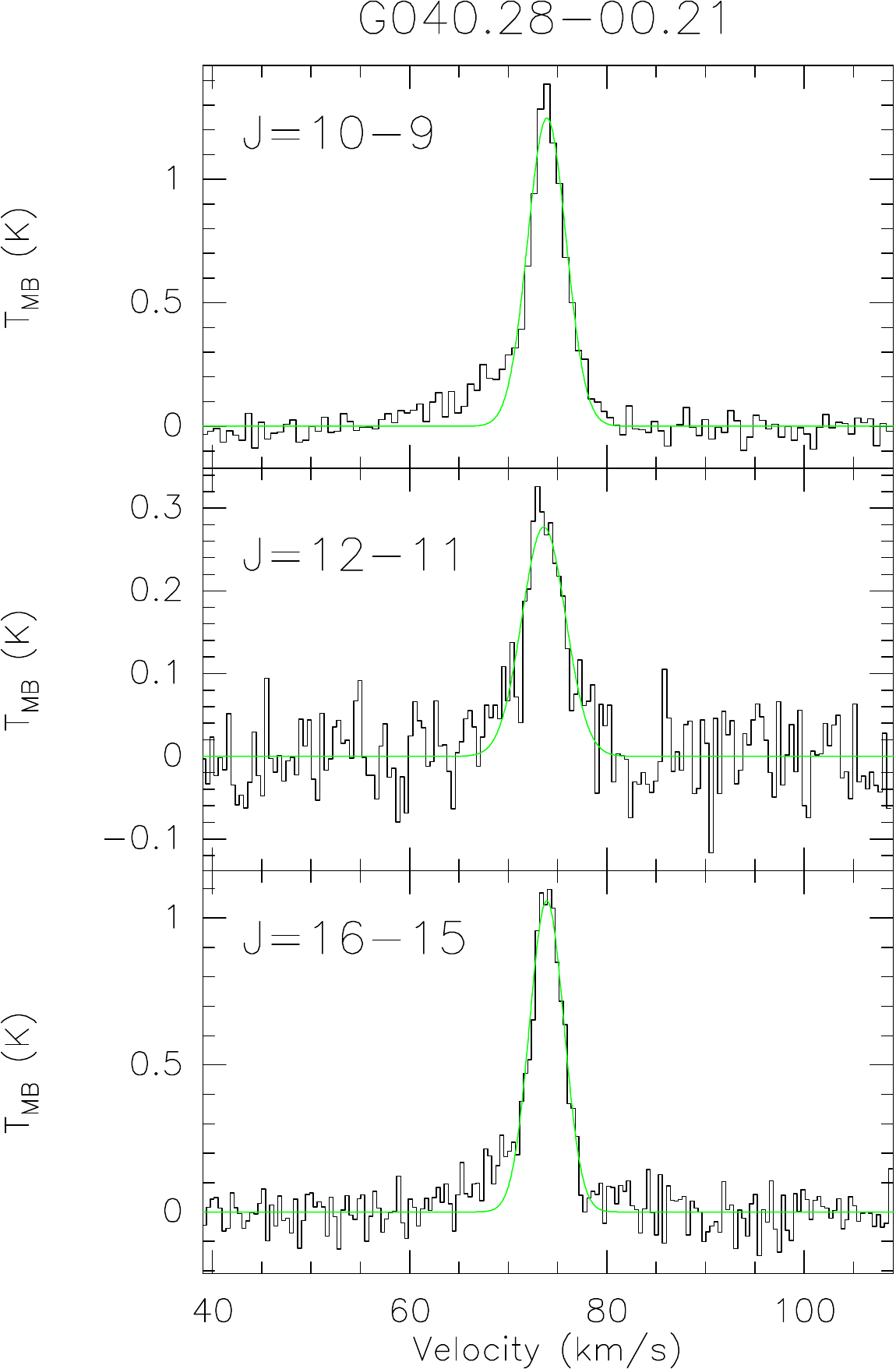}
    \includegraphics[width=0.29\textwidth]{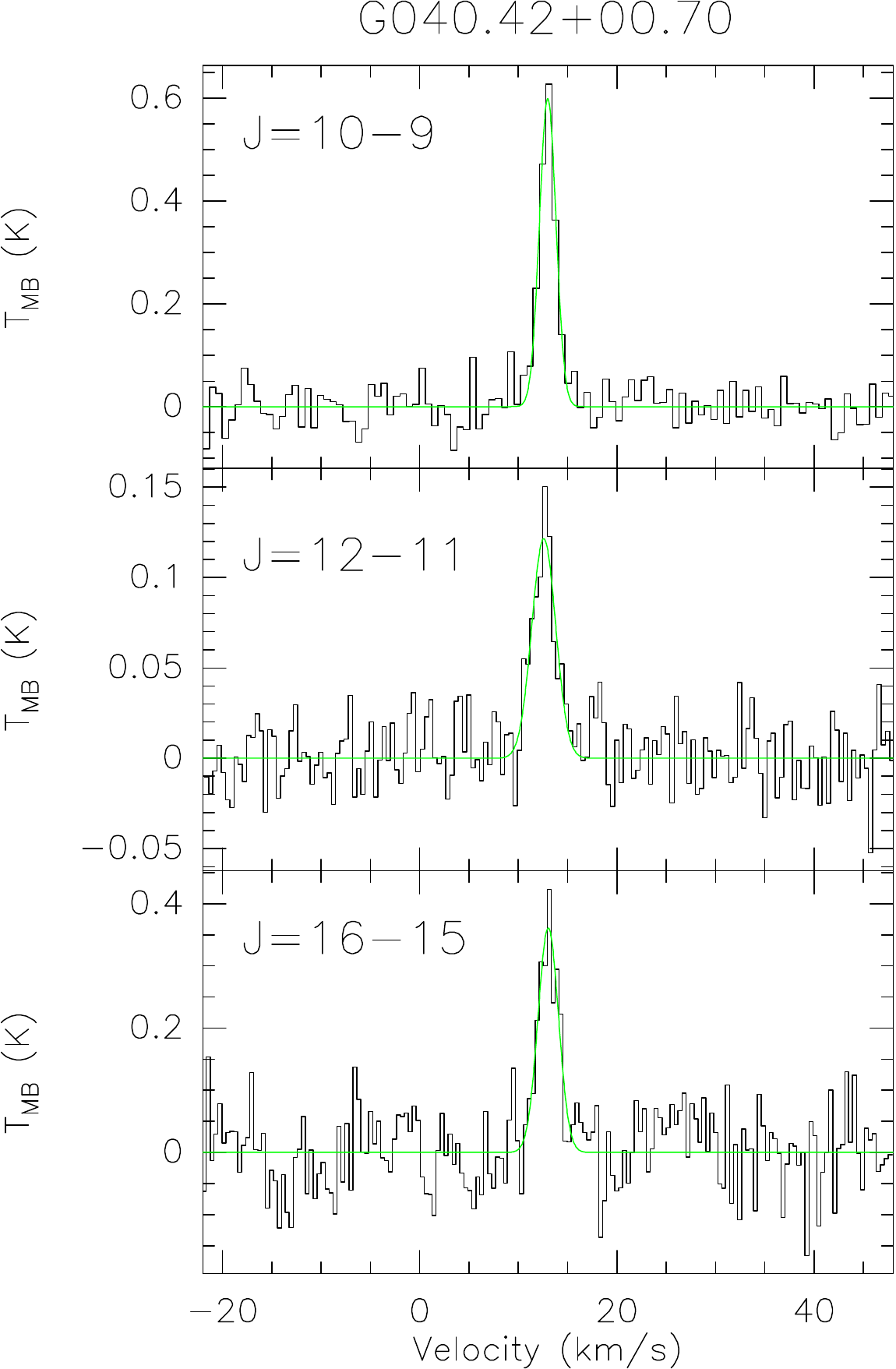}
    \includegraphics[width=0.29\textwidth]{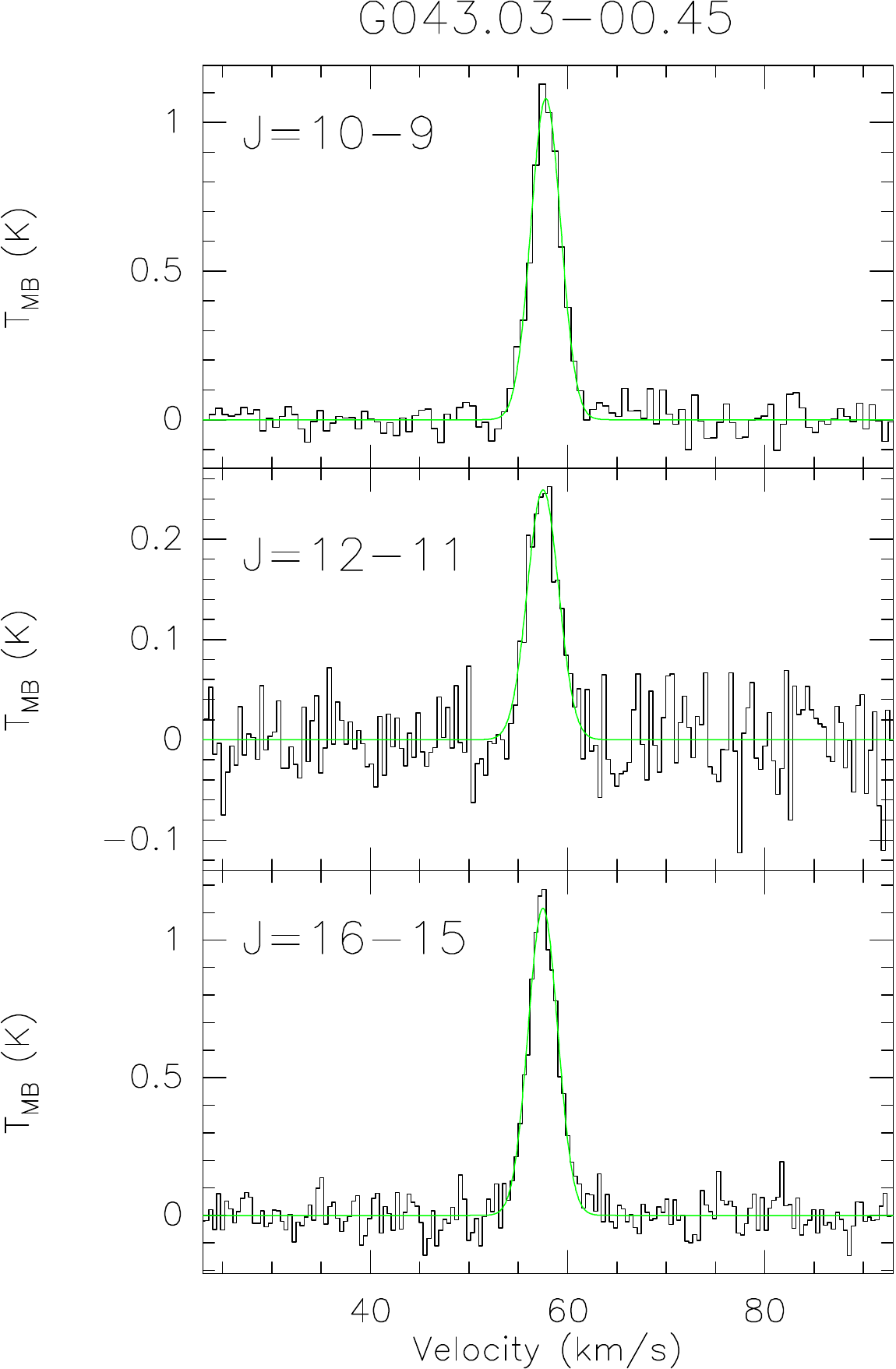} 
    \includegraphics[width=0.29\textwidth]{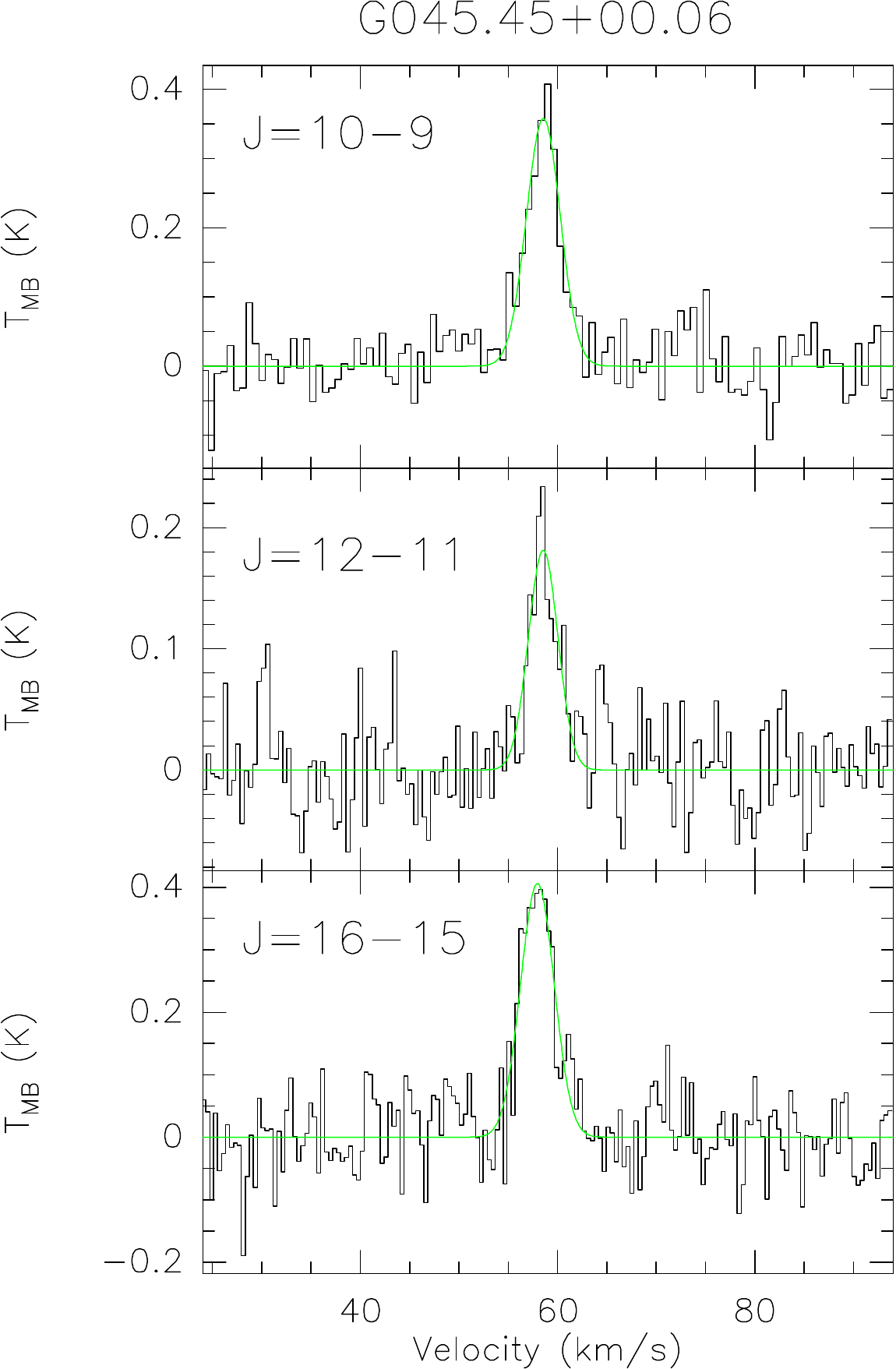}
    \includegraphics[width=0.29\textwidth]{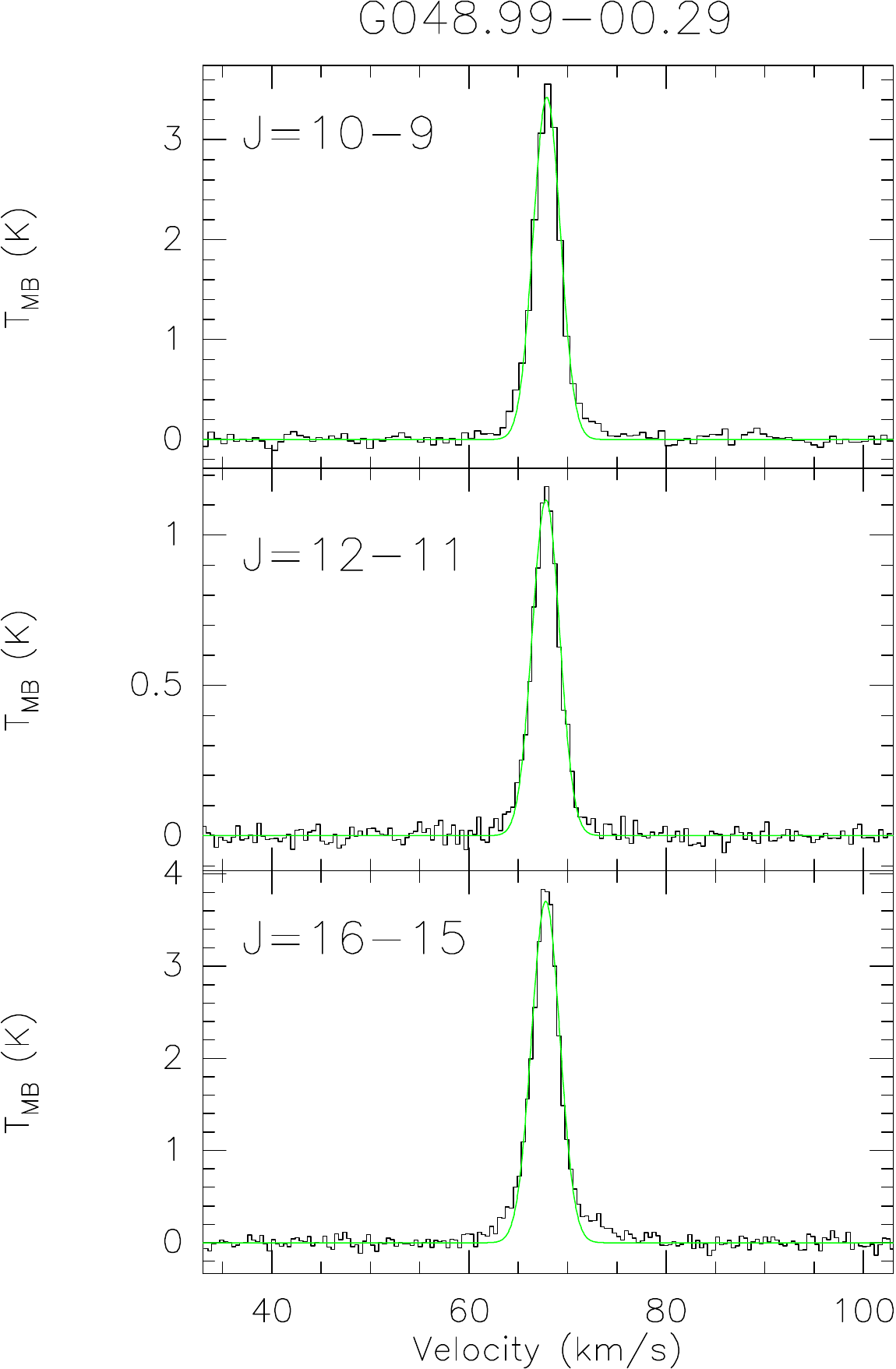}
    \includegraphics[width=0.29\textwidth]{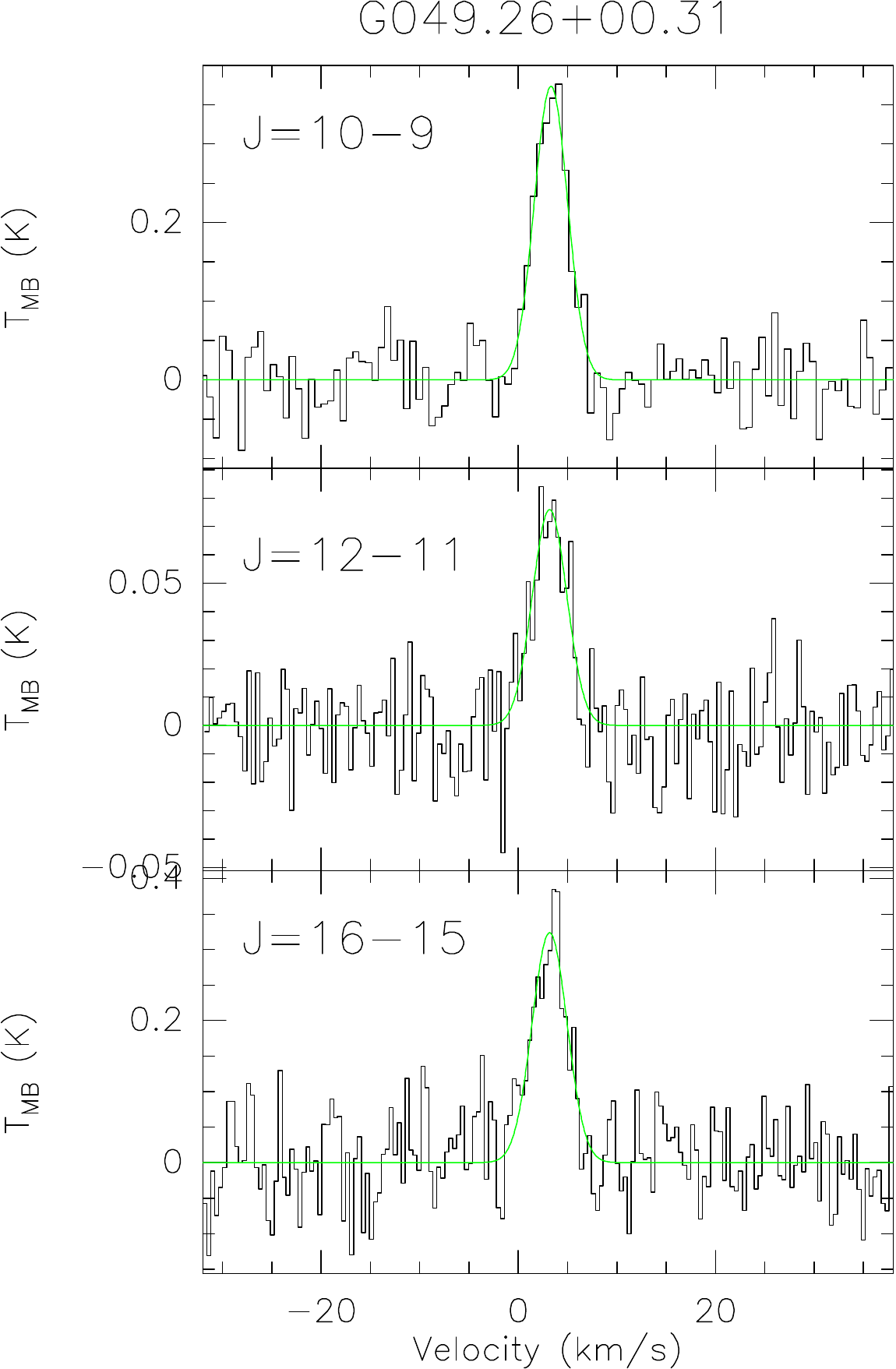}
    \caption{Continued.}
\end{figure*}

\addtocounter{figure}{-1}
\begin{figure*}
    \centering
    \includegraphics[width=0.29\textwidth]{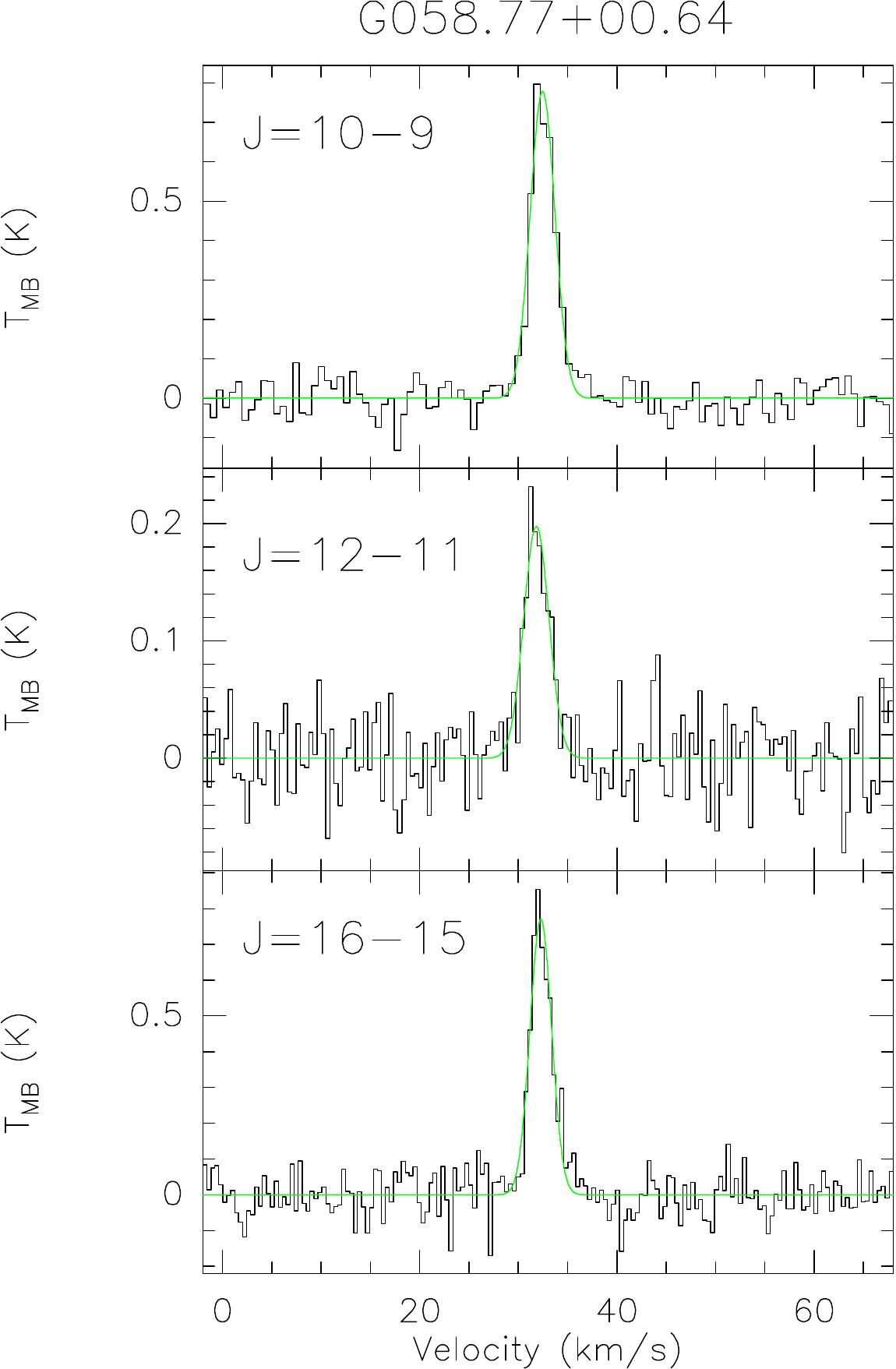}   
    \includegraphics[width=0.29\textwidth]{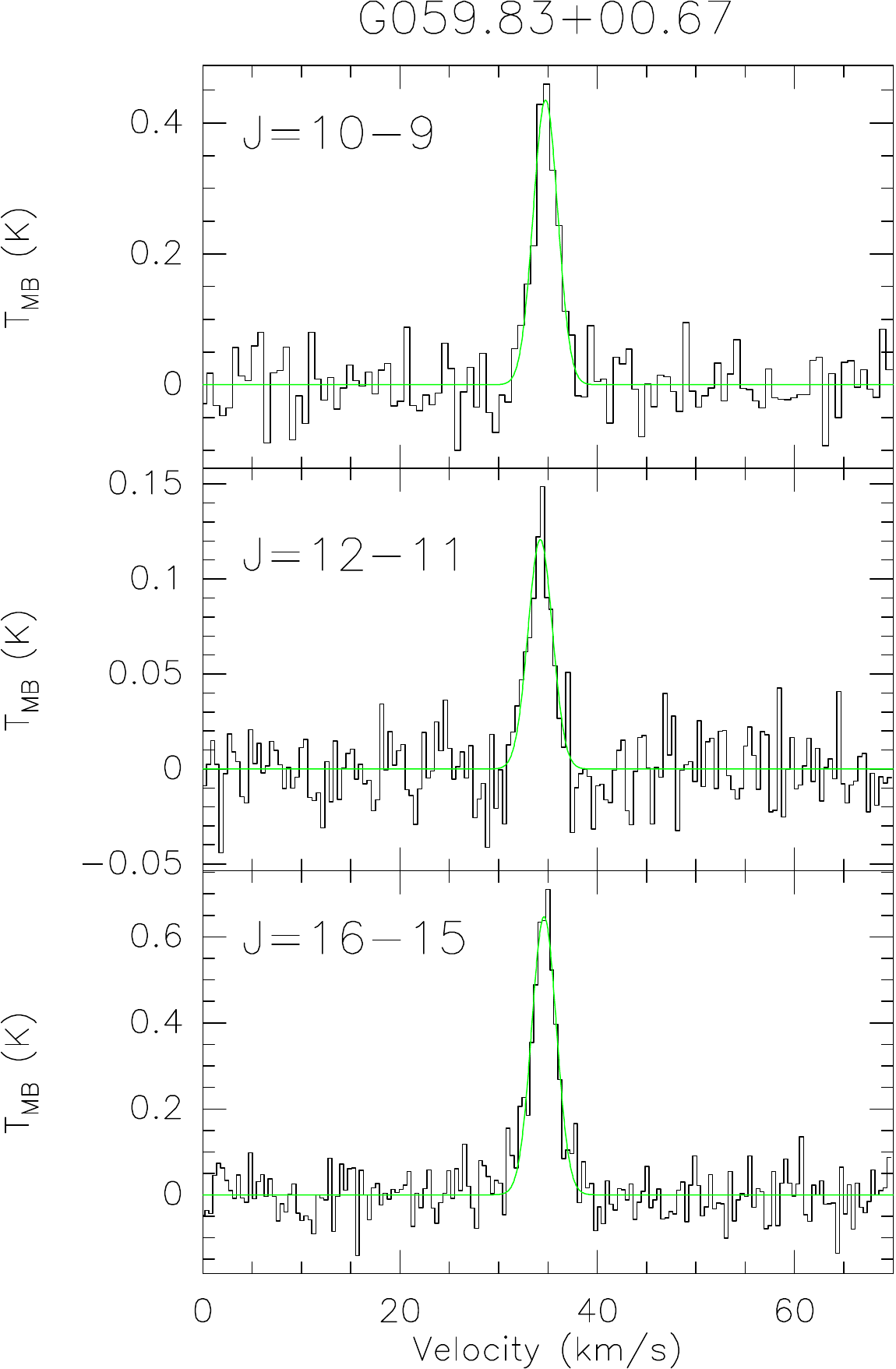}
    \includegraphics[width=0.29\textwidth]{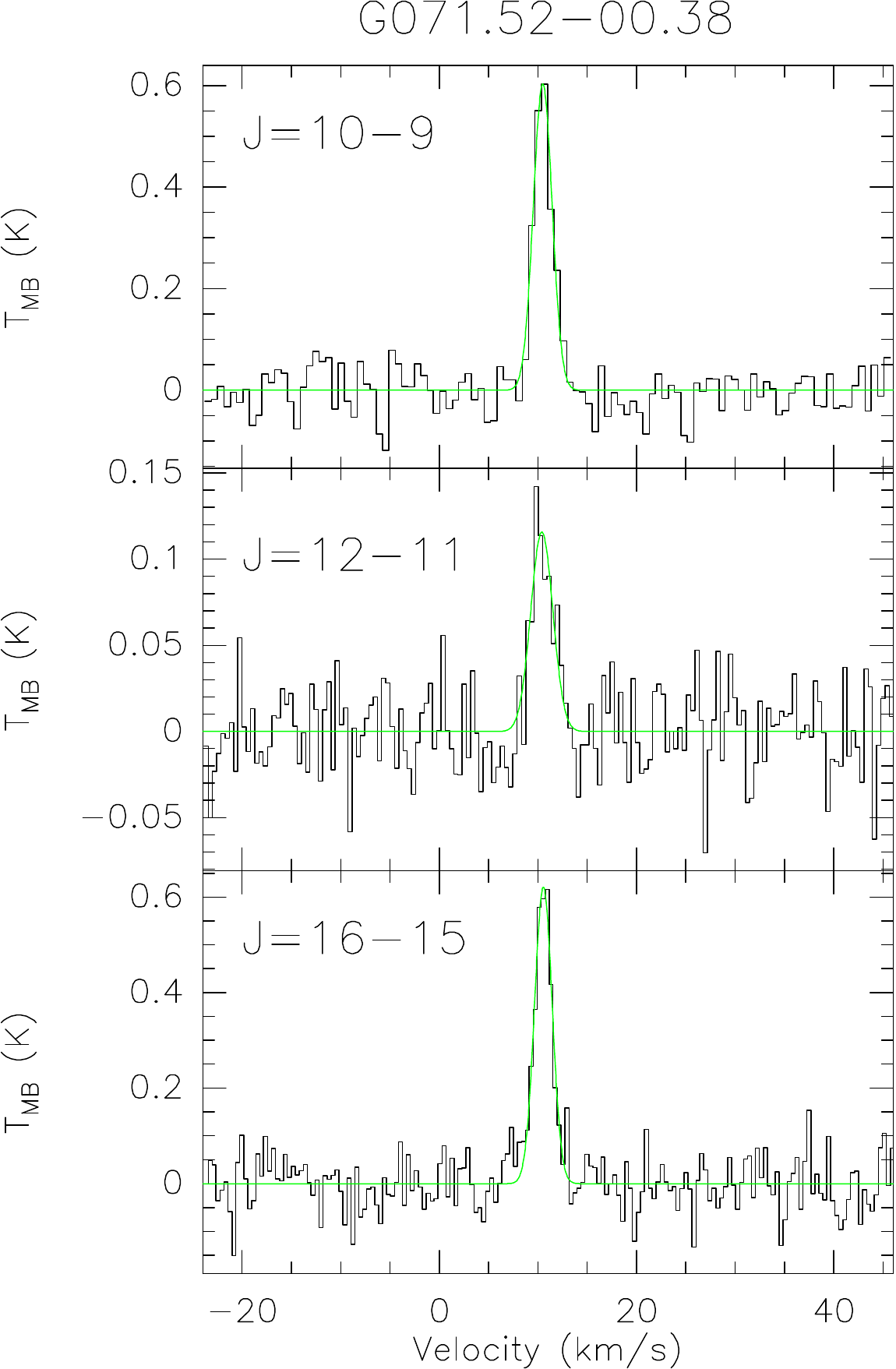}
    \includegraphics[width=0.29\textwidth]{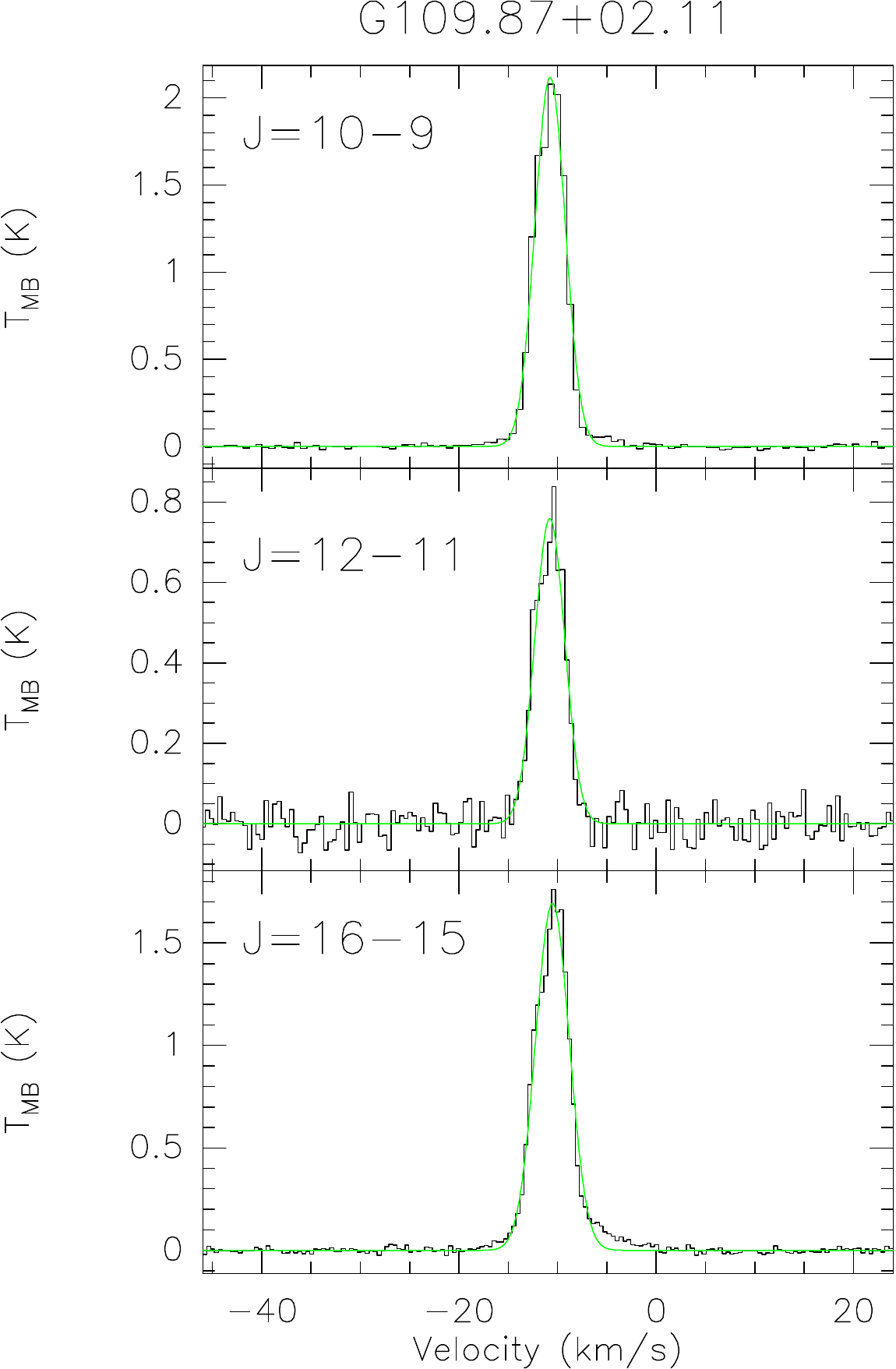} 
    \caption{Continued.}
    \label{fig1}
\end{figure*}

\begin{figure*}
    \centering
    \includegraphics[width=0.3\textwidth]{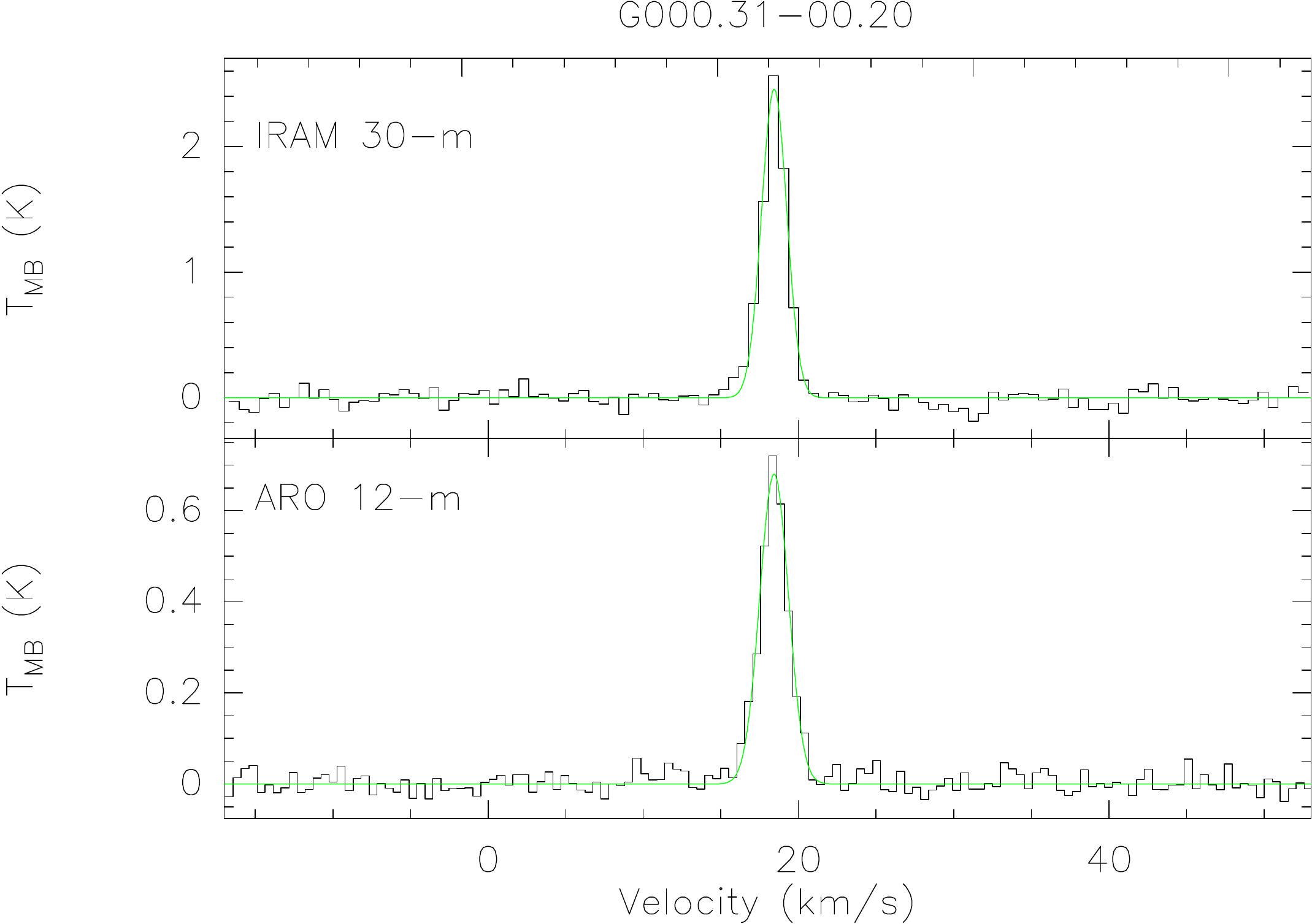}
    \includegraphics[width=0.3\textwidth]{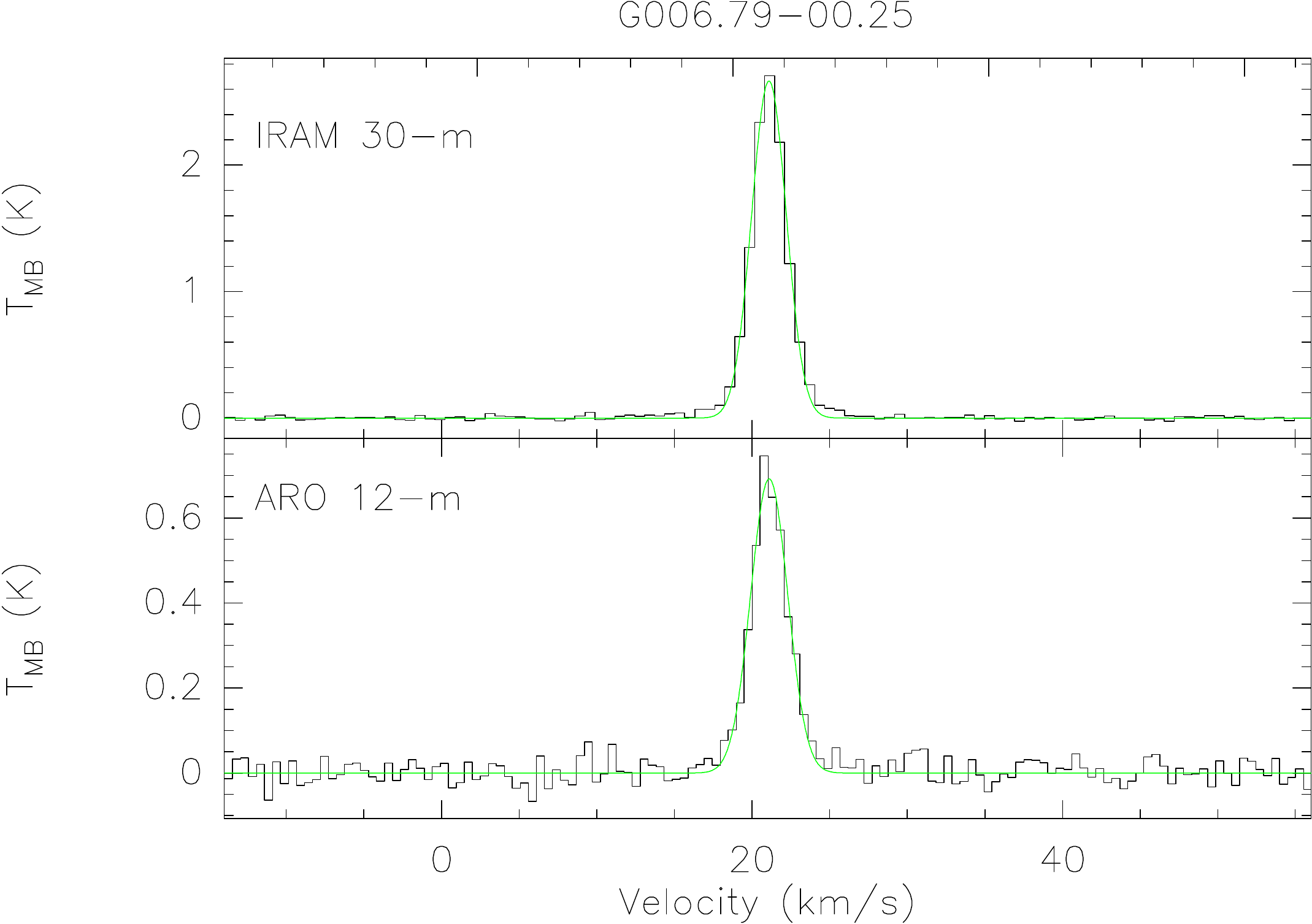}
    \includegraphics[width=0.3\textwidth]{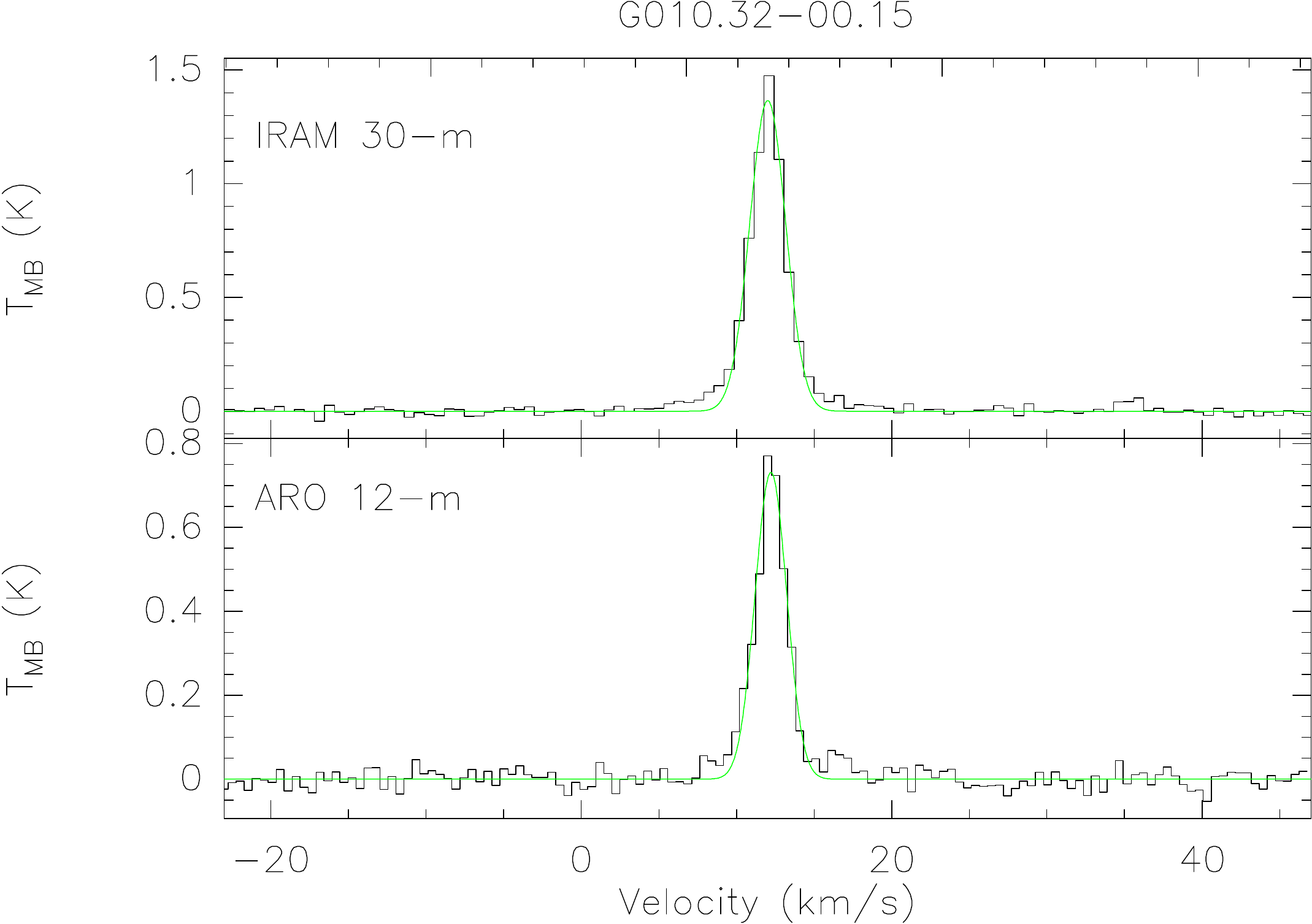} 
    \includegraphics[width=0.3\textwidth]{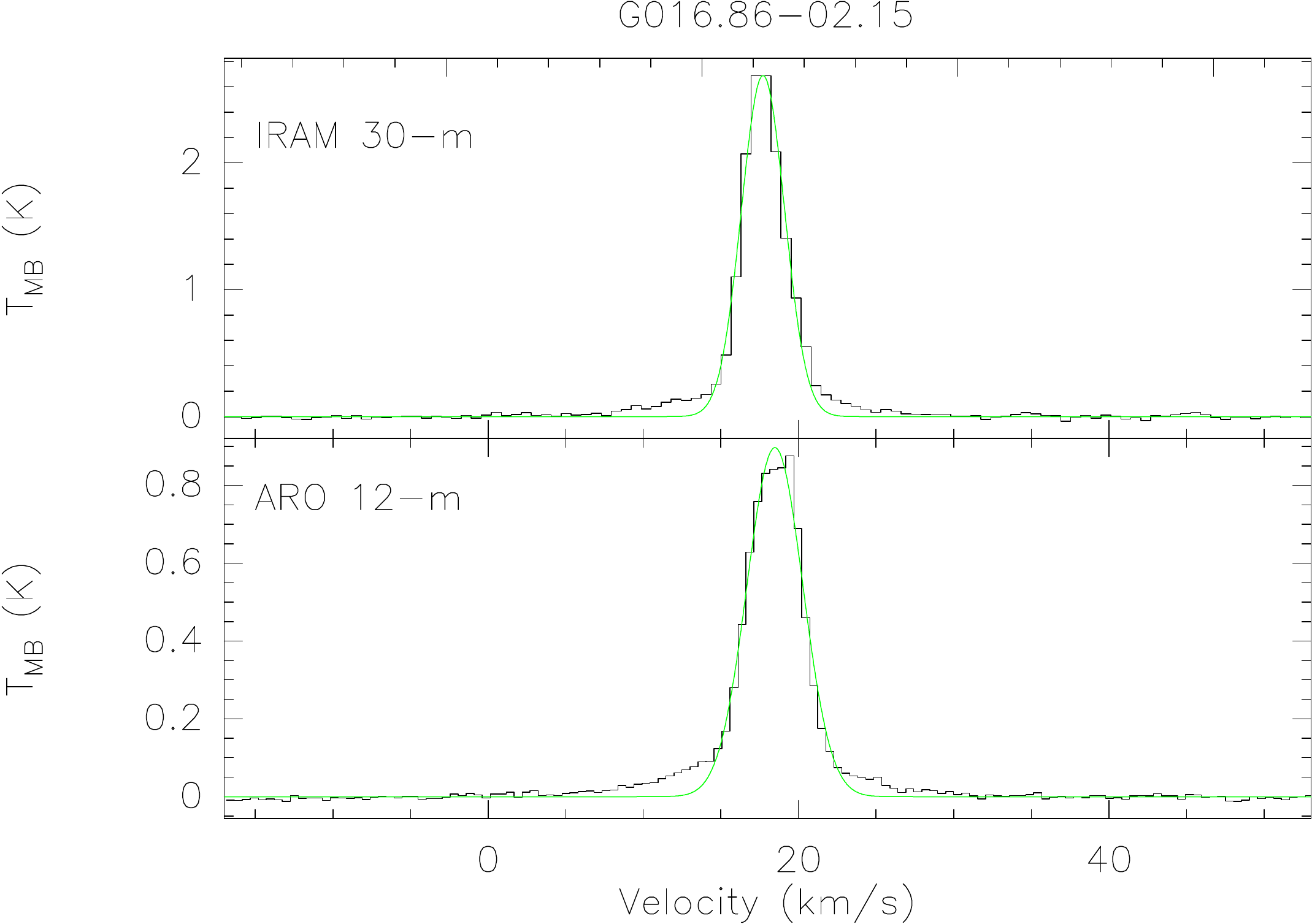}
    \includegraphics[width=0.3\textwidth]{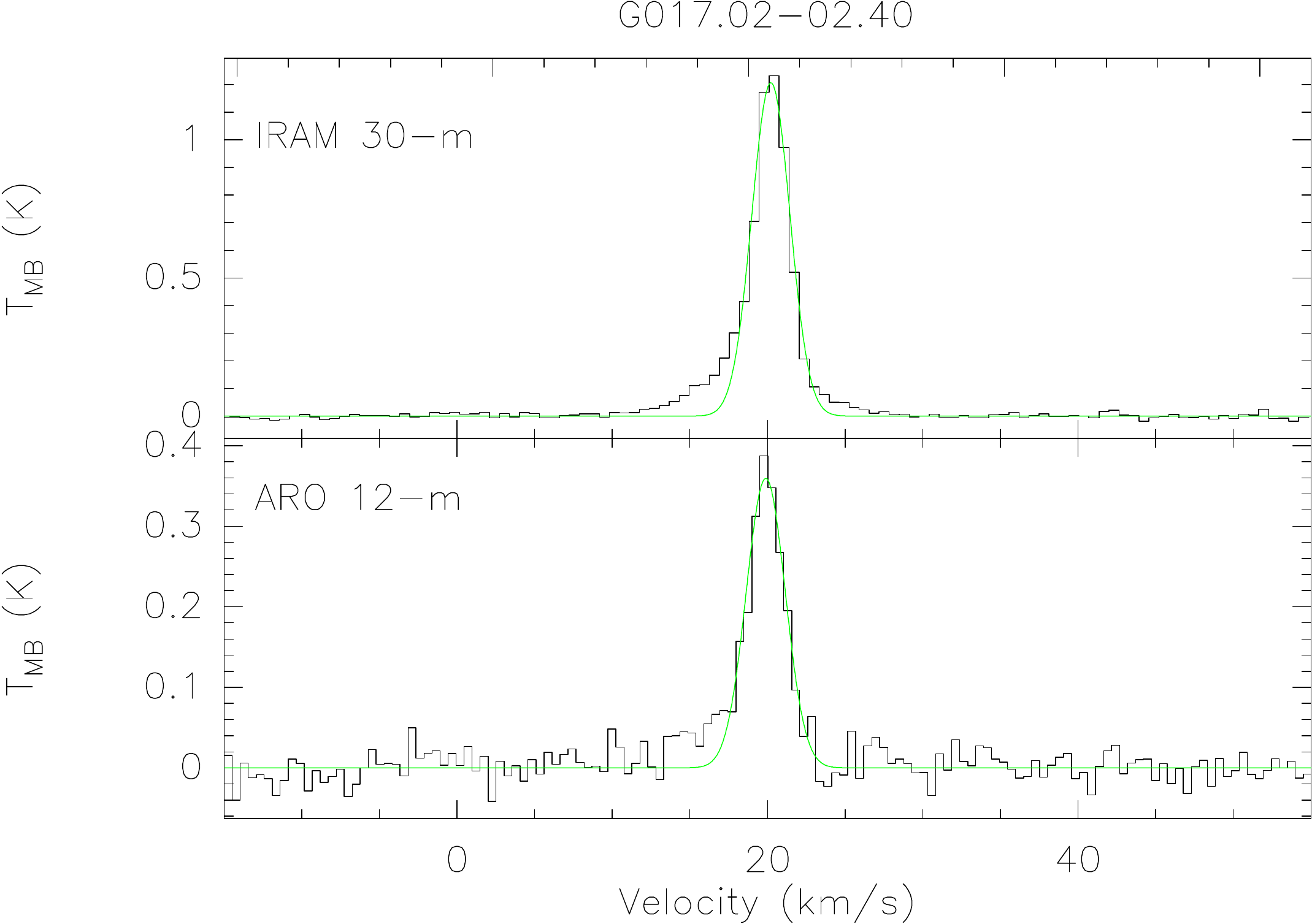}
    \includegraphics[width=0.3\textwidth]{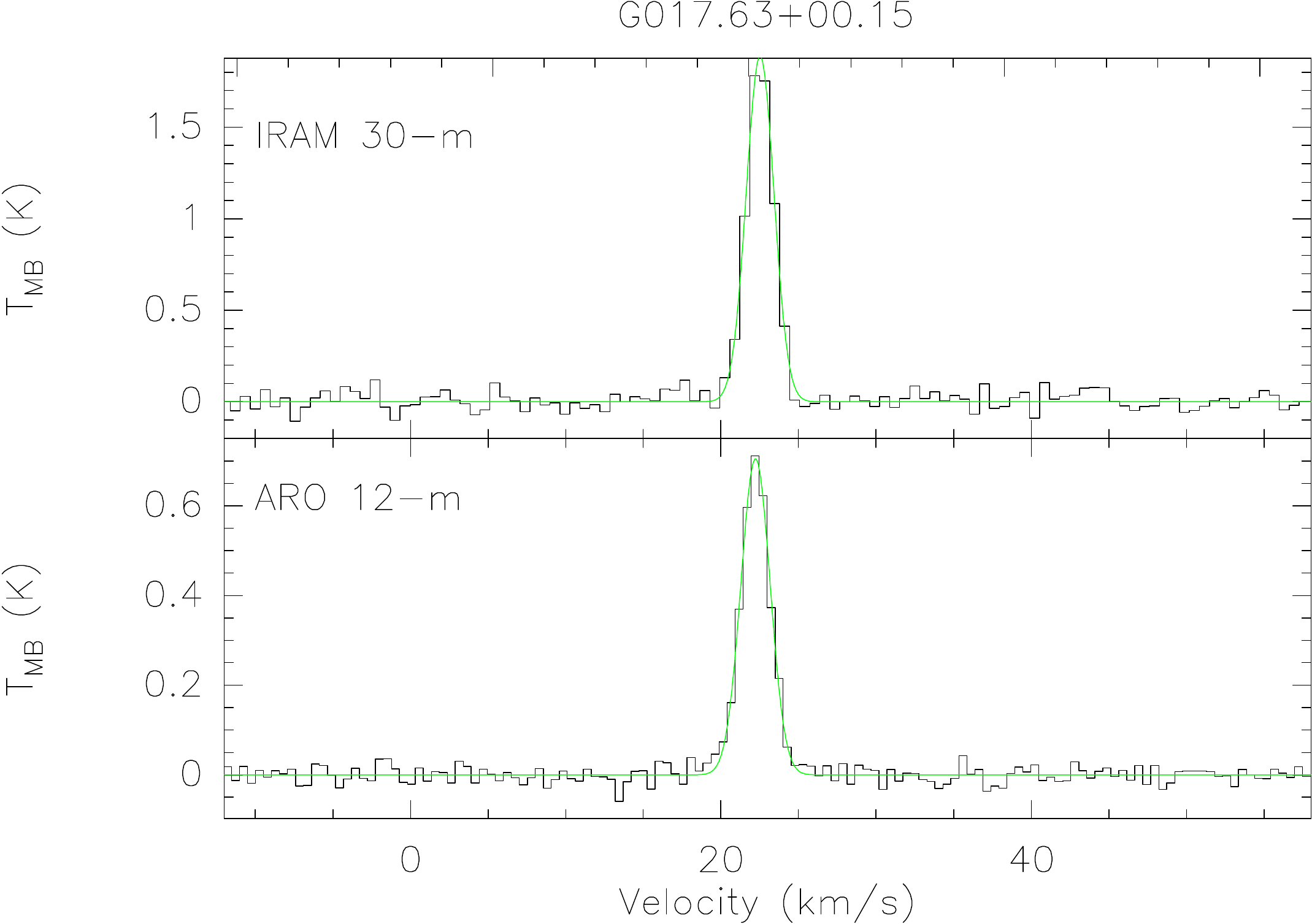} 
    \includegraphics[width=0.3\textwidth]{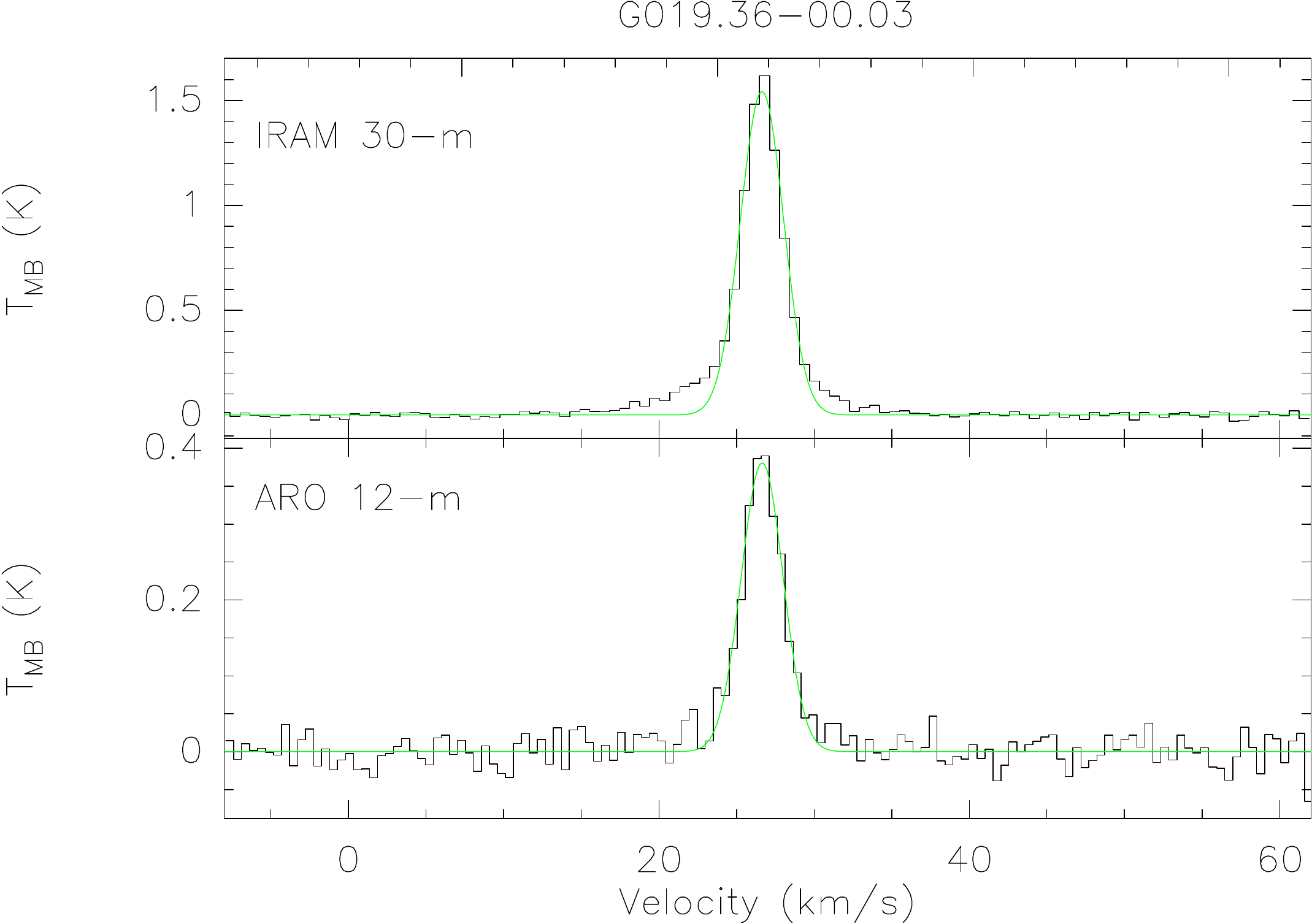}
    \includegraphics[width=0.3\textwidth]{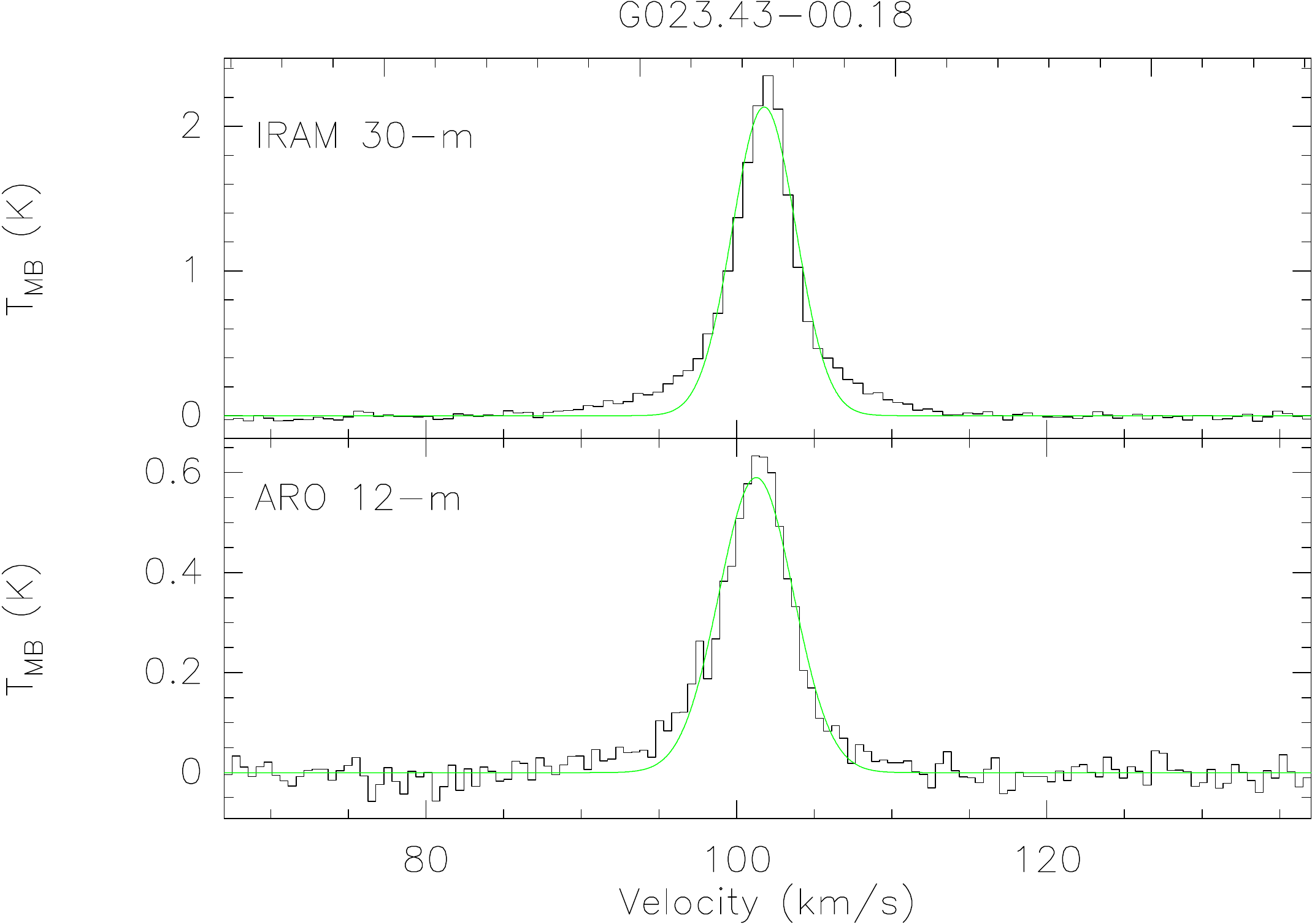}
    \includegraphics[width=0.3\textwidth]{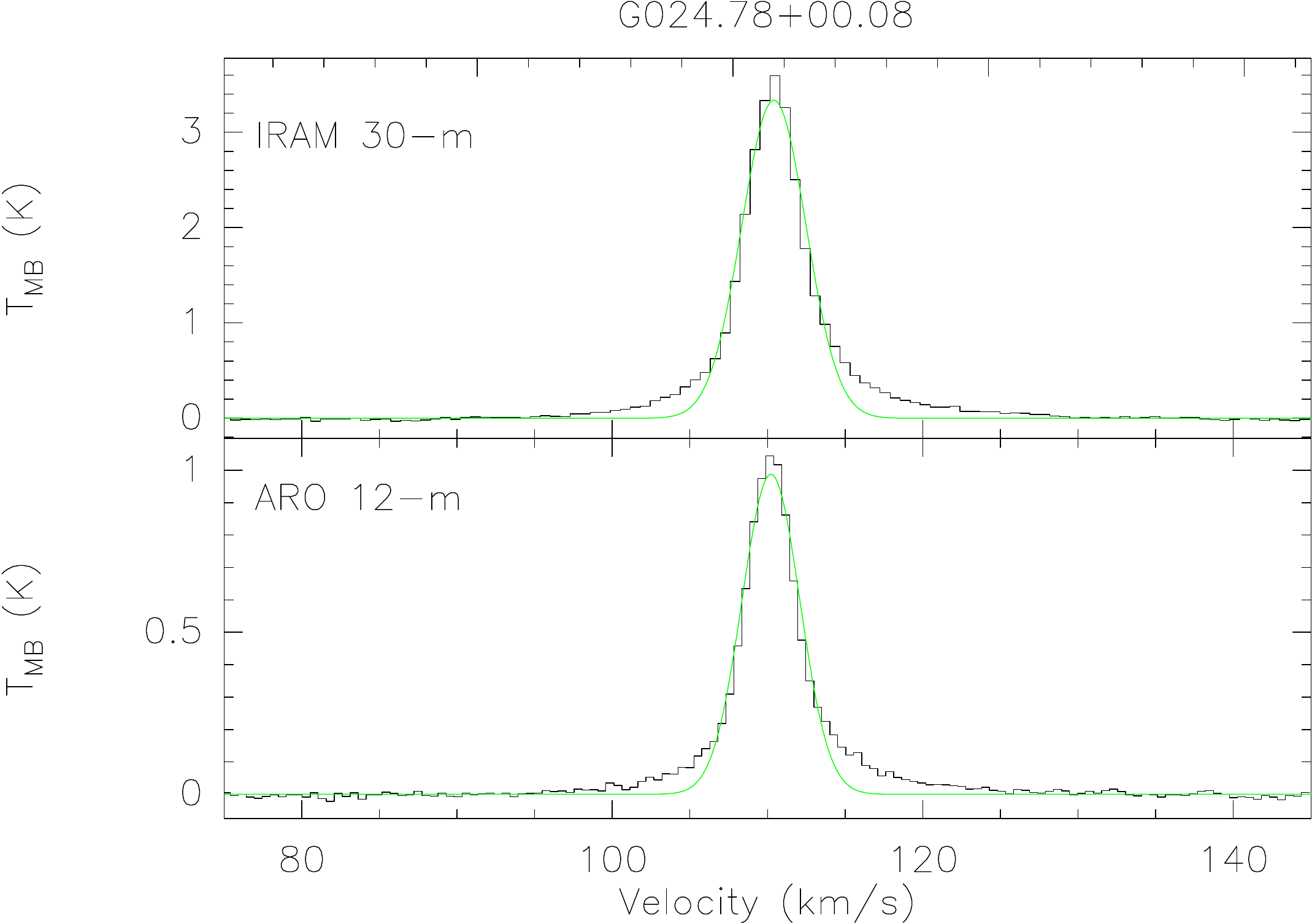} 
    \includegraphics[width=0.3\textwidth]{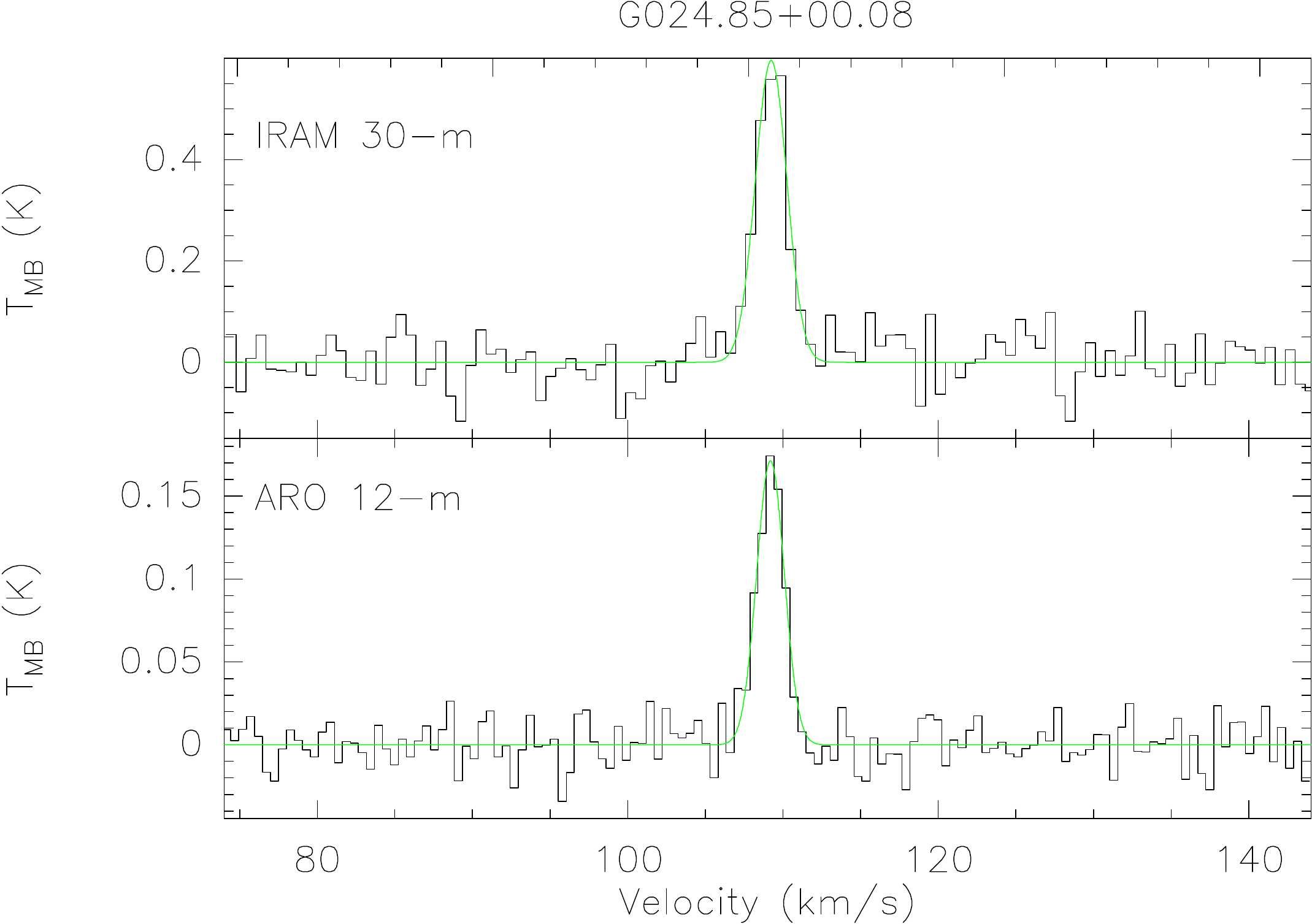}
    \includegraphics[width=0.3\textwidth]{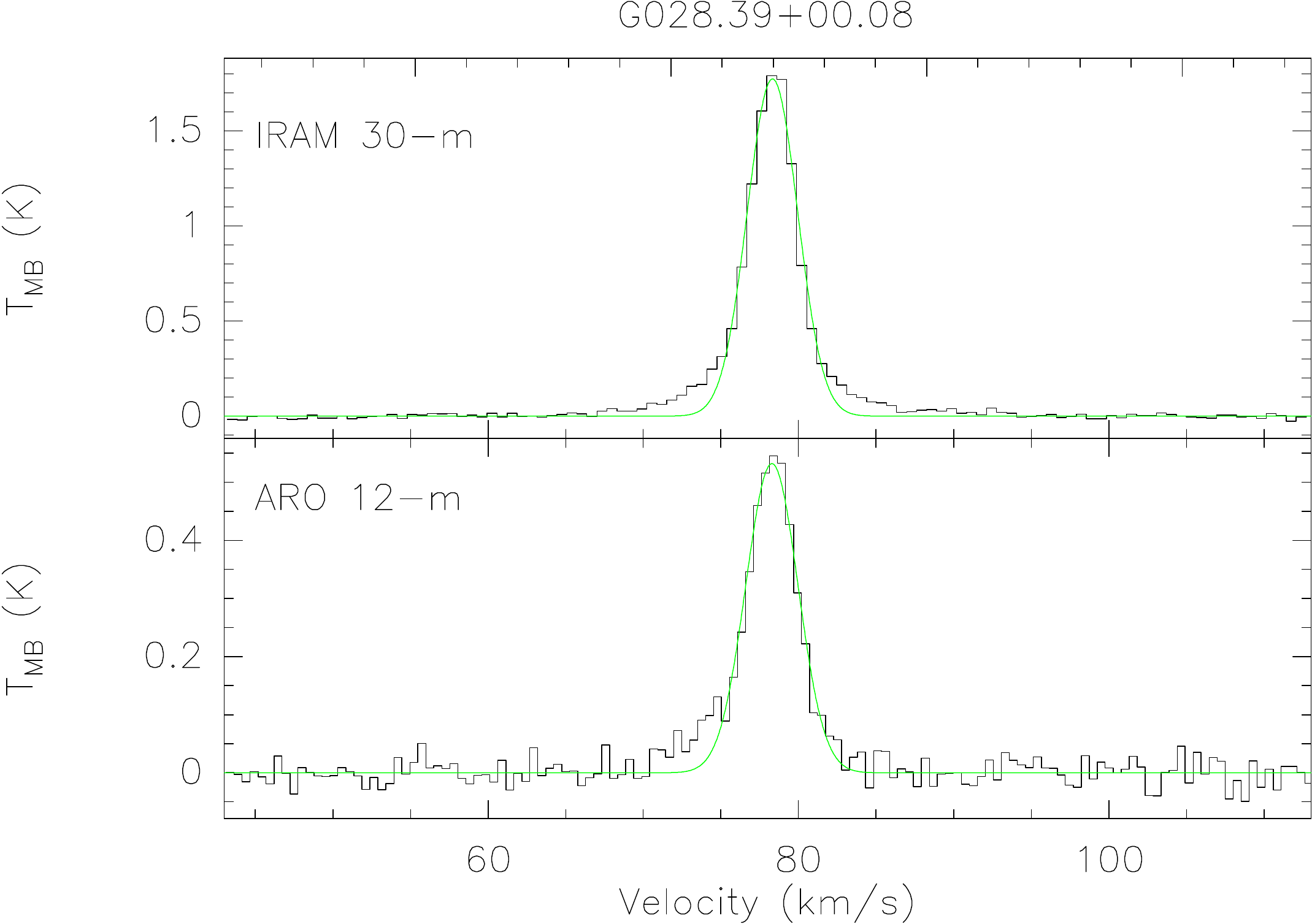}
    \includegraphics[width=0.3\textwidth]{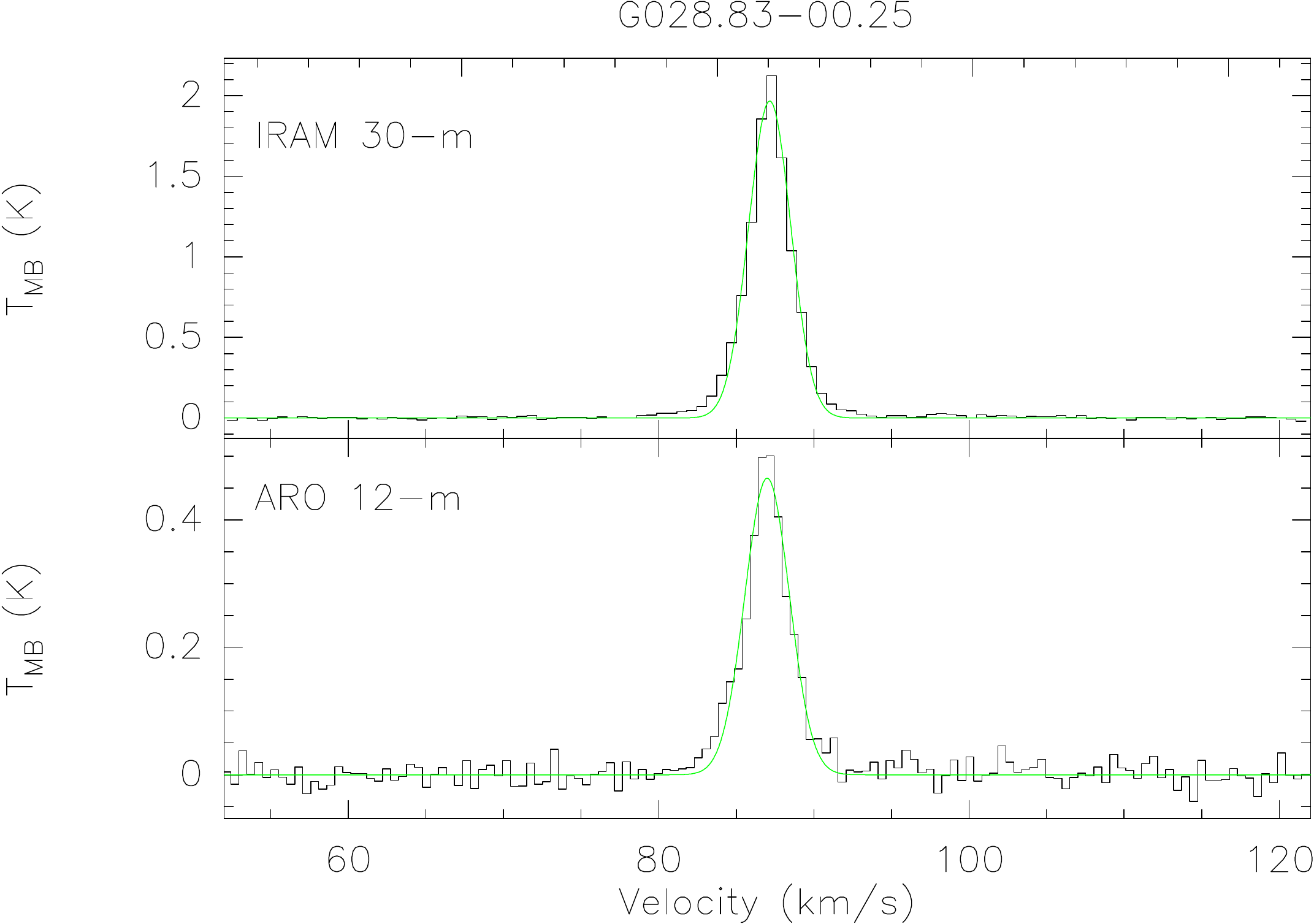} 
    \includegraphics[width=0.3\textwidth]{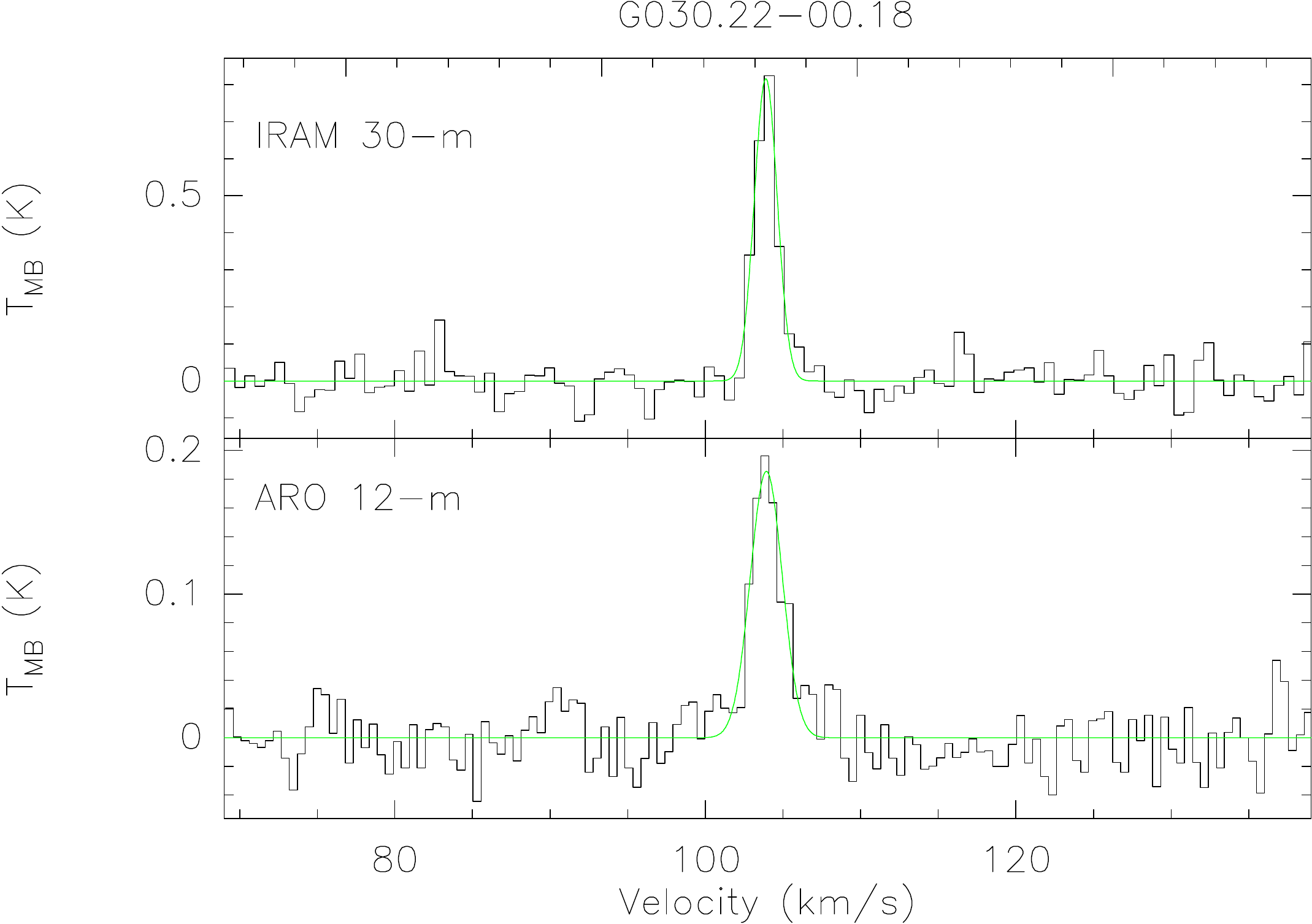}
    \includegraphics[width=0.3\textwidth]{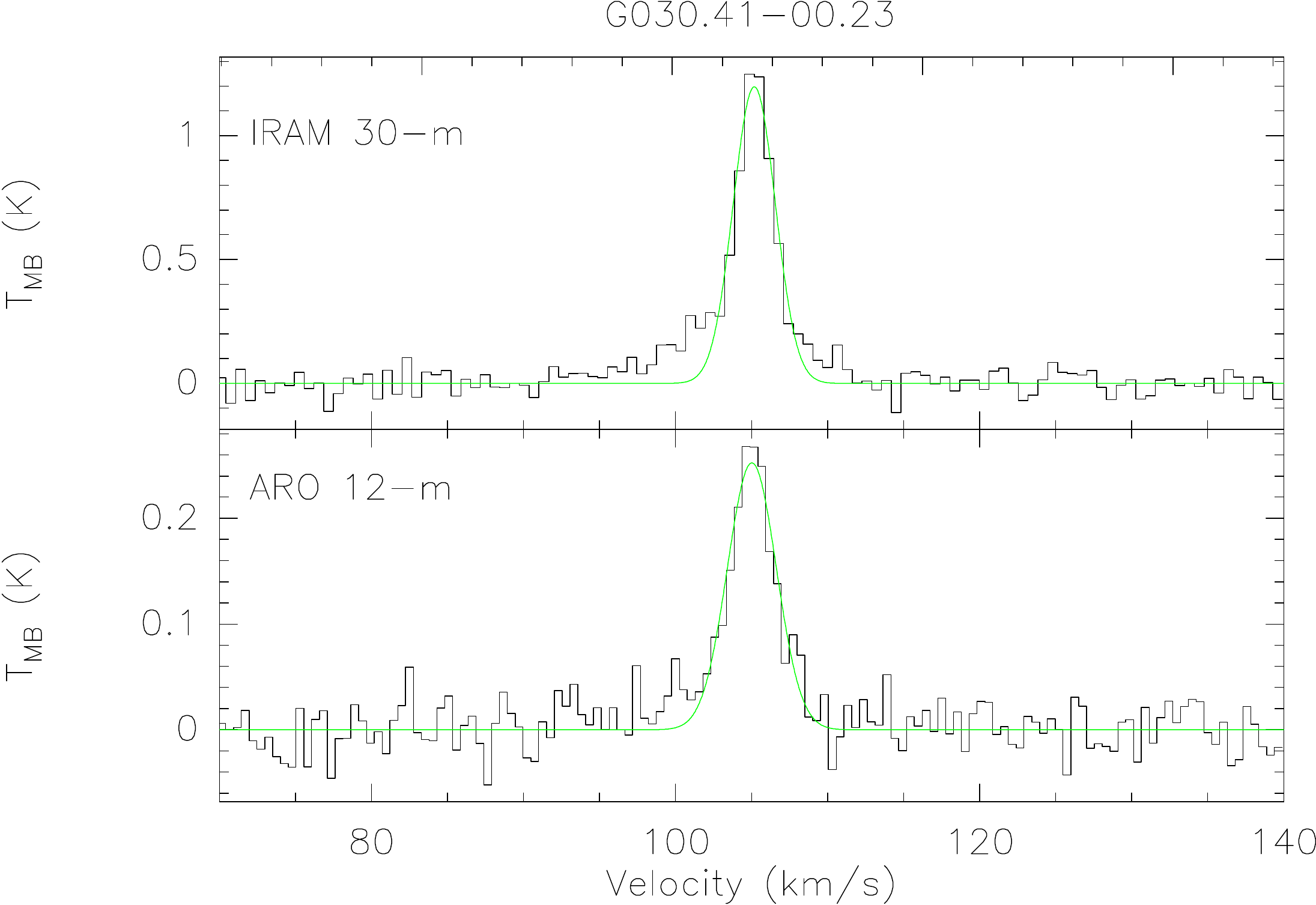}
    \includegraphics[width=0.3\textwidth]{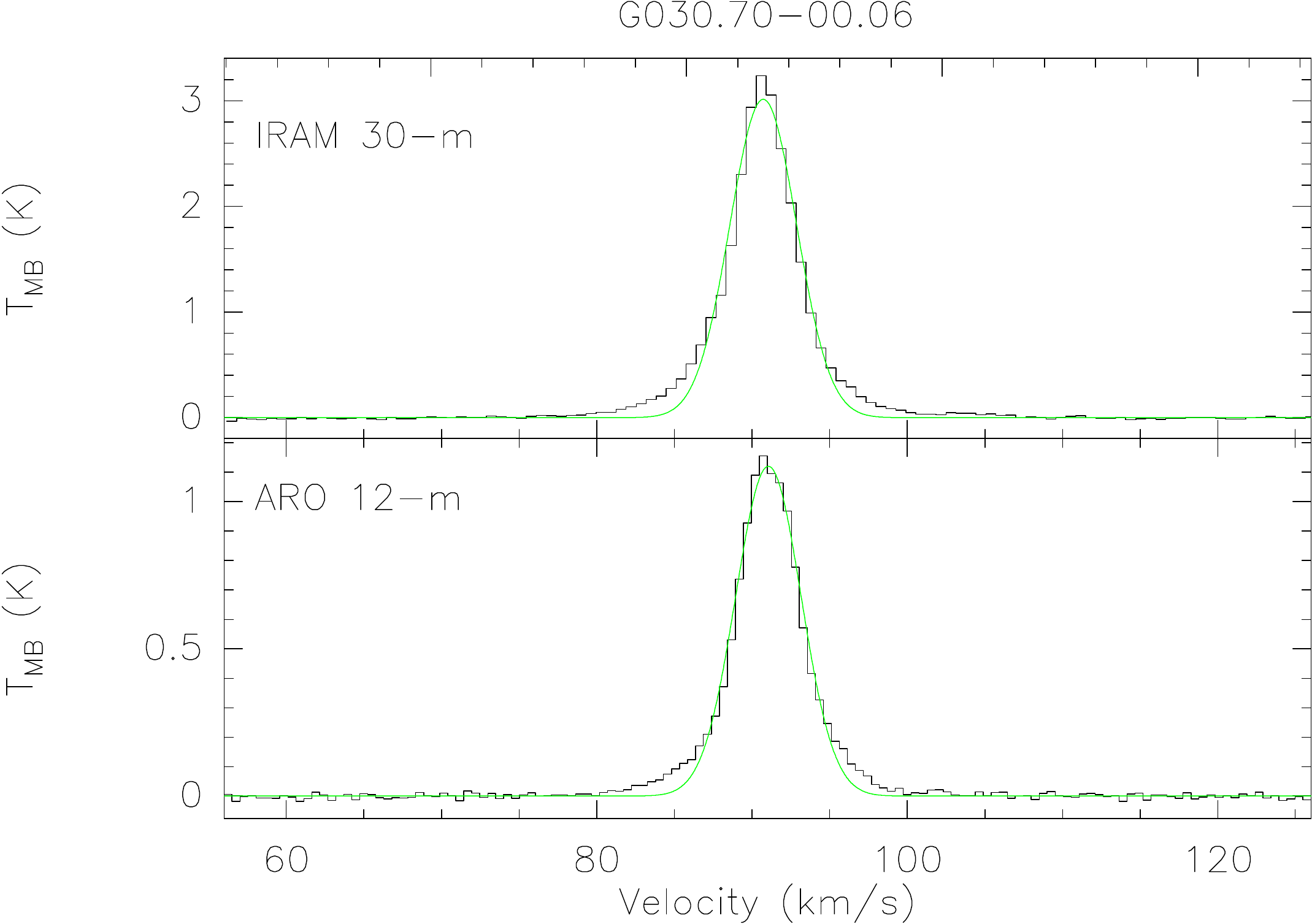} 
    \includegraphics[width=0.3\textwidth]{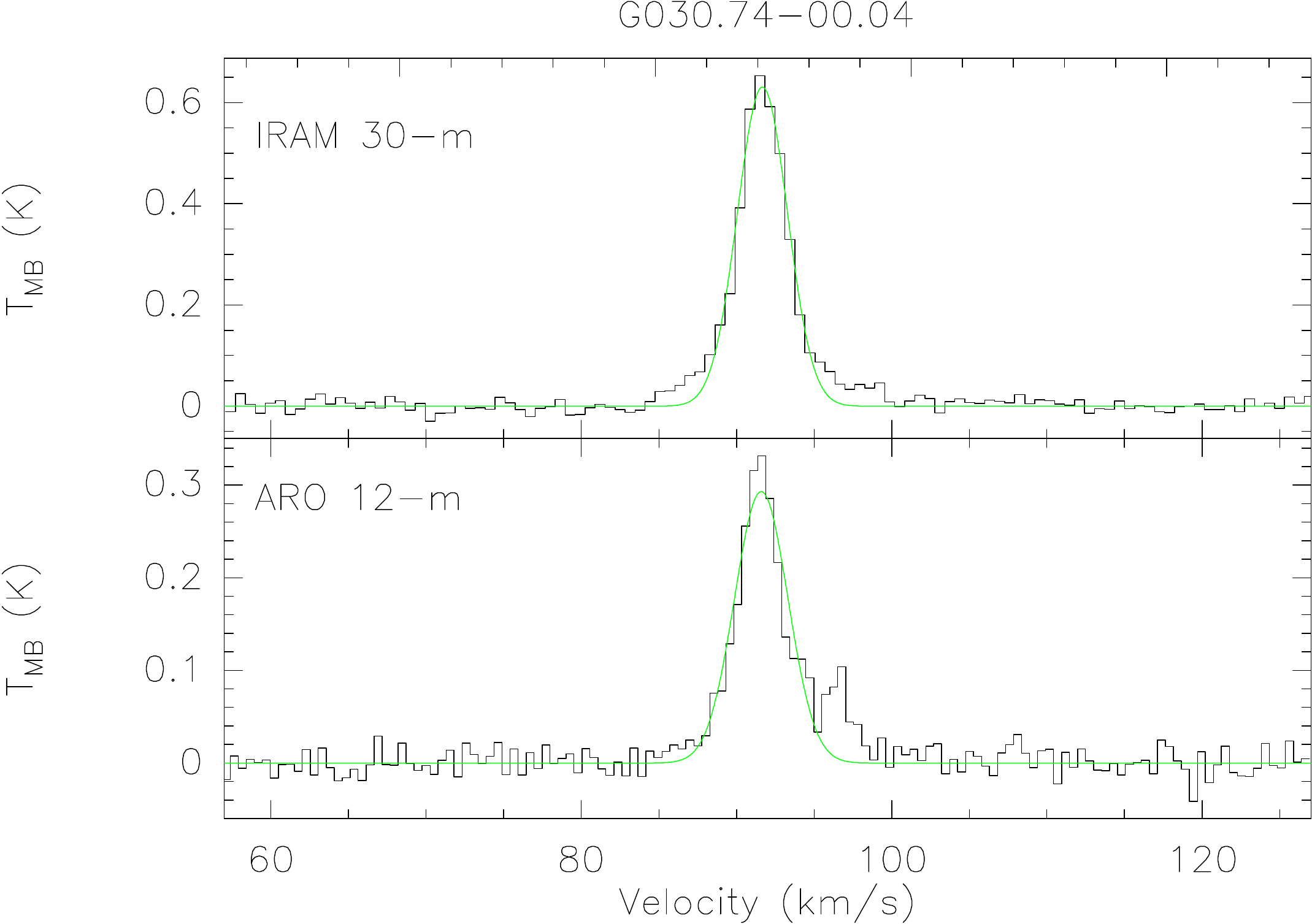}
    \includegraphics[width=0.3\textwidth]{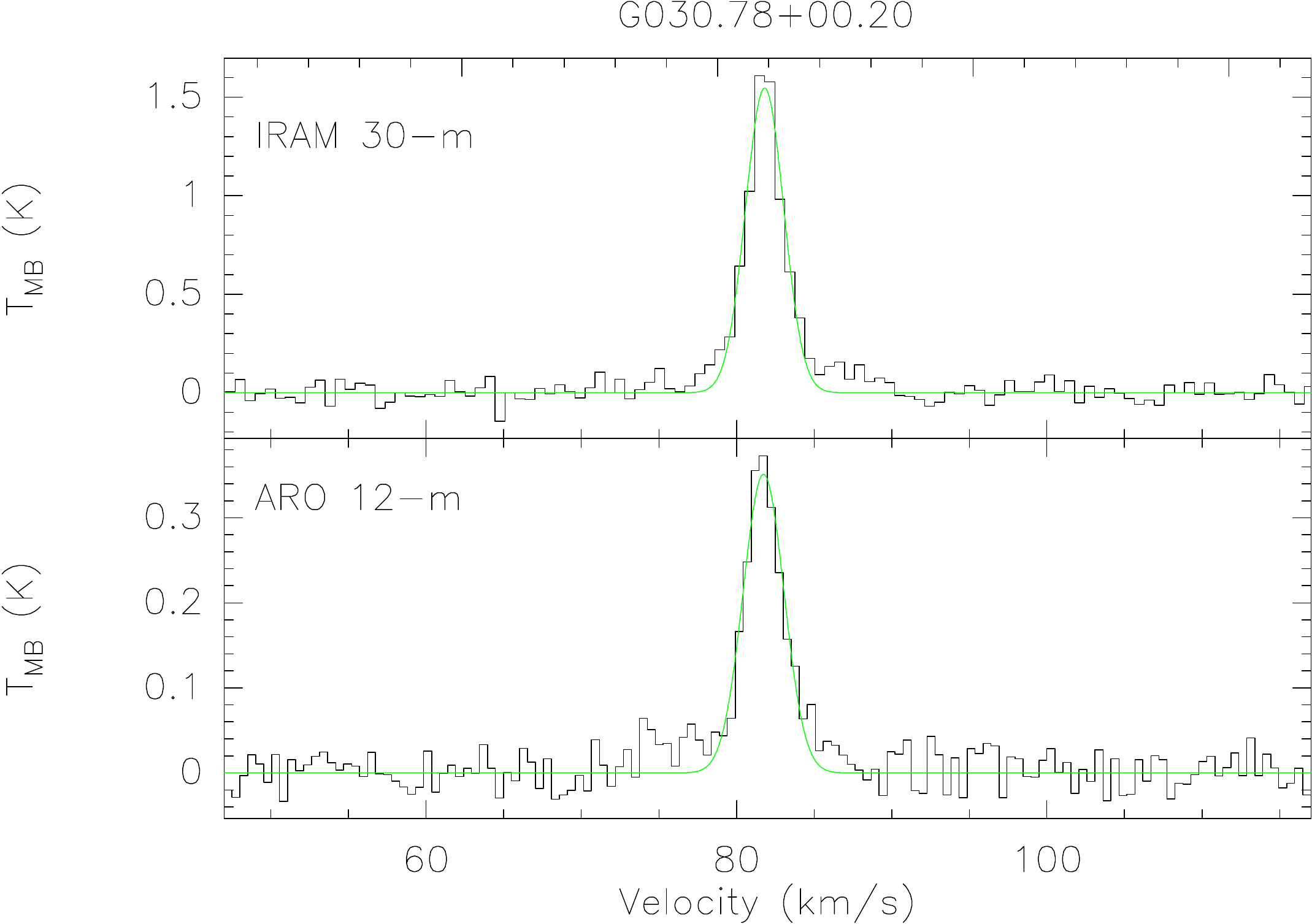}
    \includegraphics[width=0.3\textwidth]{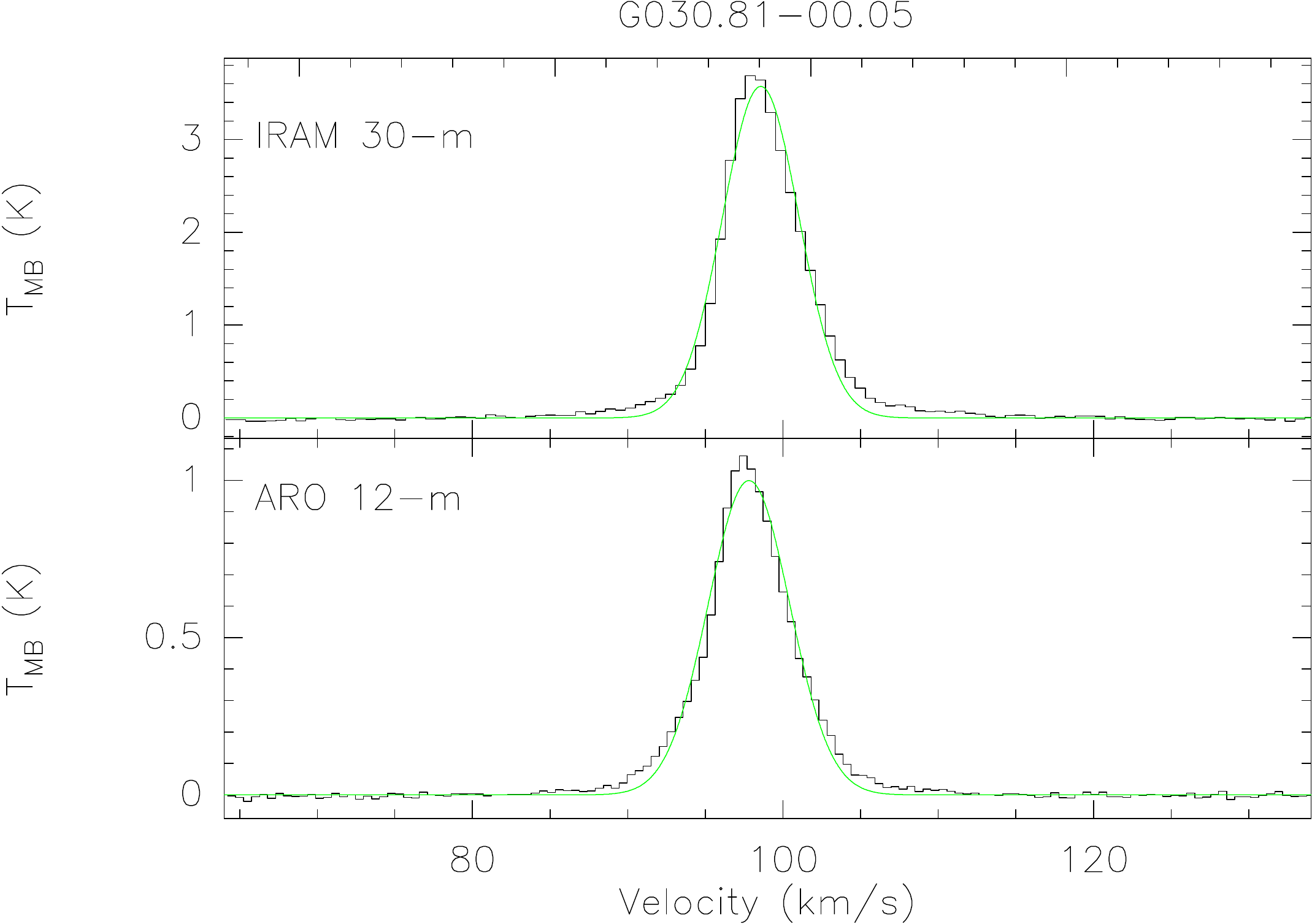} 
    \caption{HC$_{3}$N J = 10$-$9 spectra (upper panel: IRAM 30 m; lower panel: ARO 12 m) of 33 sources.}
    \label{fig2}
\end{figure*}
    
\addtocounter{figure}{-1}
\begin{figure*}    
    \centering
    \includegraphics[width=0.3\textwidth]{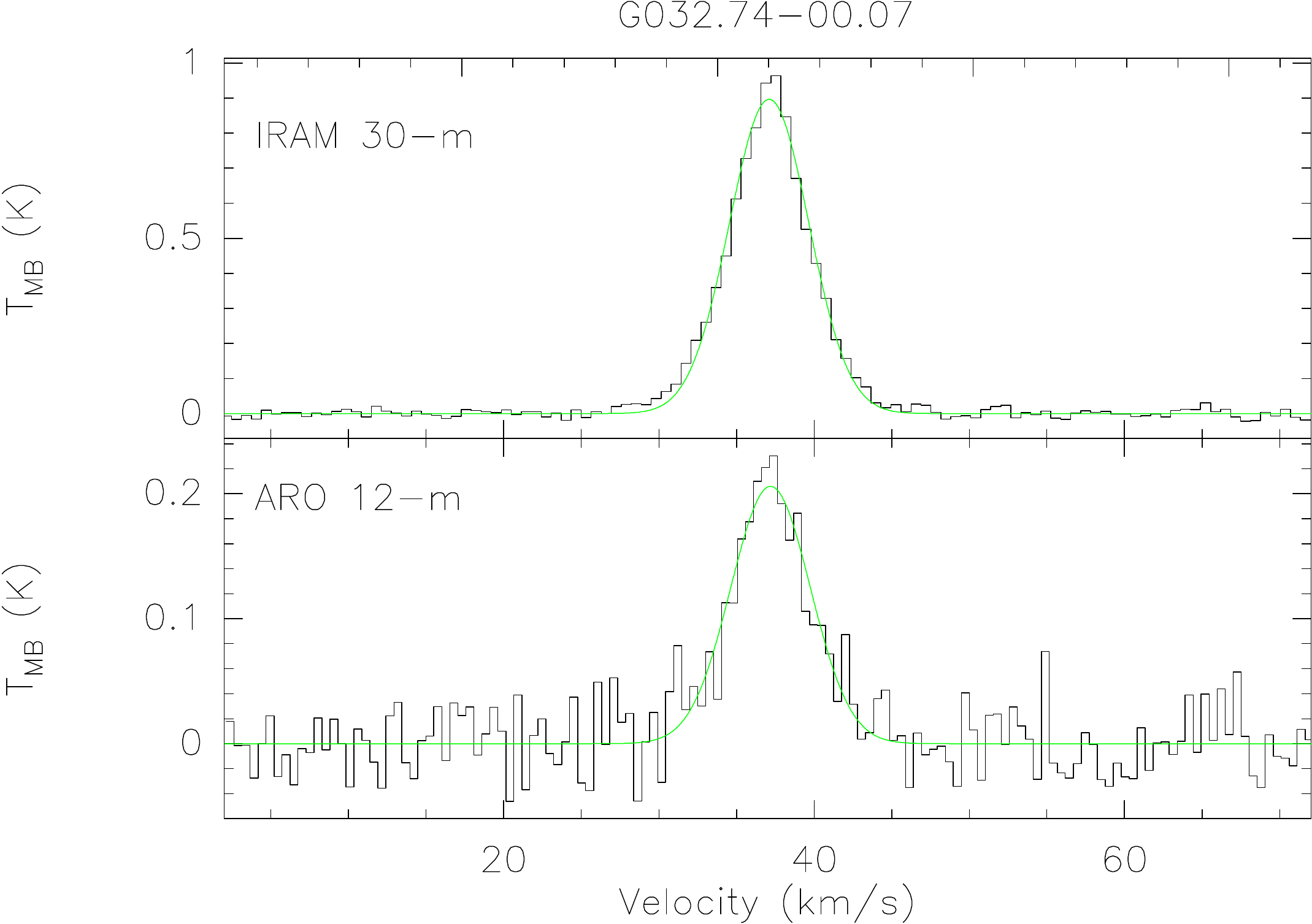}
    \includegraphics[width=0.3\textwidth]{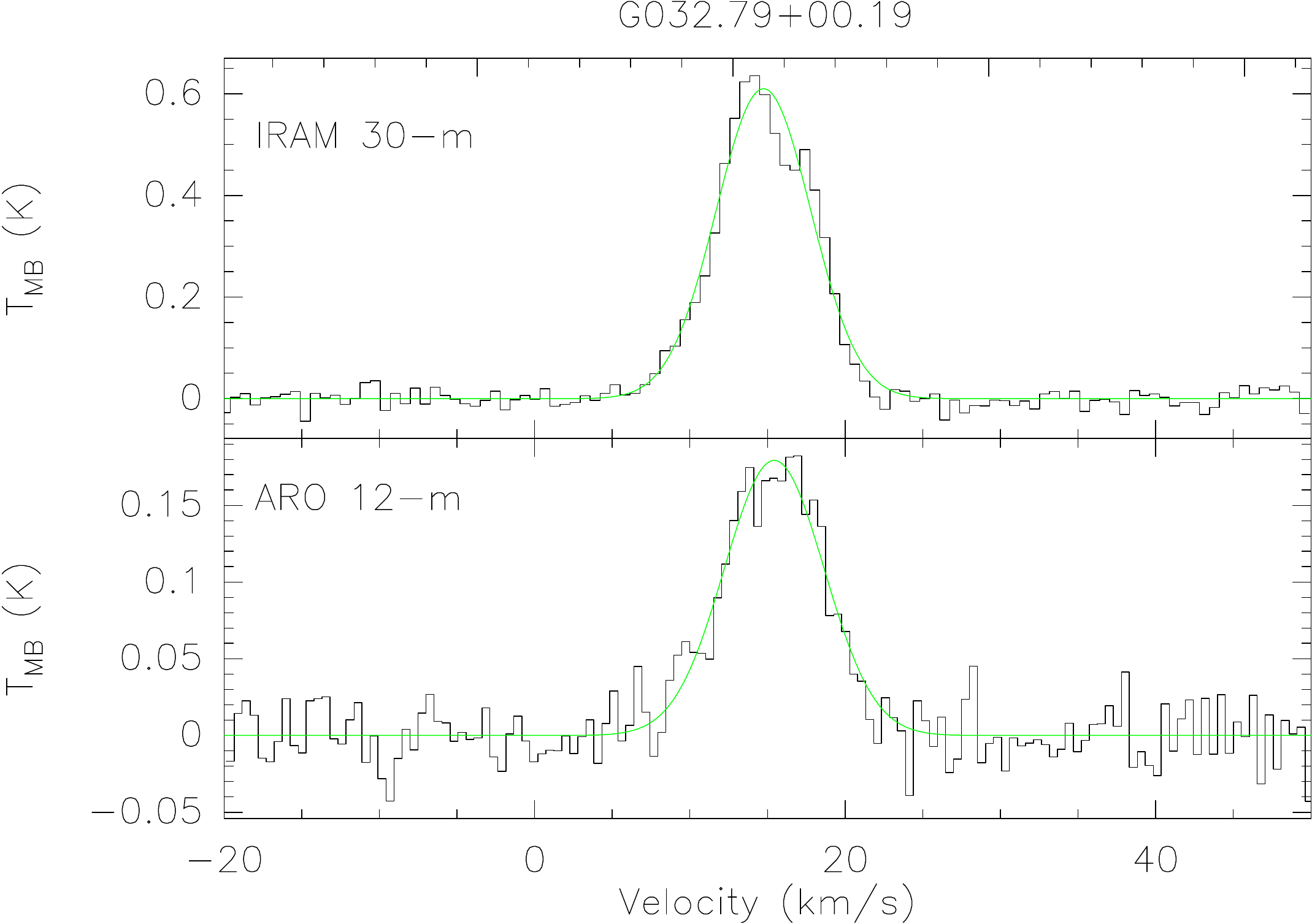}
    \includegraphics[width=0.3\textwidth]{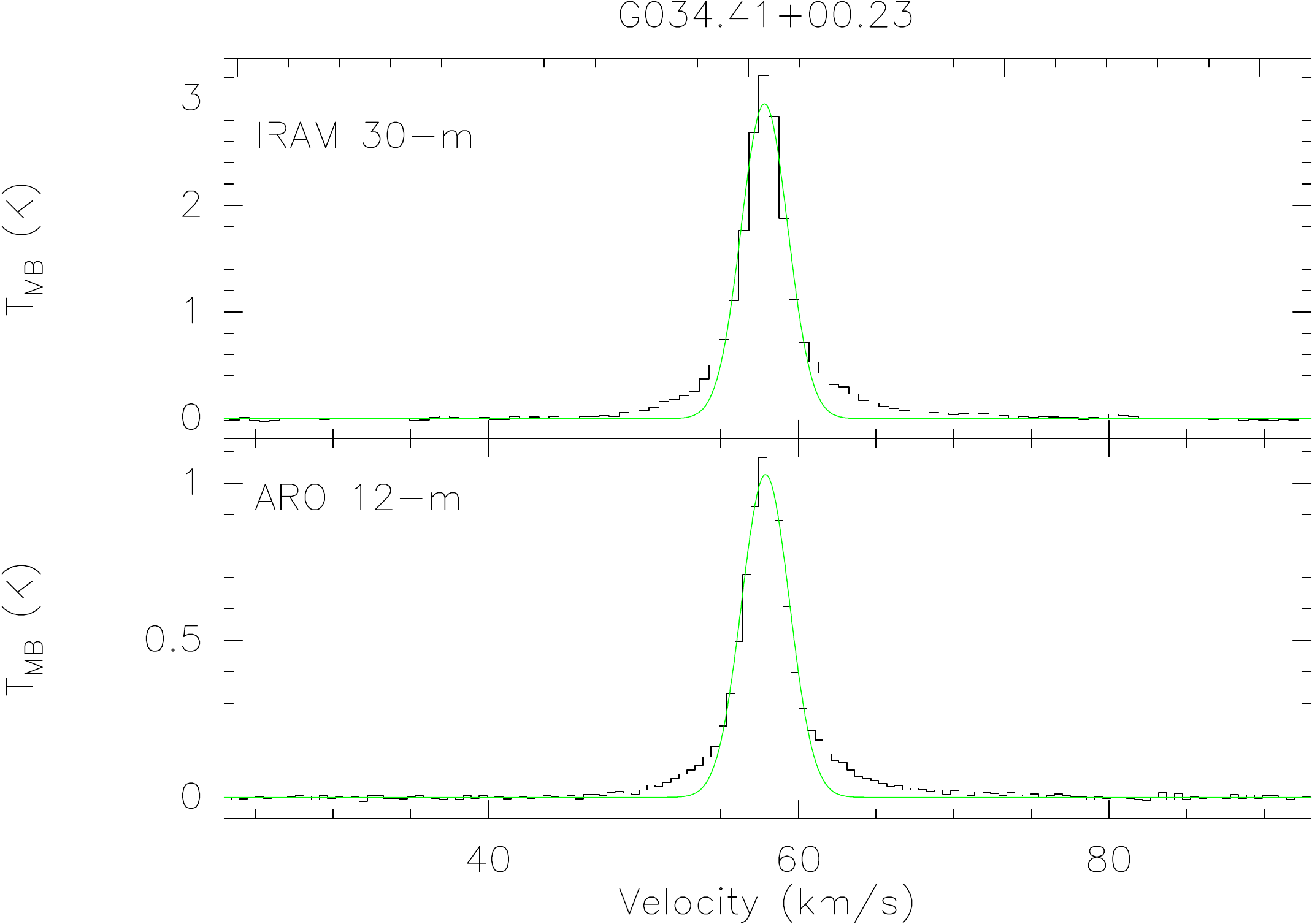} 
    \includegraphics[width=0.3\textwidth]{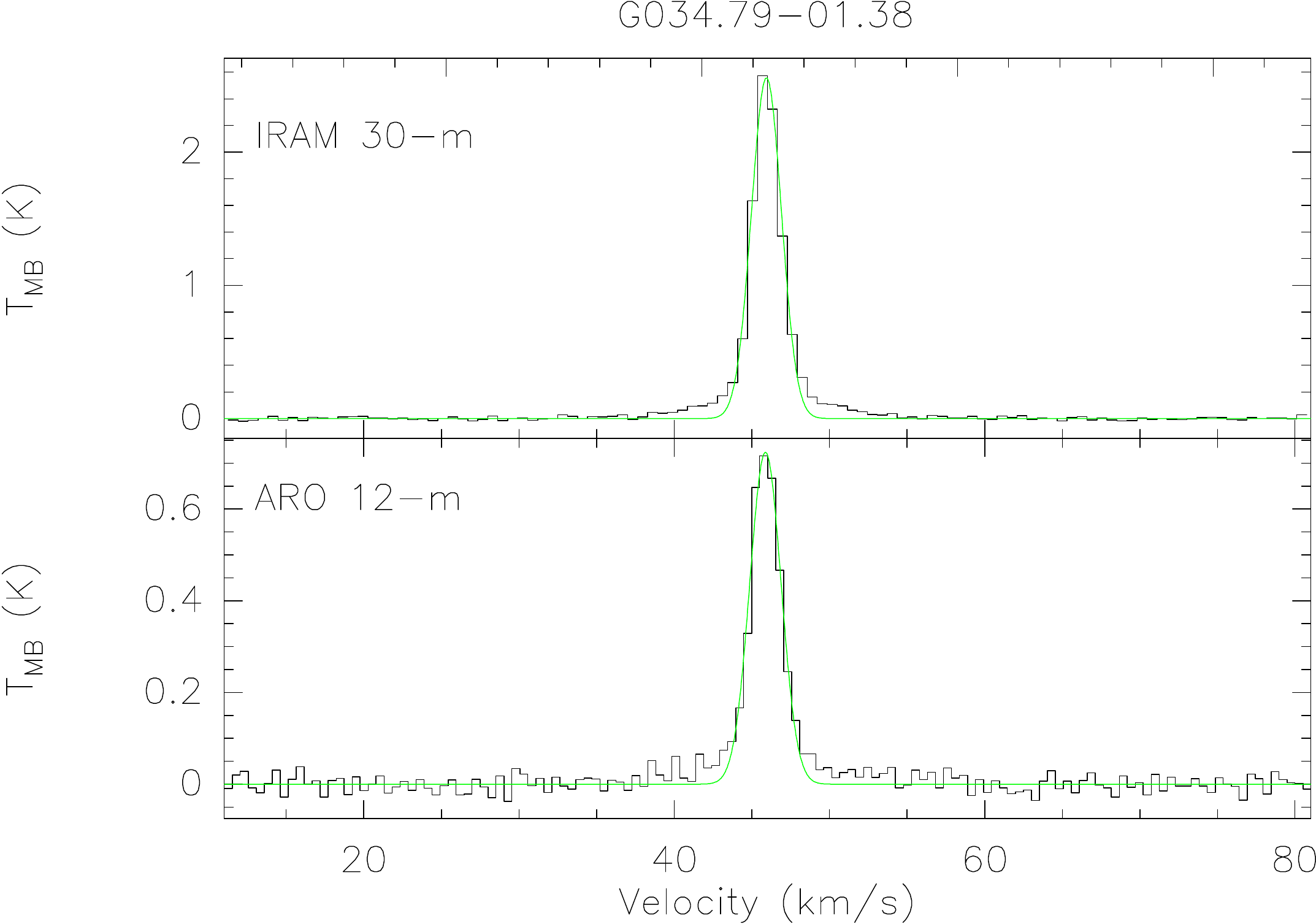}
    \includegraphics[width=0.3\textwidth]{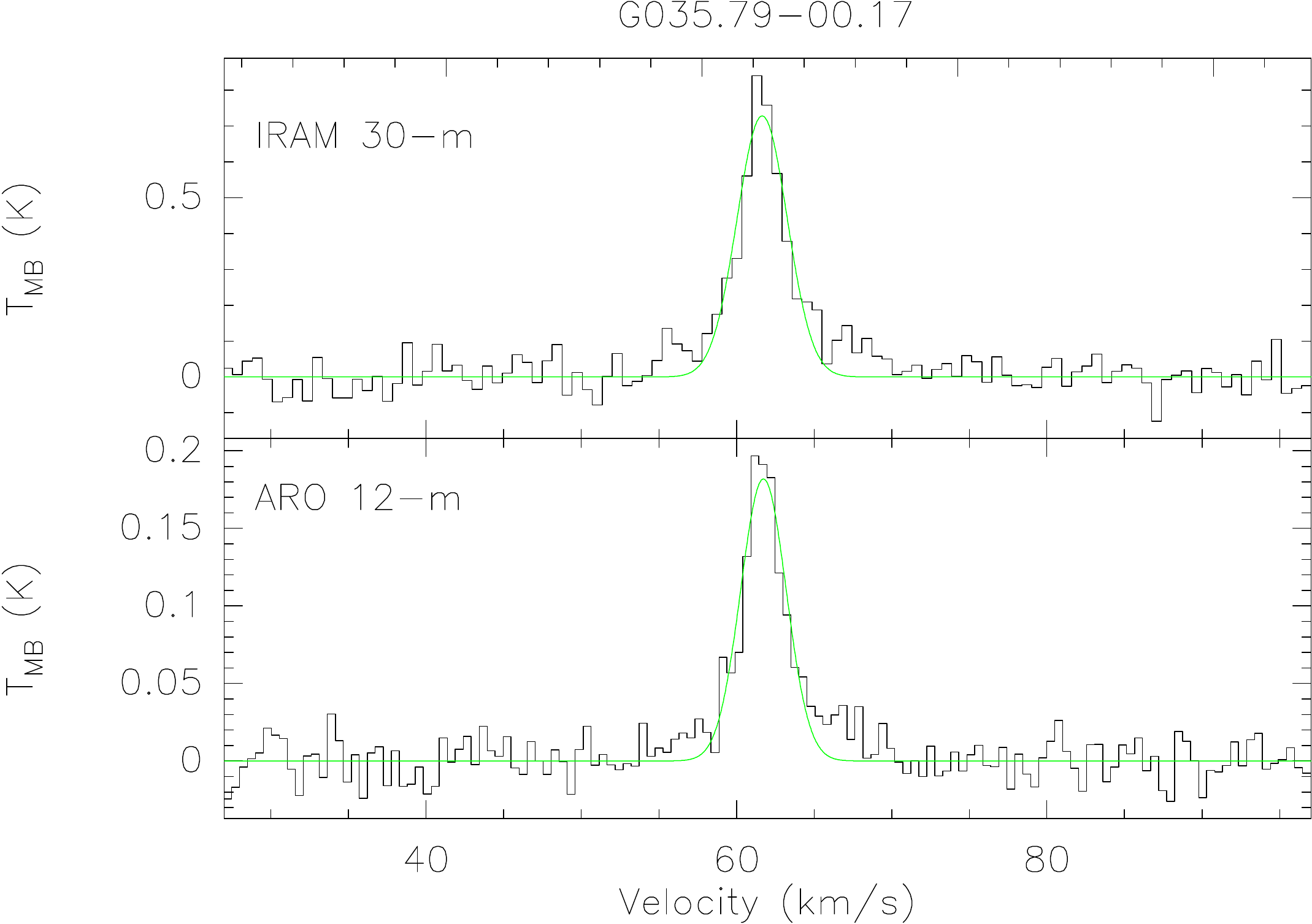}
    \includegraphics[width=0.3\textwidth]{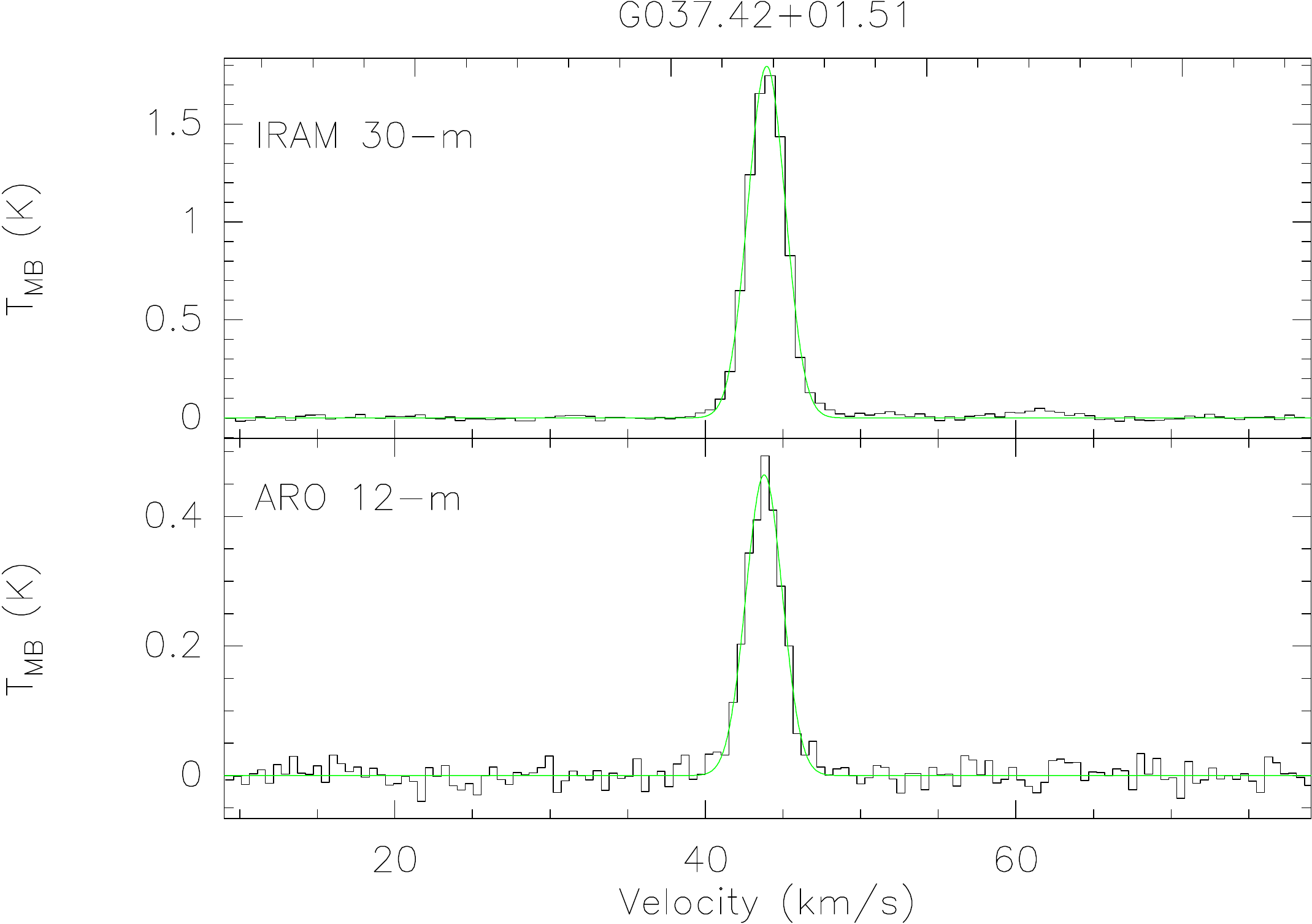}
    \includegraphics[width=0.3\textwidth]{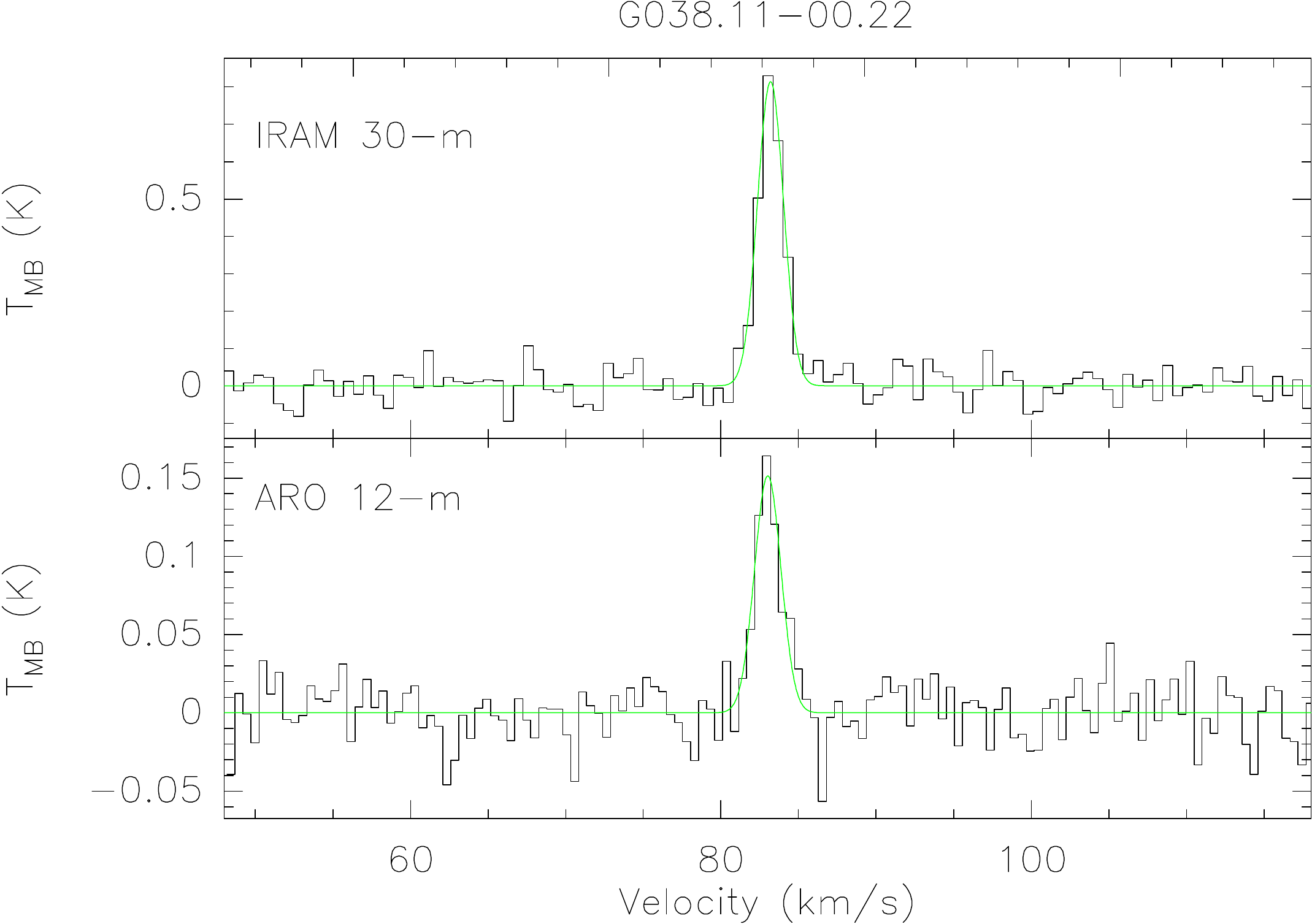}
    \includegraphics[width=0.3\textwidth]{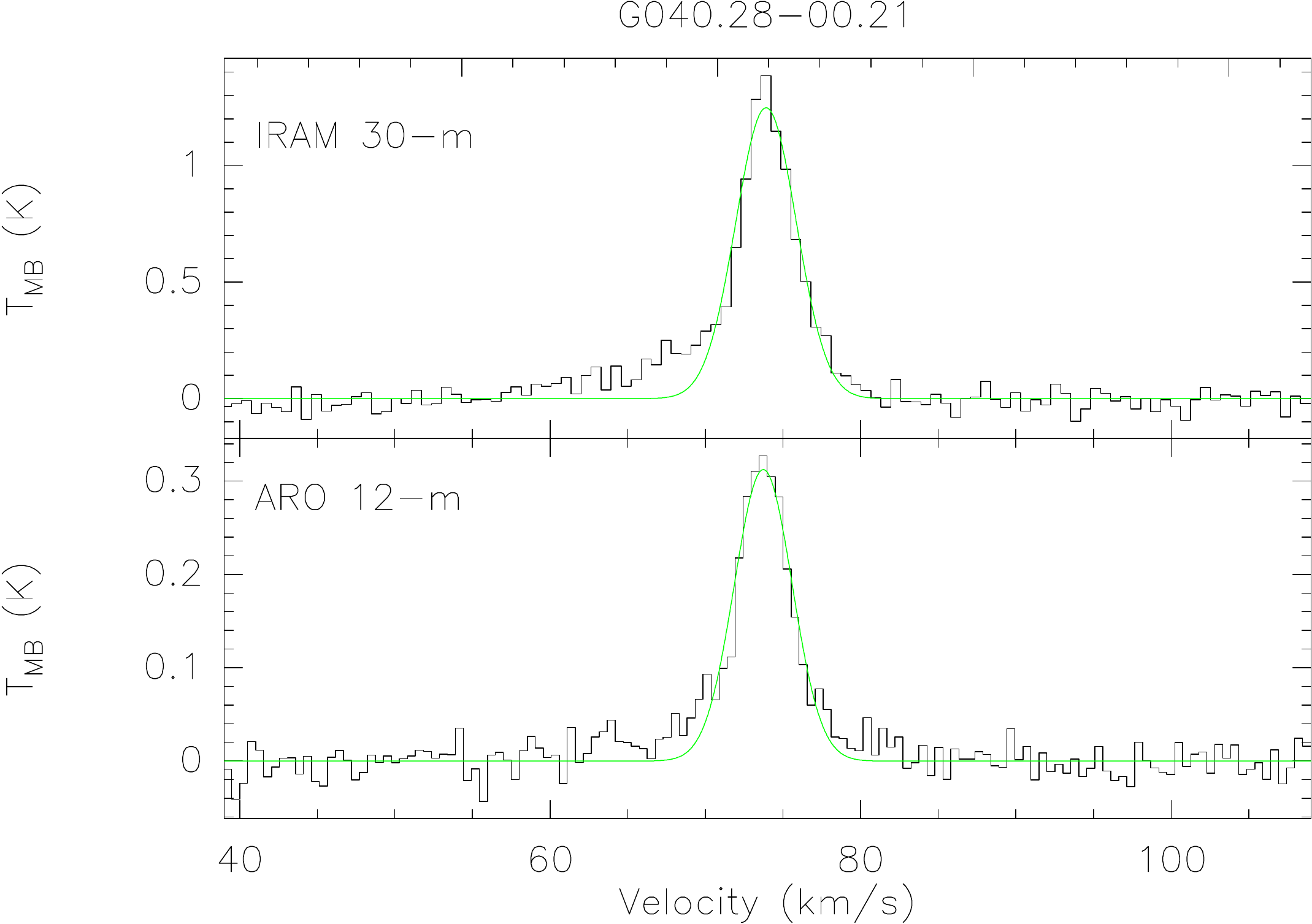}
    \includegraphics[width=0.3\textwidth]{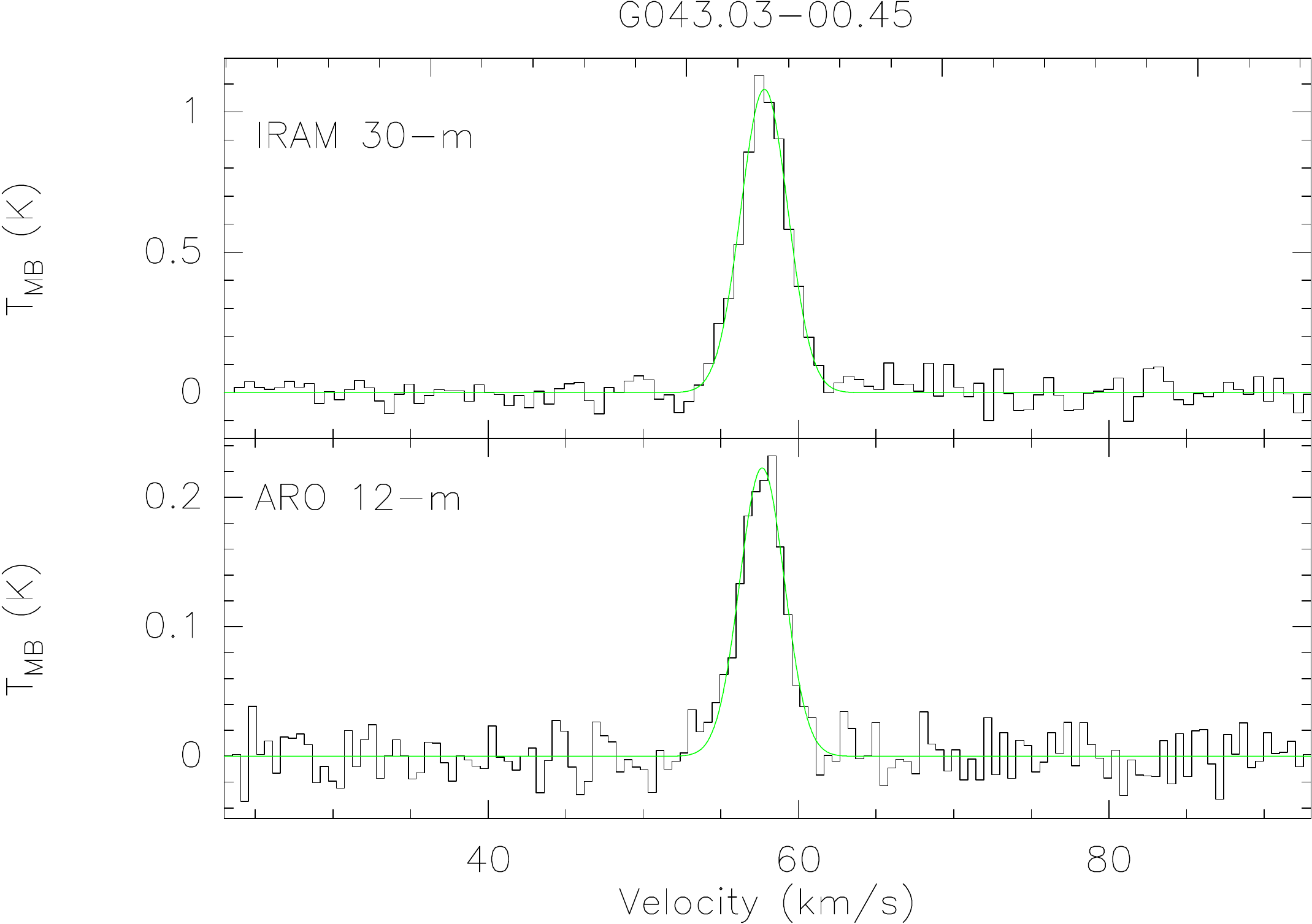}
    \includegraphics[width=0.3\textwidth]{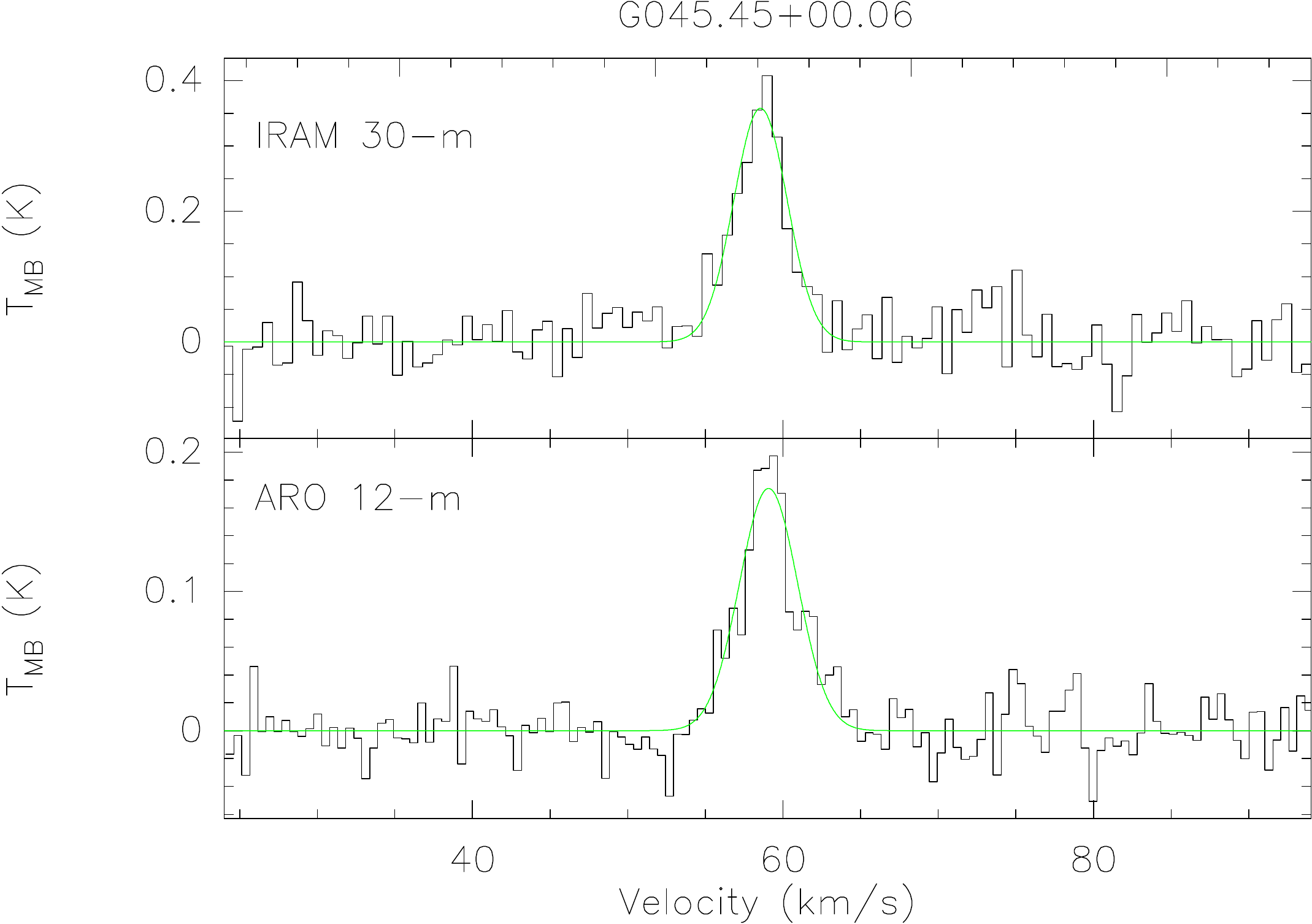}
    \includegraphics[width=0.3\textwidth]{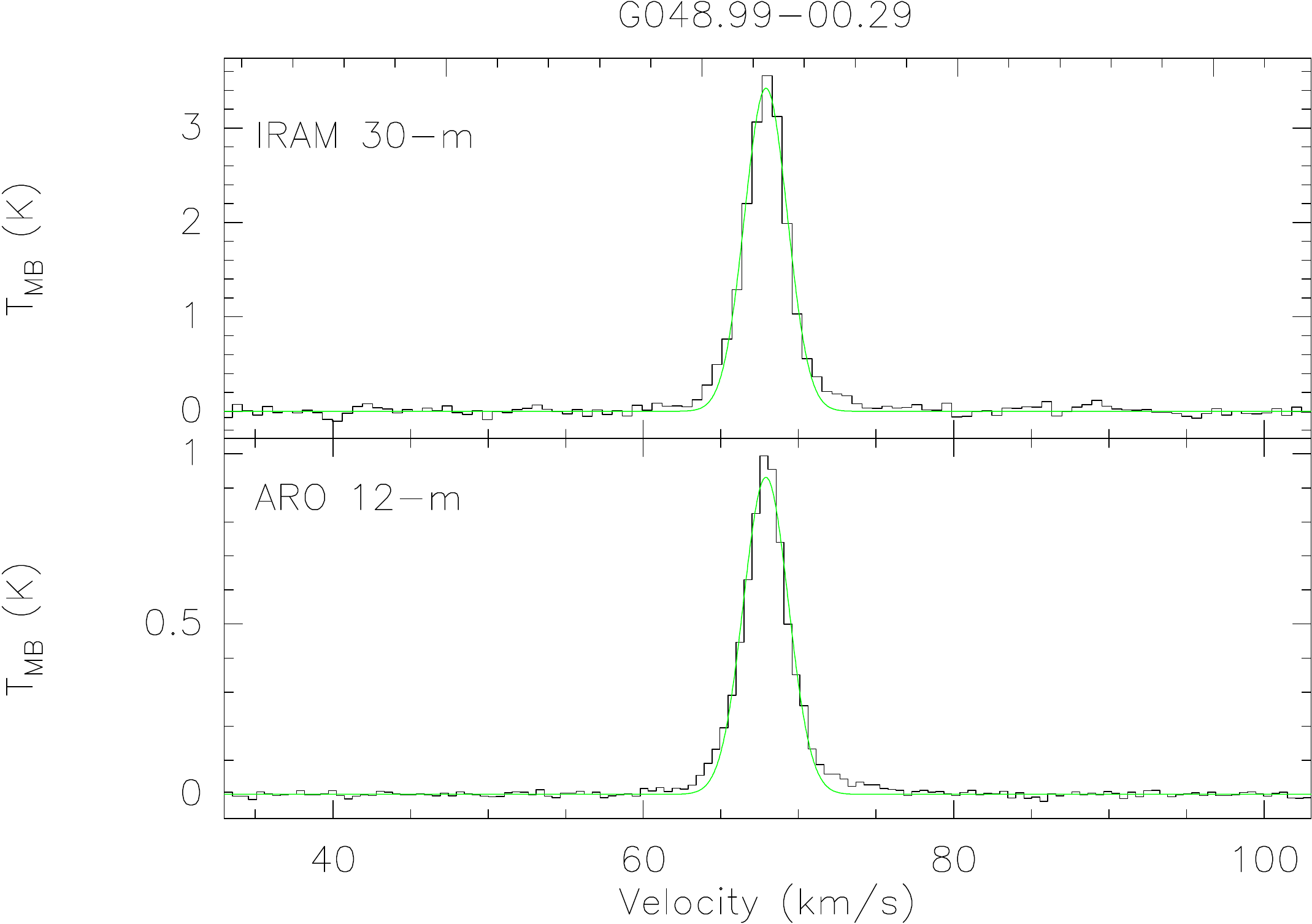}
    \includegraphics[width=0.3\textwidth]{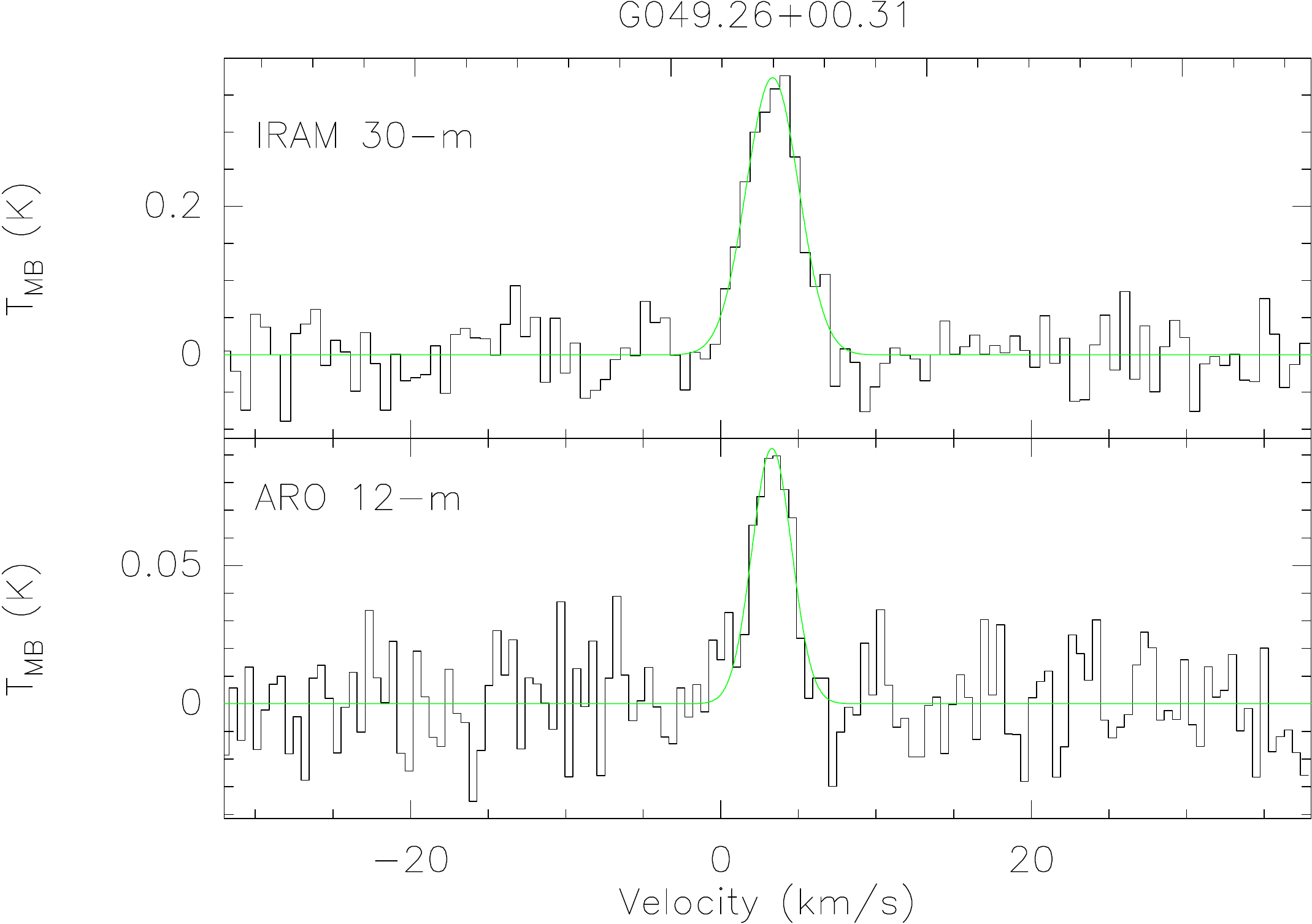}
    \includegraphics[width=0.3\textwidth]{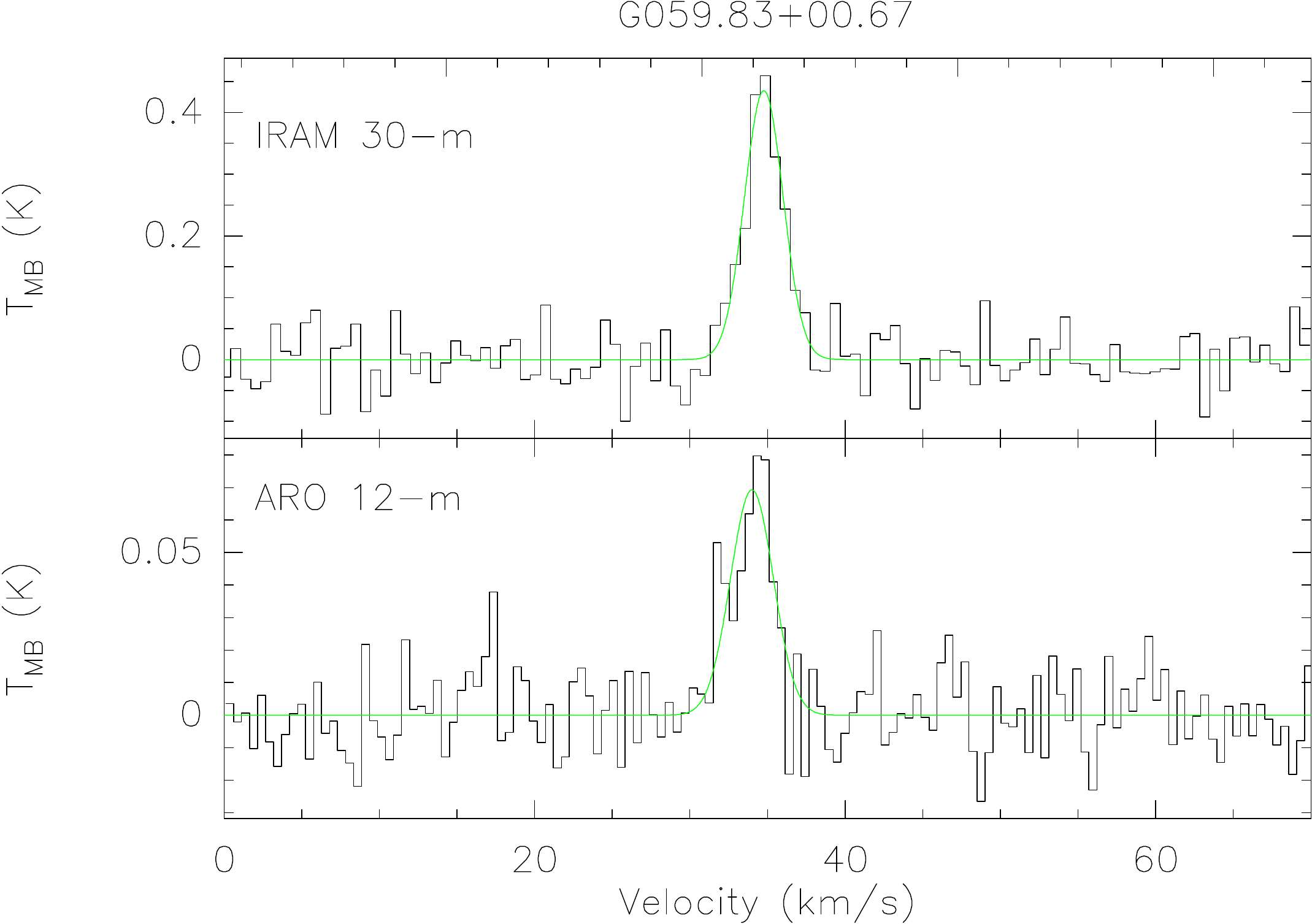}
    \includegraphics[width=0.3\textwidth]{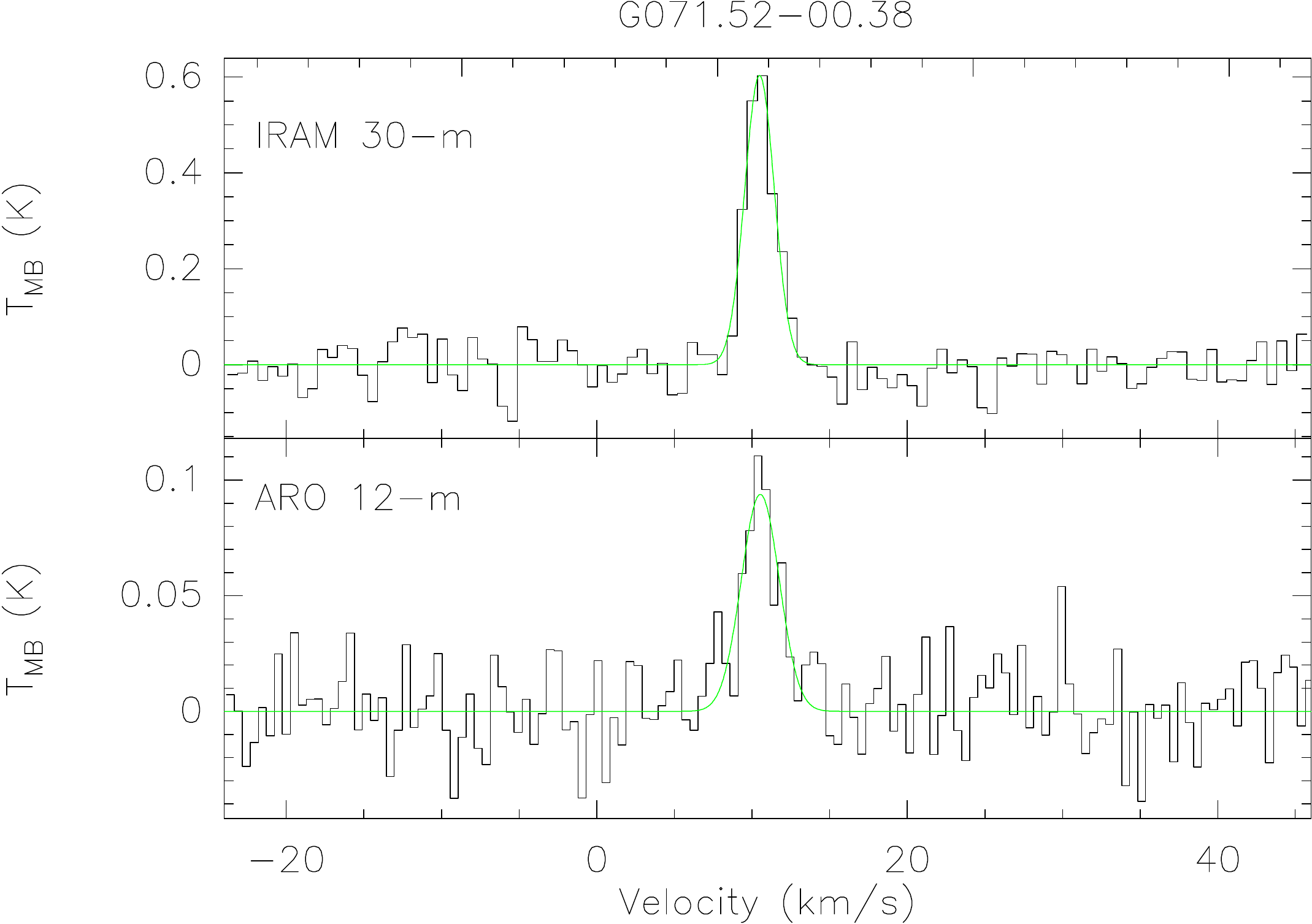}
    \includegraphics[width=0.3\textwidth]{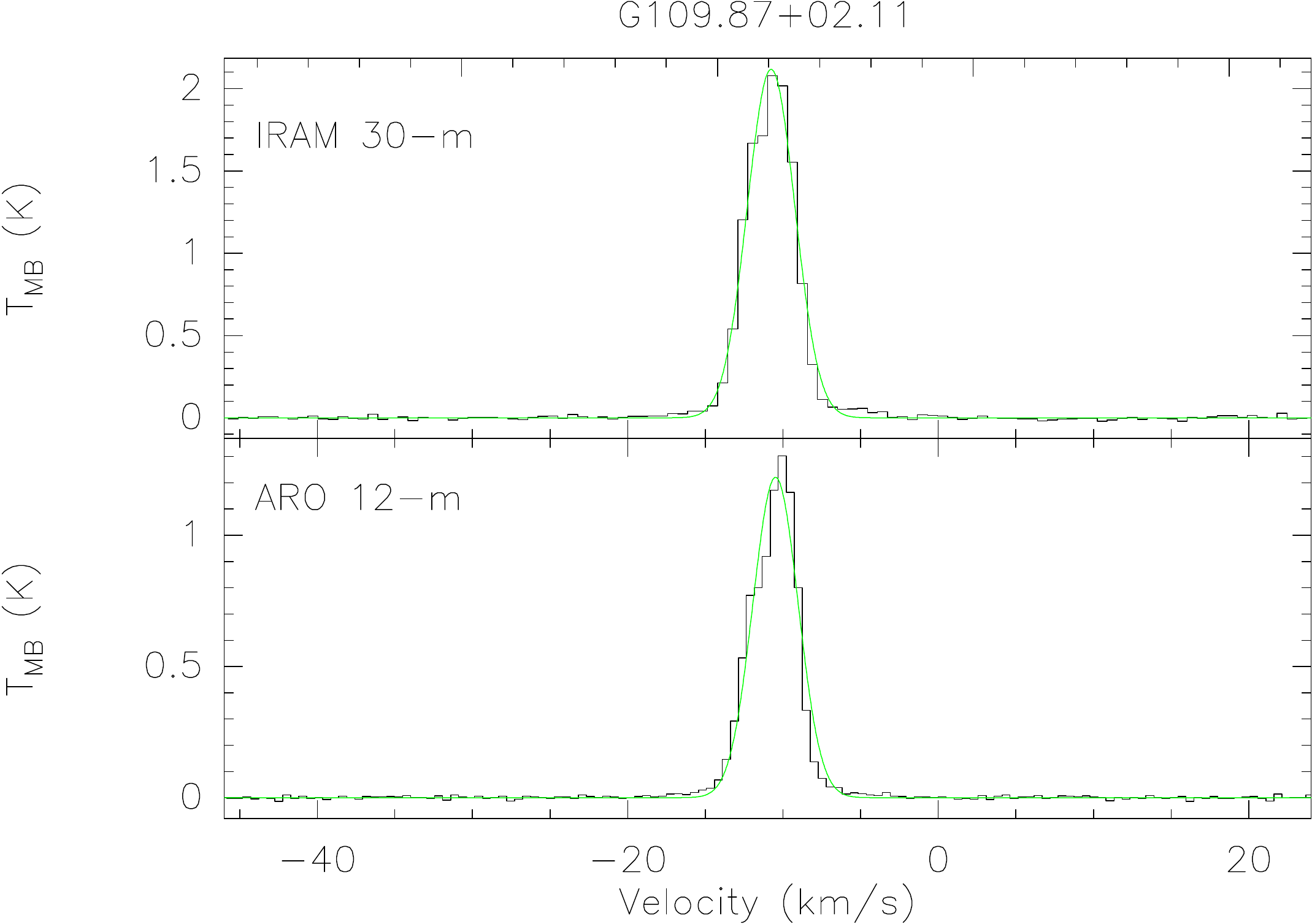}
    \caption{Continued.}
\end{figure*} 

\begin{figure*}
    \centering
    \includegraphics[width=0.3\textwidth]{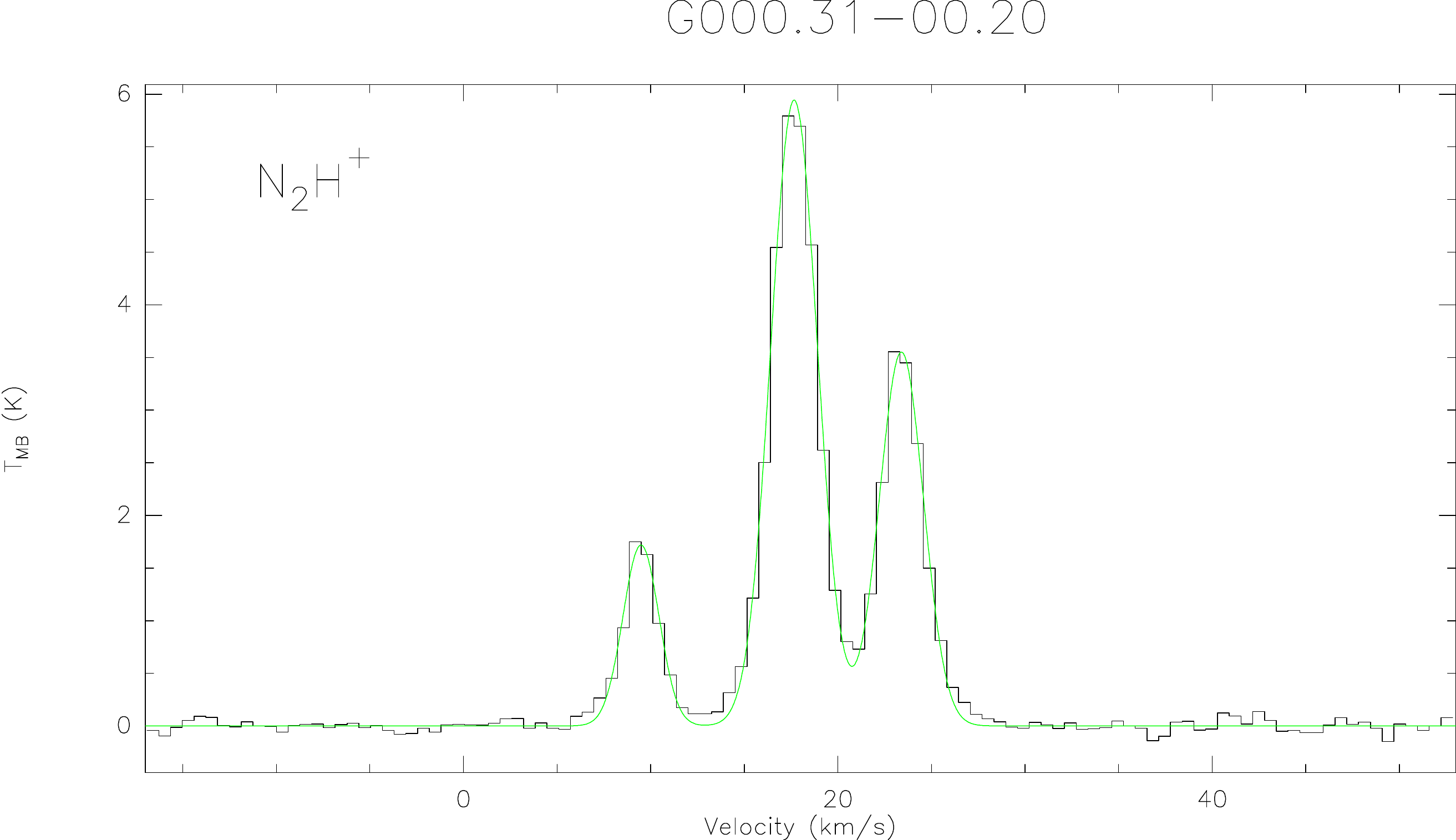}
    \includegraphics[width=0.3\textwidth]{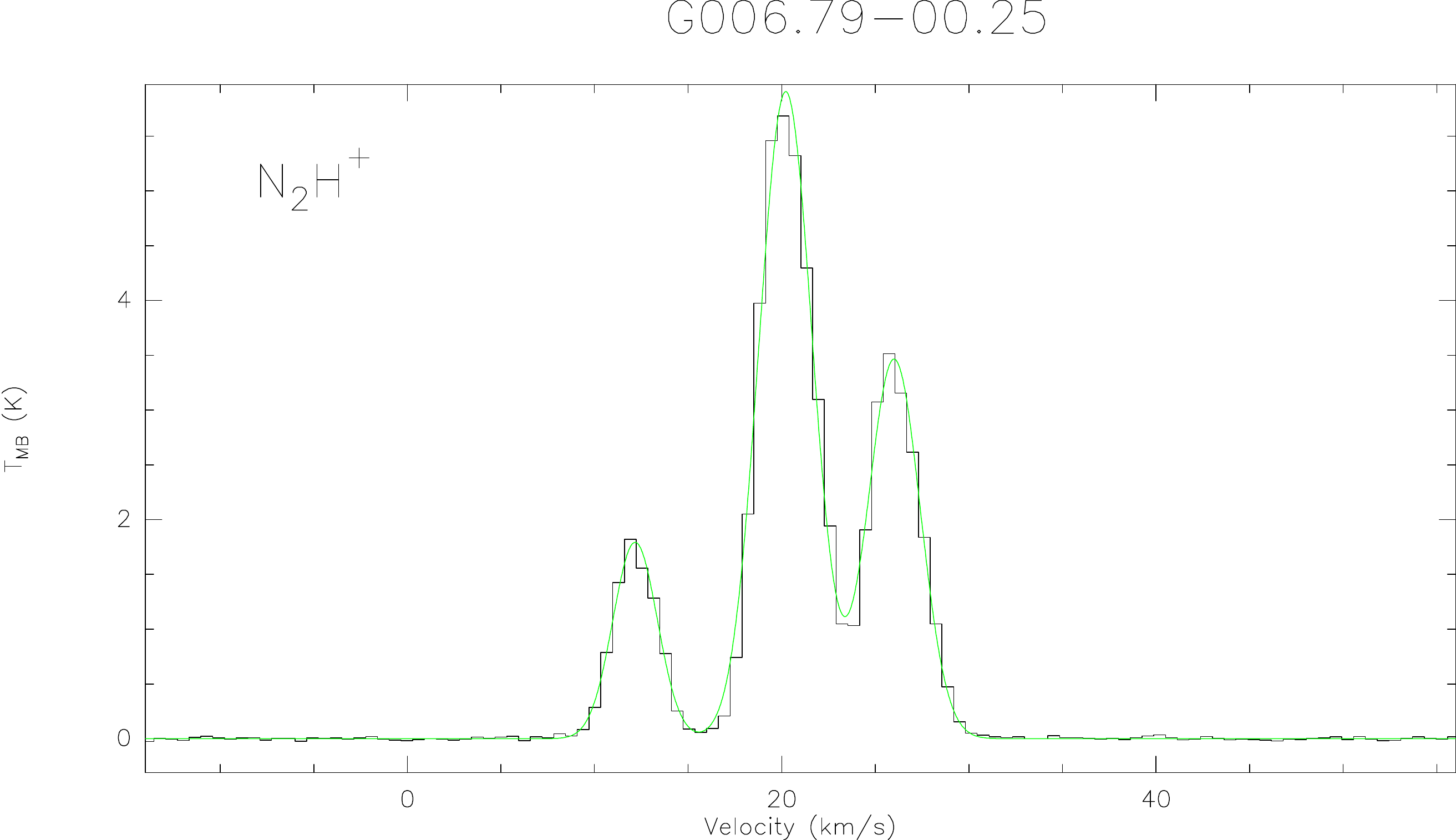}
    \includegraphics[width=0.3\textwidth]{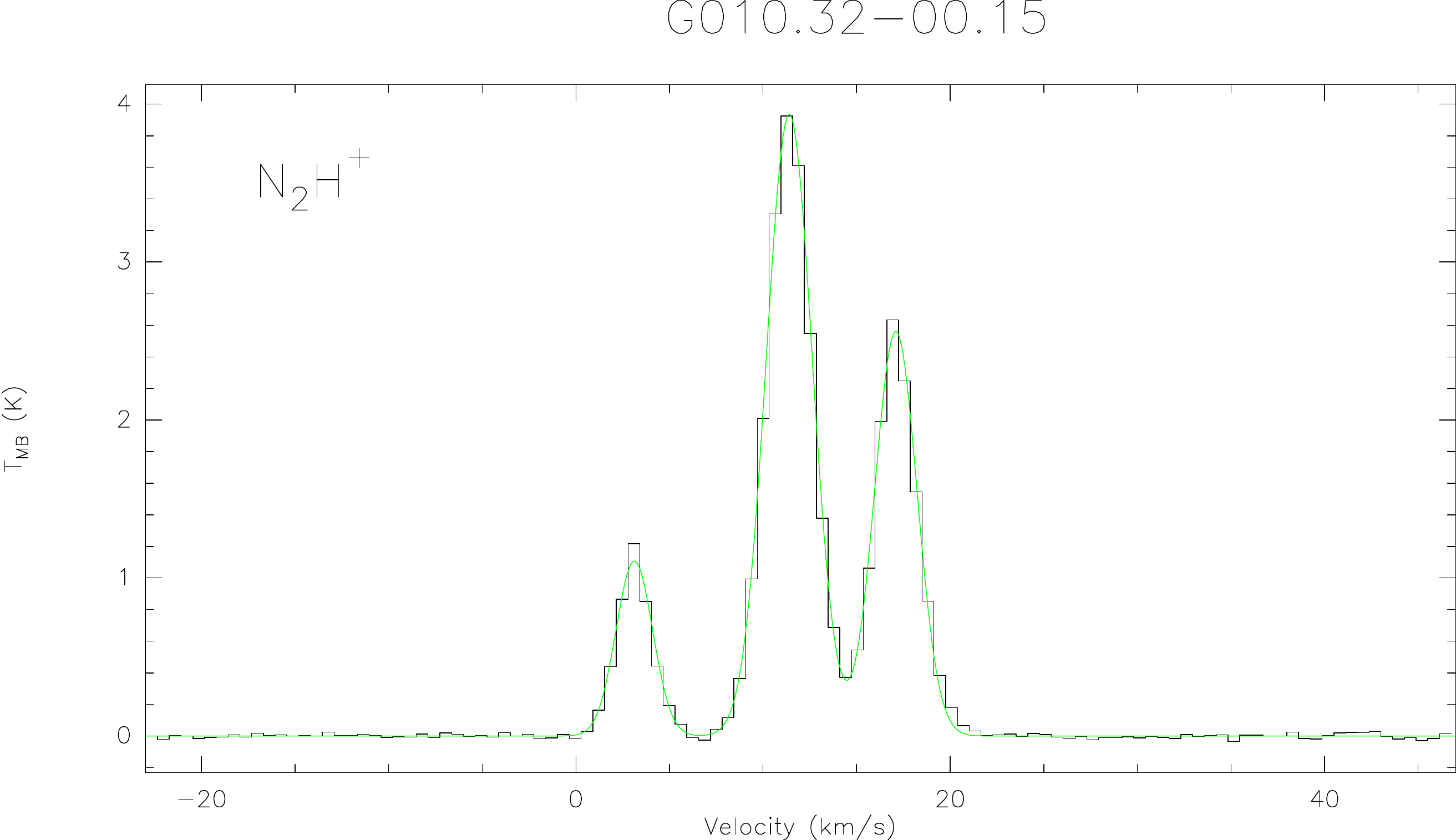} 
    \includegraphics[width=0.3\textwidth]{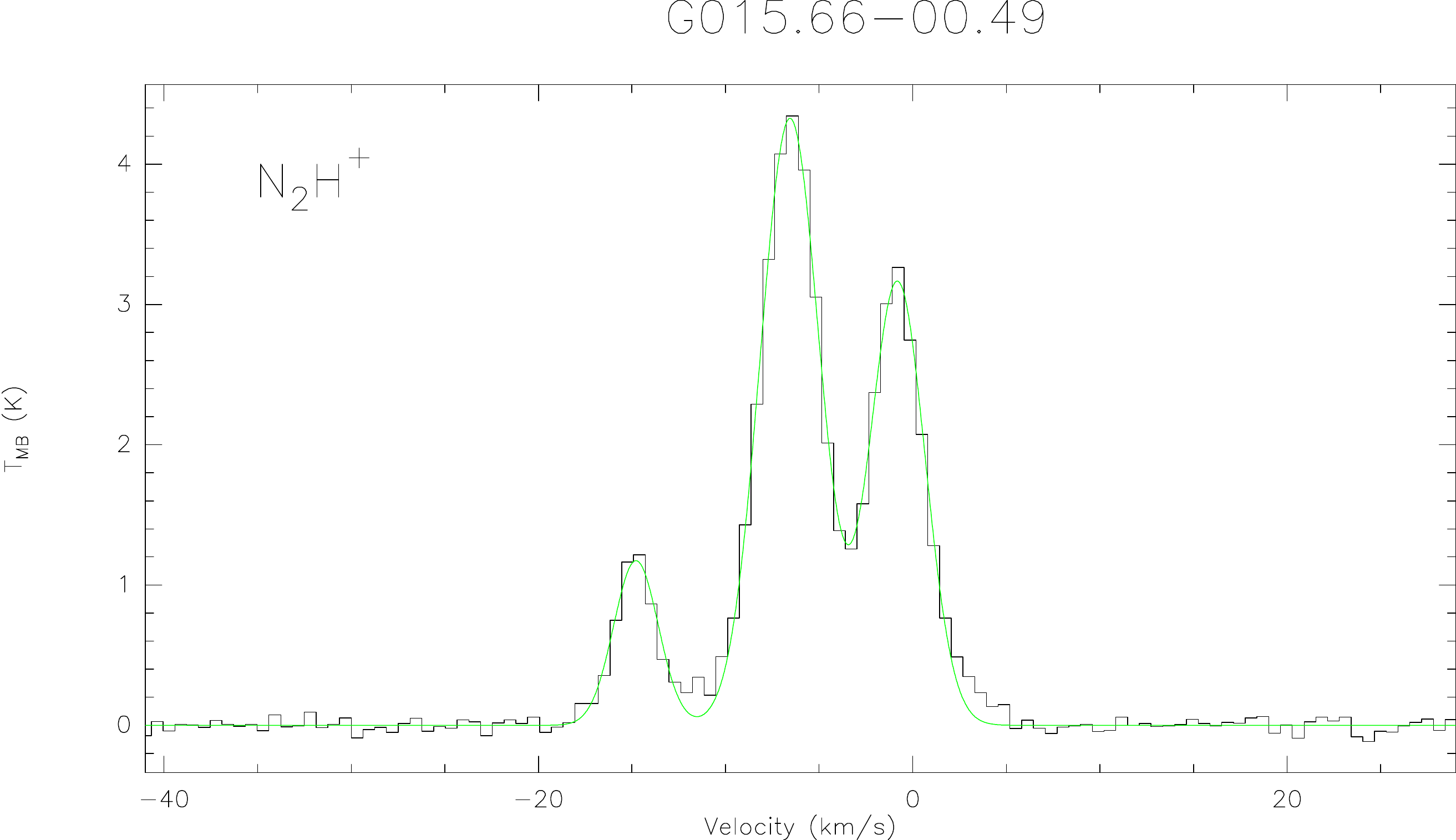}
    \includegraphics[width=0.3\textwidth]{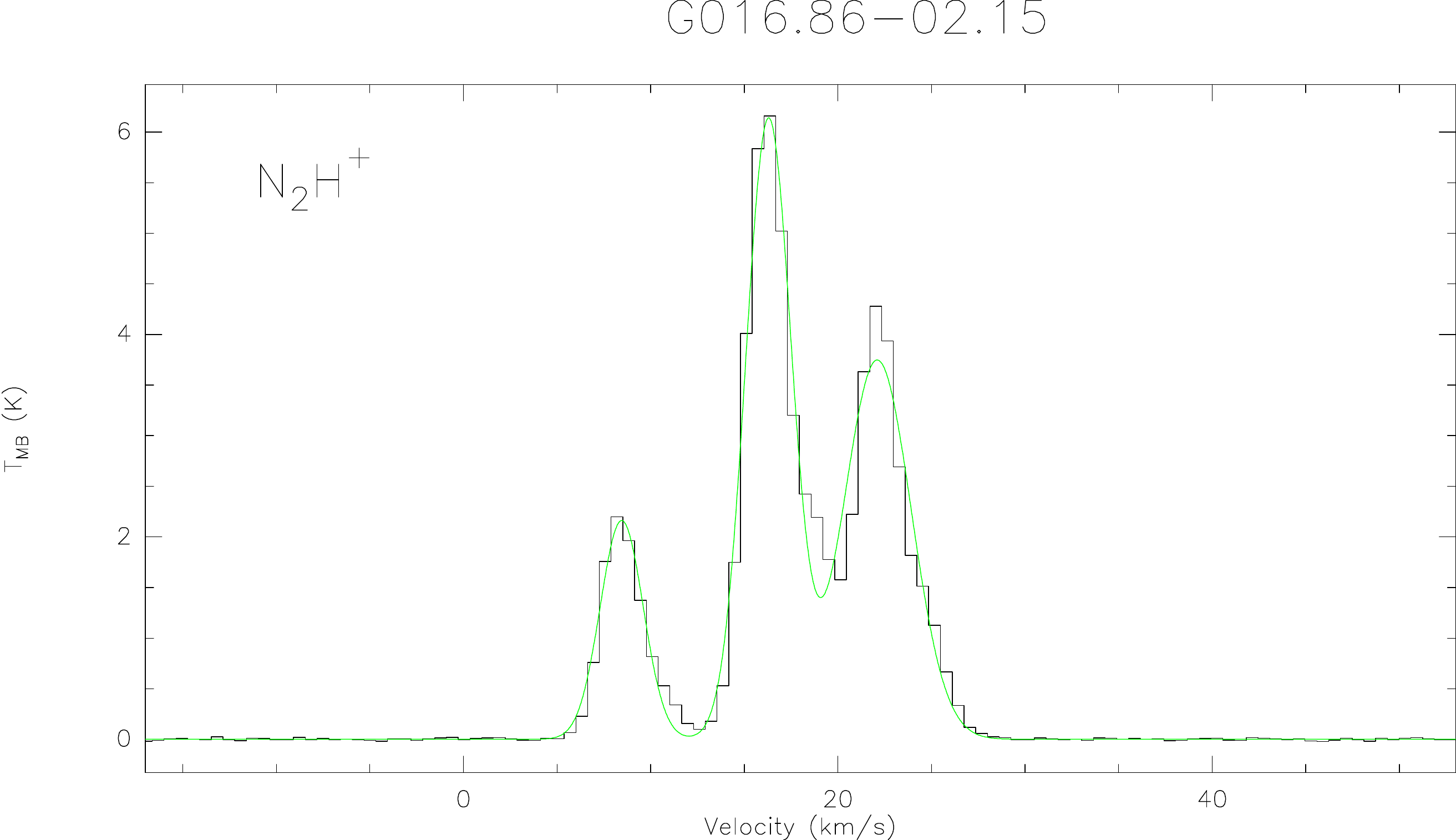}
    \includegraphics[width=0.3\textwidth]{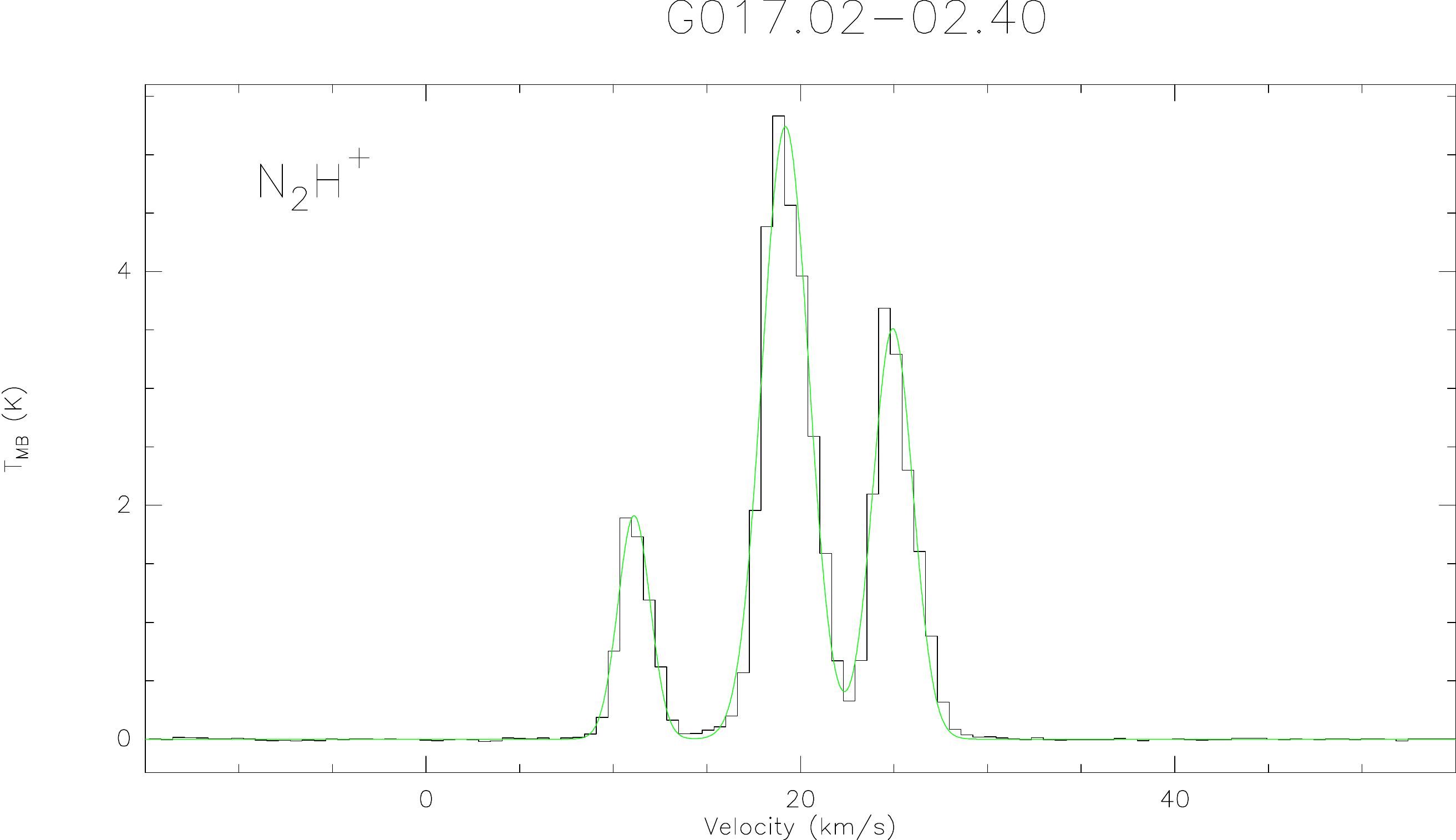} 
    \includegraphics[width=0.3\textwidth]{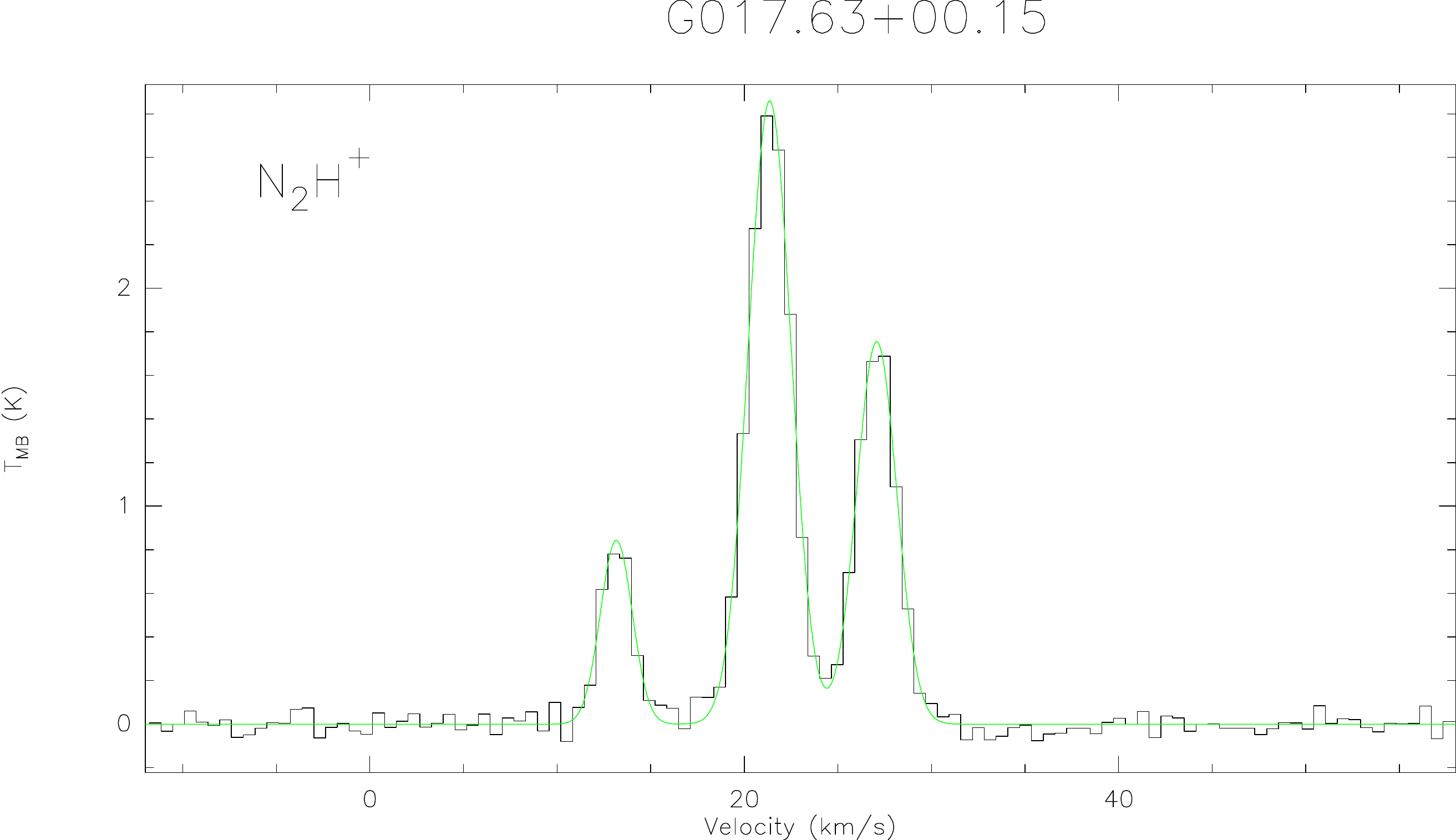}
    \includegraphics[width=0.3\textwidth]{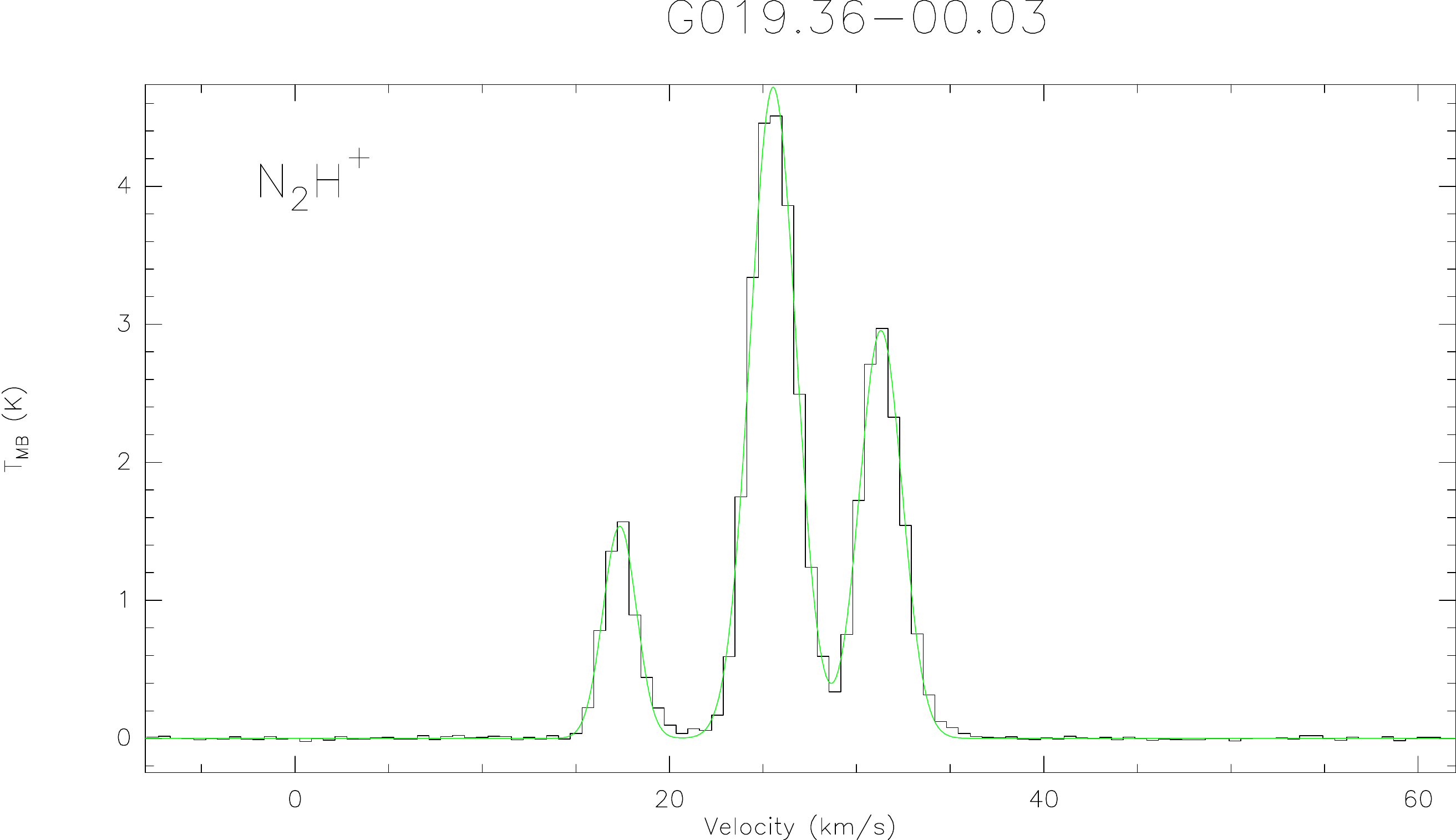}
    \includegraphics[width=0.3\textwidth]{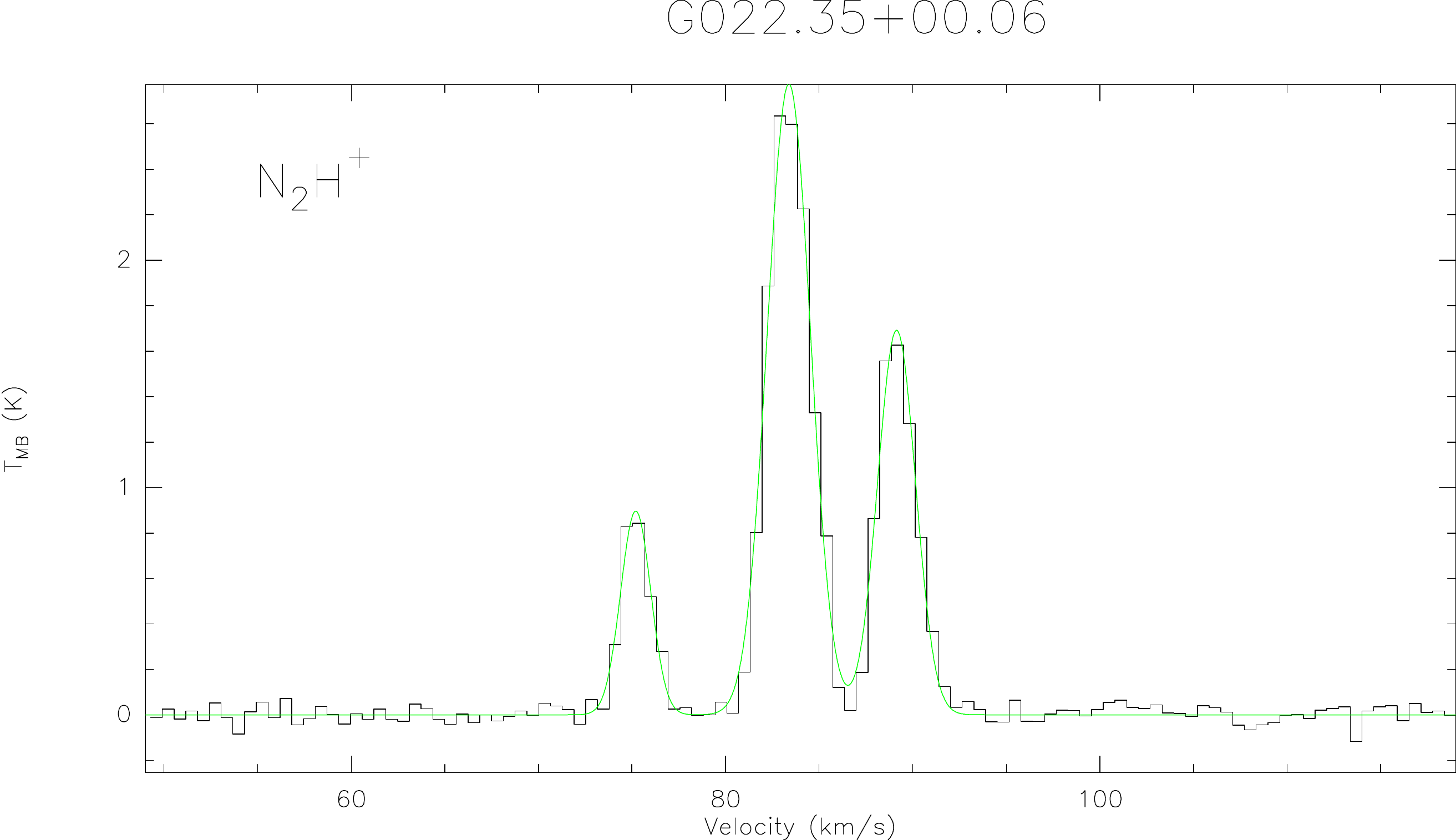} 
    \includegraphics[width=0.3\textwidth]{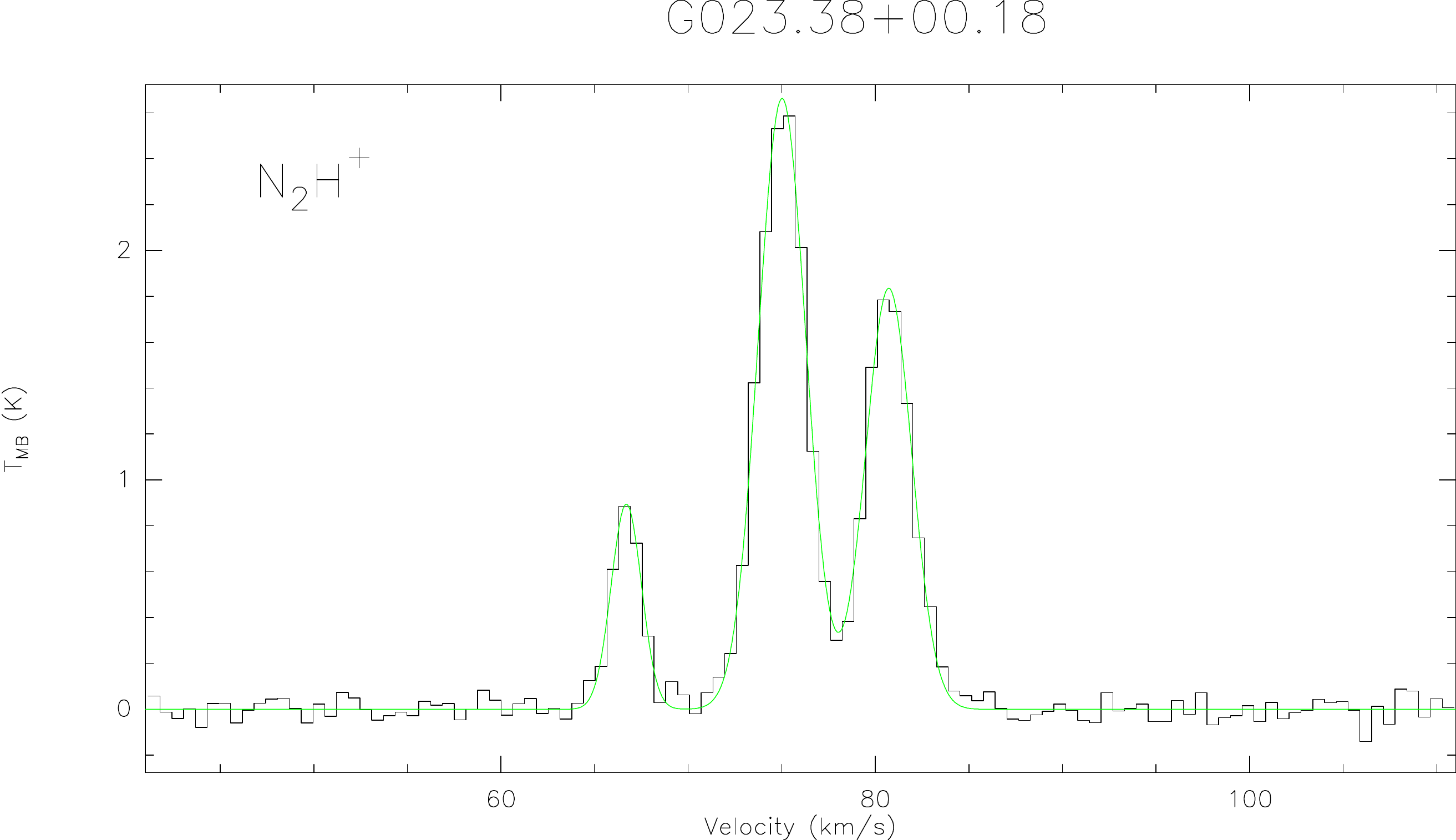}
    \includegraphics[width=0.3\textwidth]{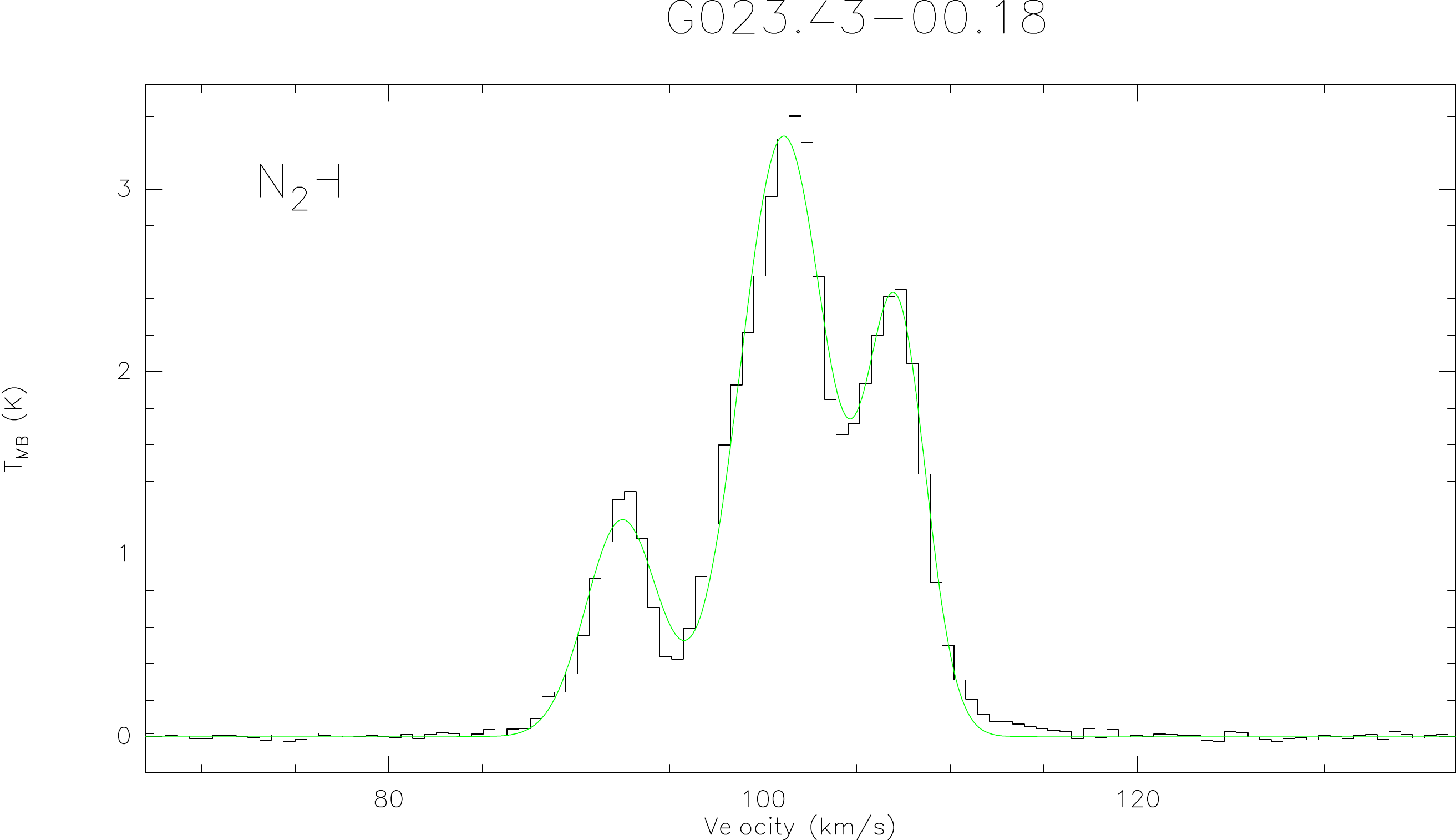}
    \includegraphics[width=0.3\textwidth]{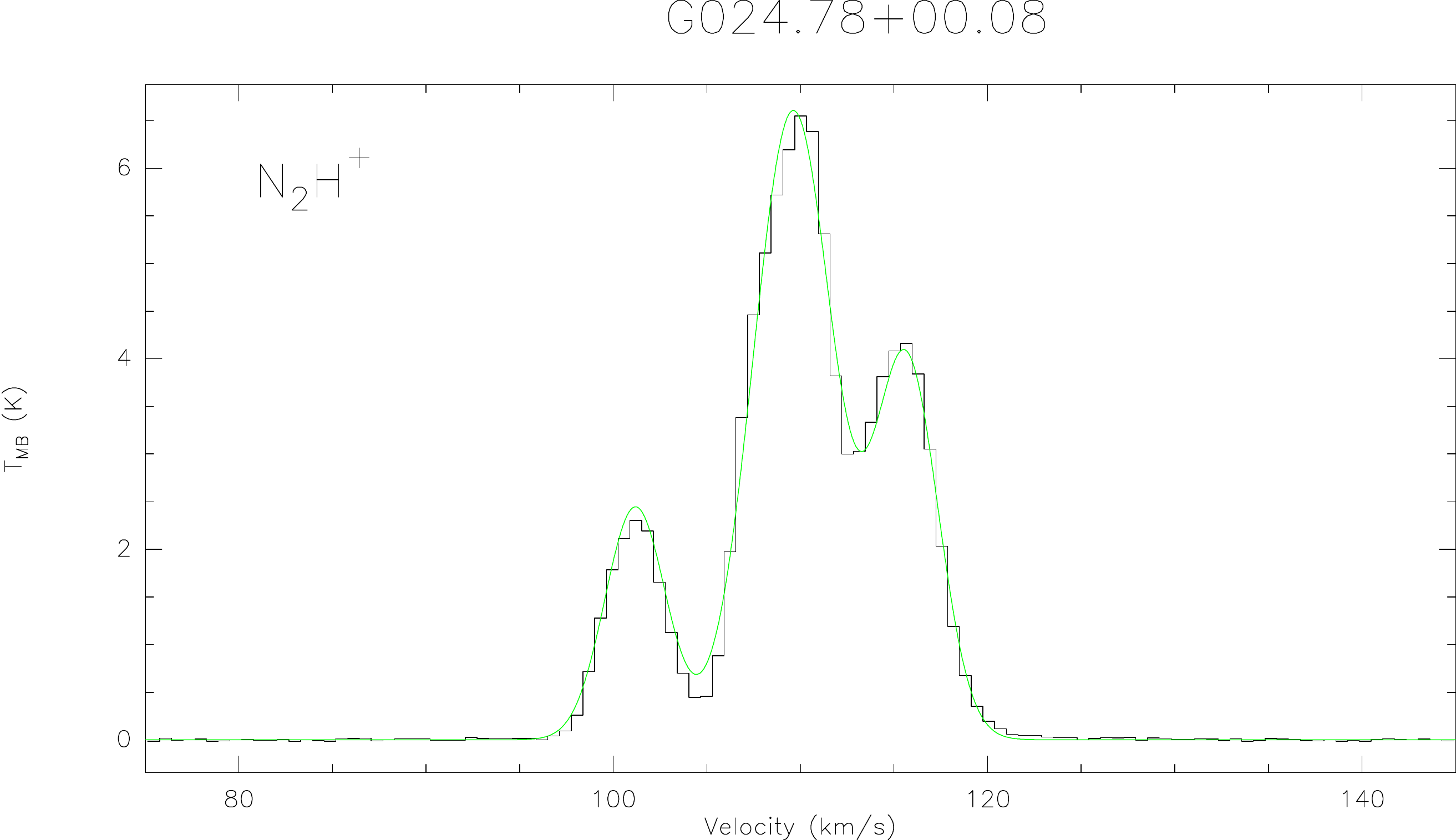} 
    \includegraphics[width=0.3\textwidth]{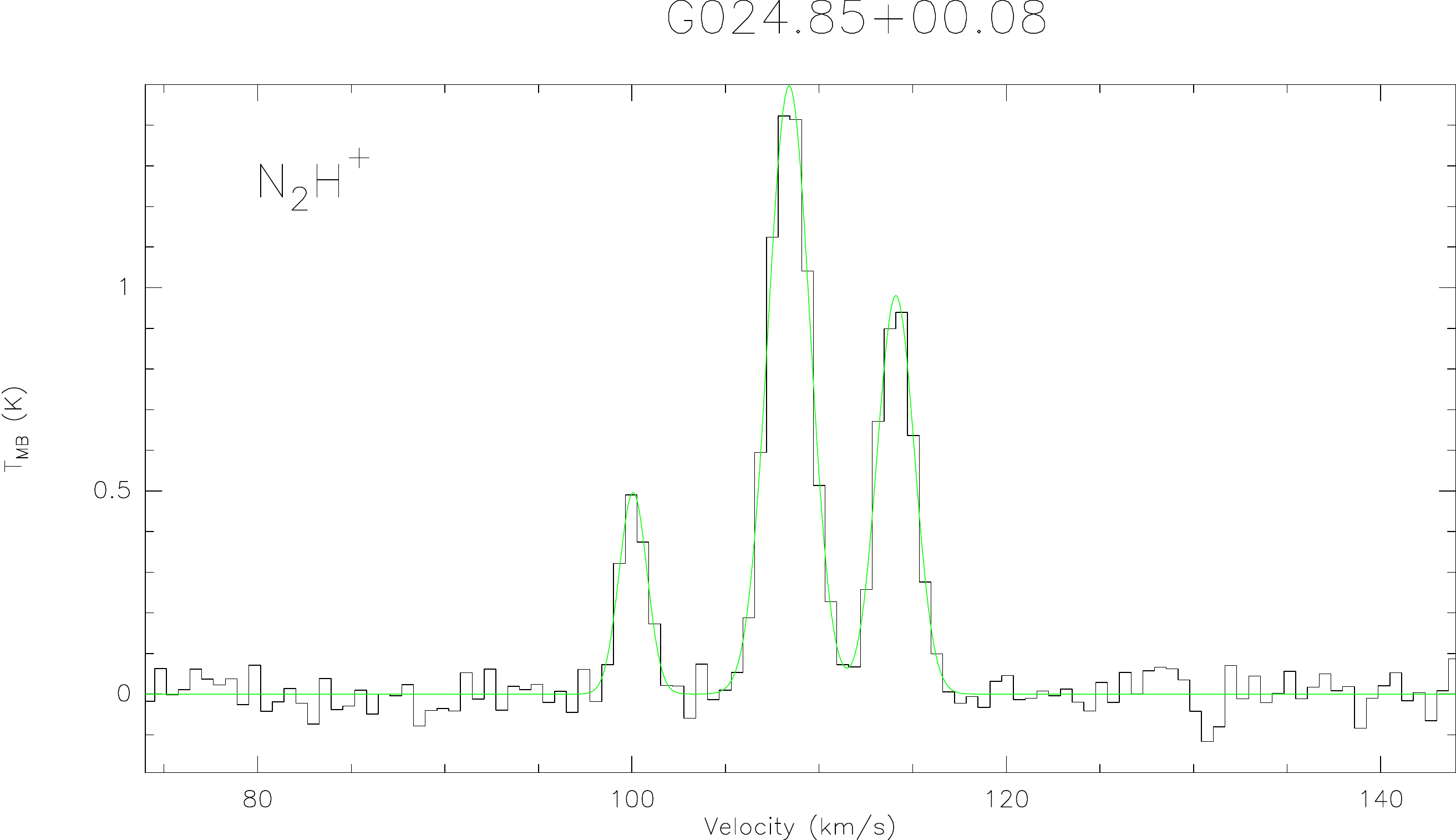}
    \includegraphics[width=0.3\textwidth]{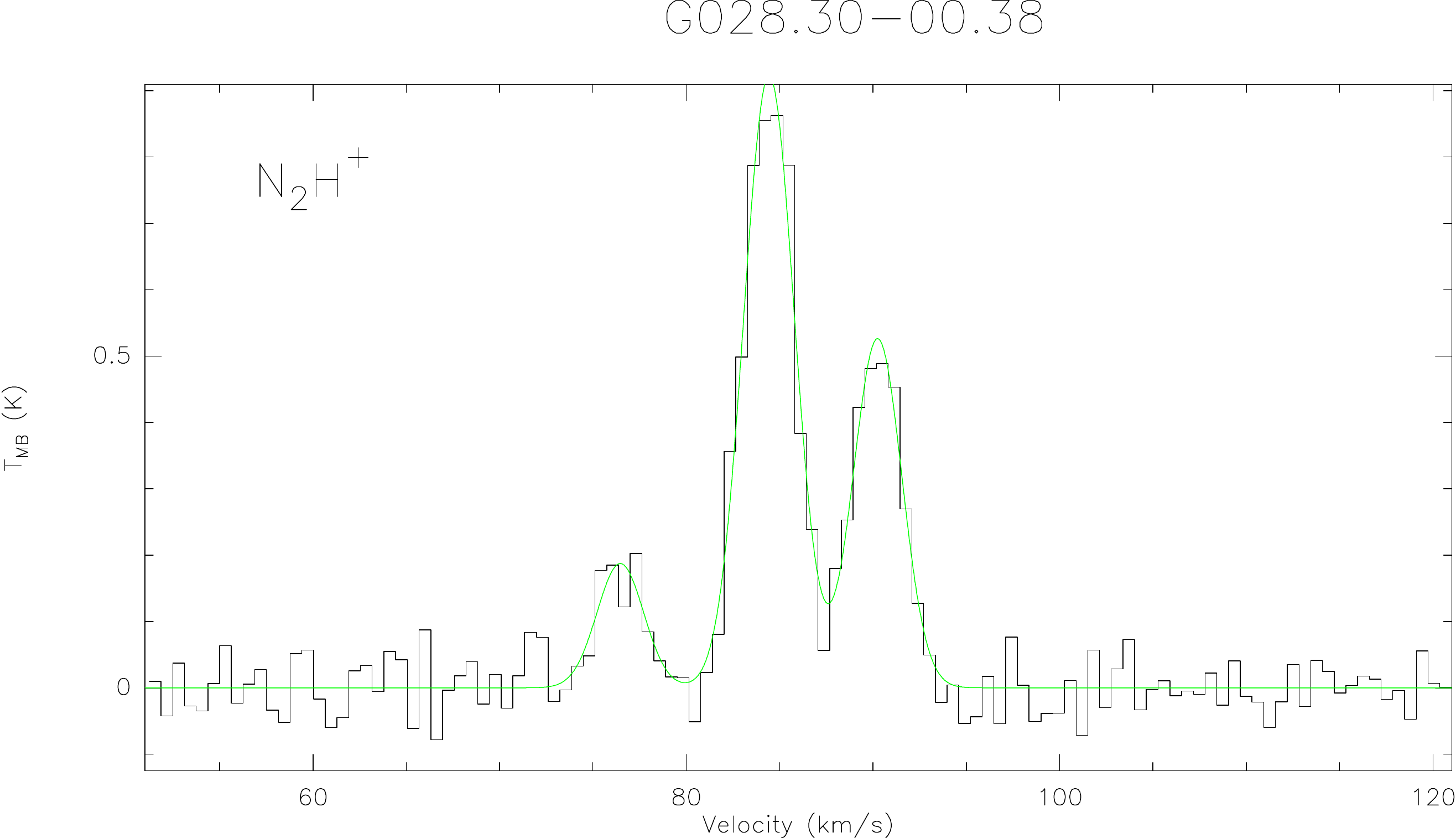}
    \includegraphics[width=0.3\textwidth]{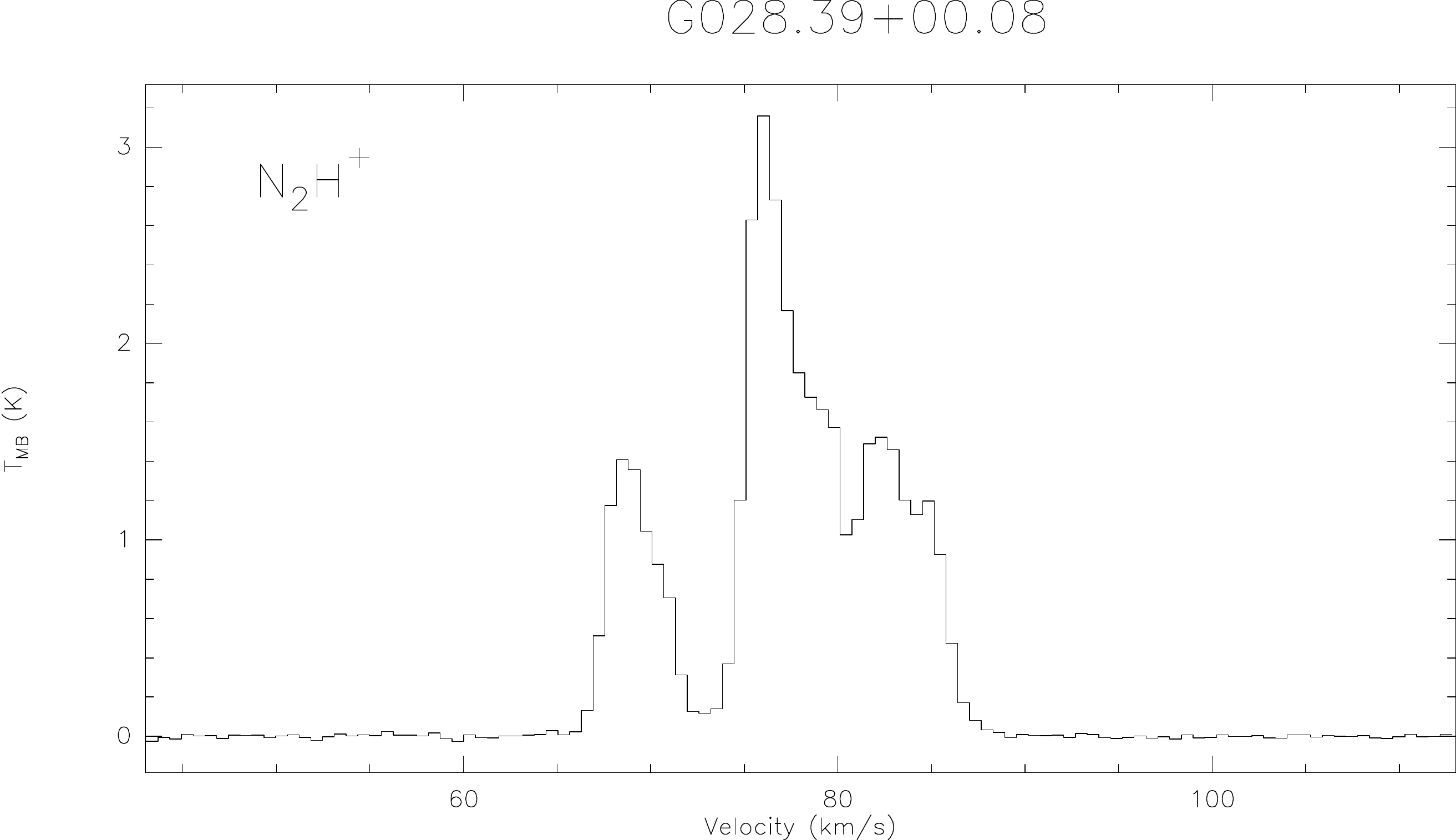} 
    \includegraphics[width=0.3\textwidth]{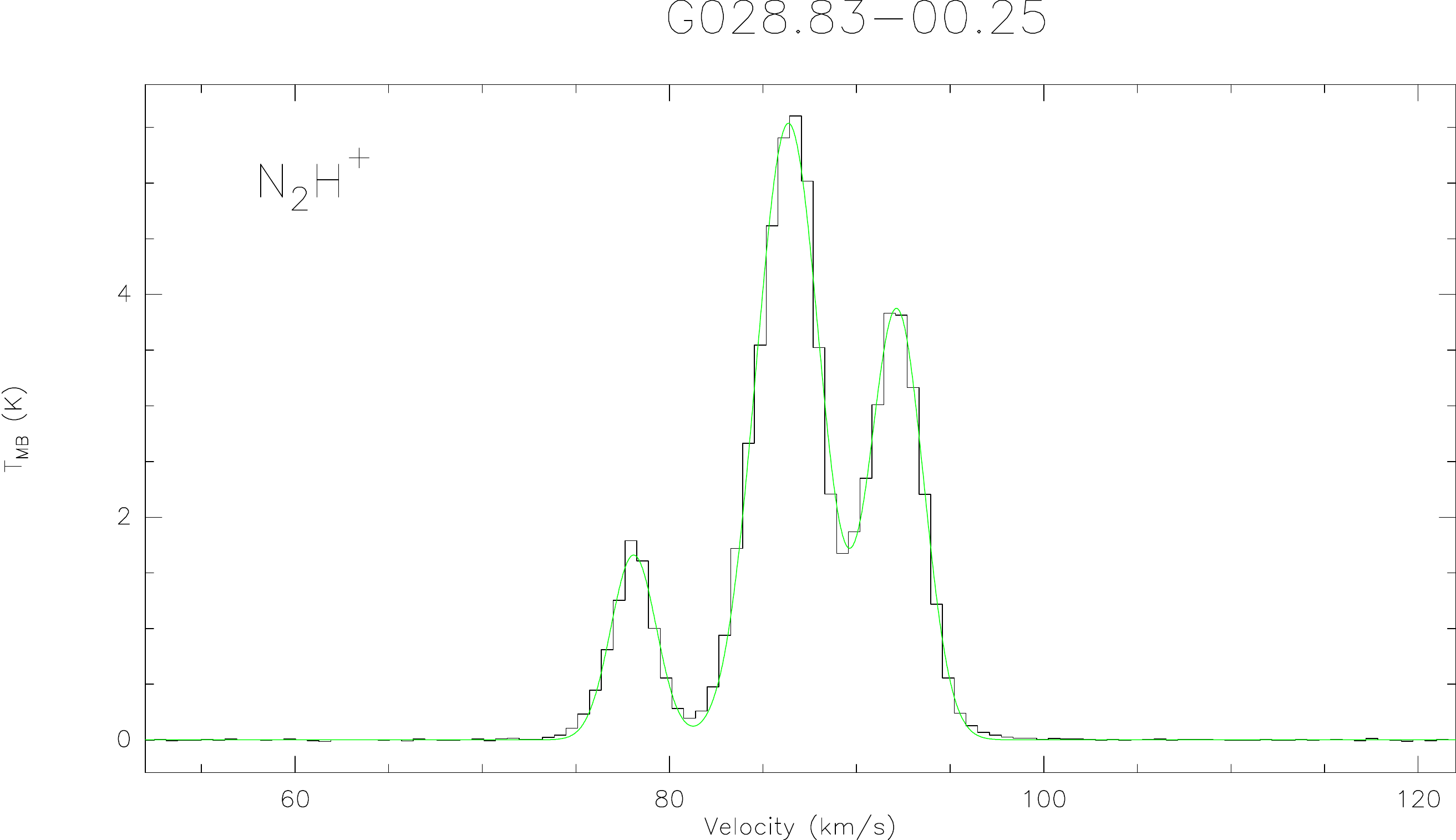}
    \includegraphics[width=0.3\textwidth]{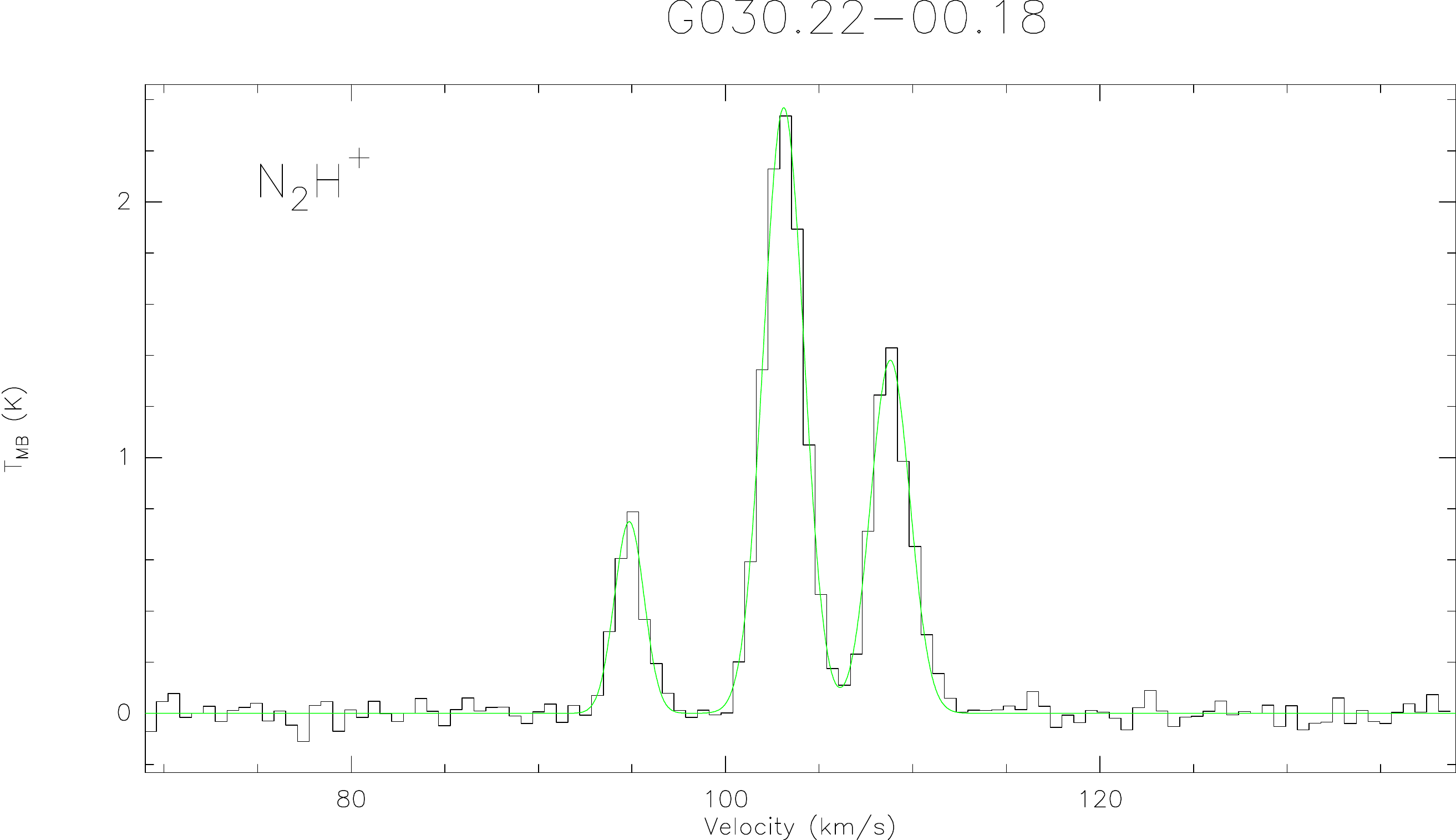}
    \includegraphics[width=0.3\textwidth]{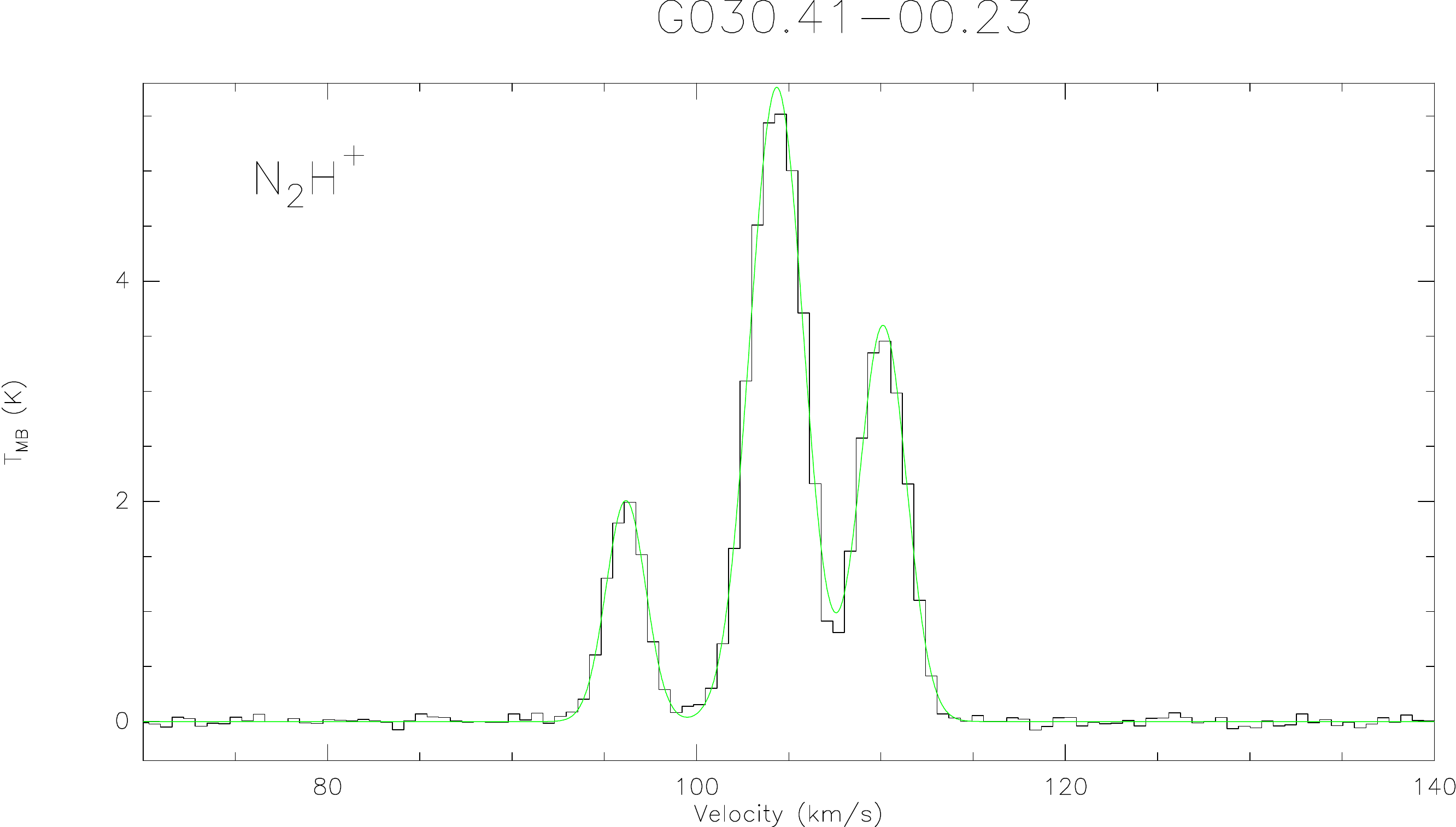} 
    \includegraphics[width=0.3\textwidth]{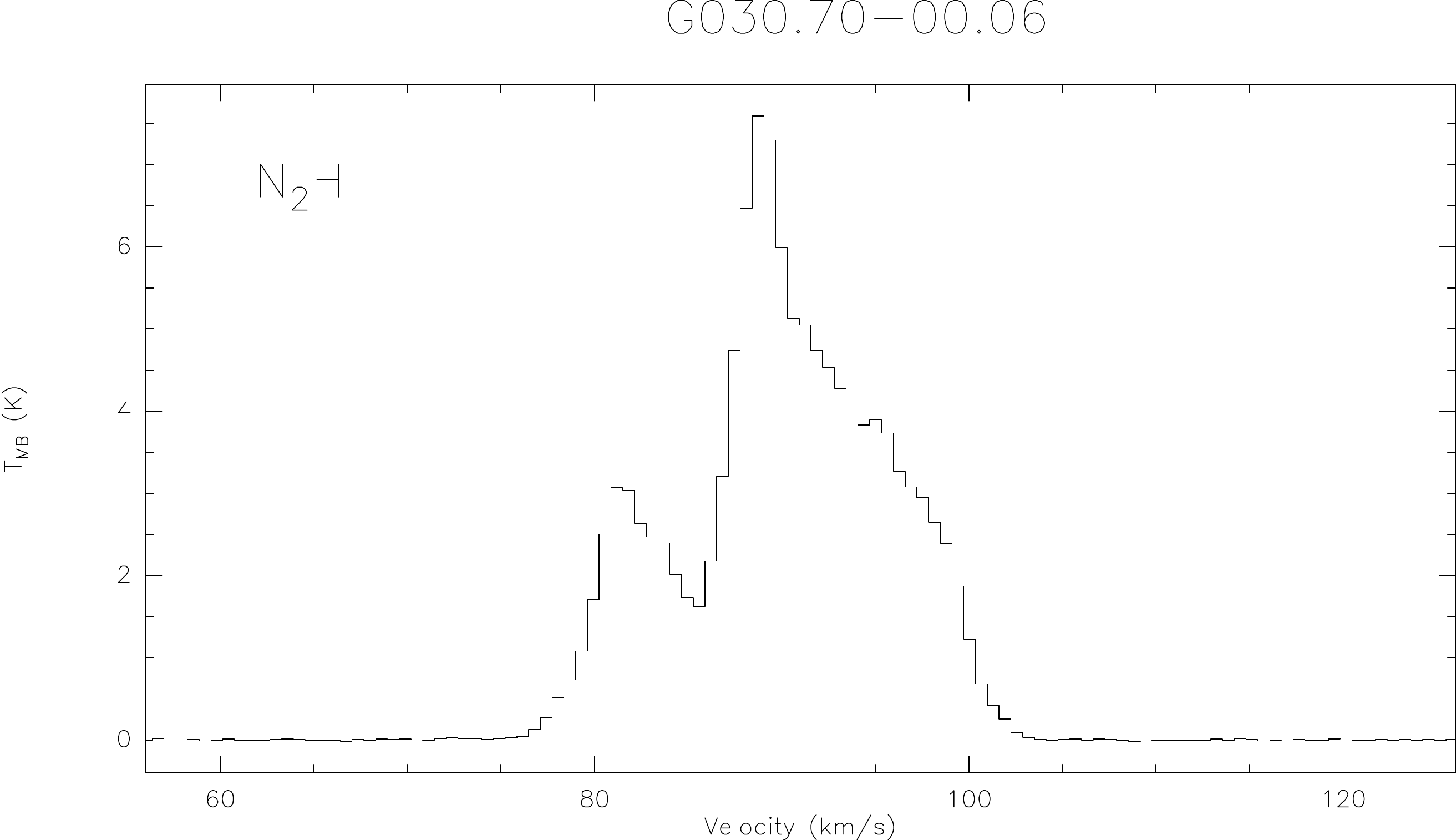}
    \includegraphics[width=0.3\textwidth]{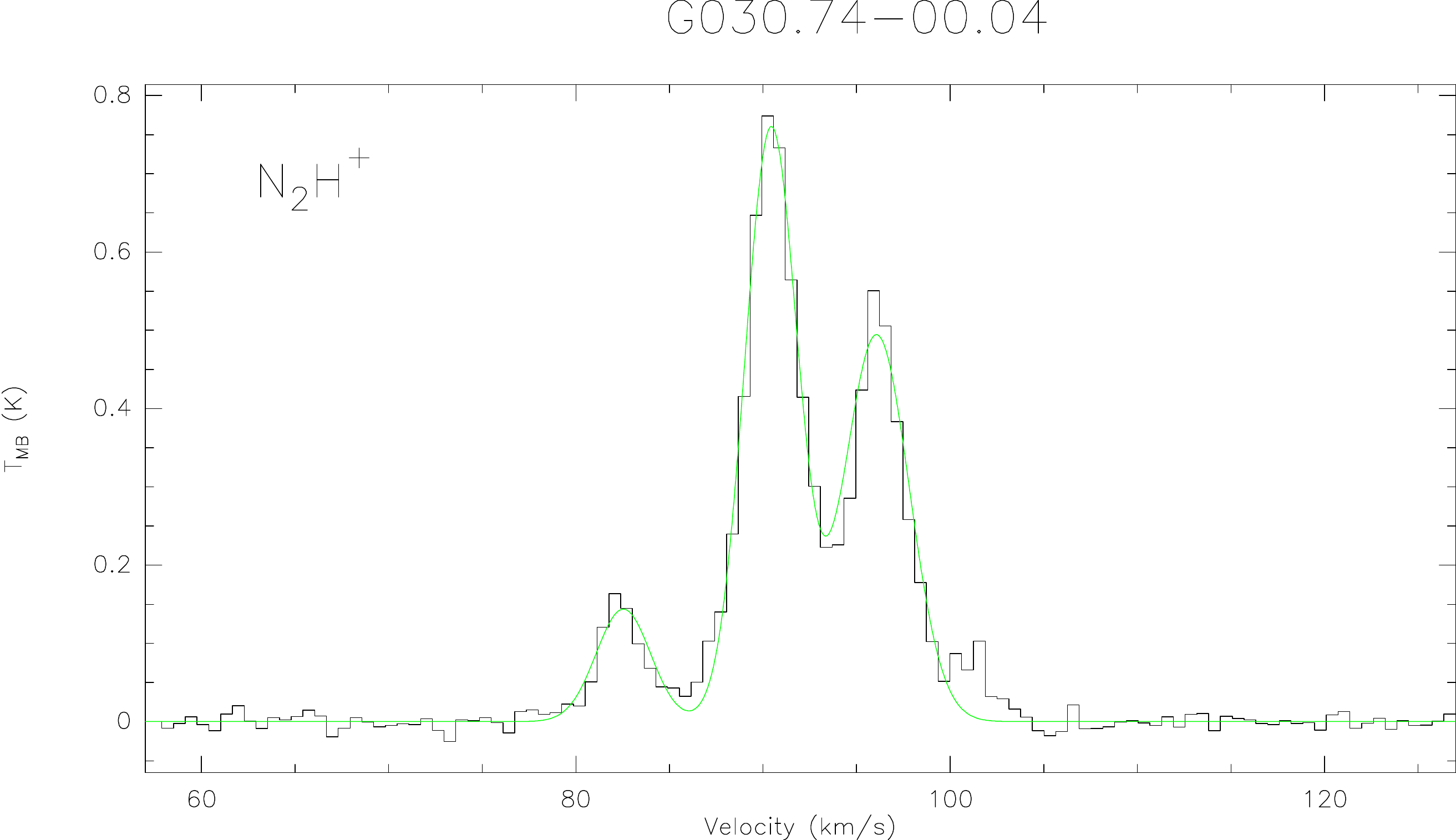}
    \includegraphics[width=0.3\textwidth]{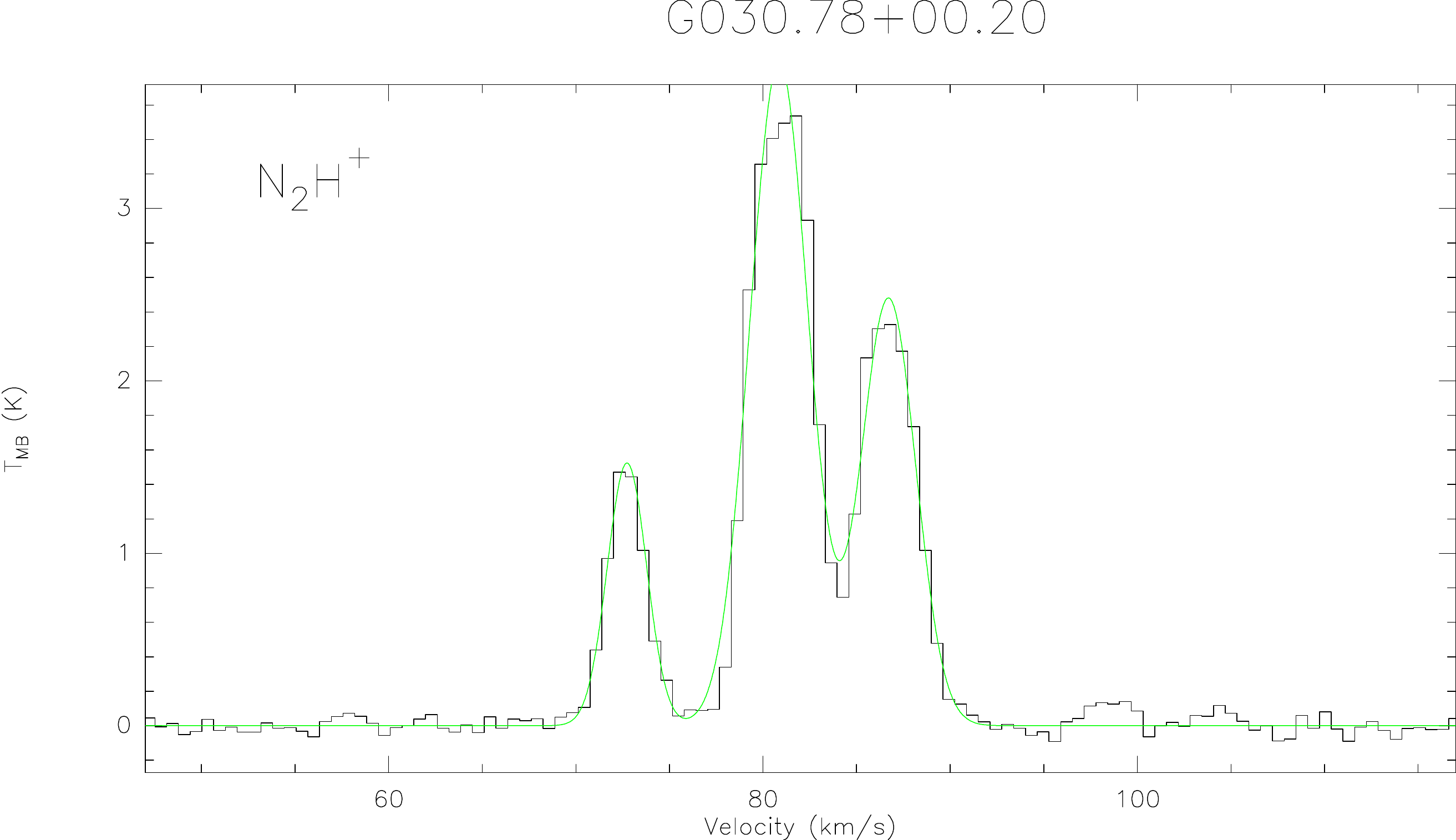} 
    \caption{N$_{2}$H$^{+}$ (J = 1$-$0) spectra detected by the IRAM 30 m telescope.}
\end{figure*}
    
\addtocounter{figure}{-1}
\begin{figure*}    
    \centering
    \includegraphics[width=0.3\textwidth]{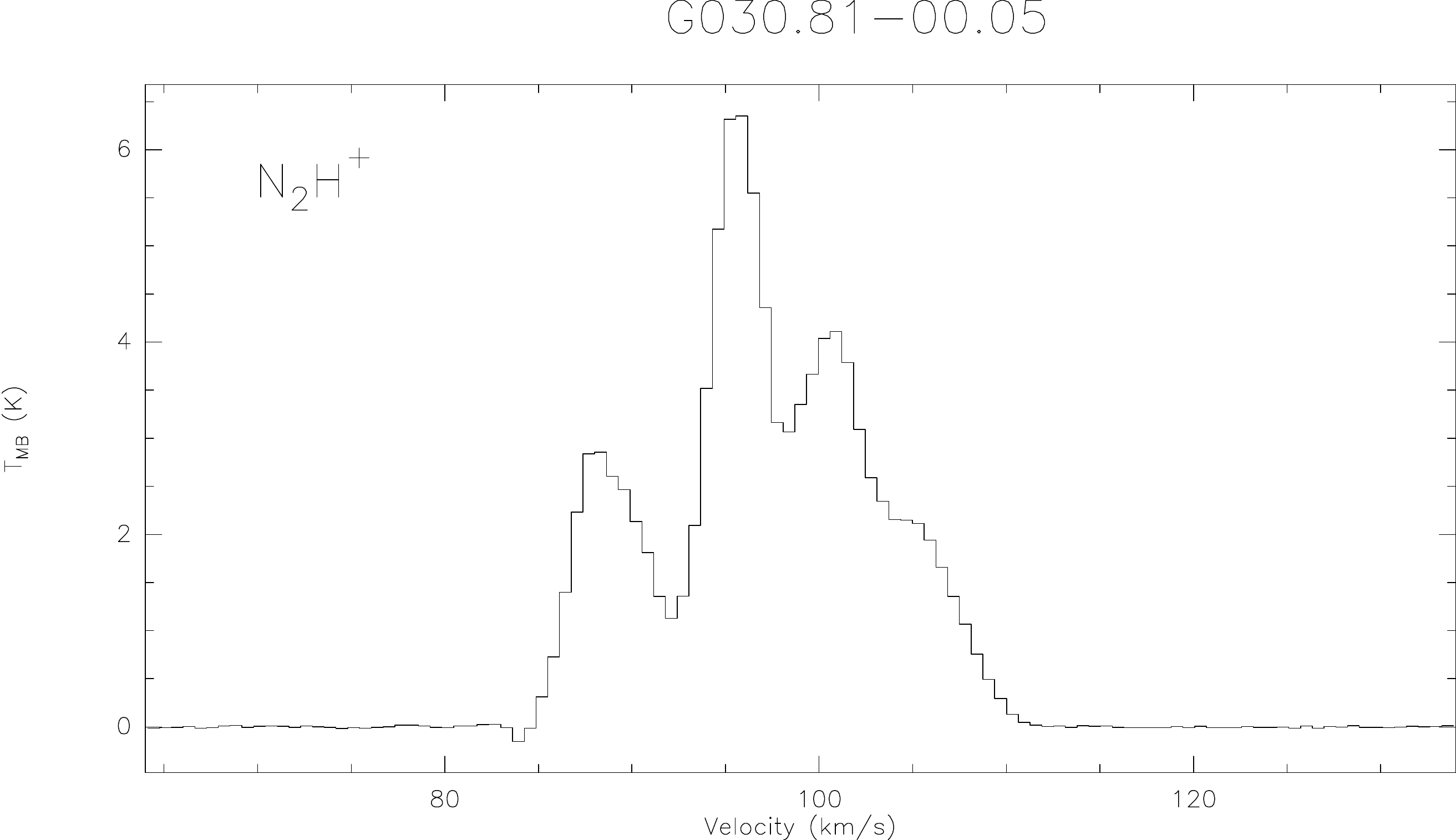}
    \includegraphics[width=0.3\textwidth]{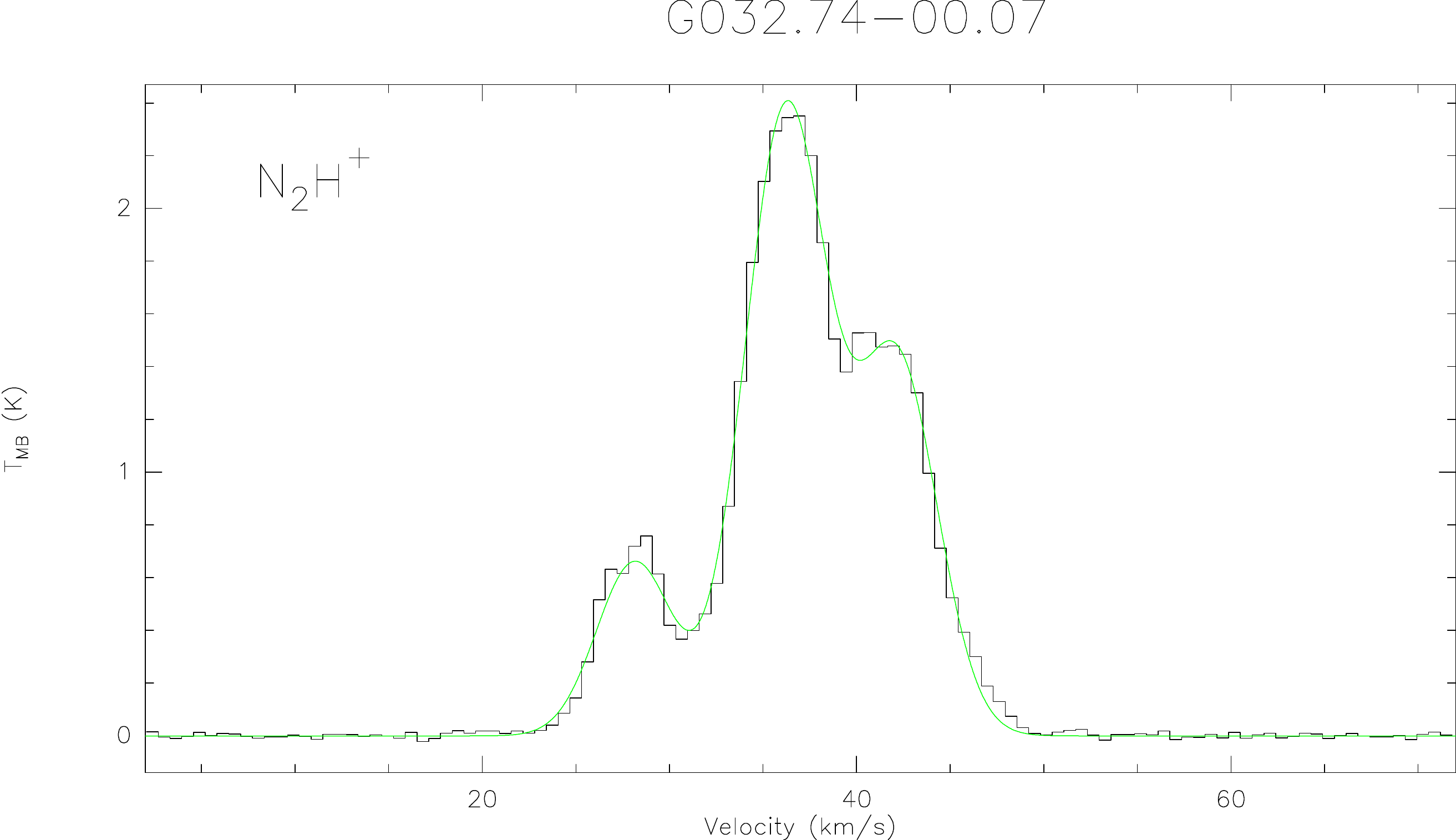}
    \includegraphics[width=0.3\textwidth]{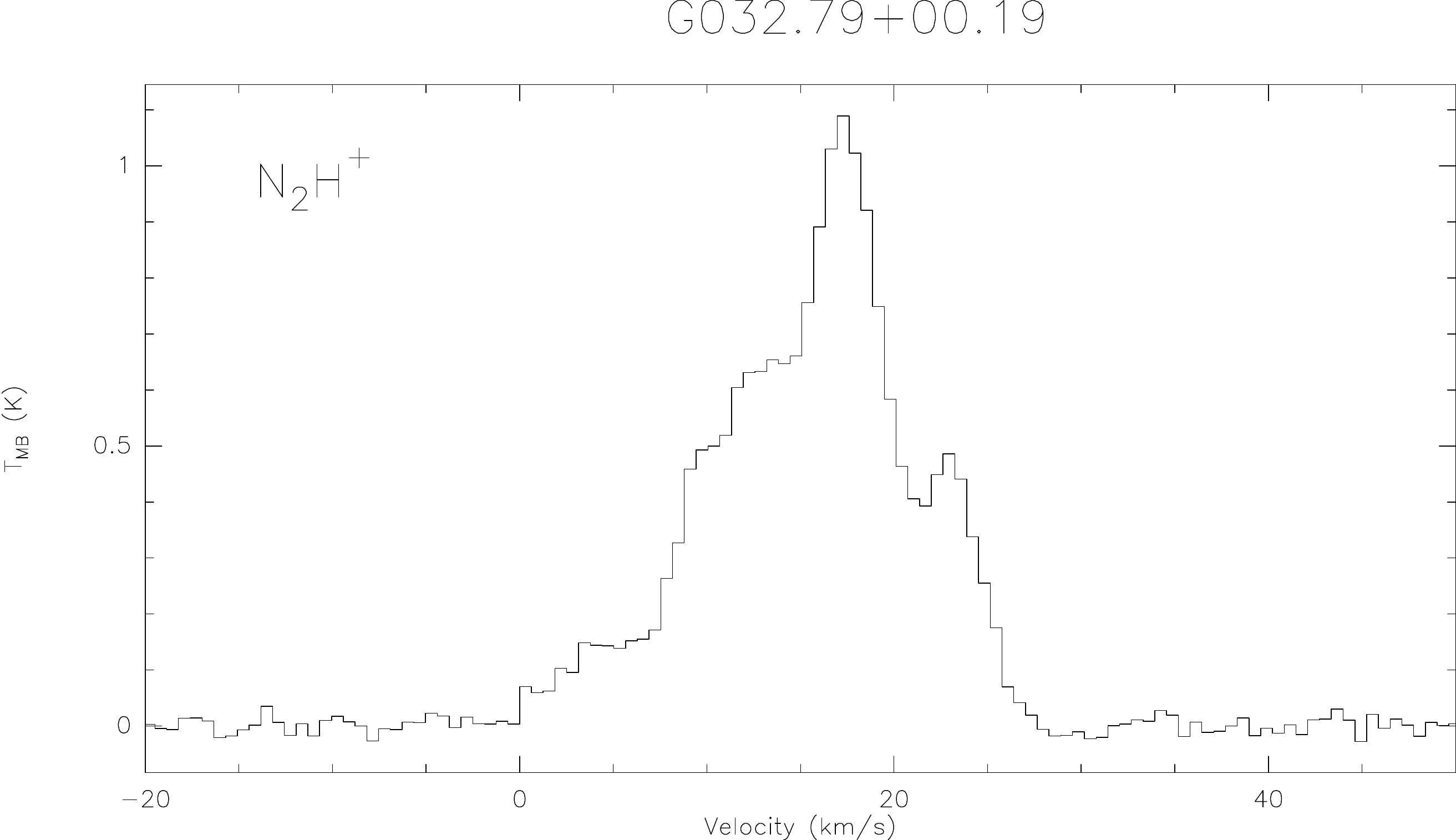} 
    \includegraphics[width=0.3\textwidth]{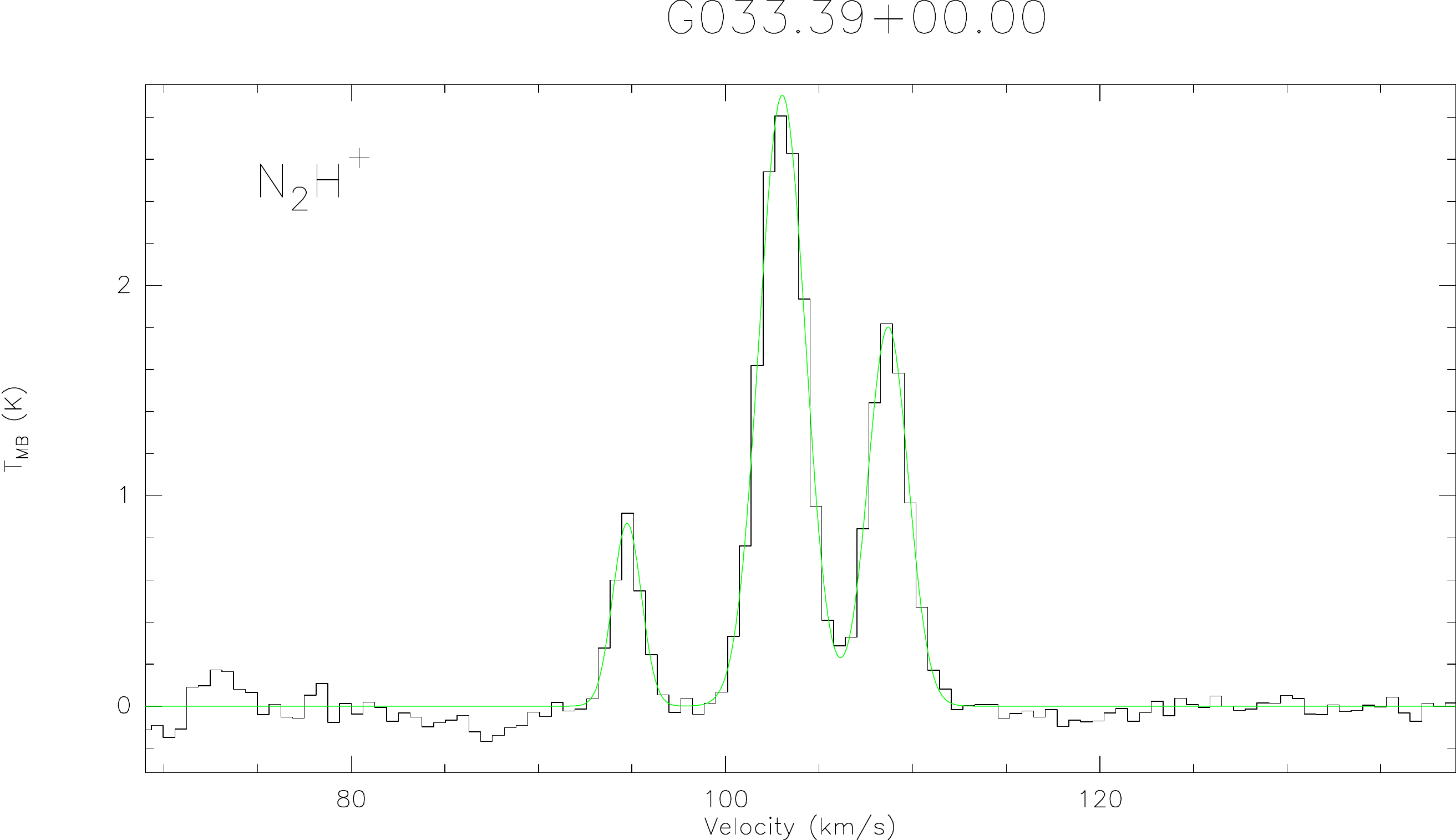}
    \includegraphics[width=0.3\textwidth]{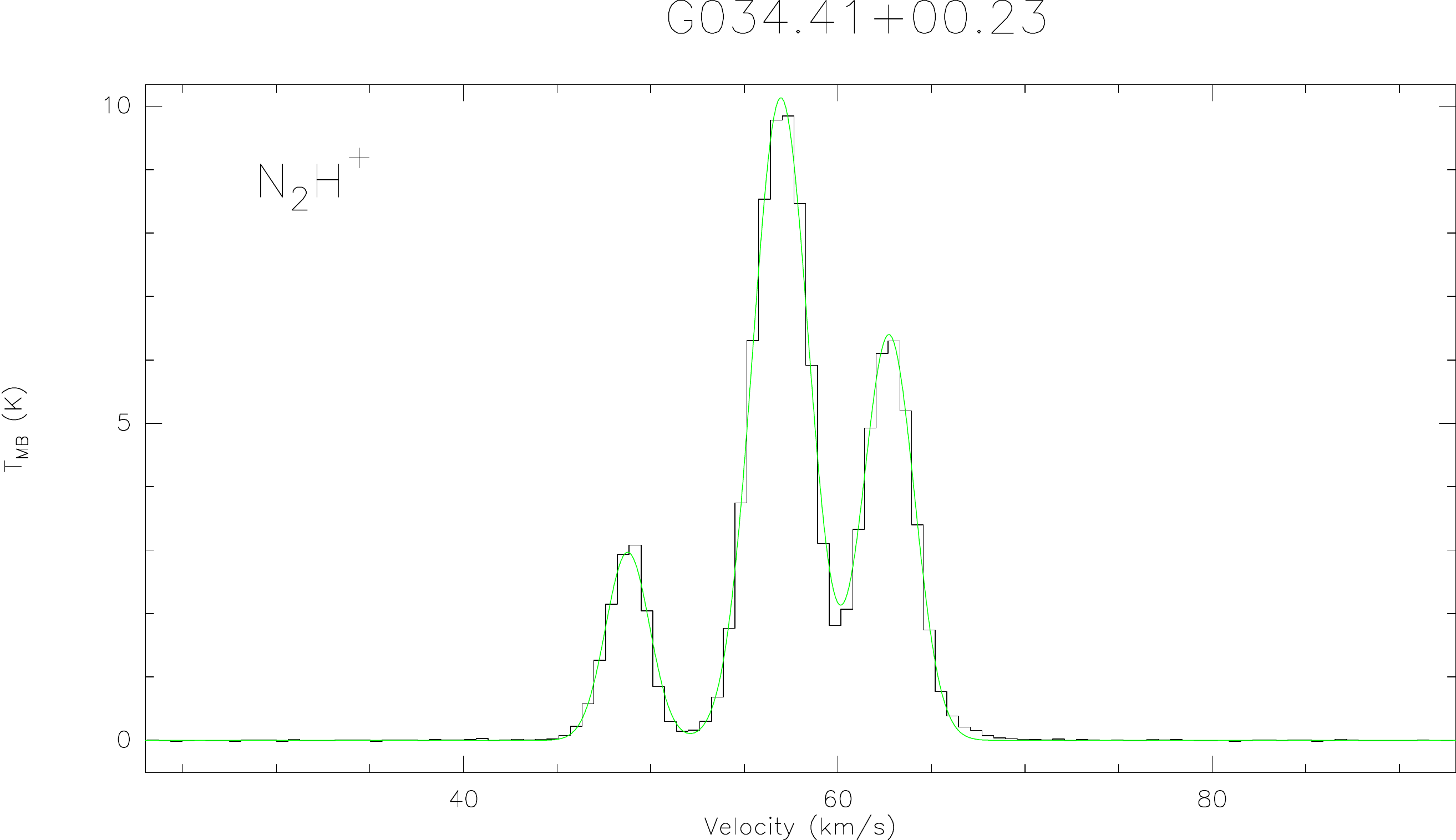}
    \includegraphics[width=0.3\textwidth]{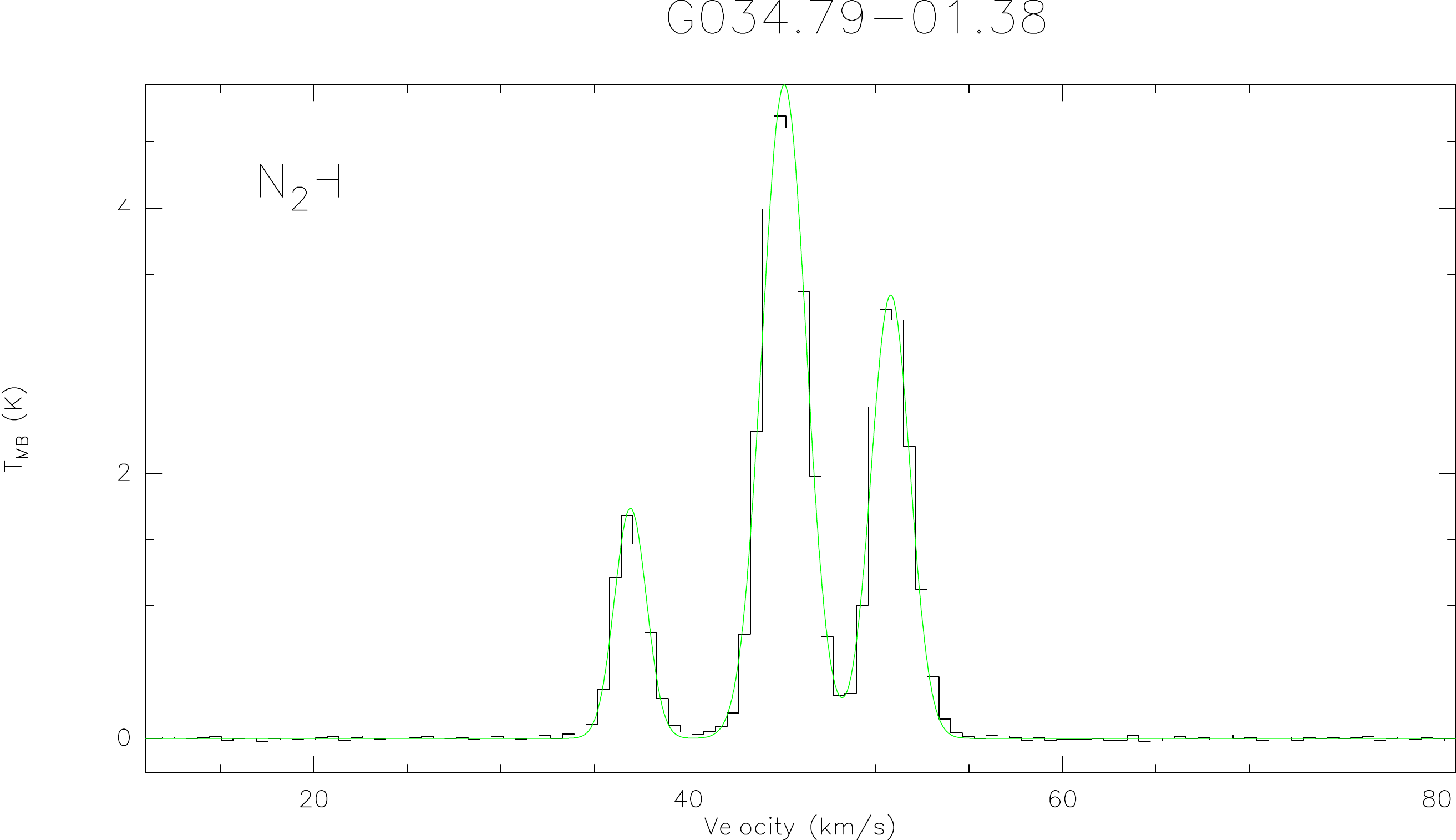} 
    \includegraphics[width=0.3\textwidth]{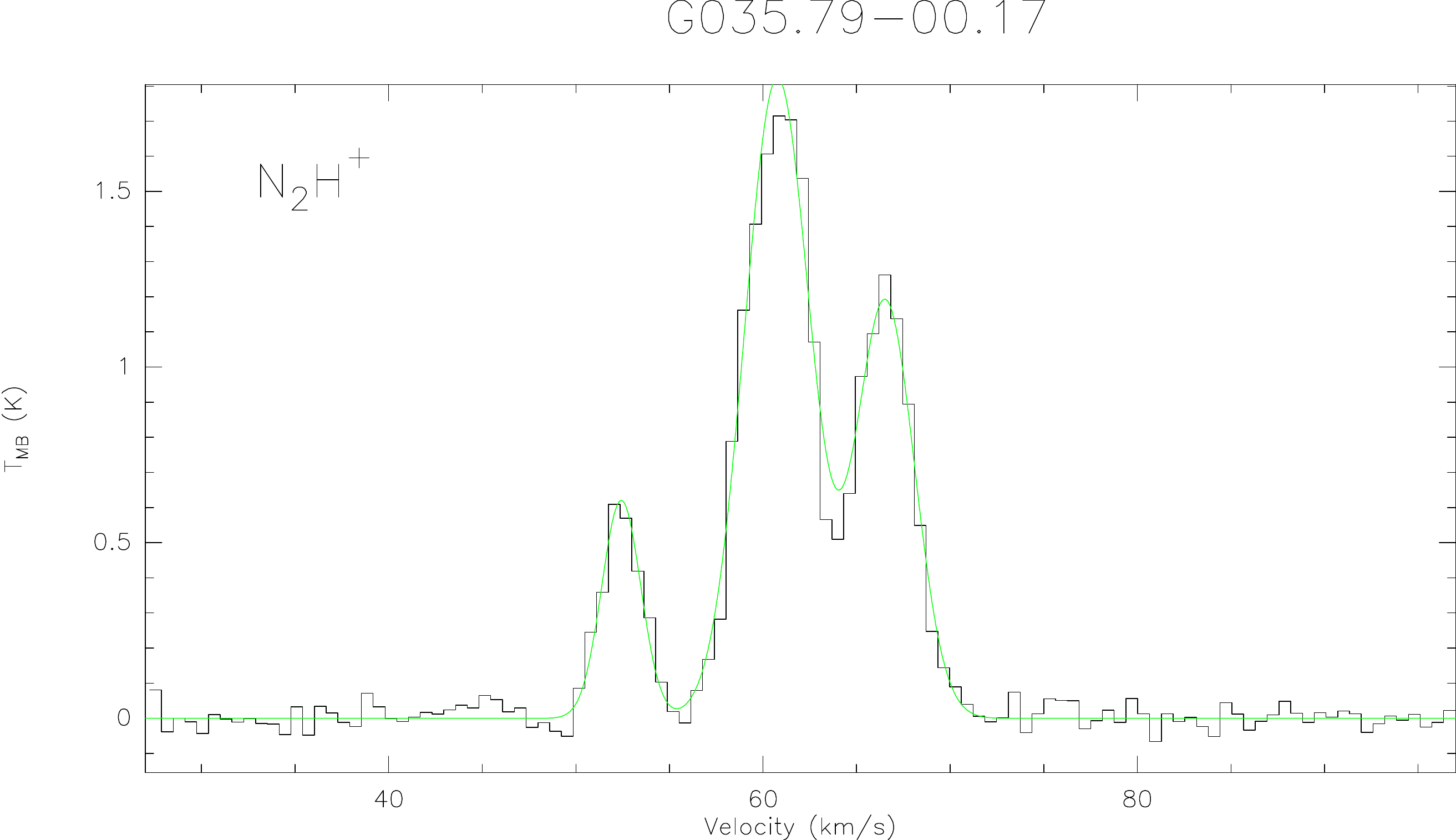}
    \includegraphics[width=0.3\textwidth]{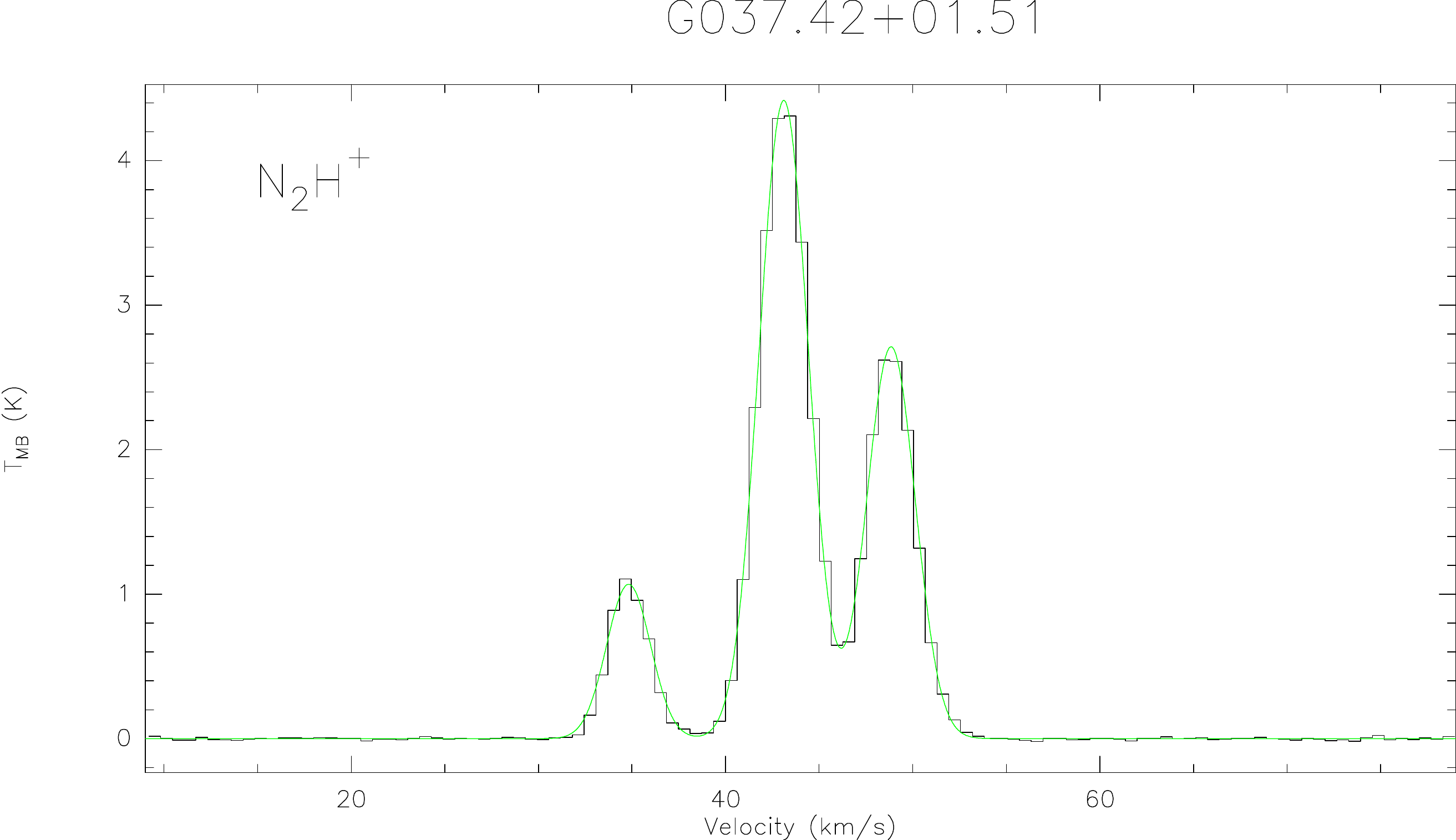} 
    \includegraphics[width=0.3\textwidth]{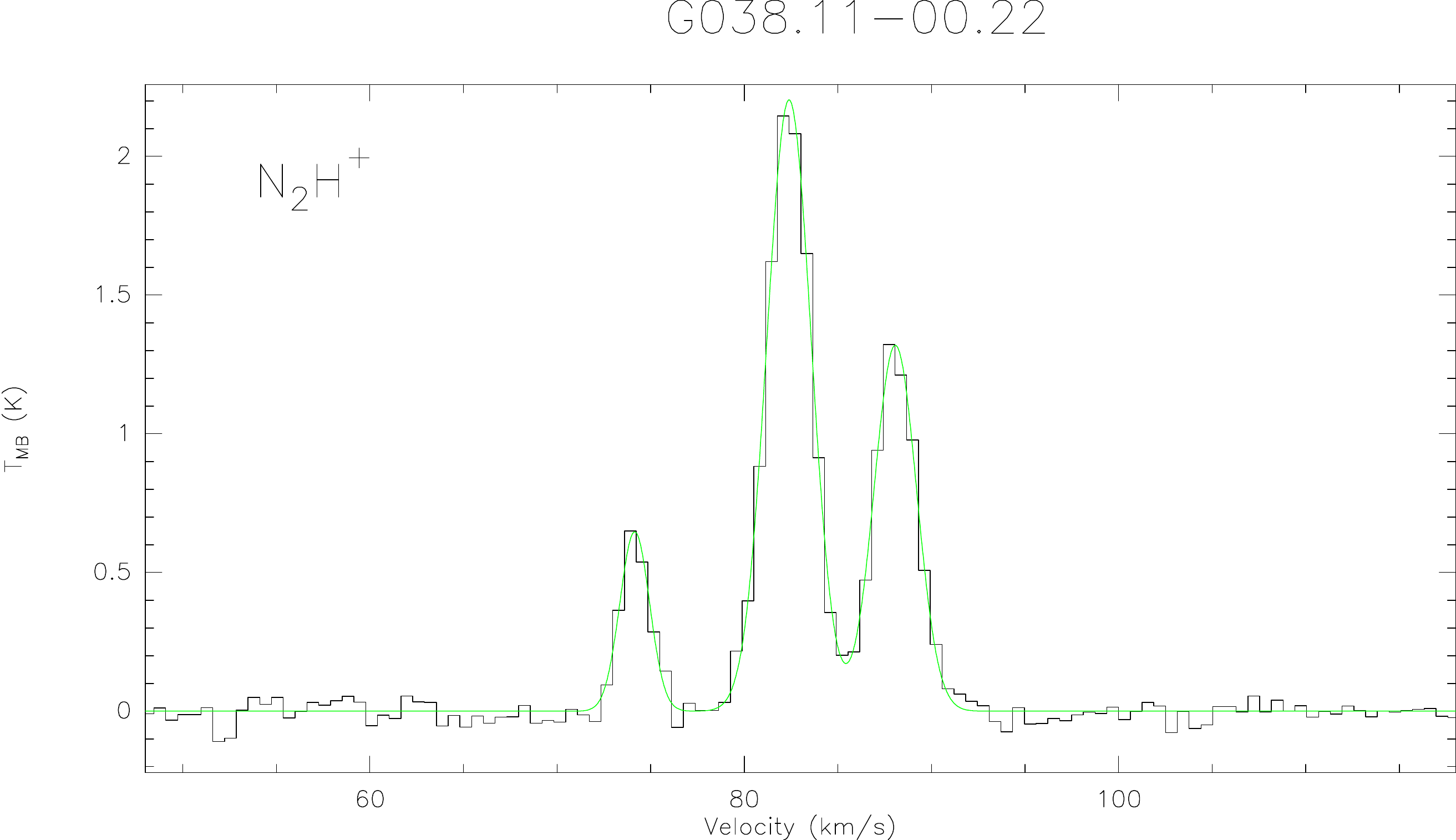} 
    \includegraphics[width=0.3\textwidth]{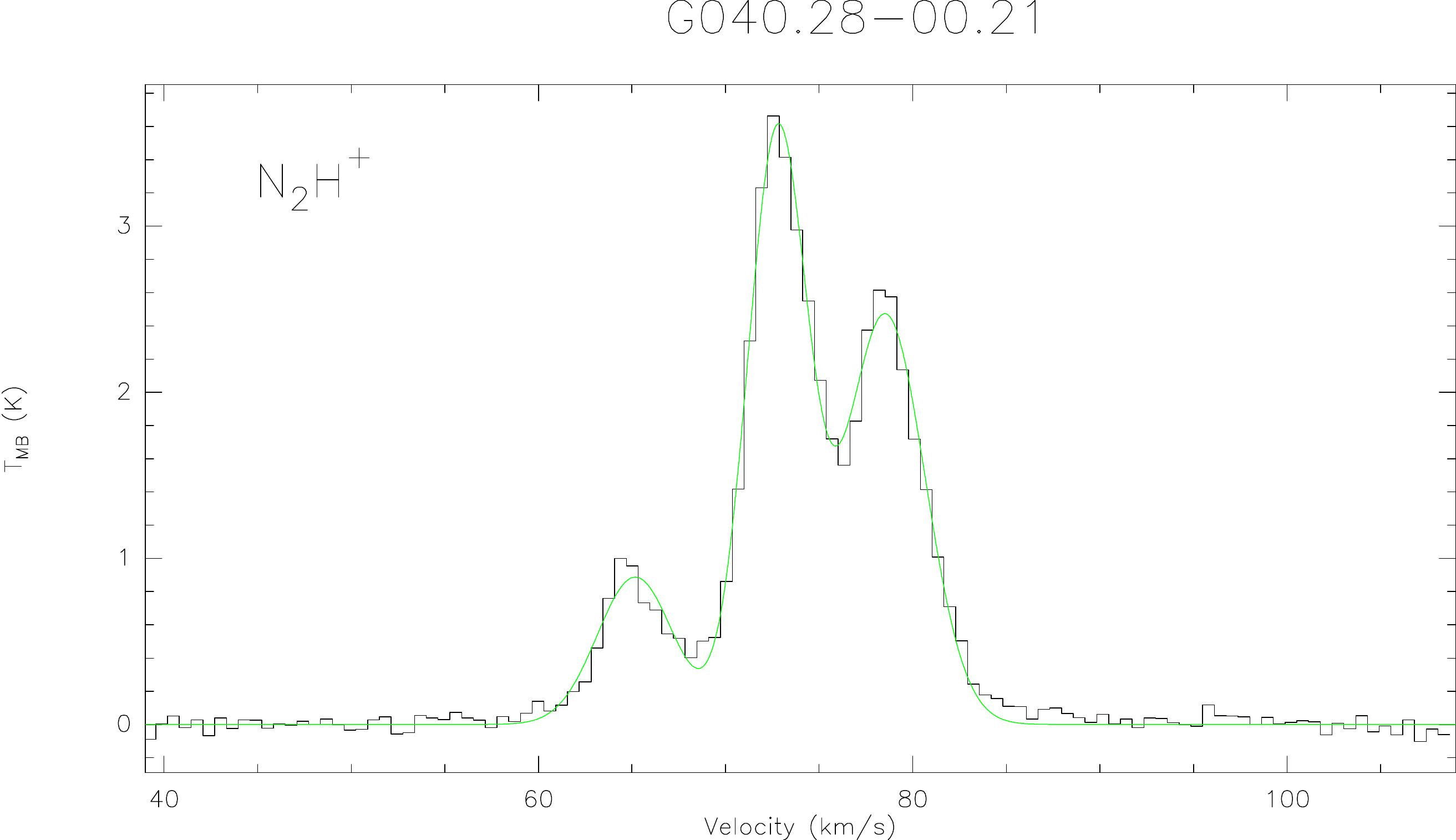}
    \includegraphics[width=0.3\textwidth]{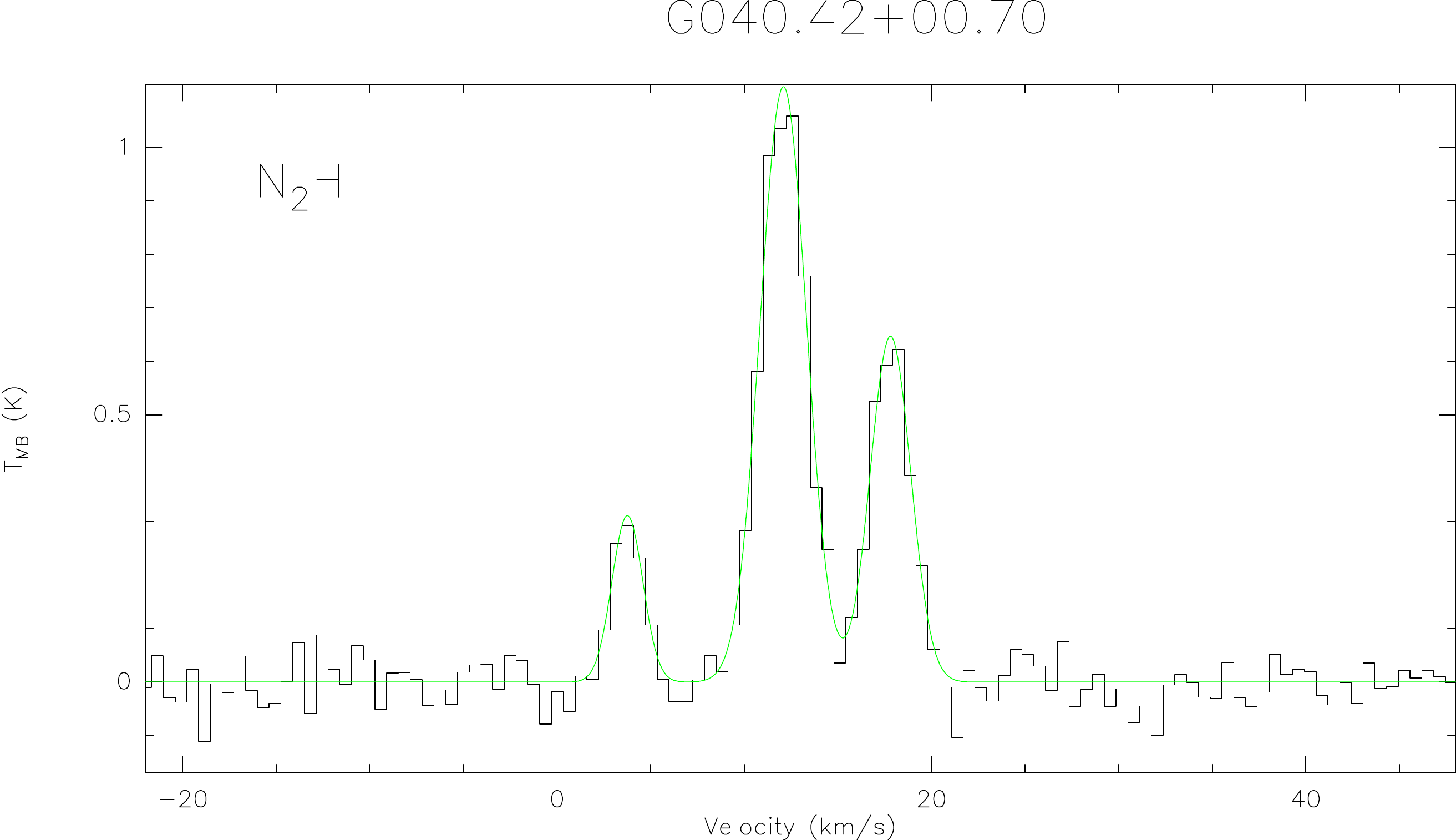}
    \includegraphics[width=0.3\textwidth]{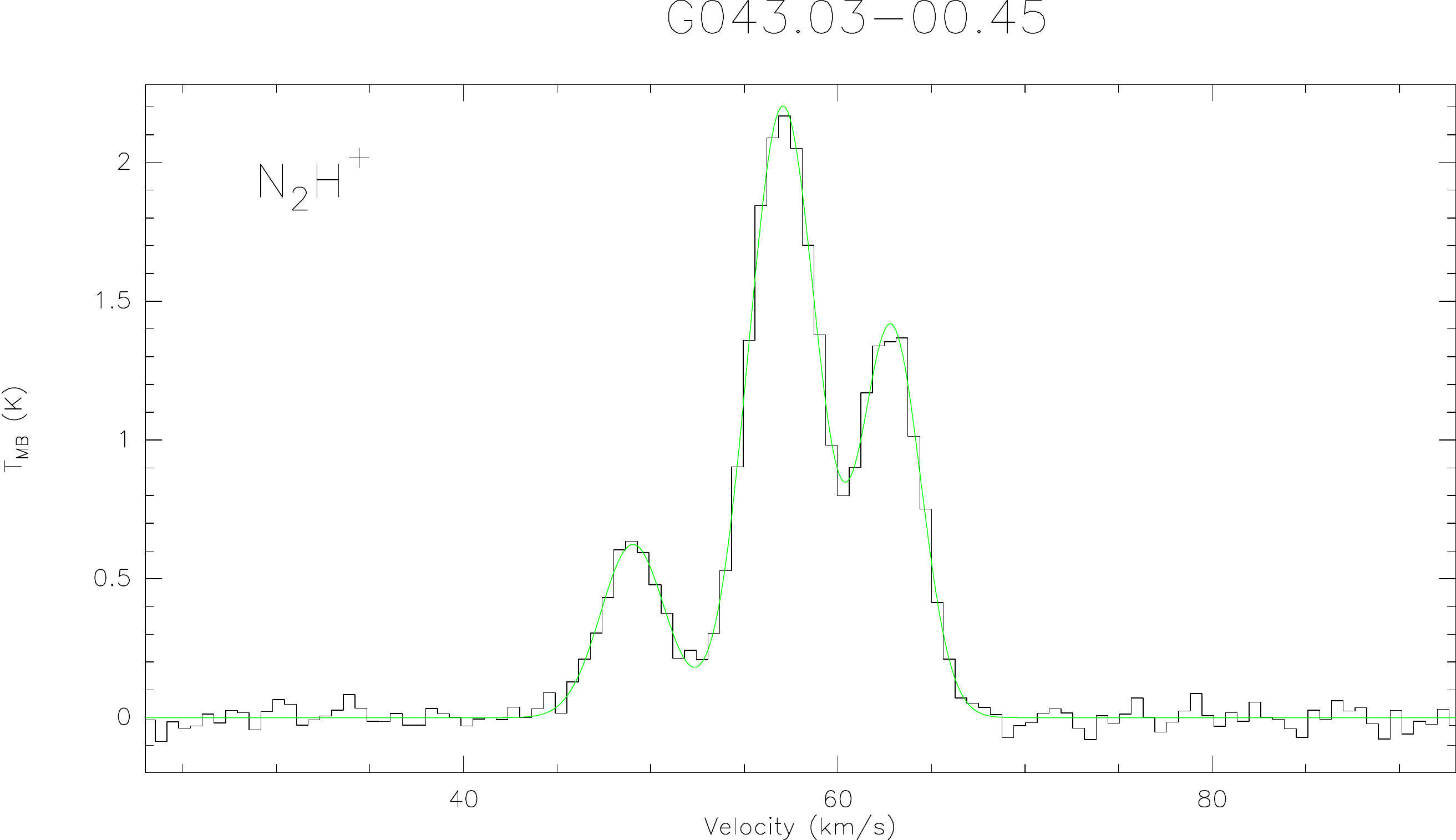} 
    \includegraphics[width=0.3\textwidth]{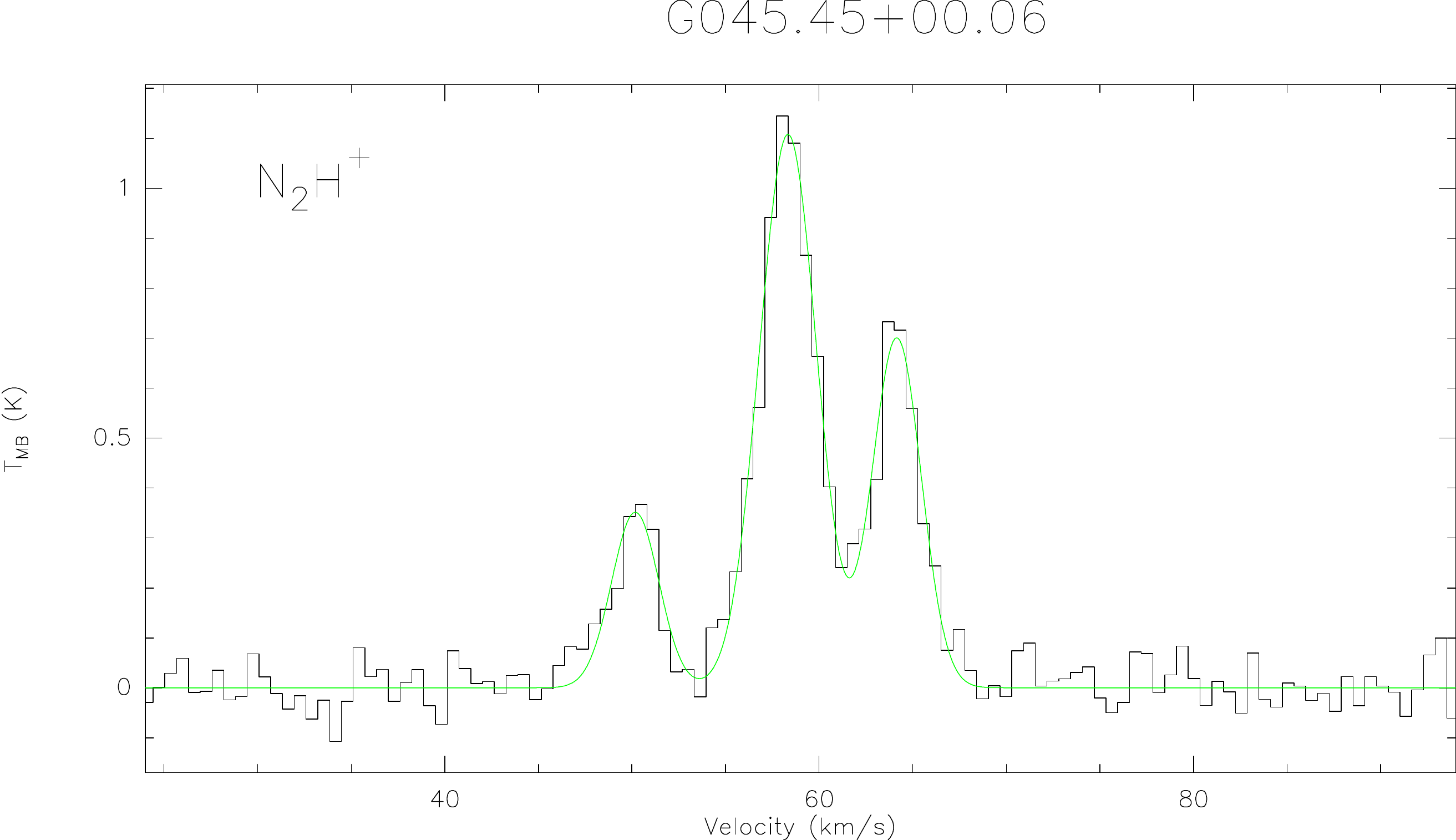}
    \includegraphics[width=0.3\textwidth]{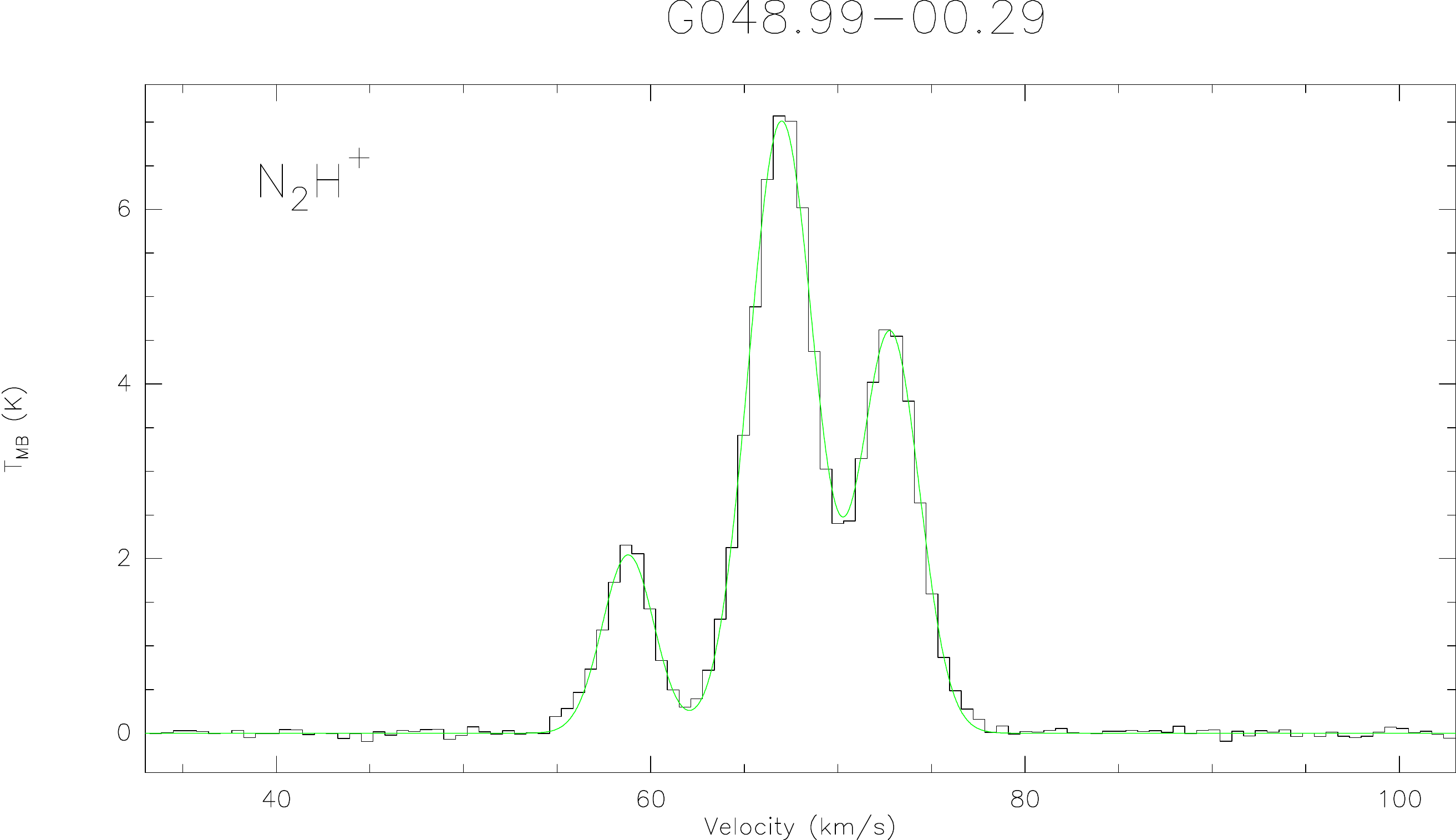}
    \includegraphics[width=0.3\textwidth]{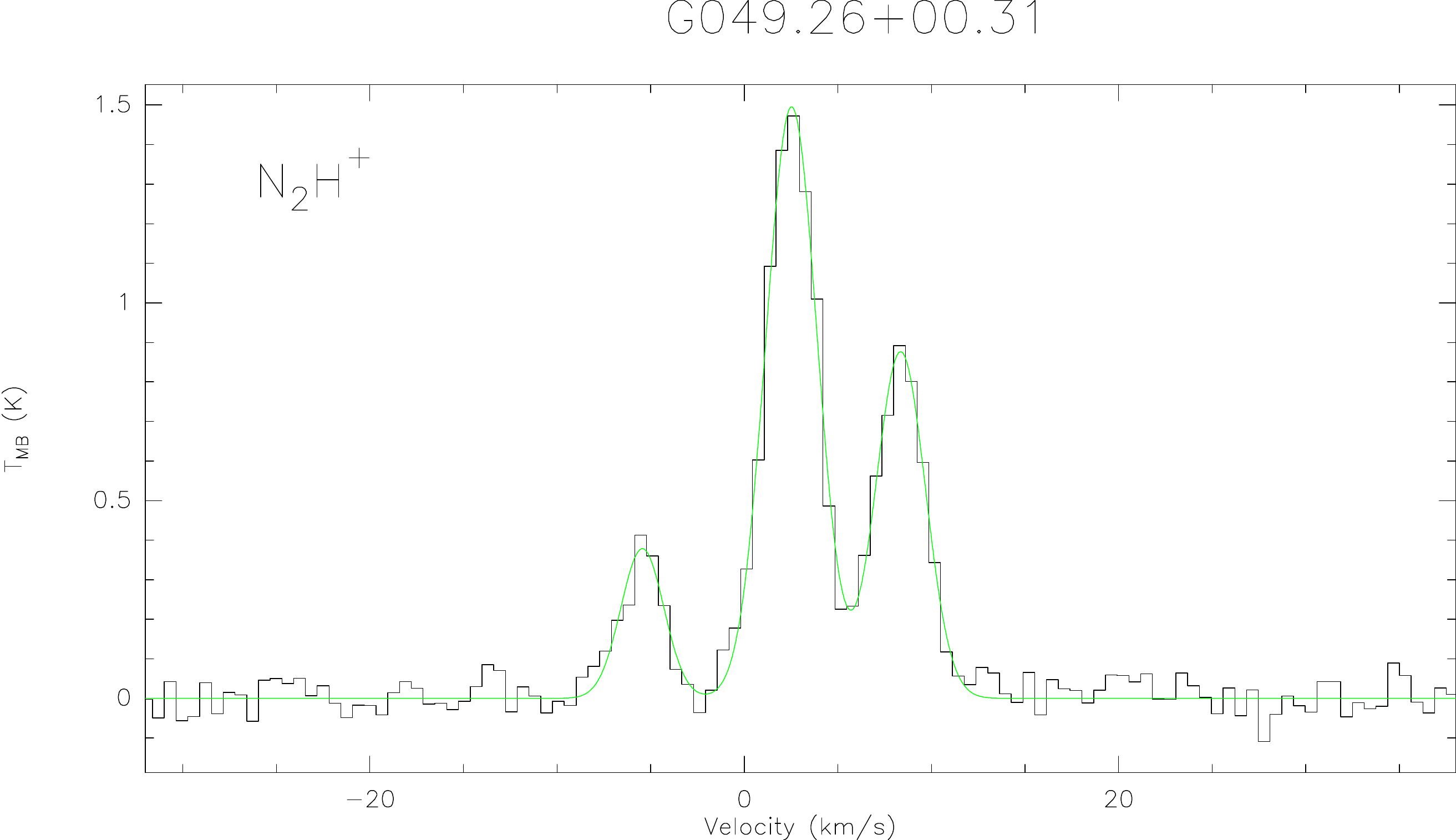} 
    
    \includegraphics[width=0.3\textwidth]{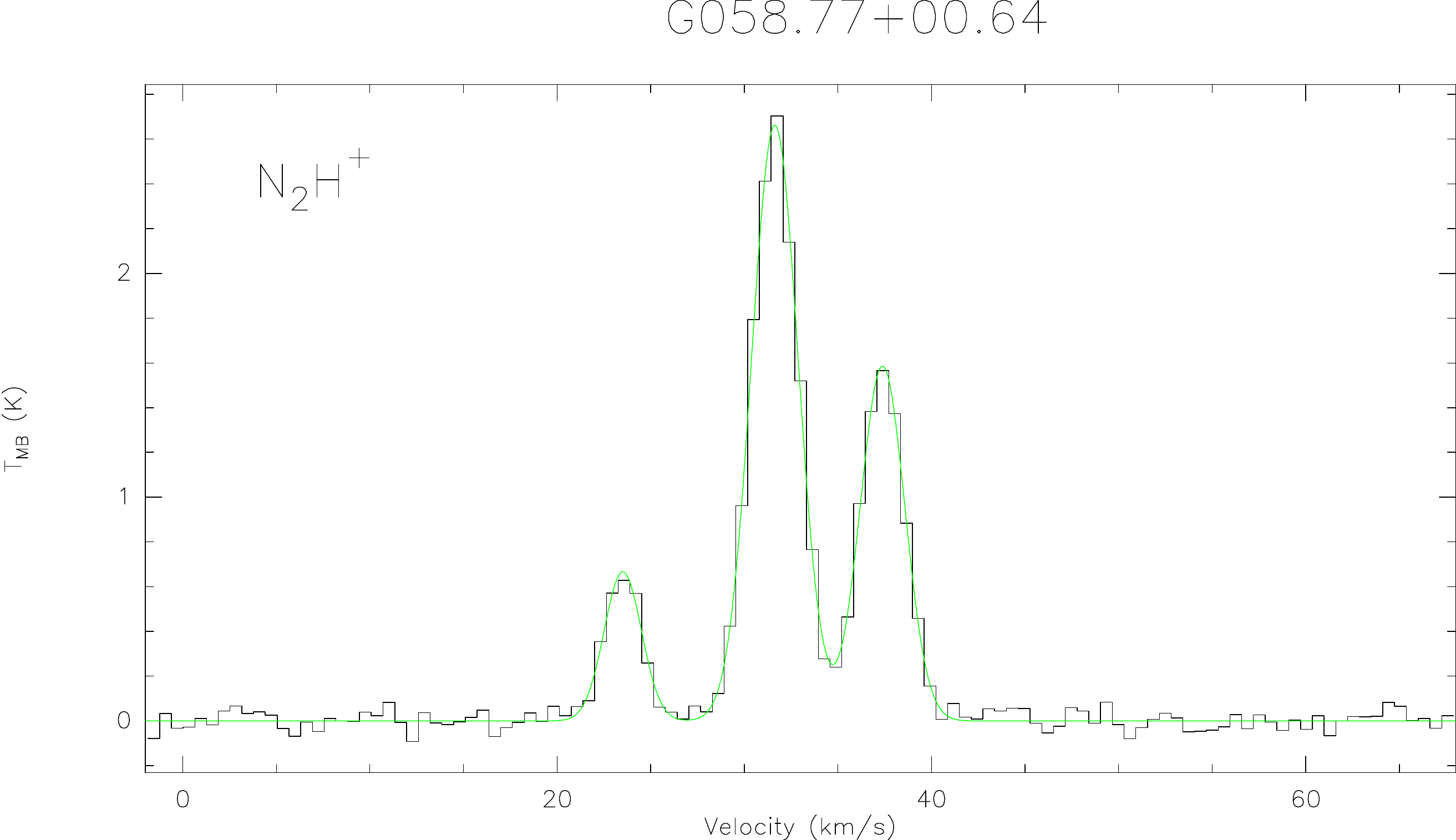}
    \includegraphics[width=0.3\textwidth]{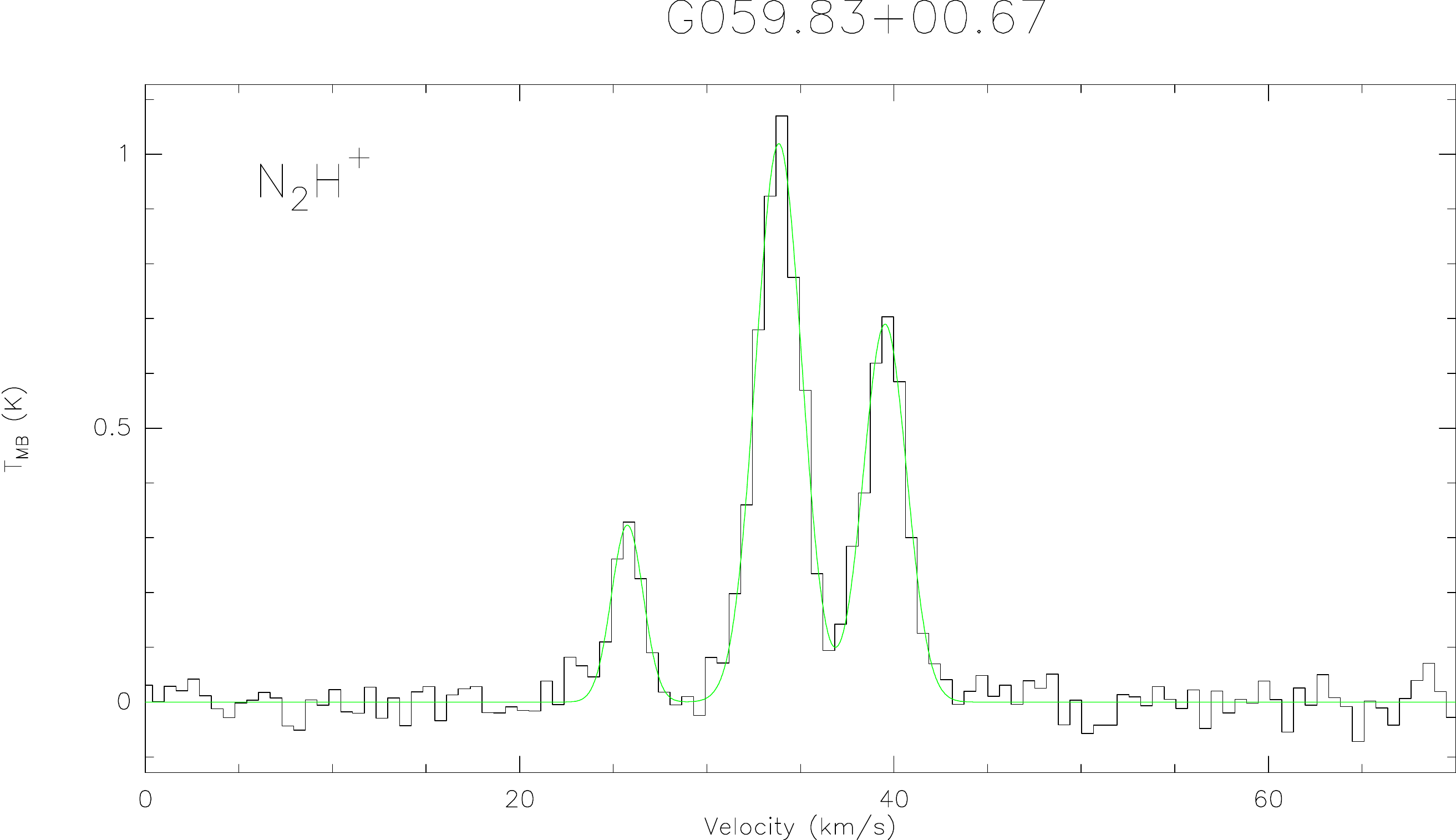}
    \includegraphics[width=0.3\textwidth]{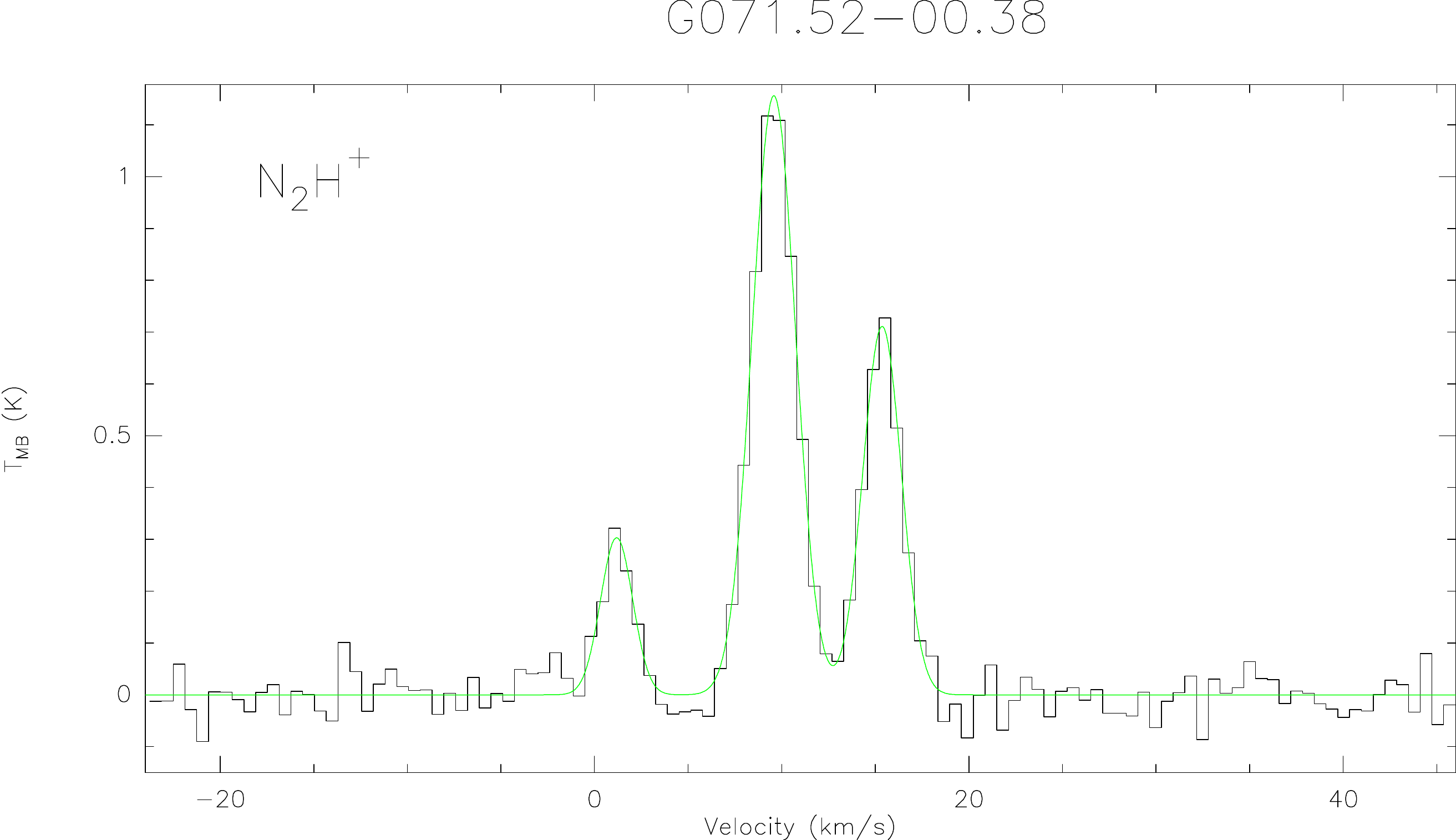} 
    \includegraphics[width=0.3\textwidth]{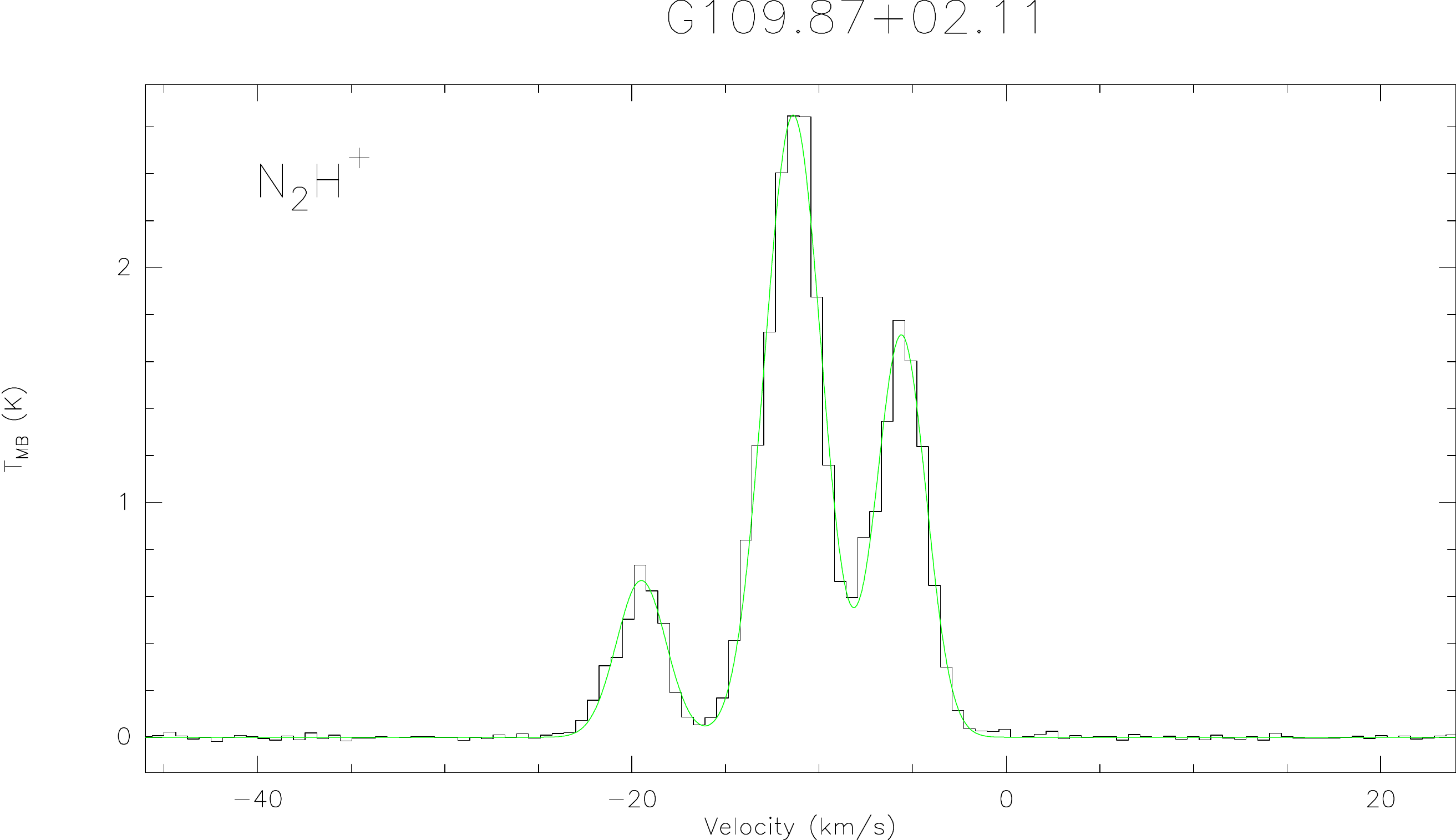}
    \caption{Continued.}
    \label{fig3}
\end{figure*}

\begin{figure}
\centering
    \includegraphics[width=0.5\textwidth]{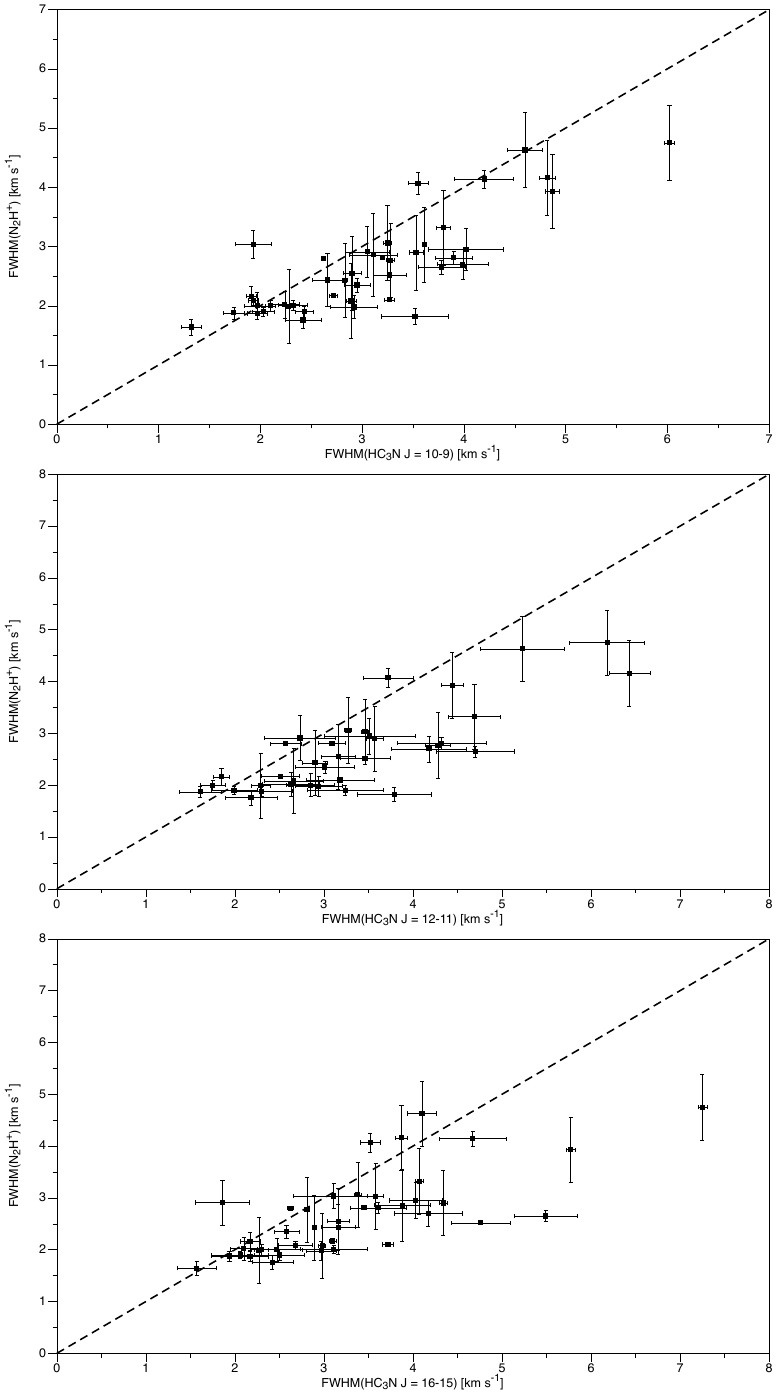}
    \caption{These panels show the comparison of line widths between HC$_{3}$N (J = 10$-$9, 12$-$11 and 16$-$15) and N$_{2}$H$^{+}$ (J = 1$-$0). The black dashed line means that both lines have the same line width.}
    \label{fig4}
\end{figure}

\begin{figure*}
    \centering
    \includegraphics[width=0.3\textwidth]{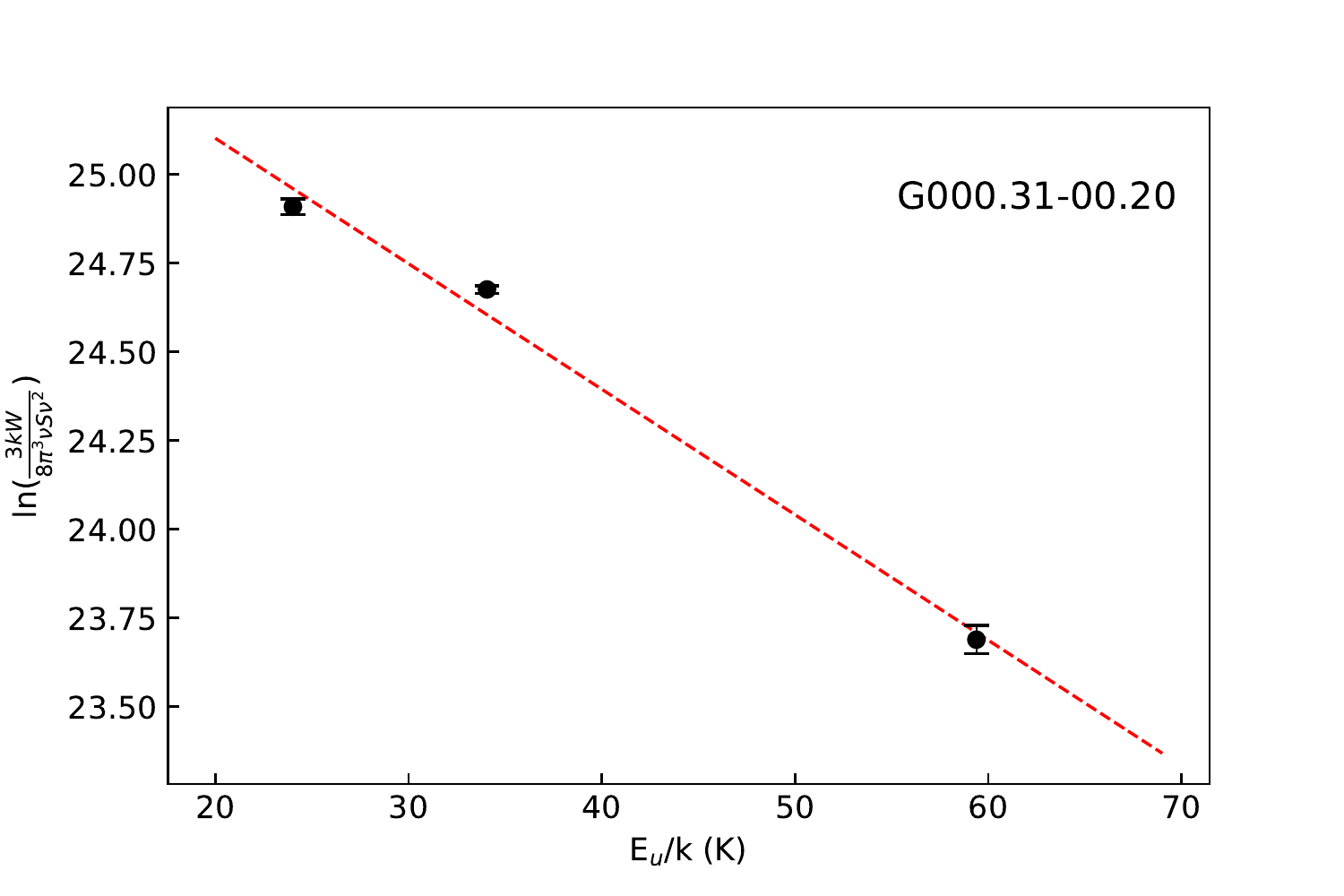}
    \includegraphics[width=0.3\textwidth]{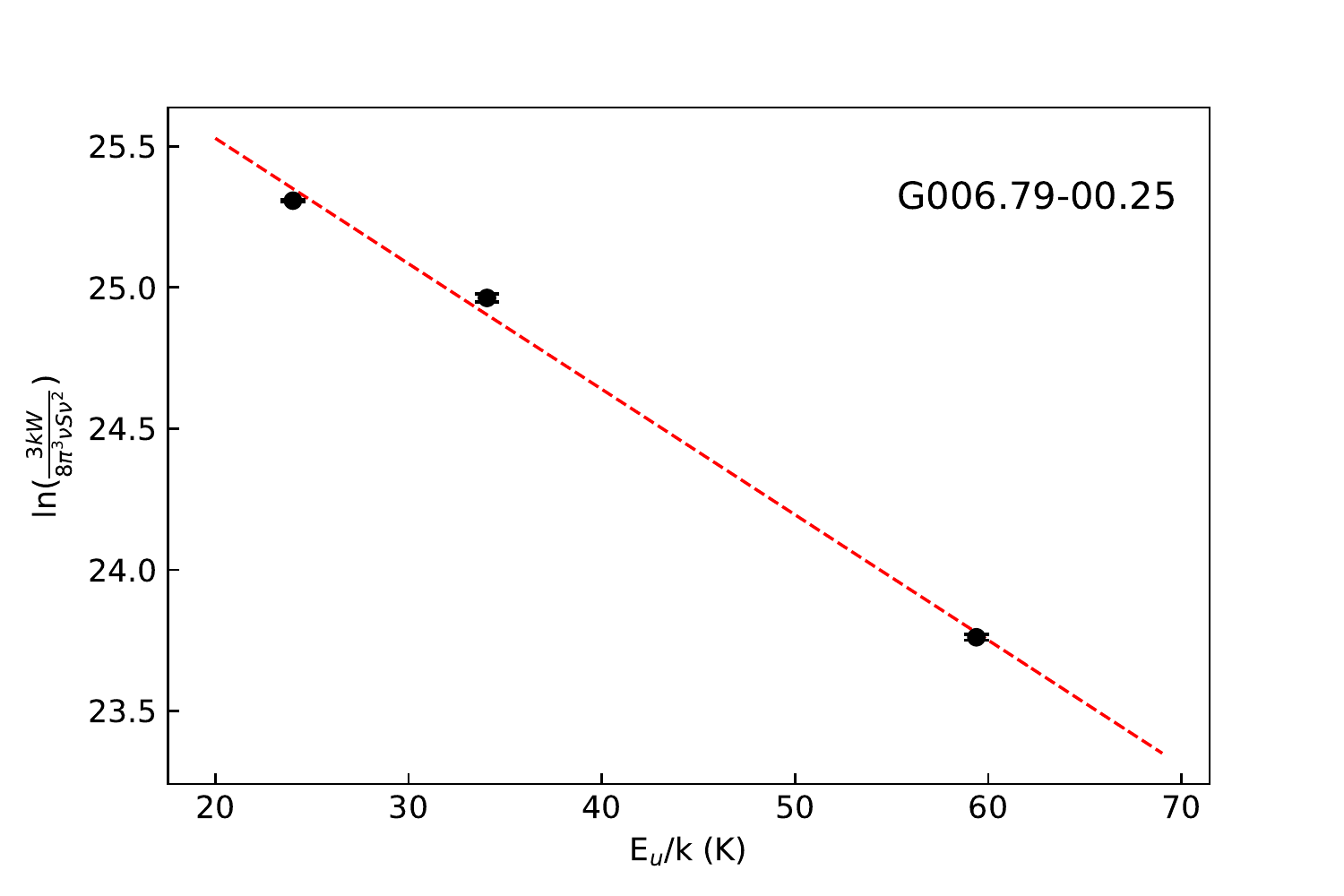}
    \includegraphics[width=0.3\textwidth]{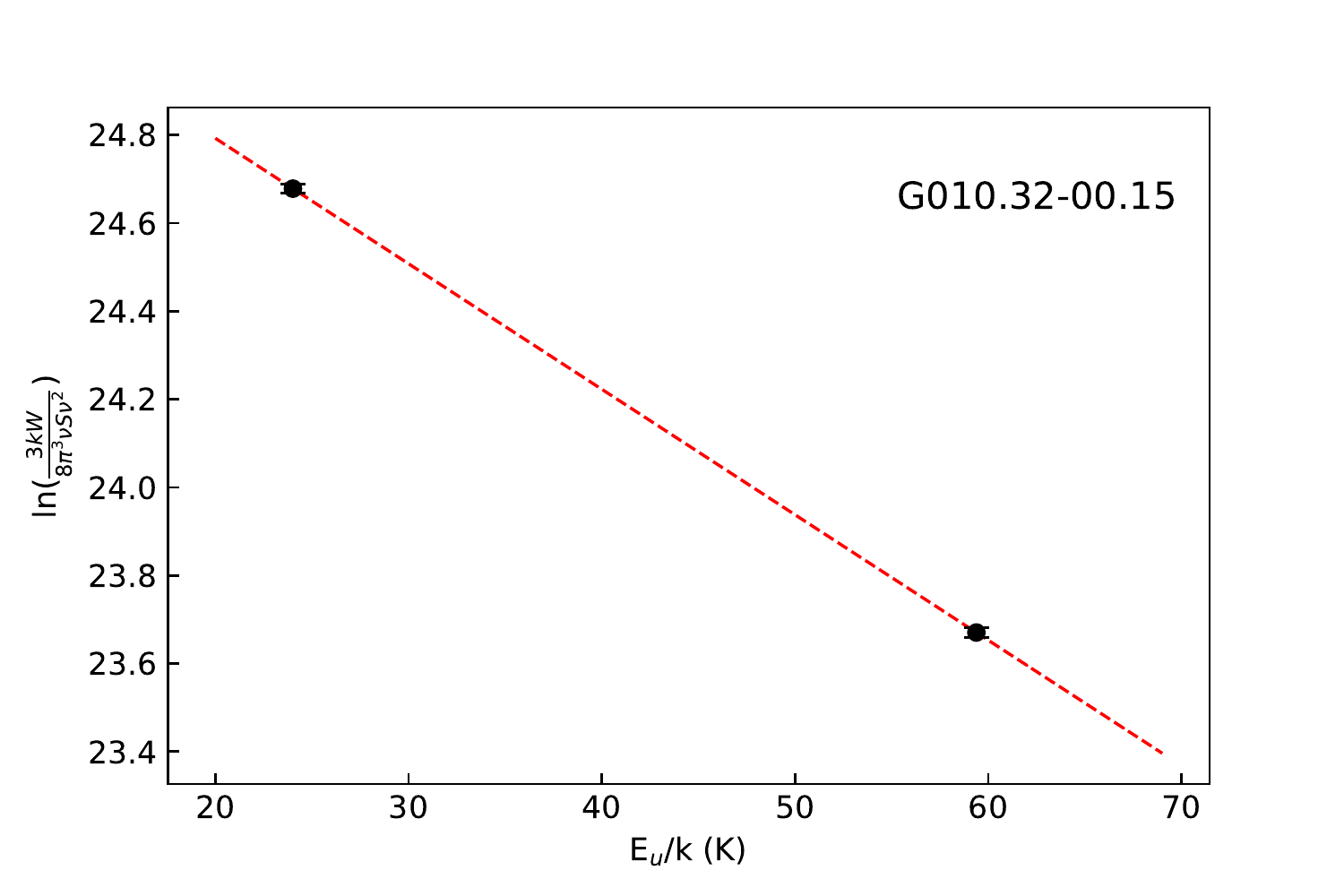} 
    \includegraphics[width=0.3\textwidth]{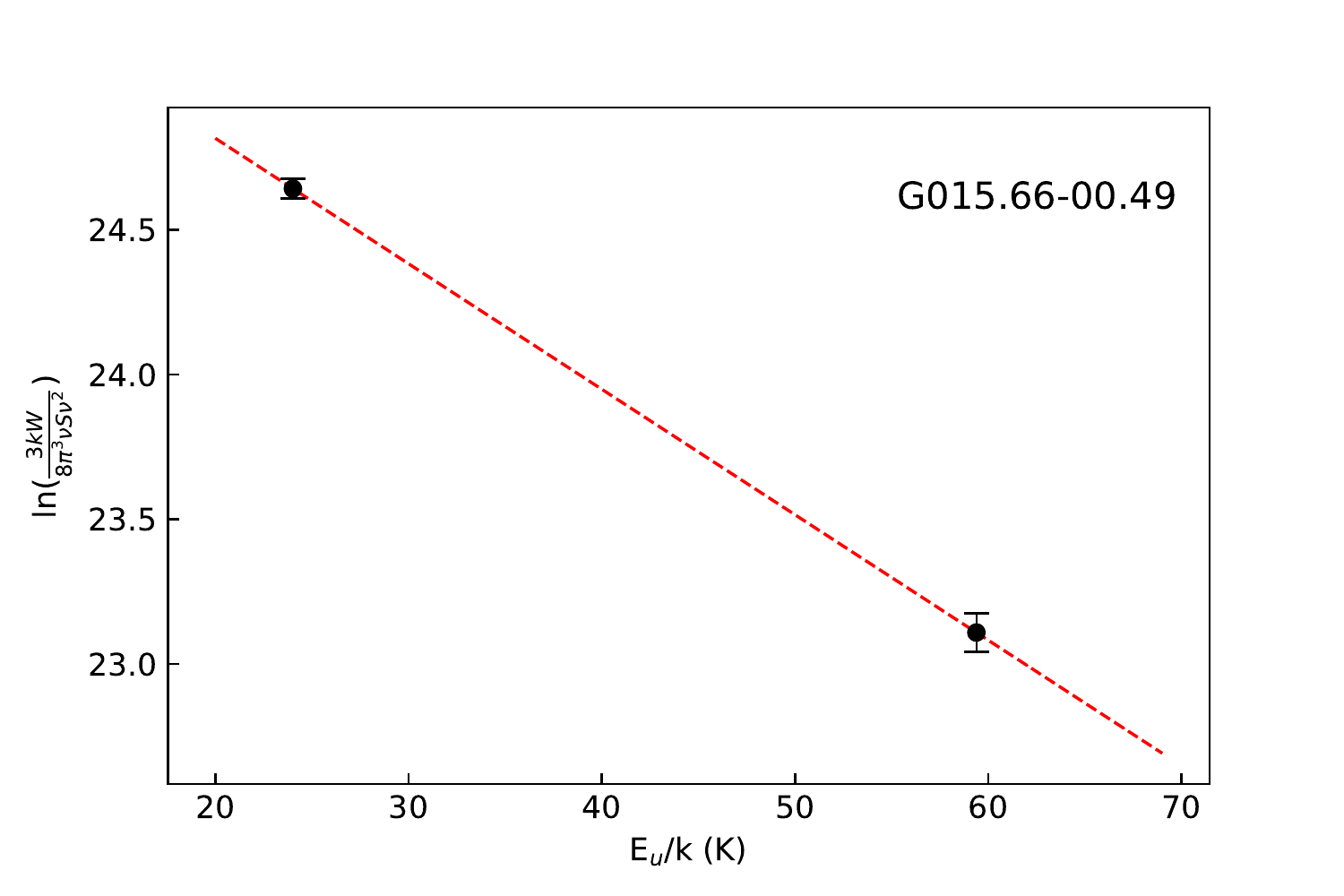}
    \includegraphics[width=0.3\textwidth]{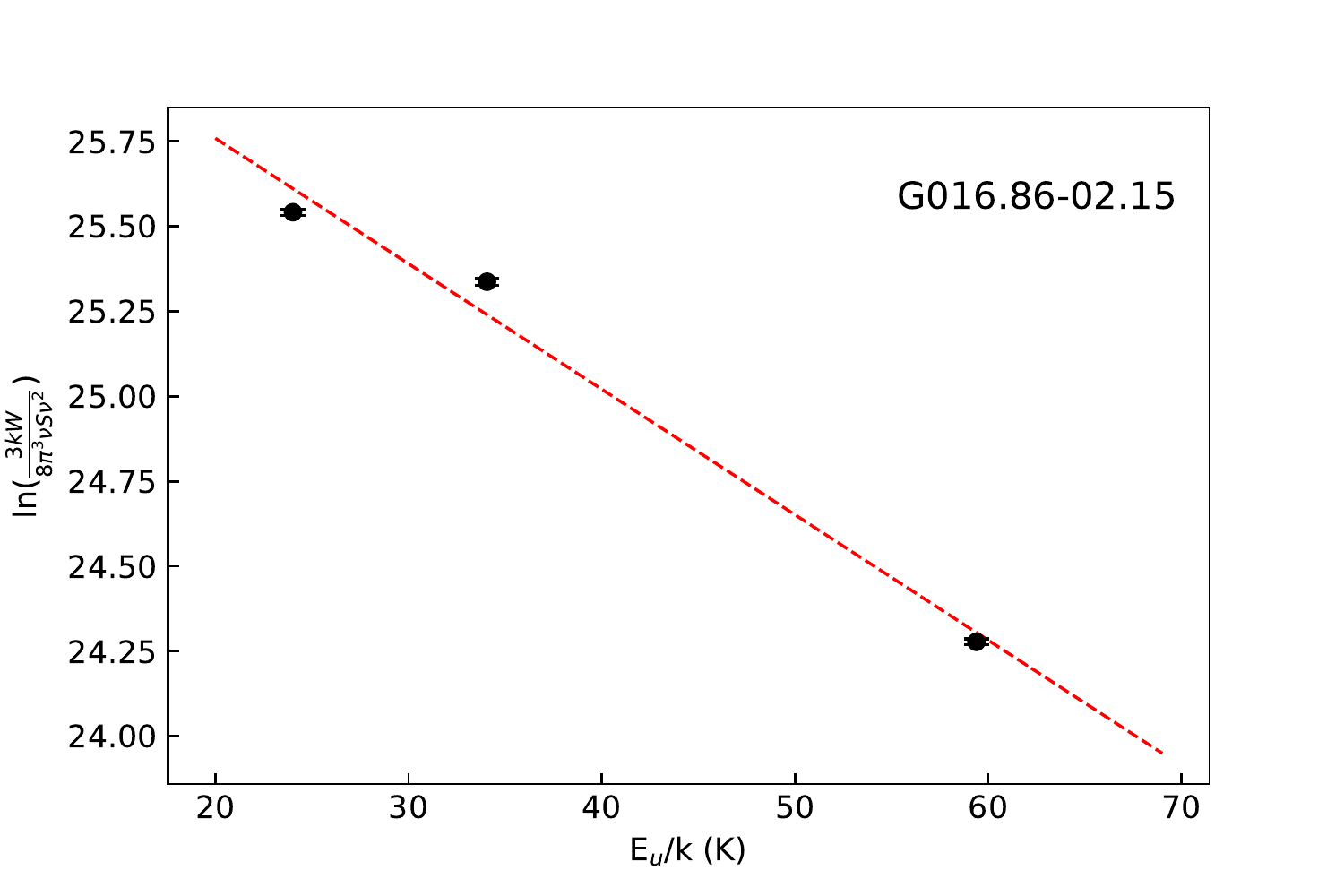}
    \includegraphics[width=0.3\textwidth]{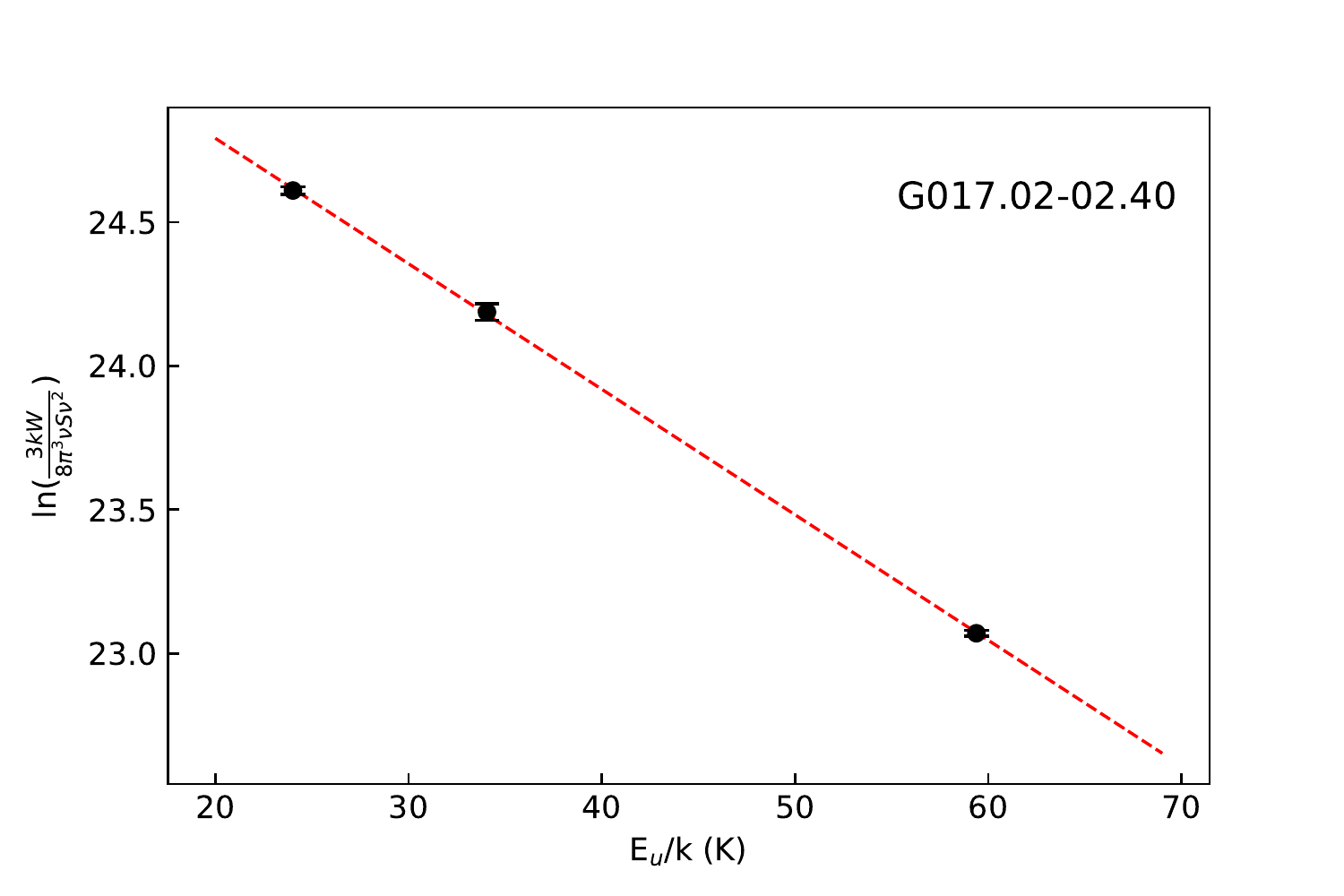} 
    \includegraphics[width=0.3\textwidth]{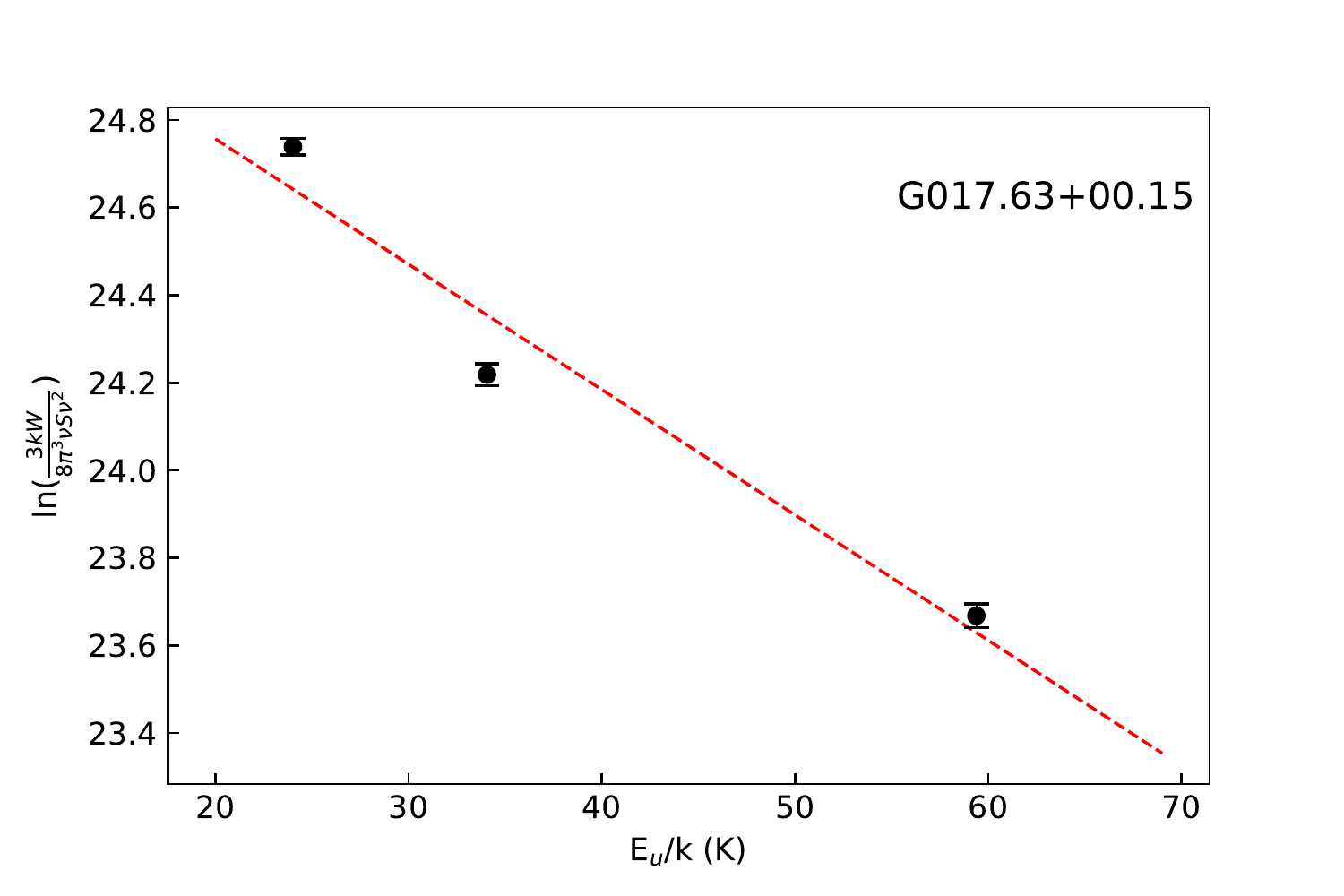}
    \includegraphics[width=0.3\textwidth]{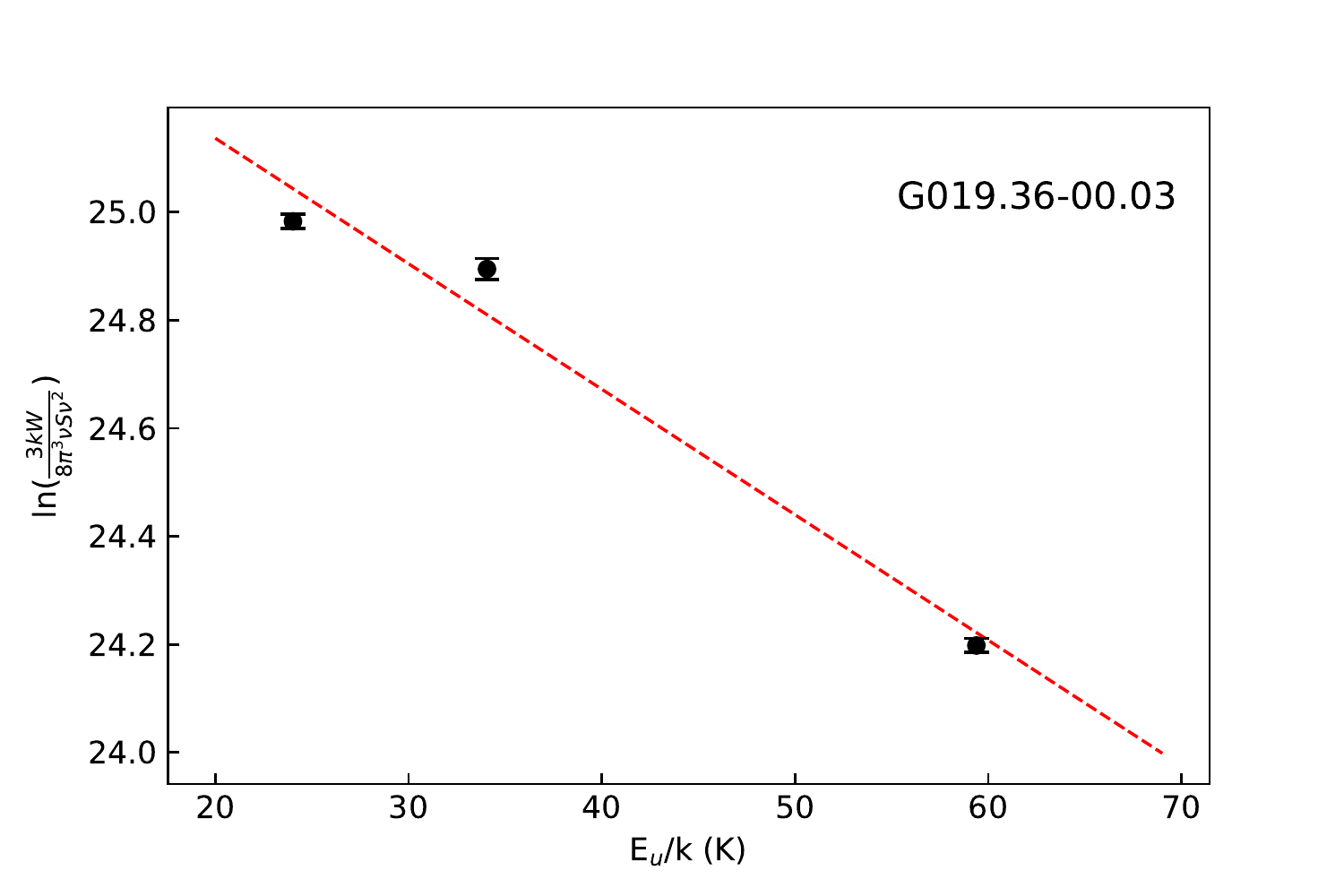}
    \includegraphics[width=0.3\textwidth]{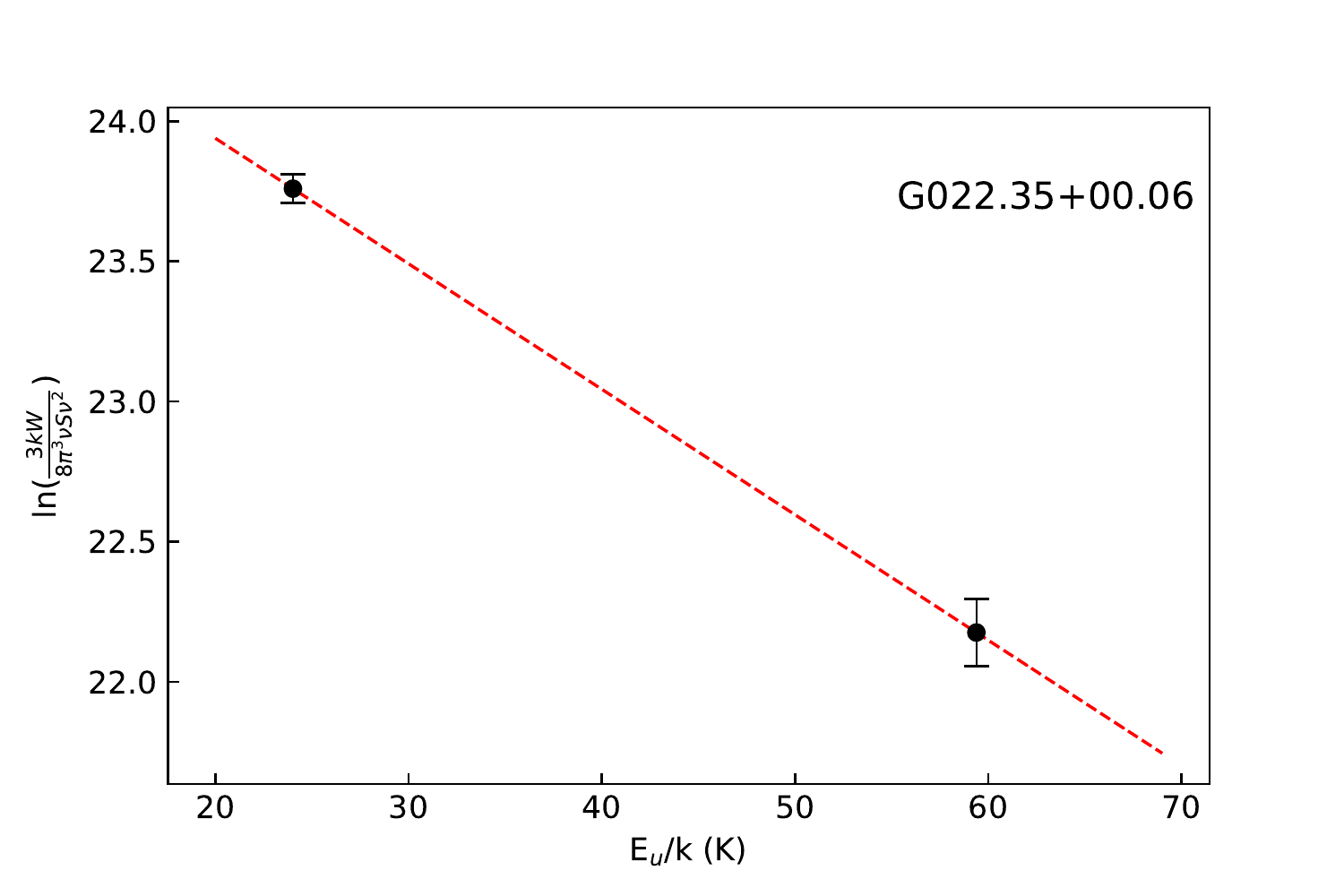} 
    \includegraphics[width=0.3\textwidth]{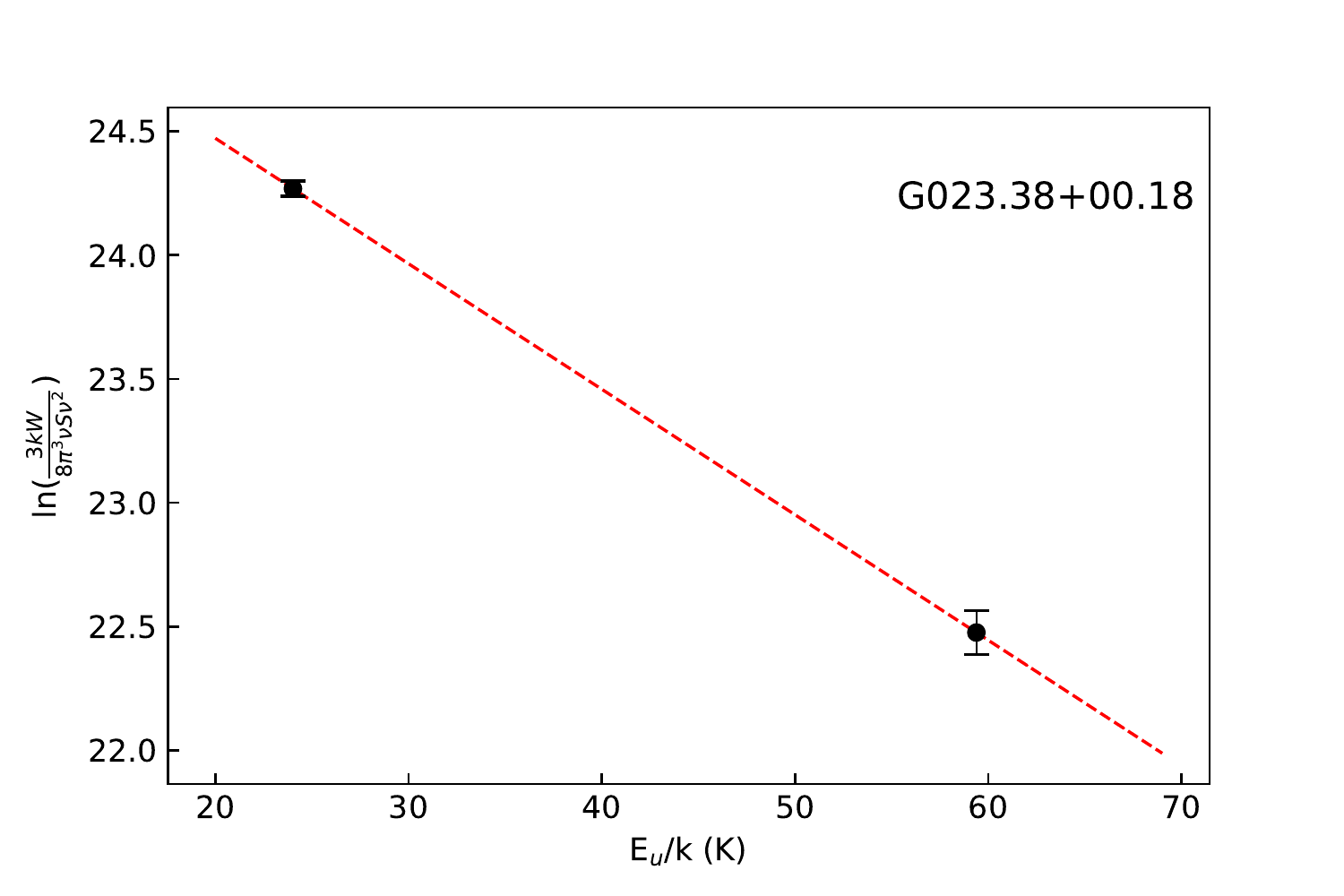}
    \includegraphics[width=0.3\textwidth]{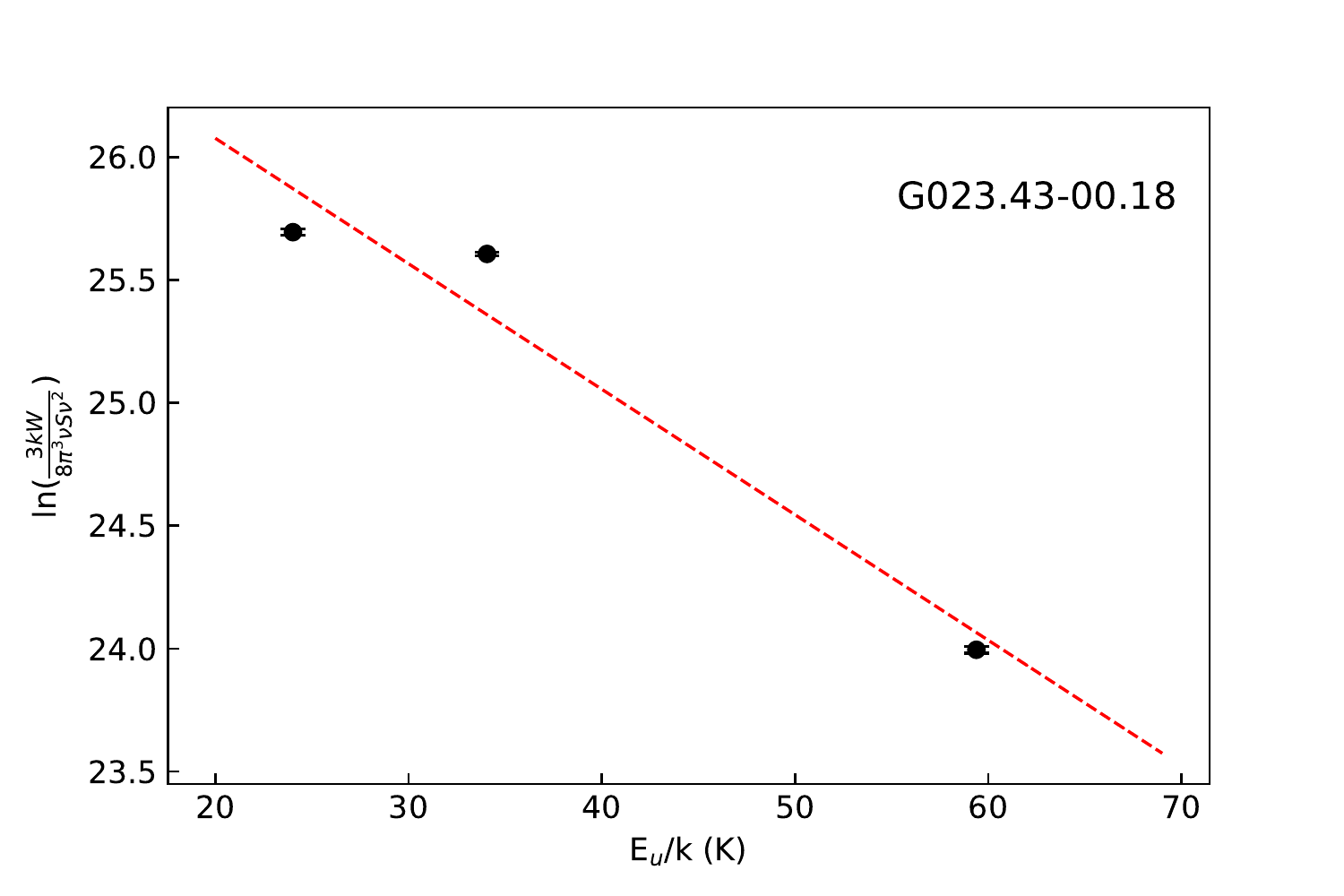}
    \includegraphics[width=0.3\textwidth]{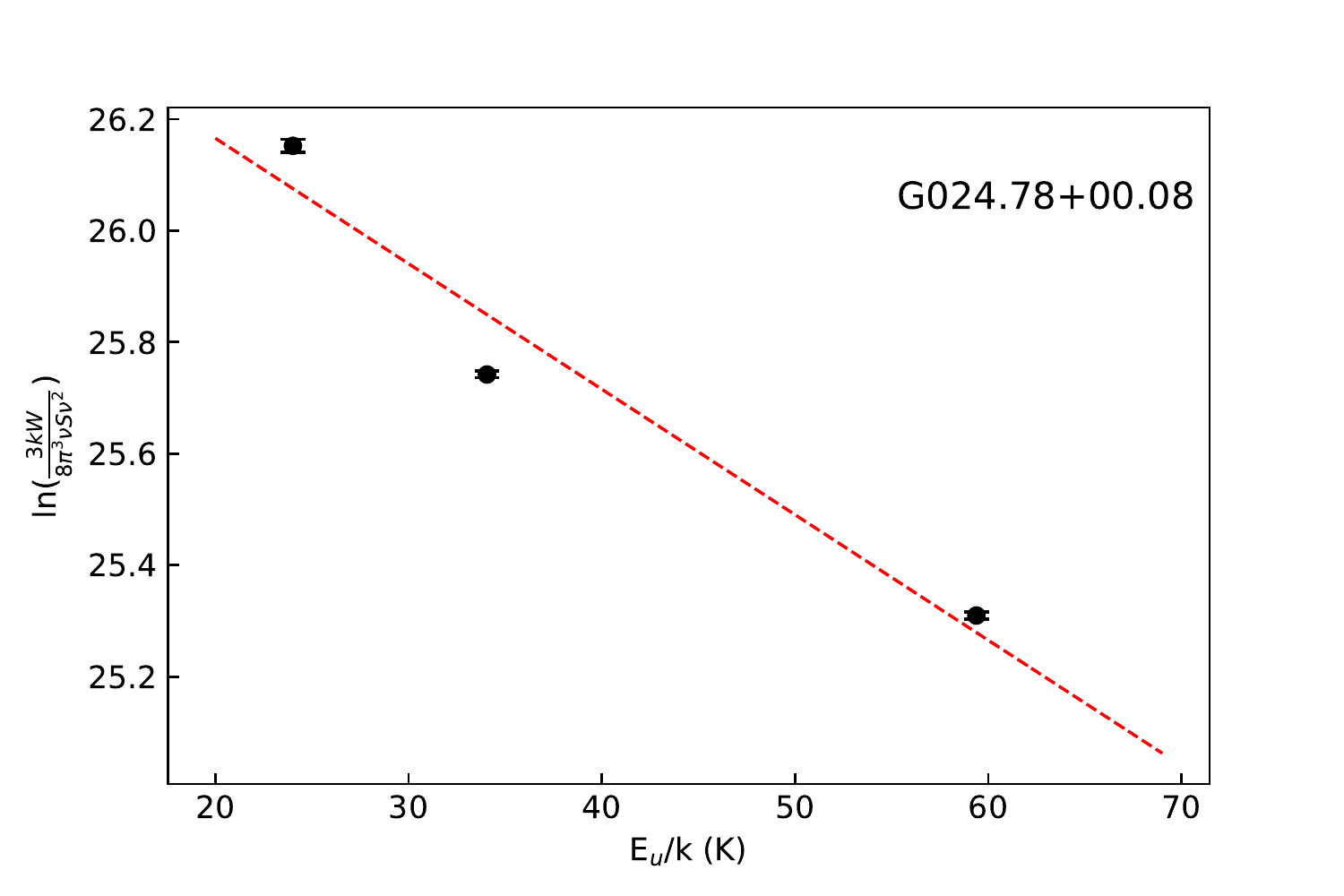} 
    \includegraphics[width=0.3\textwidth]{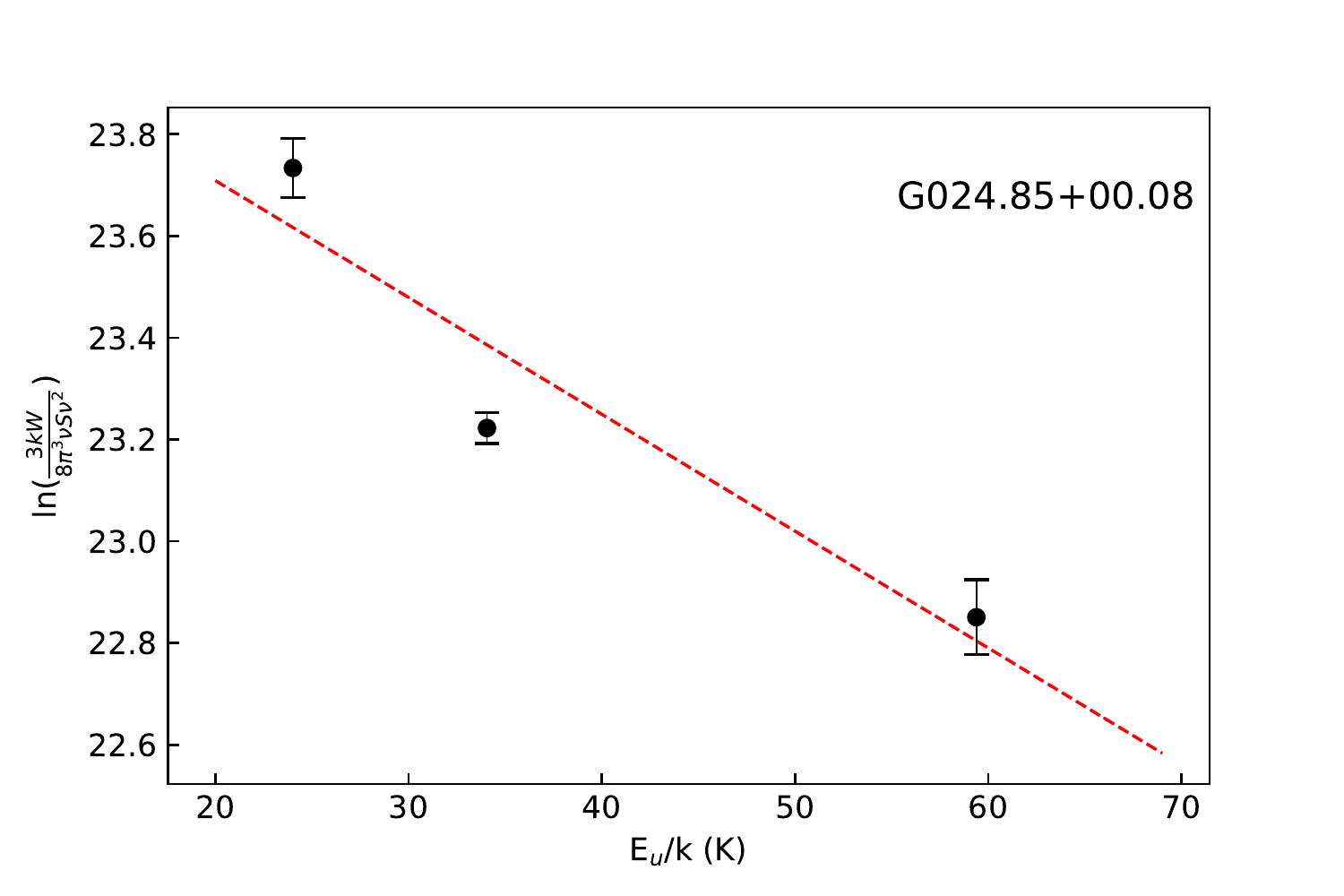}
    \includegraphics[width=0.3\textwidth]{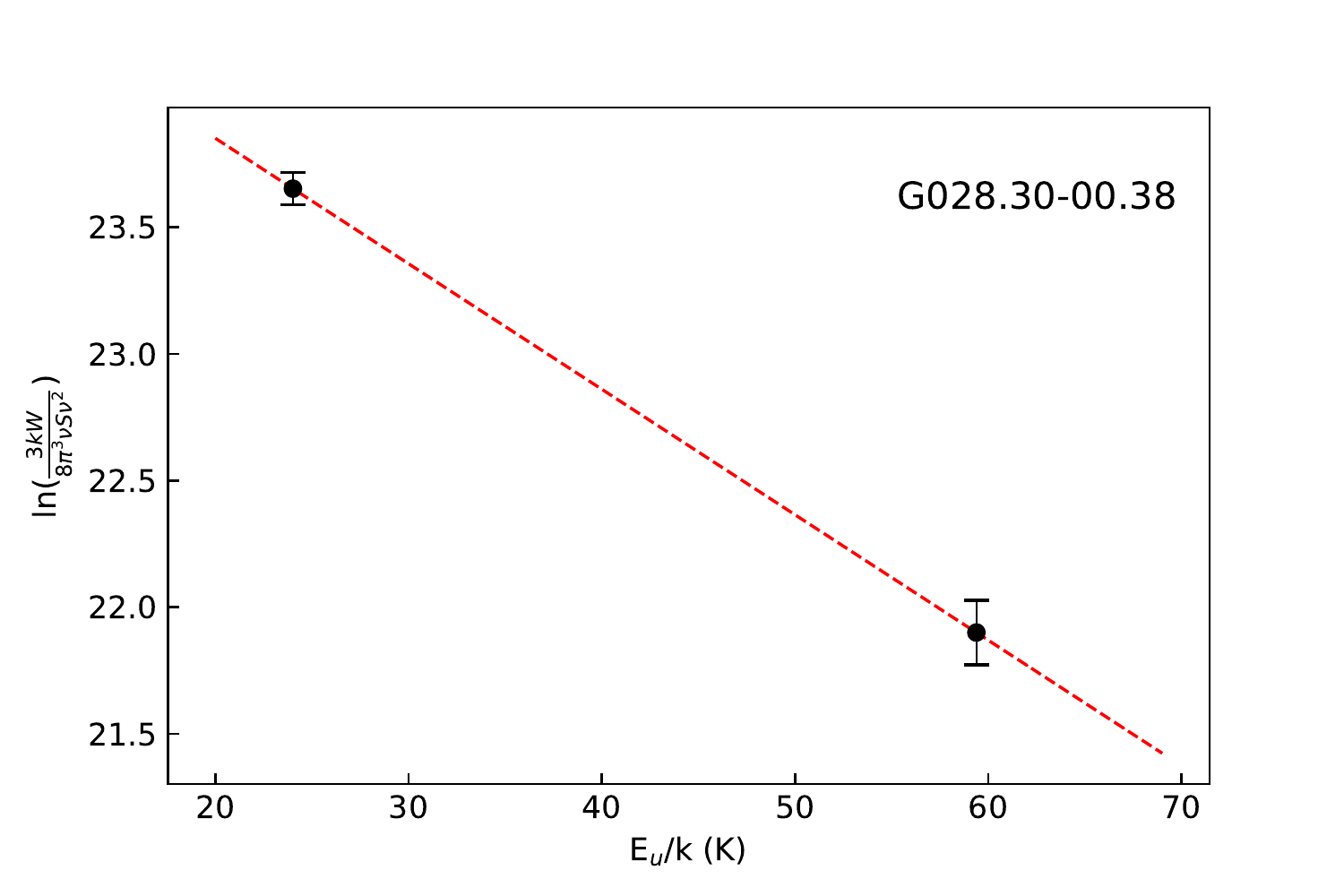}
    \includegraphics[width=0.3\textwidth]{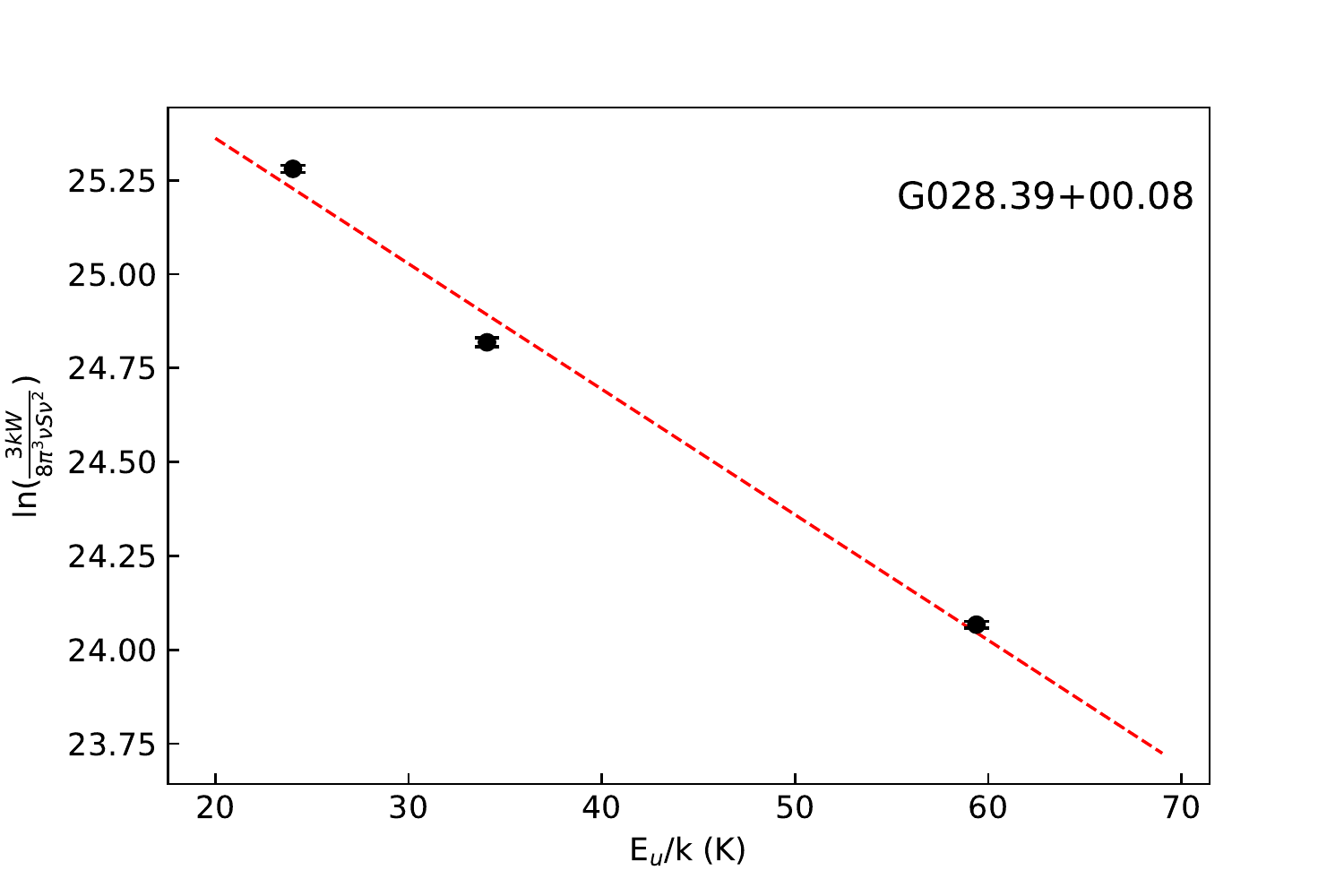} 
    \includegraphics[width=0.3\textwidth]{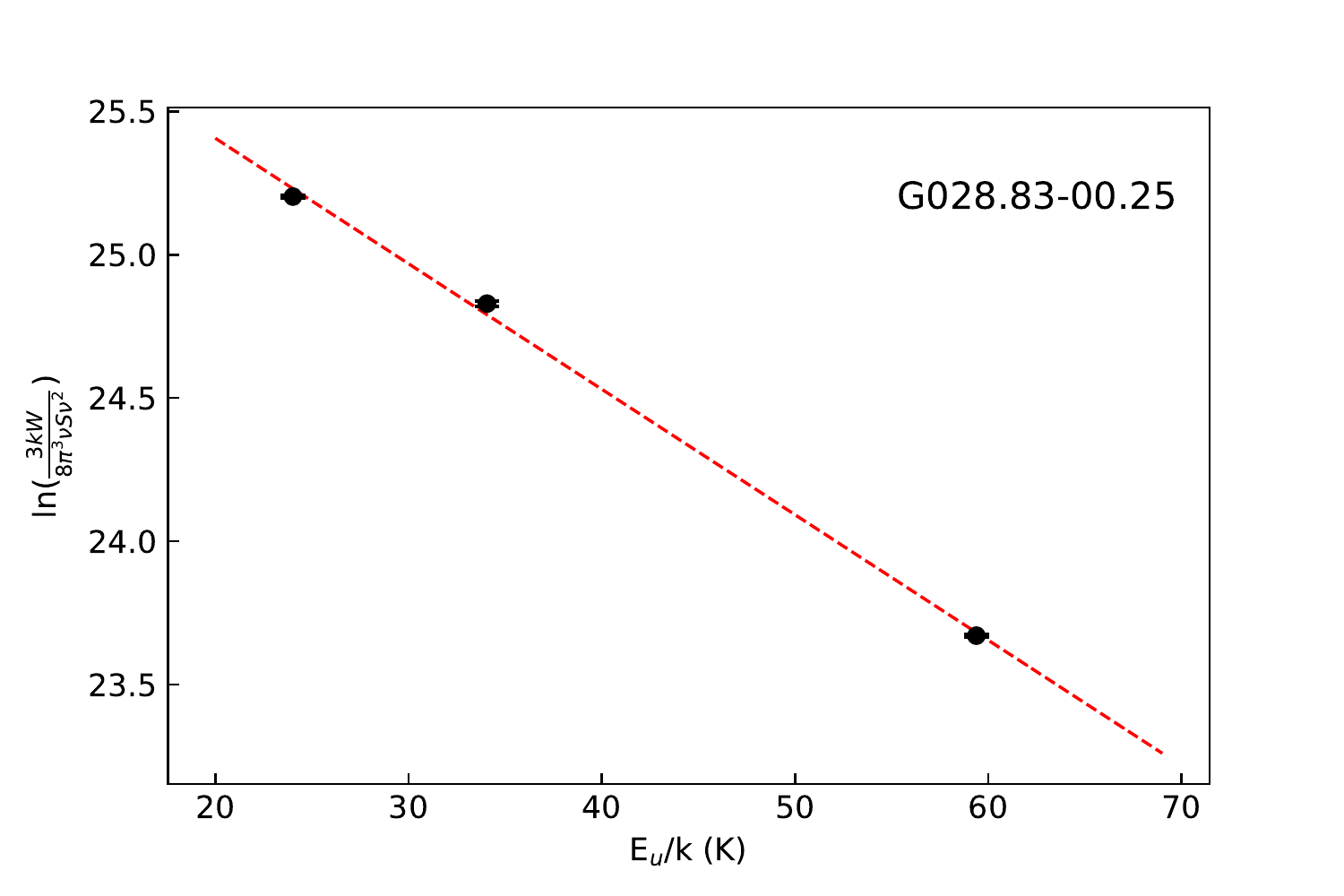}
    \includegraphics[width=0.3\textwidth]{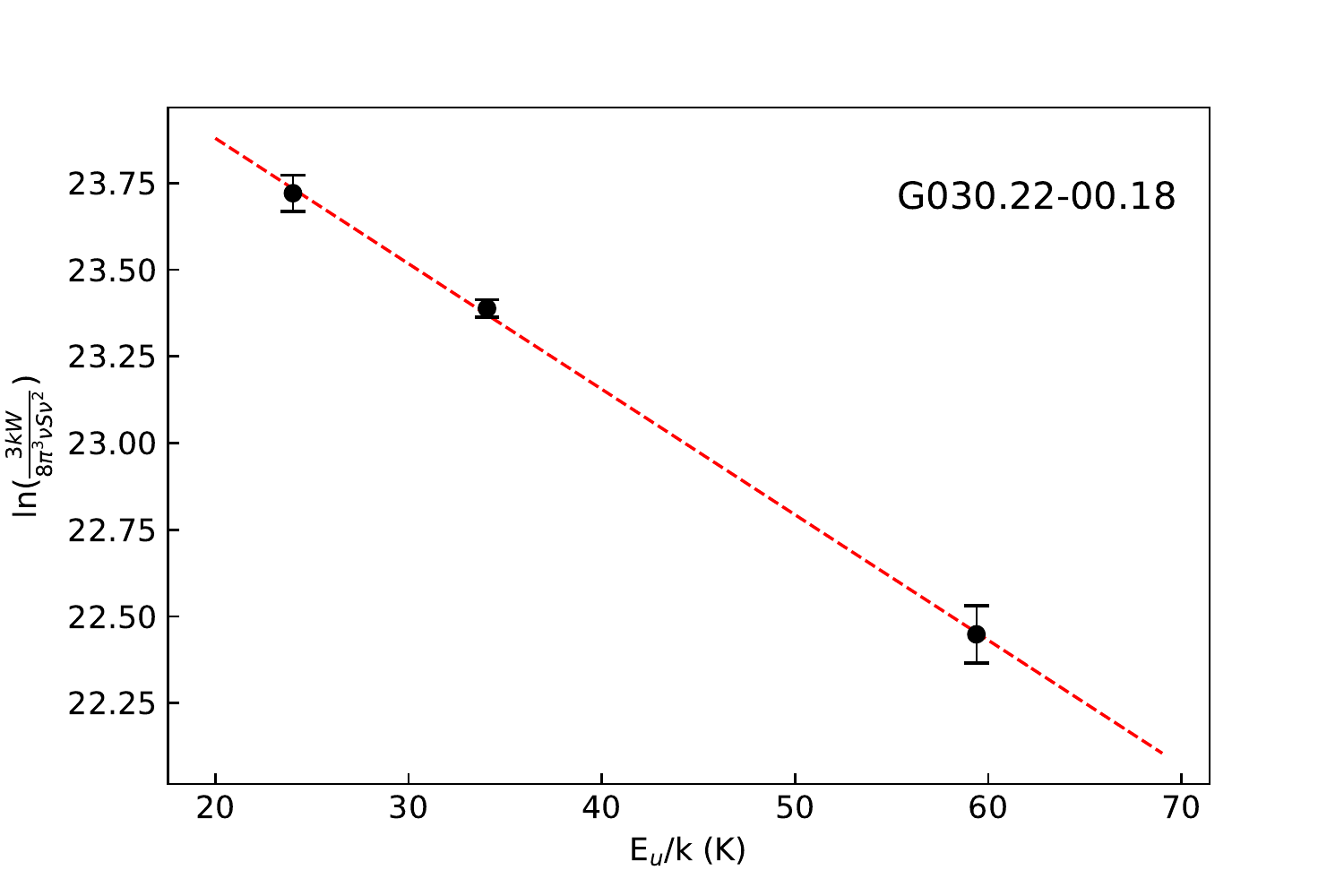}
    \includegraphics[width=0.3\textwidth]{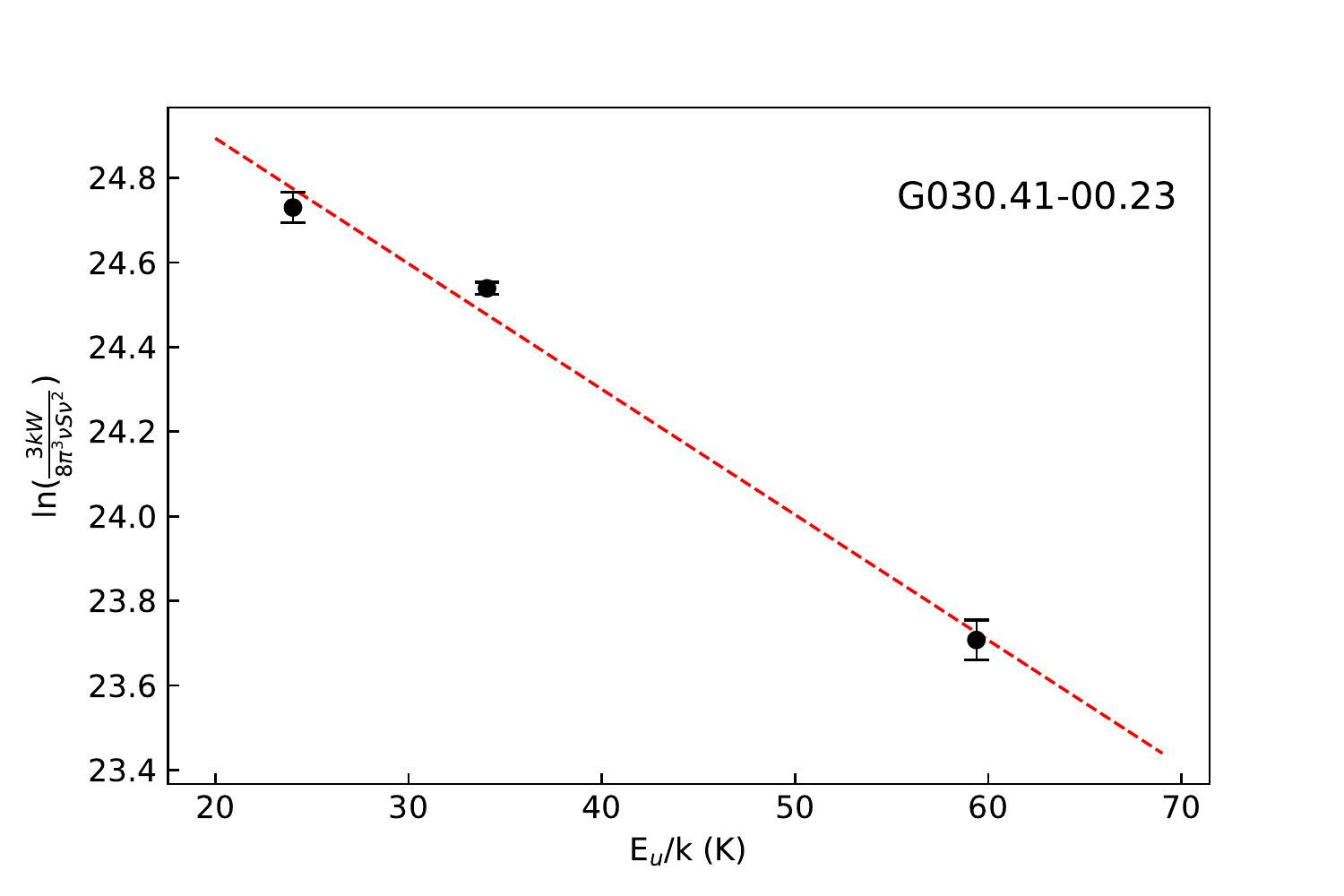} 
    \caption{Rotational diagrams of HC$_{3}$N for 40 sources with detections of all three HC$_{3}$N lines, where HC$_{3}$N (J = 12$-11$) was corrected by the beam dilution.}
\end{figure*}
    
\addtocounter{figure}{-1}
\begin{figure*}    
    \centering
    \includegraphics[width=0.3\textwidth]{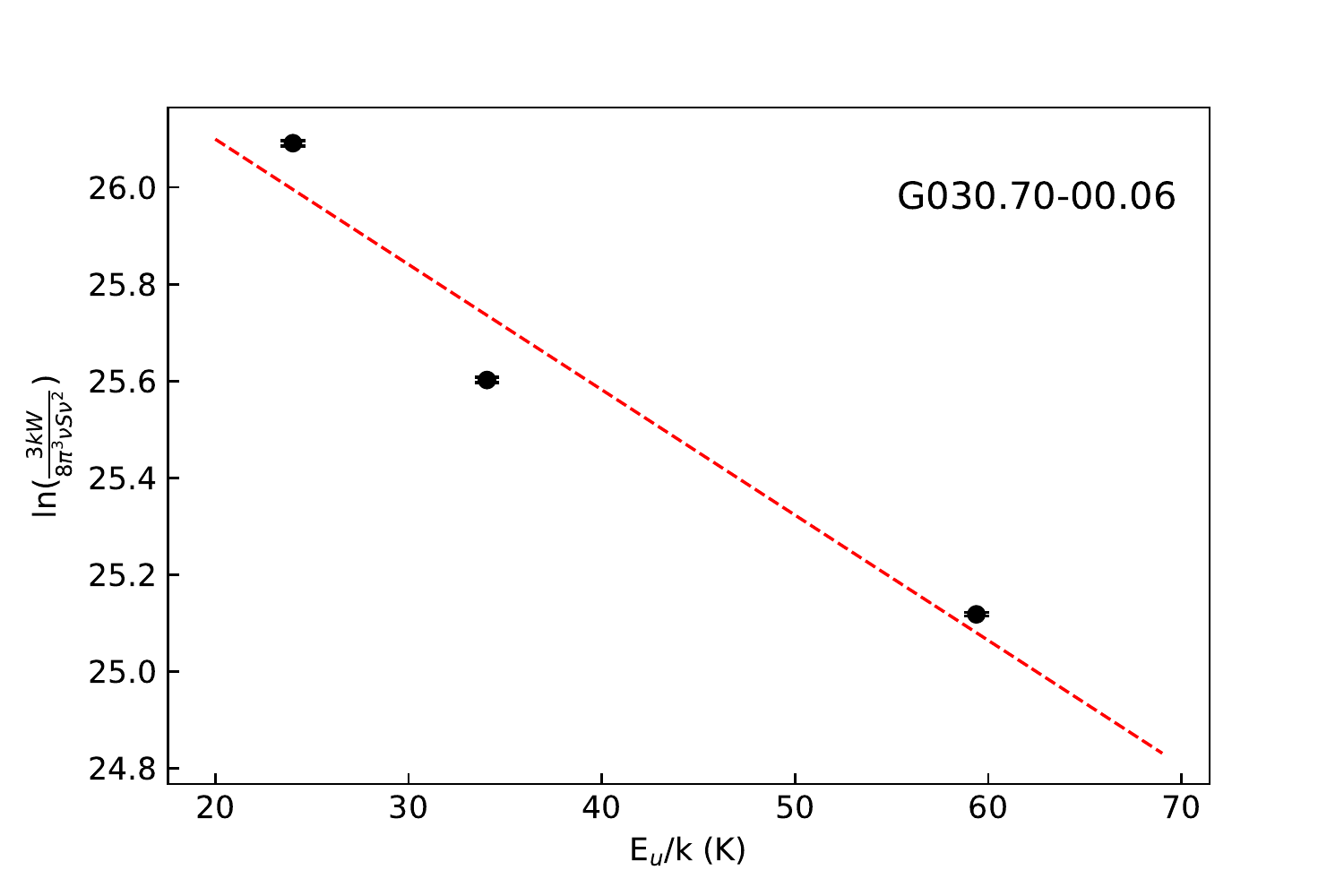}
    \includegraphics[width=0.3\textwidth]{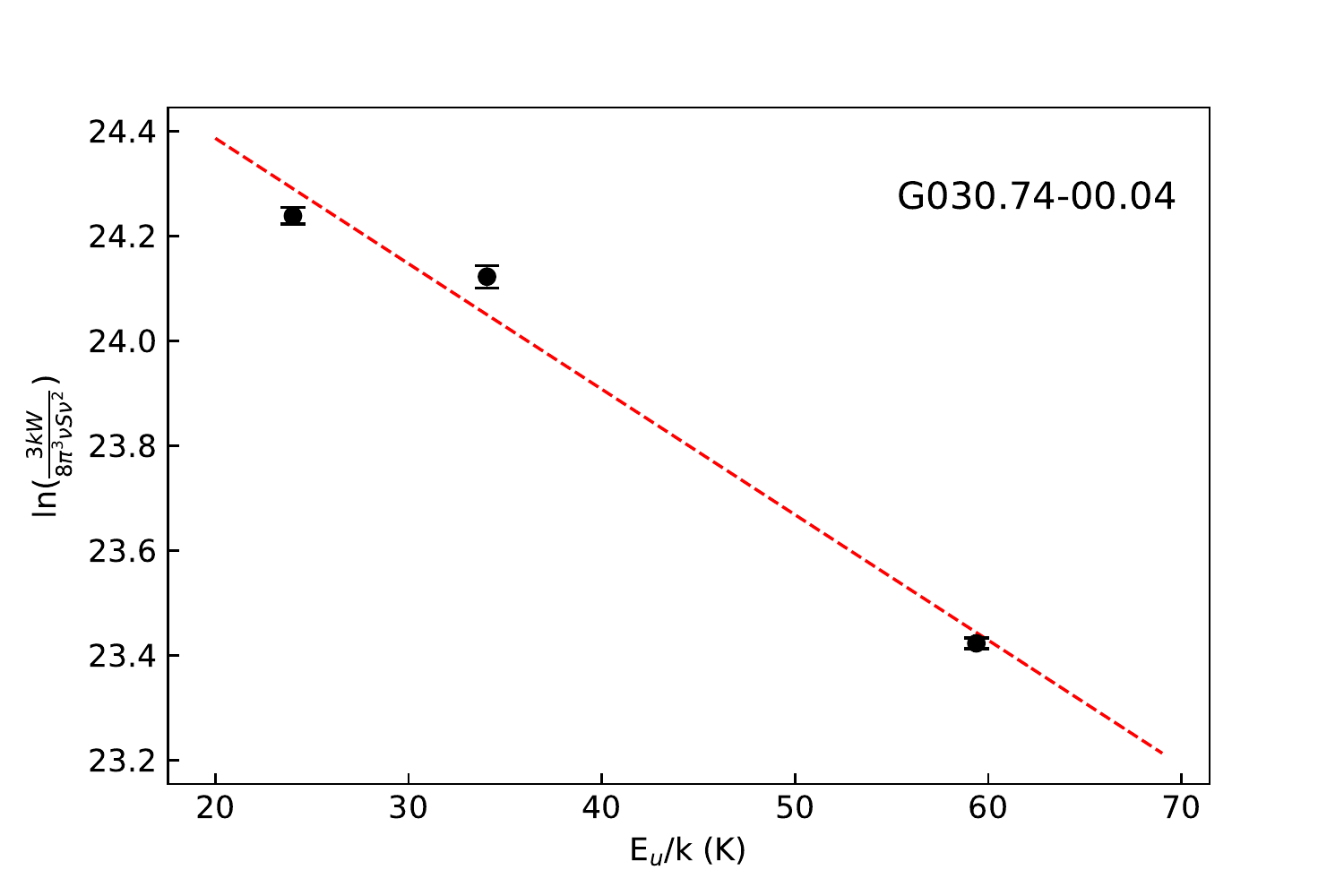}
    \includegraphics[width=0.3\textwidth]{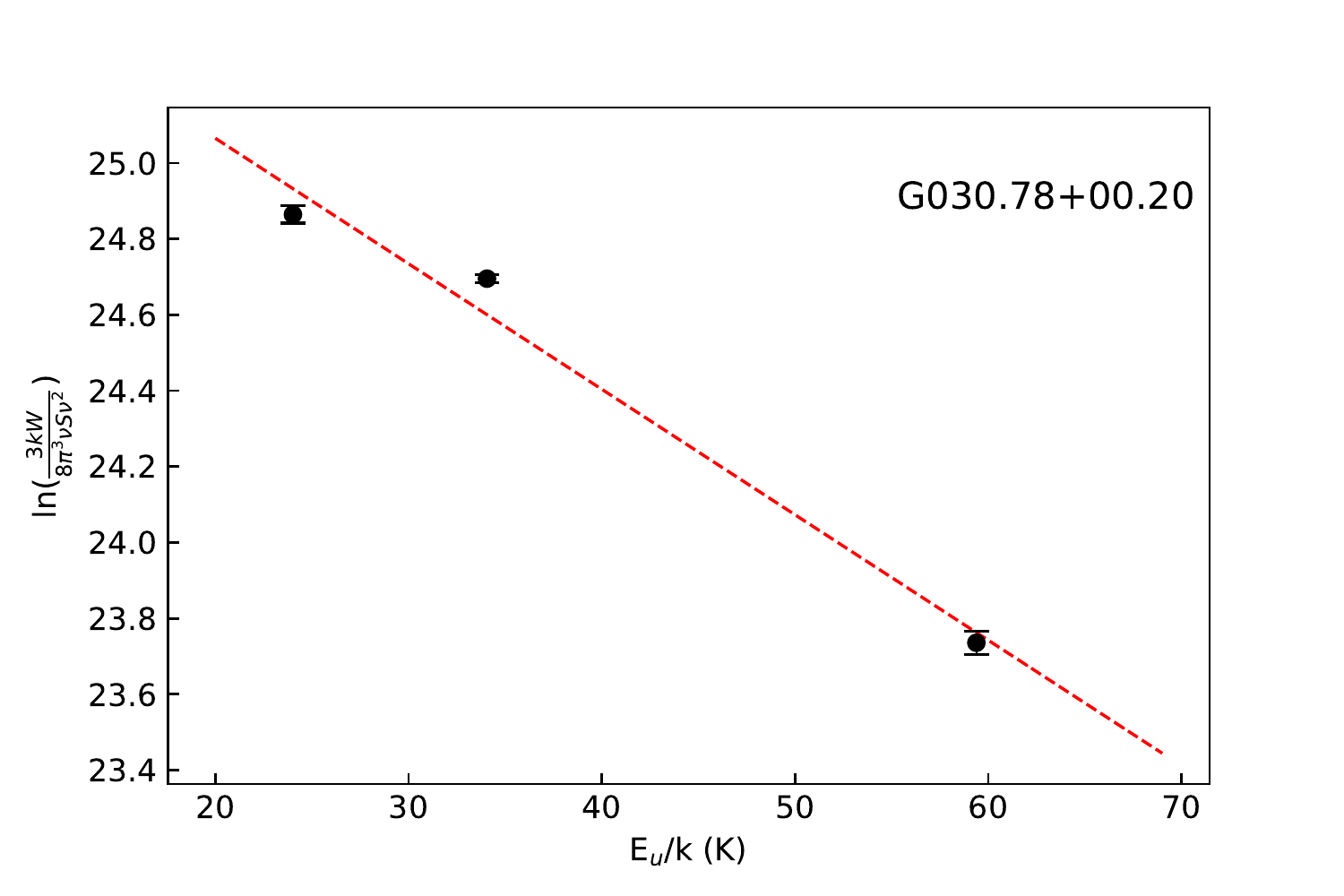} 
    \includegraphics[width=0.3\textwidth]{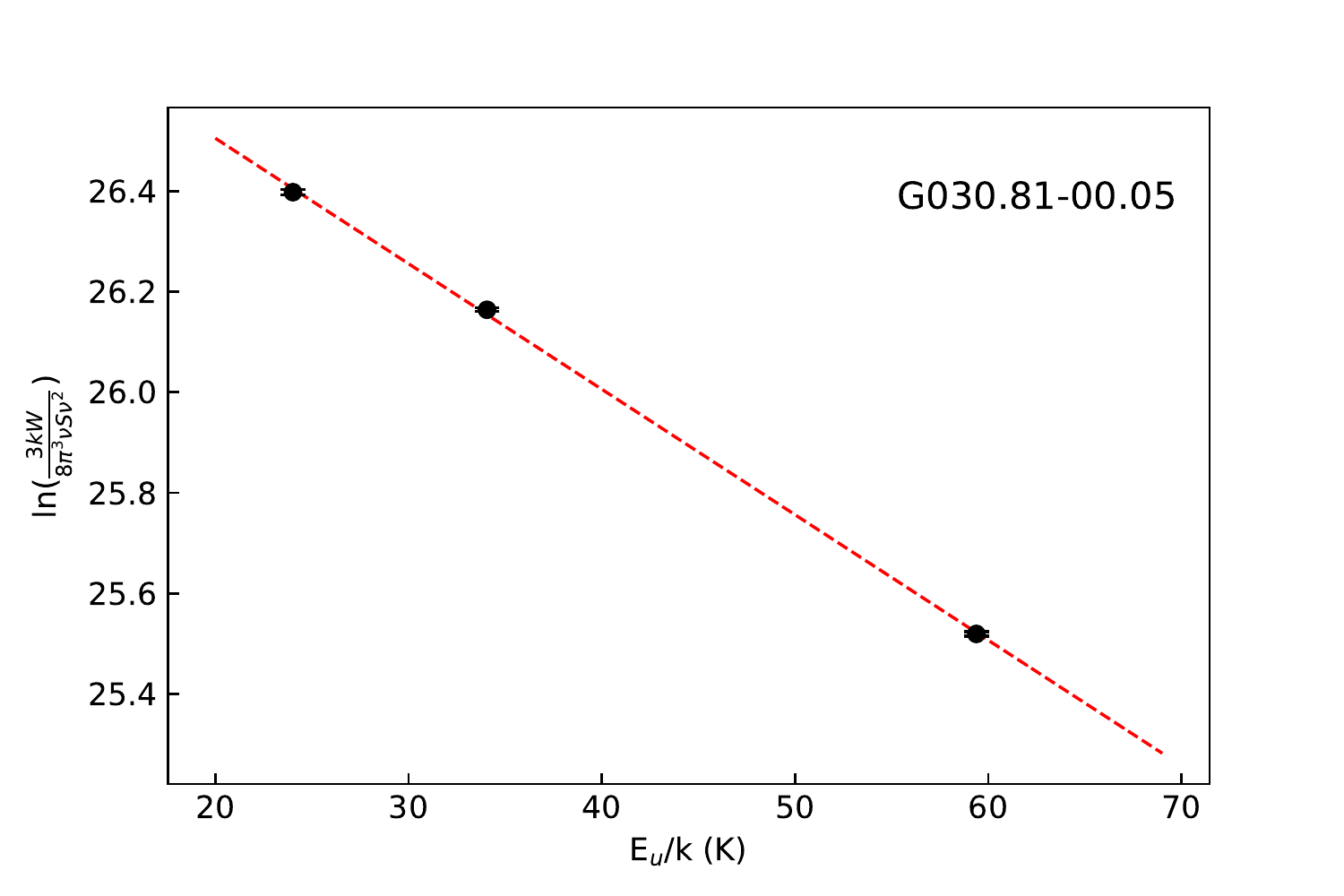}
    \includegraphics[width=0.3\textwidth]{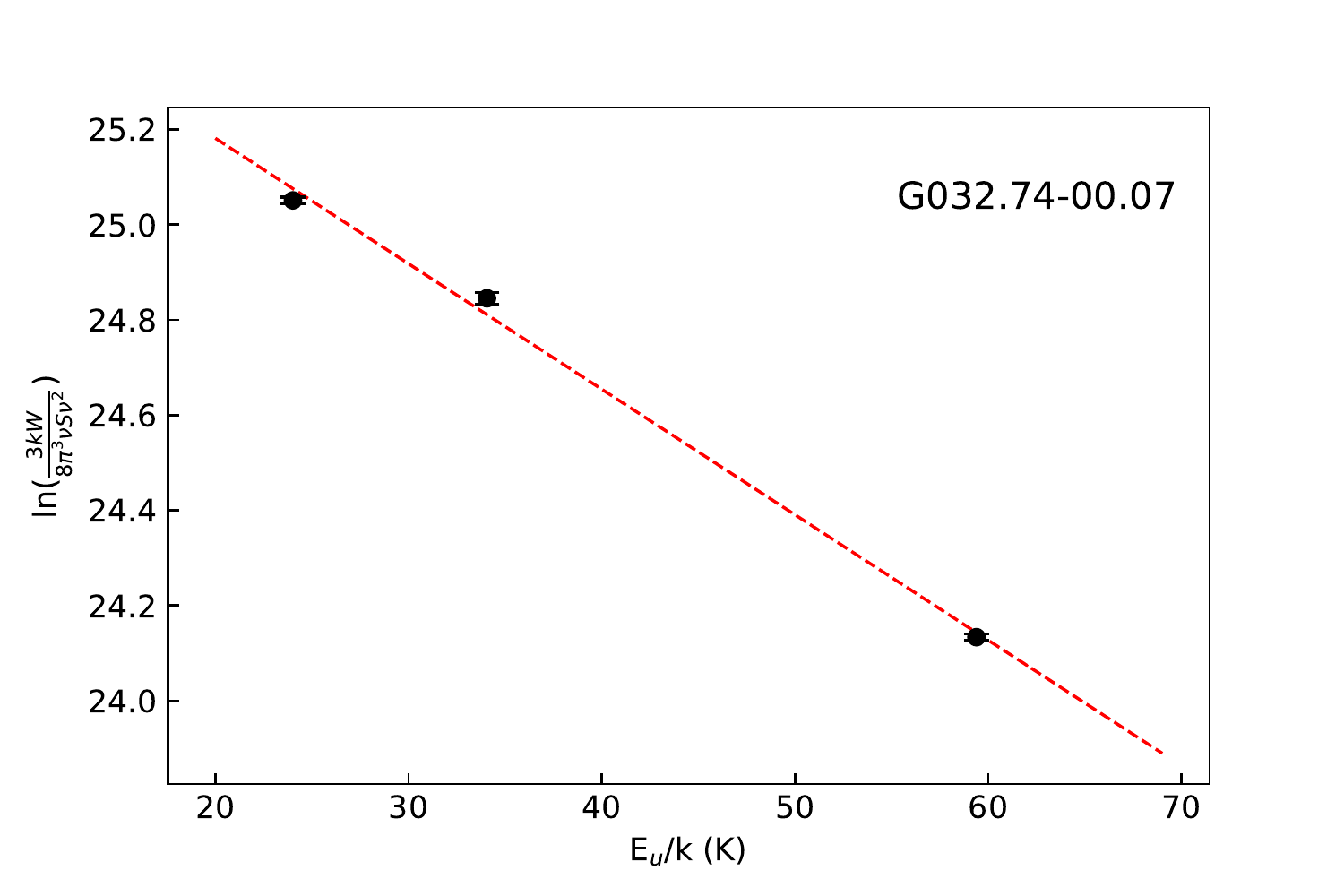} 
    \includegraphics[width=0.3\textwidth]{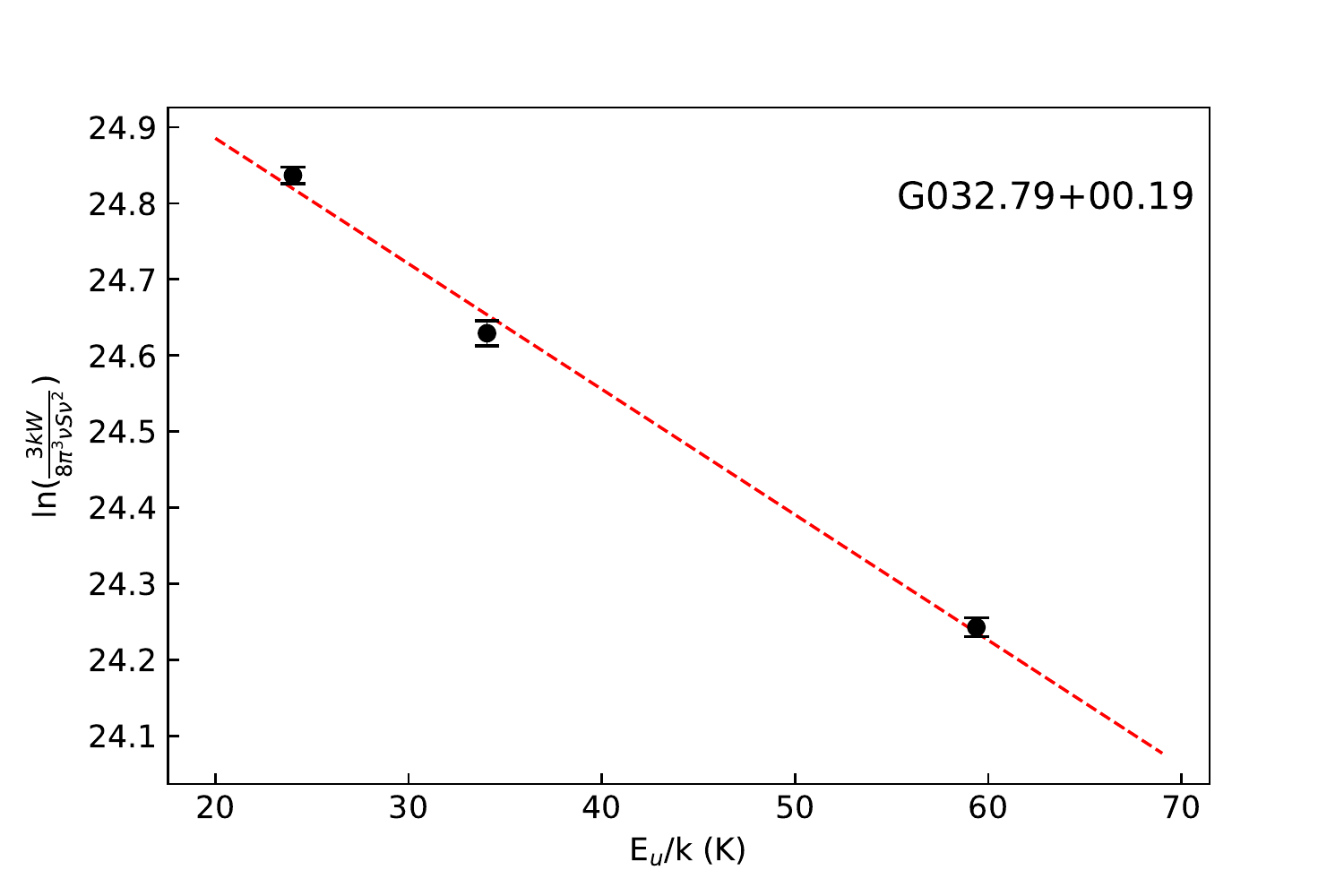}
    \includegraphics[width=0.3\textwidth]{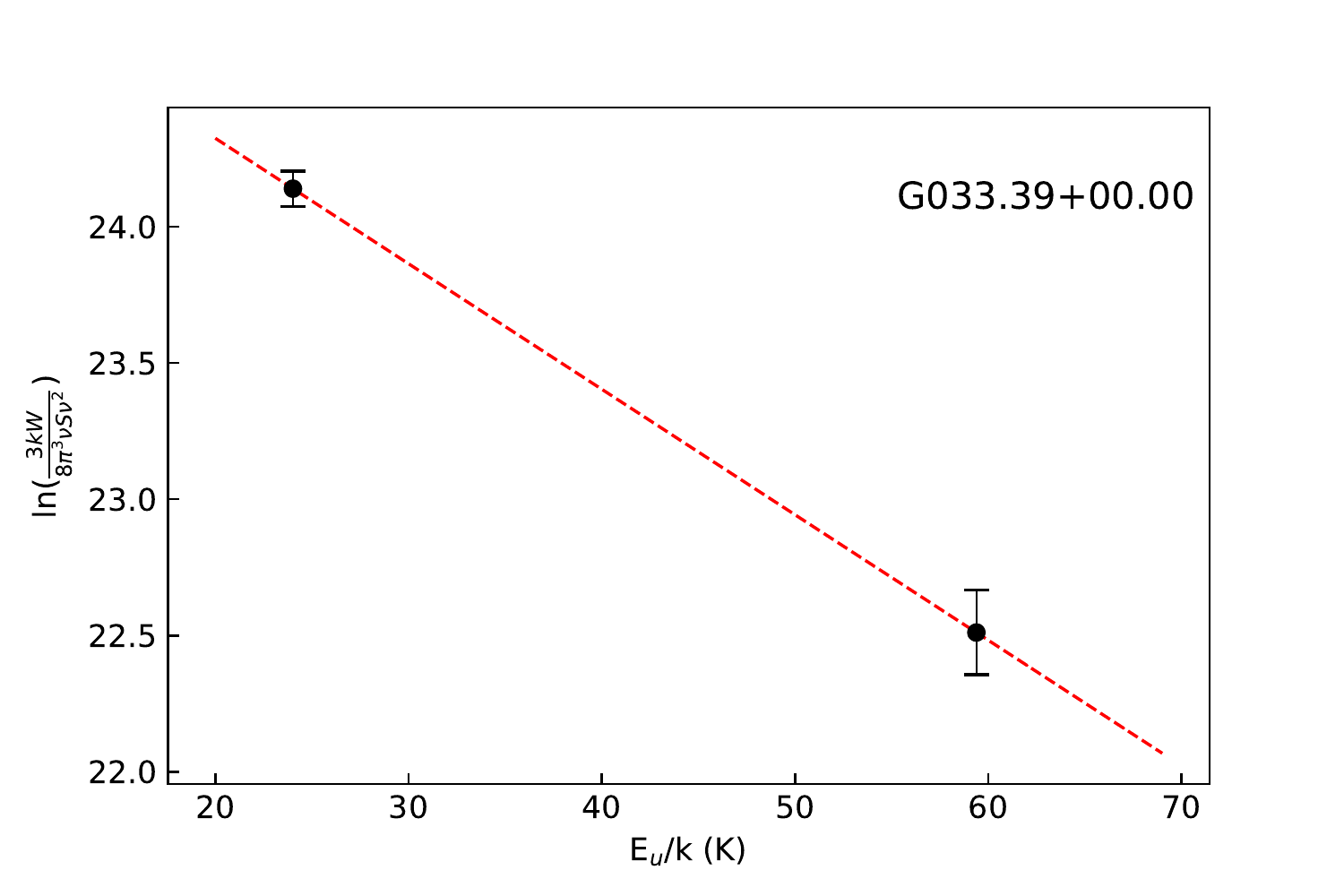}
    \includegraphics[width=0.3\textwidth]{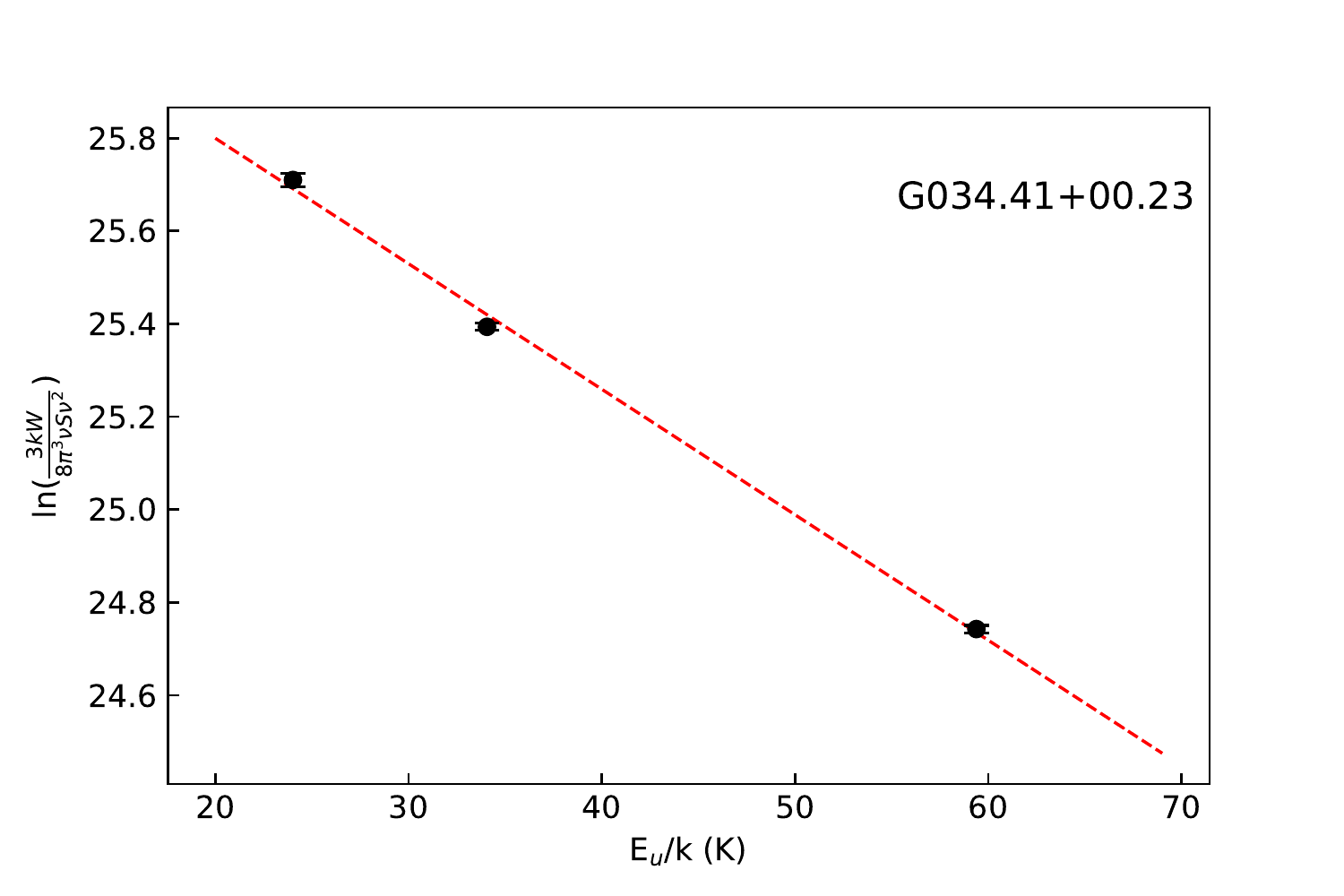} 
    \includegraphics[width=0.3\textwidth]{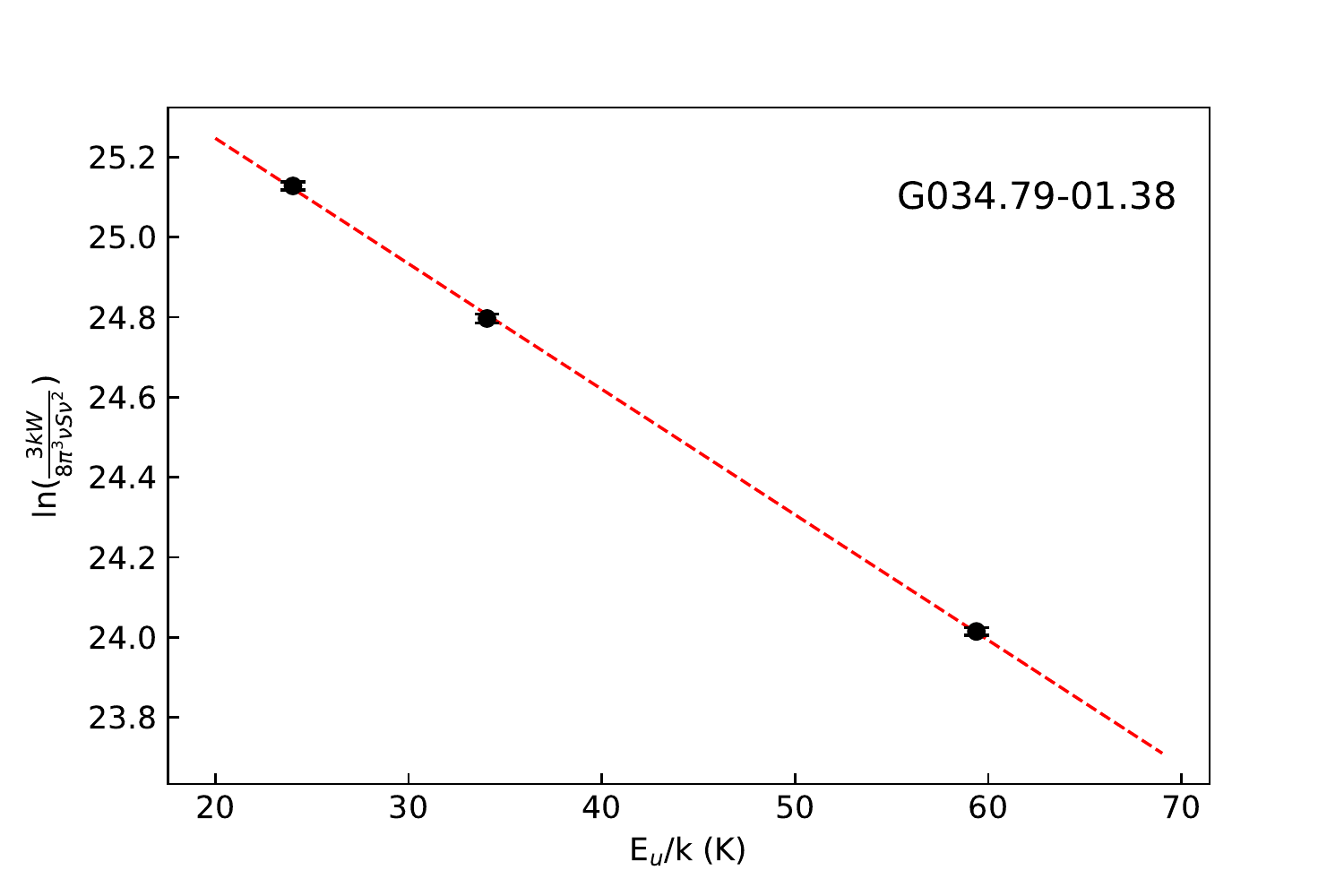}
    \includegraphics[width=0.3\textwidth]{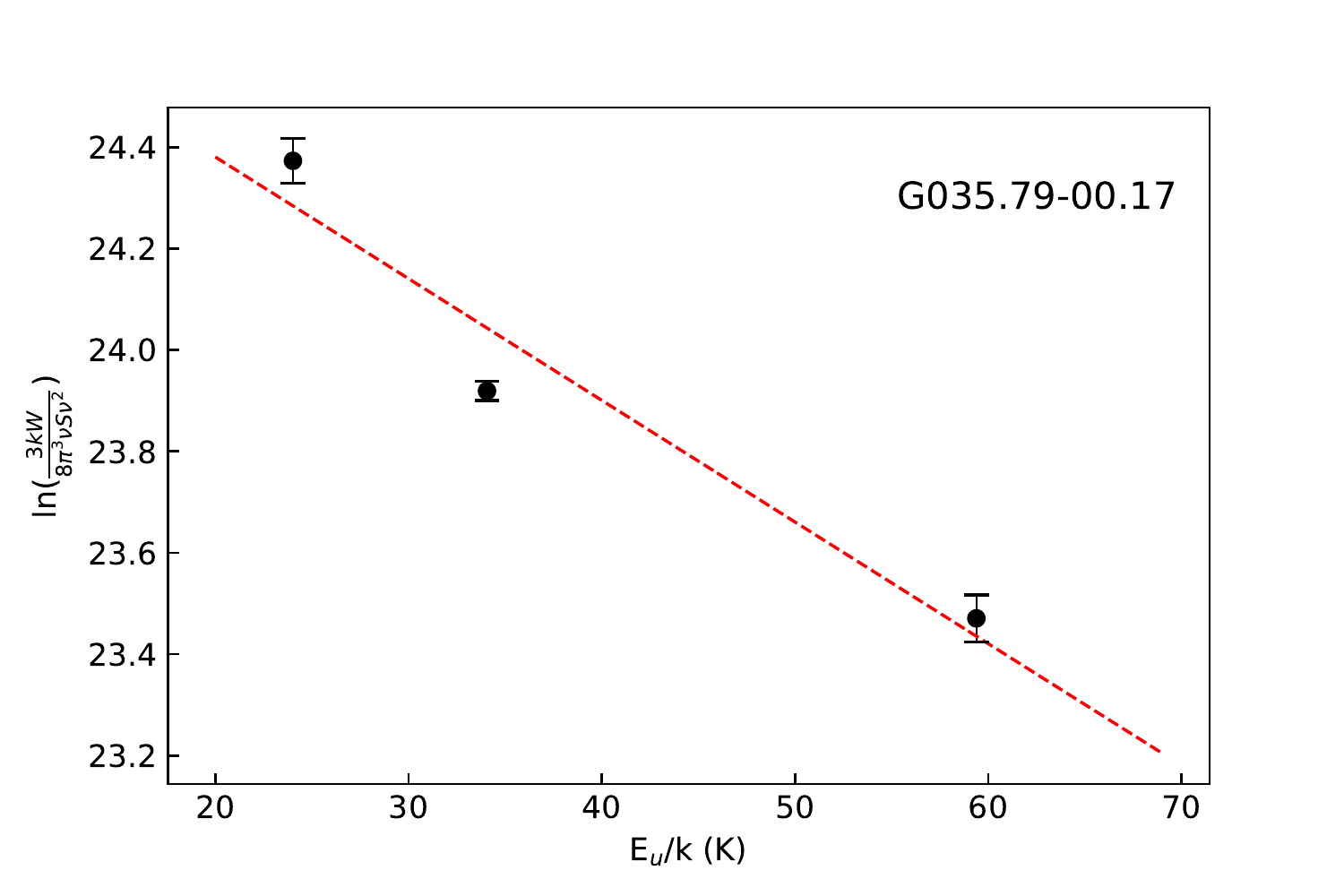}
    \includegraphics[width=0.3\textwidth]{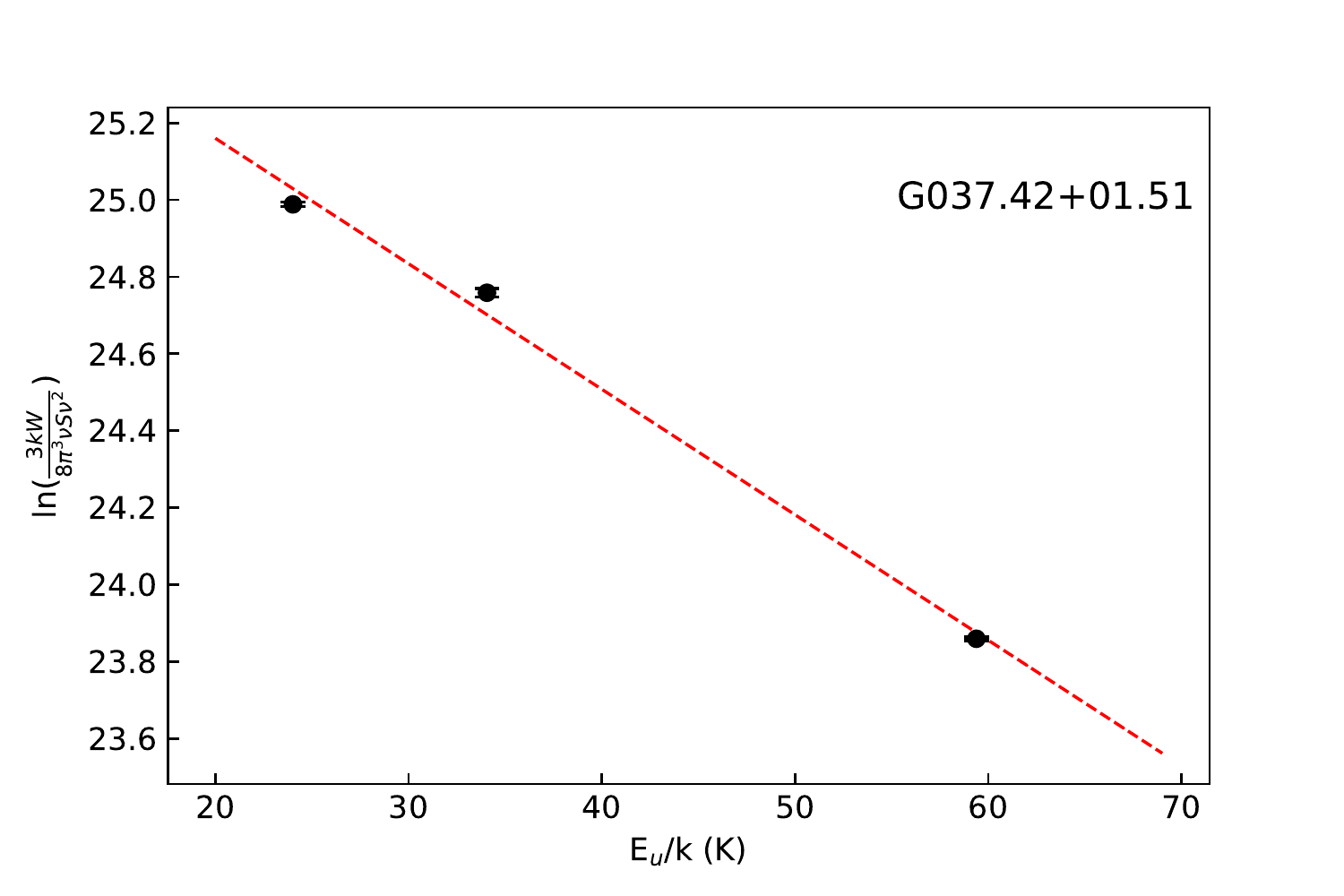} 
    \includegraphics[width=0.3\textwidth]{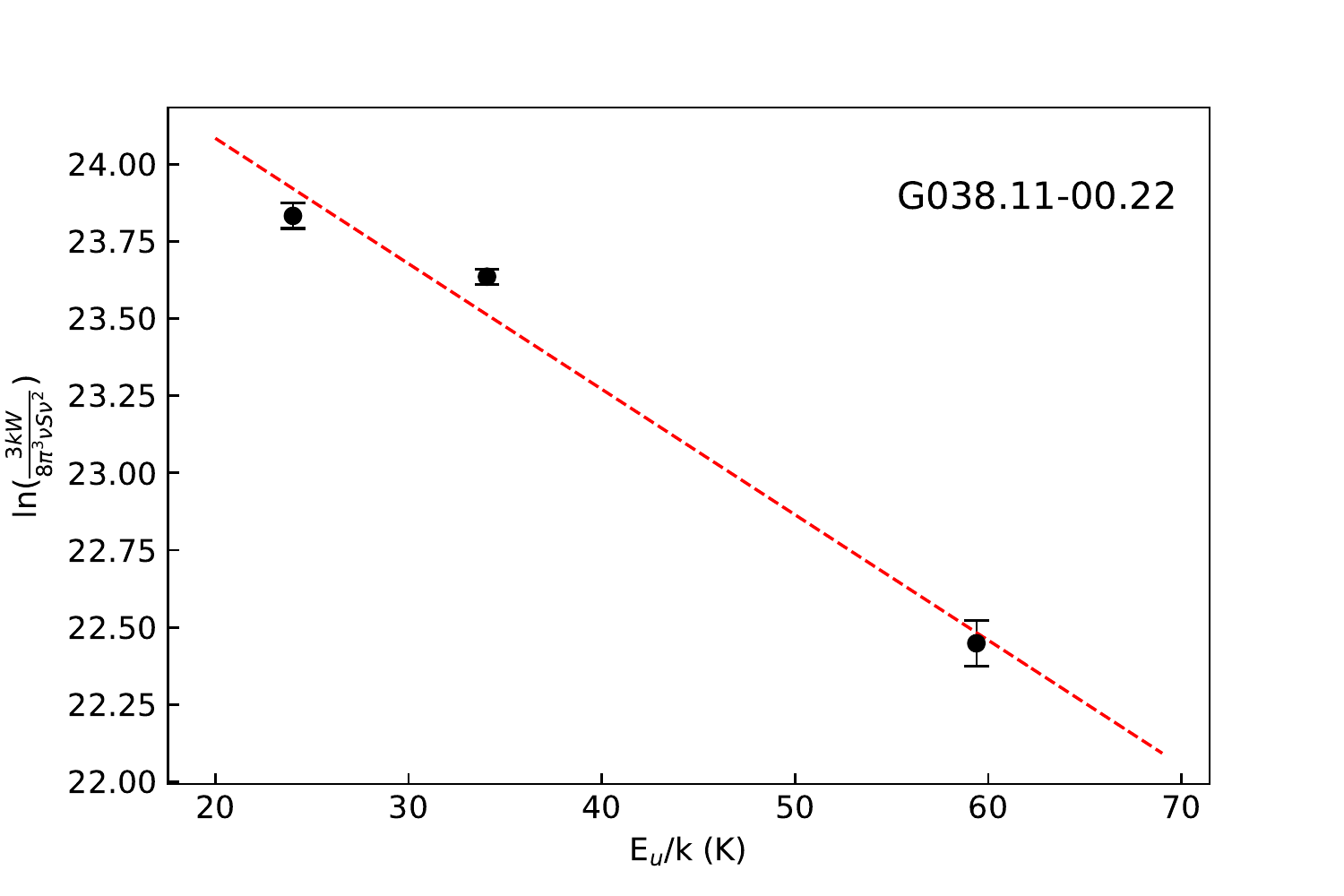}
    \includegraphics[width=0.3\textwidth]{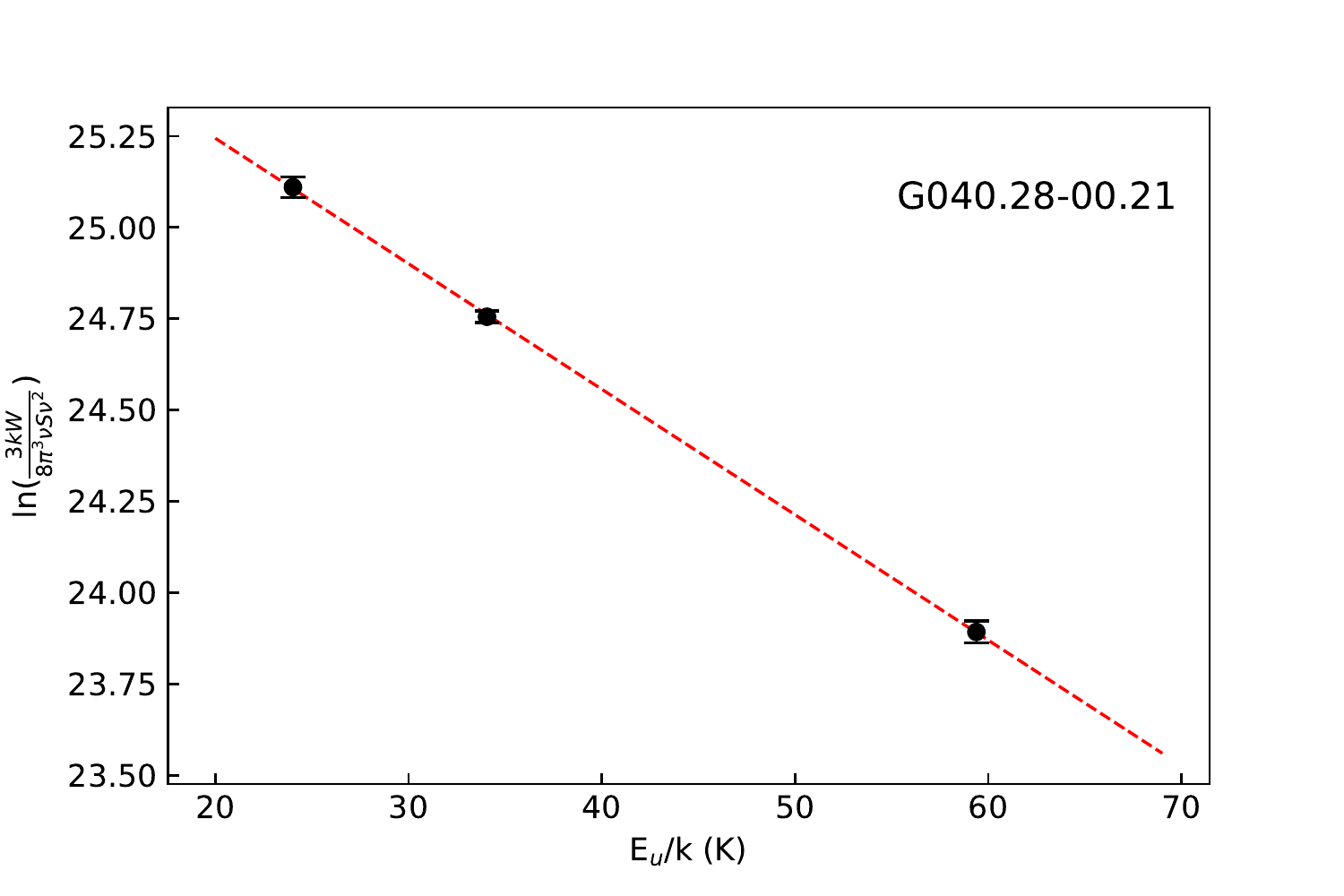} 
    \includegraphics[width=0.3\textwidth]{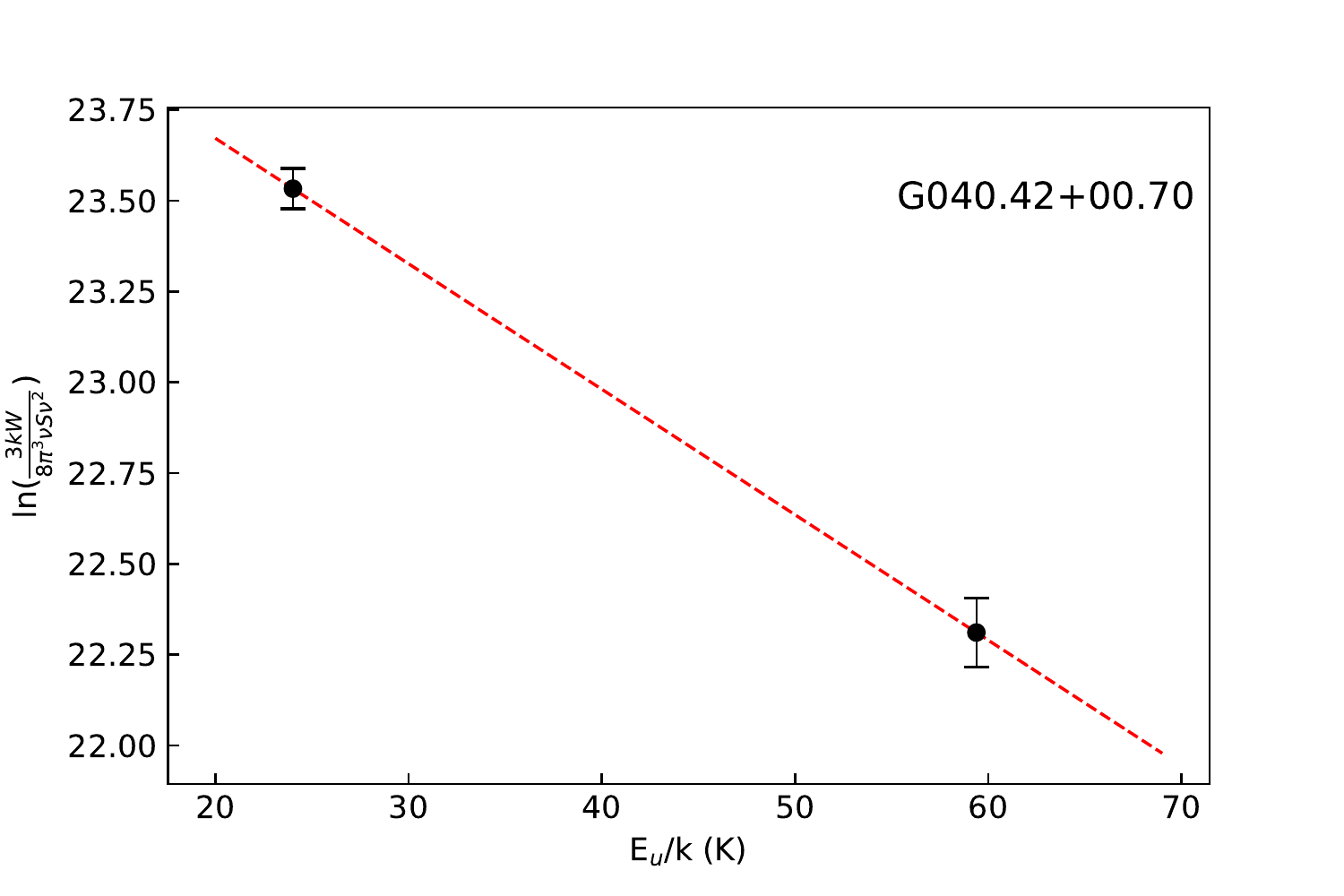} 
    \includegraphics[width=0.3\textwidth]{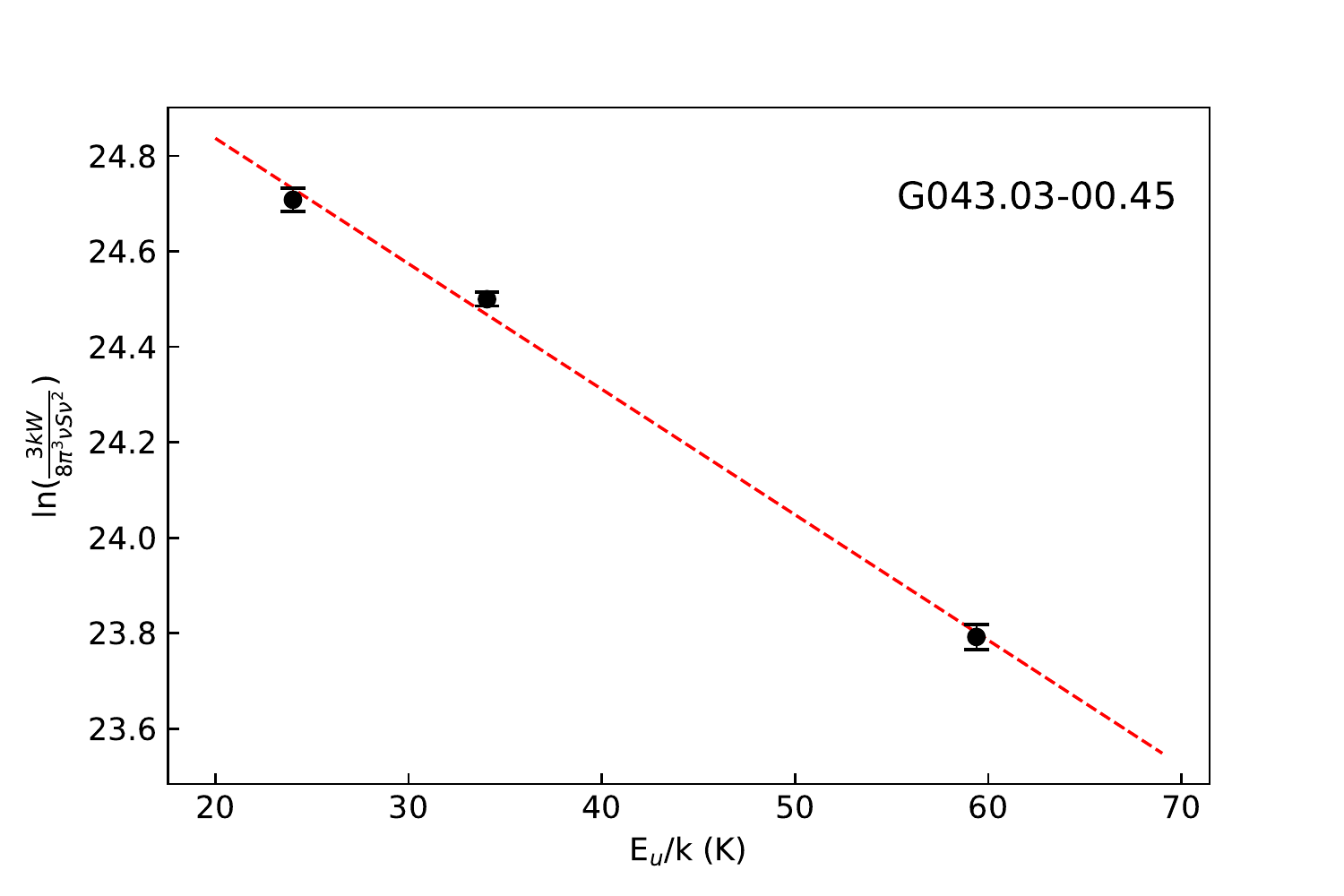} 
    \includegraphics[width=0.3\textwidth]{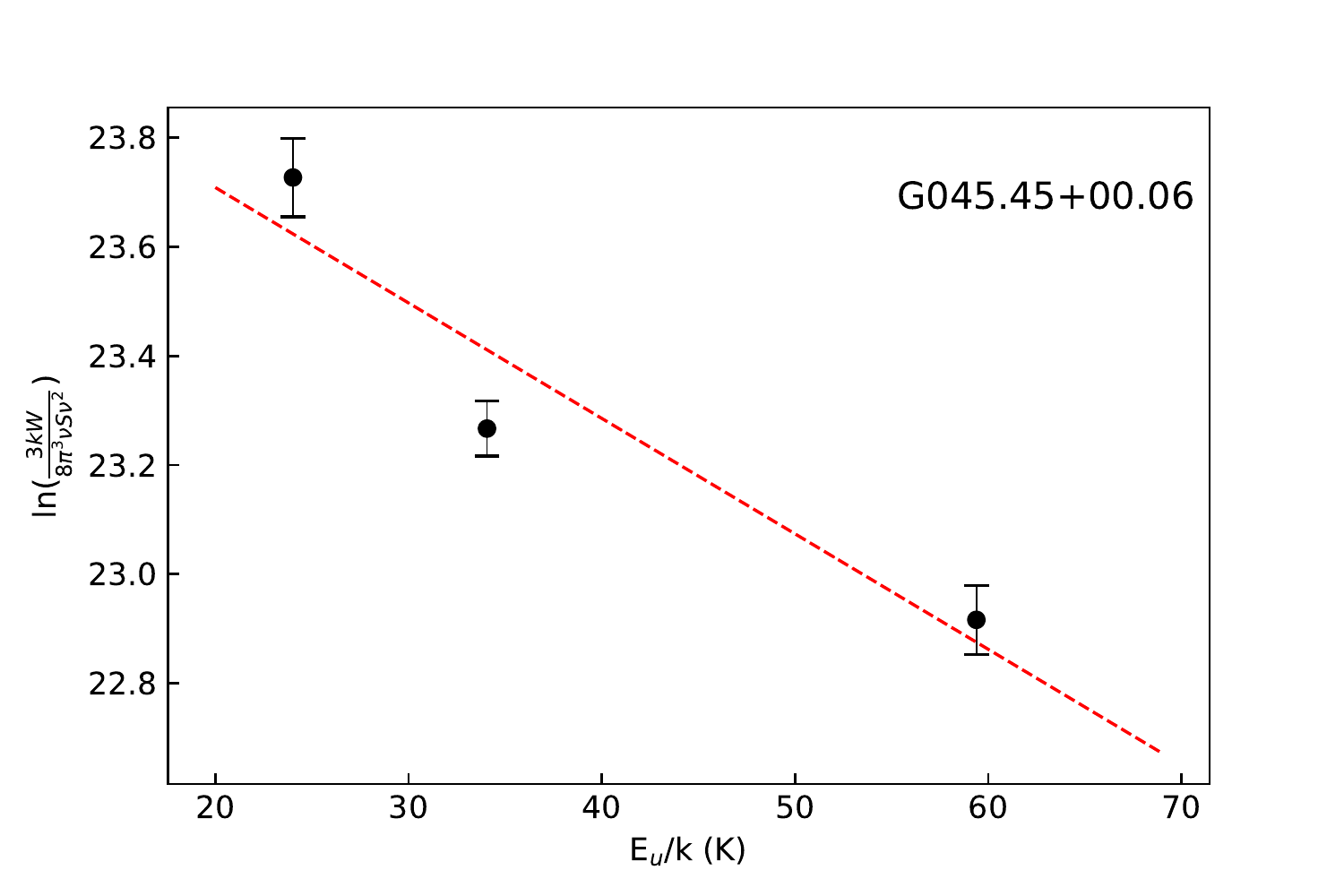} 
    \includegraphics[width=0.3\textwidth]{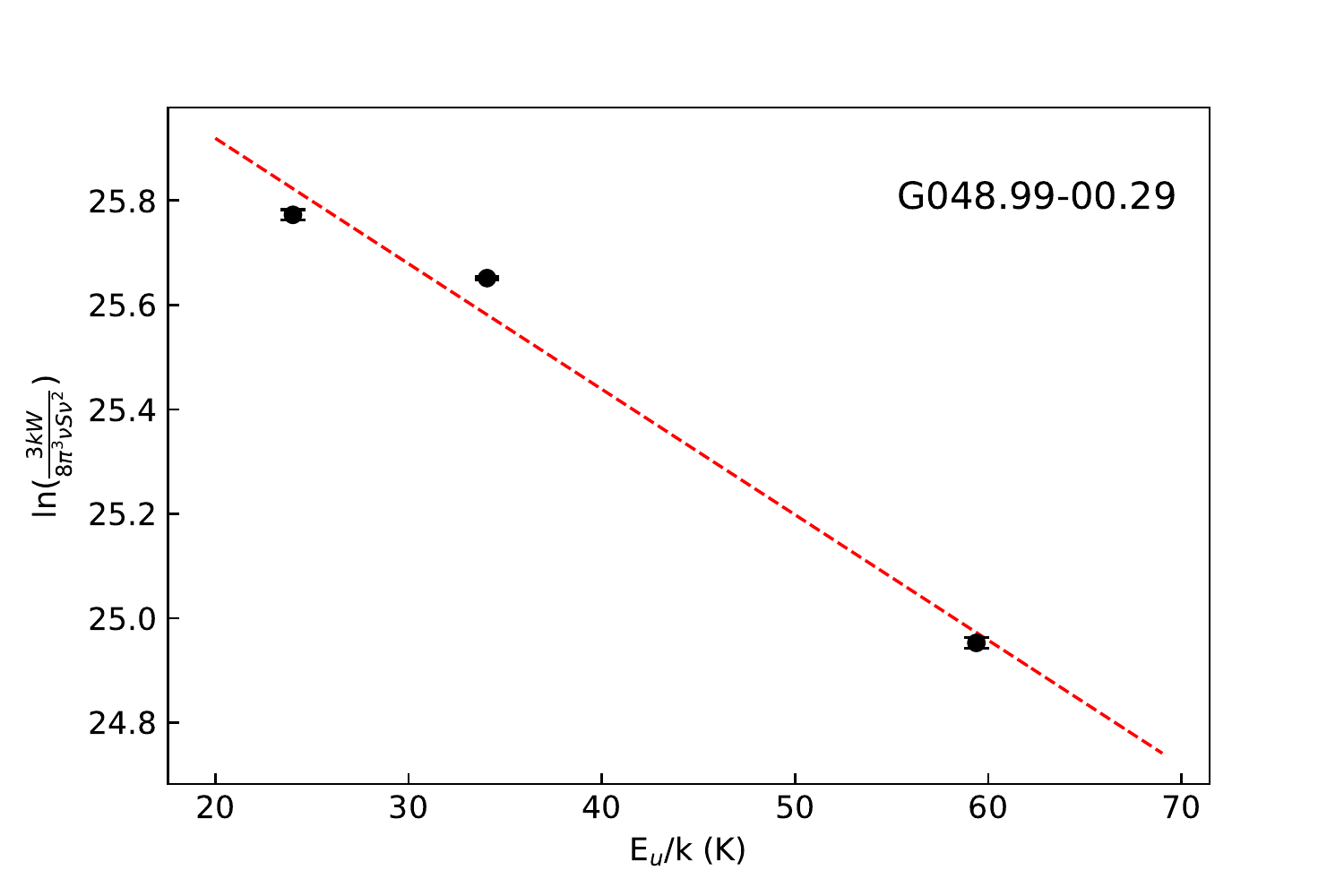} 
    \includegraphics[width=0.3\textwidth]{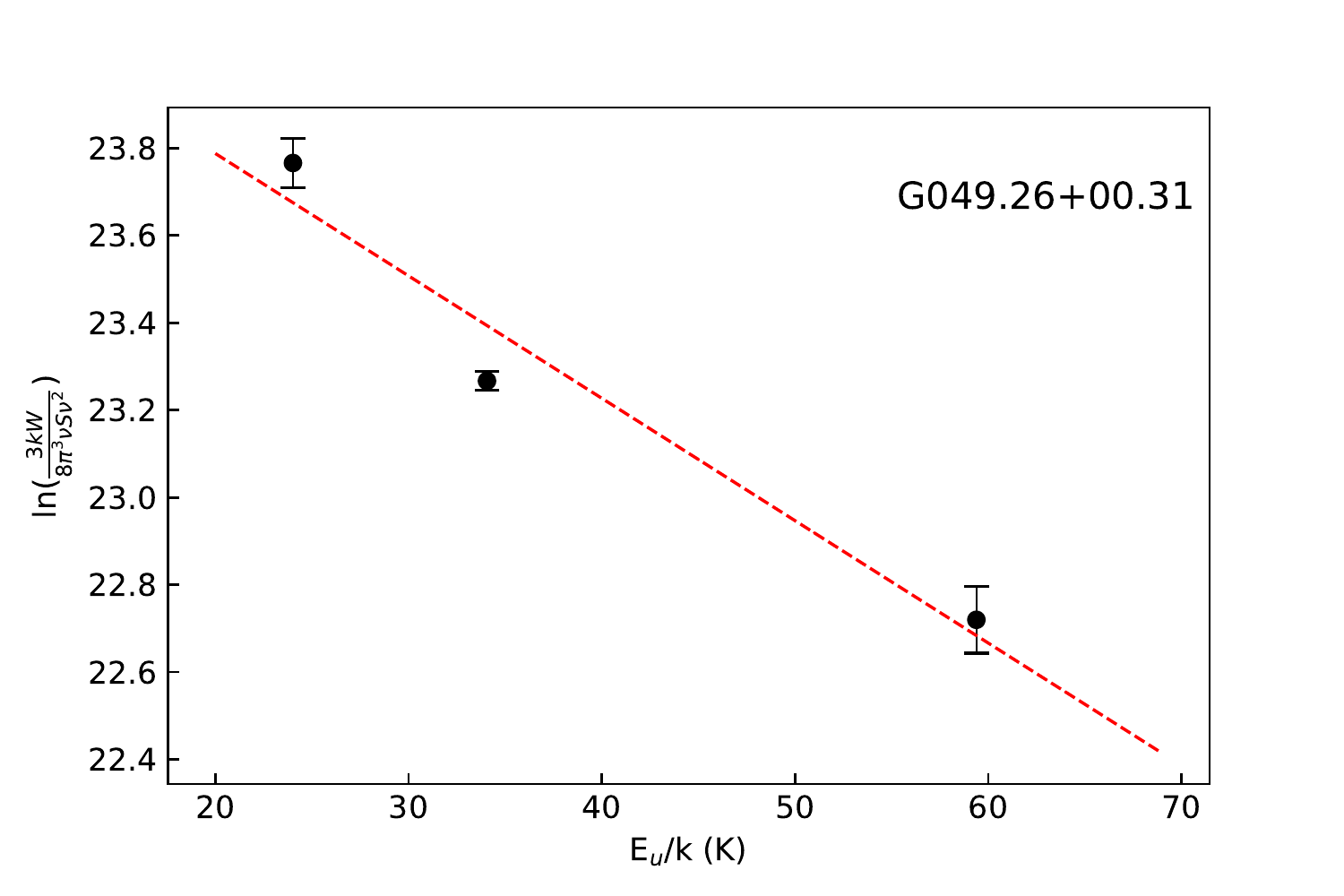}
    \caption{Continued.}
\end{figure*}
    
\addtocounter{figure}{-1}
\begin{figure*}    
    \centering
    \includegraphics[width=0.3\textwidth]{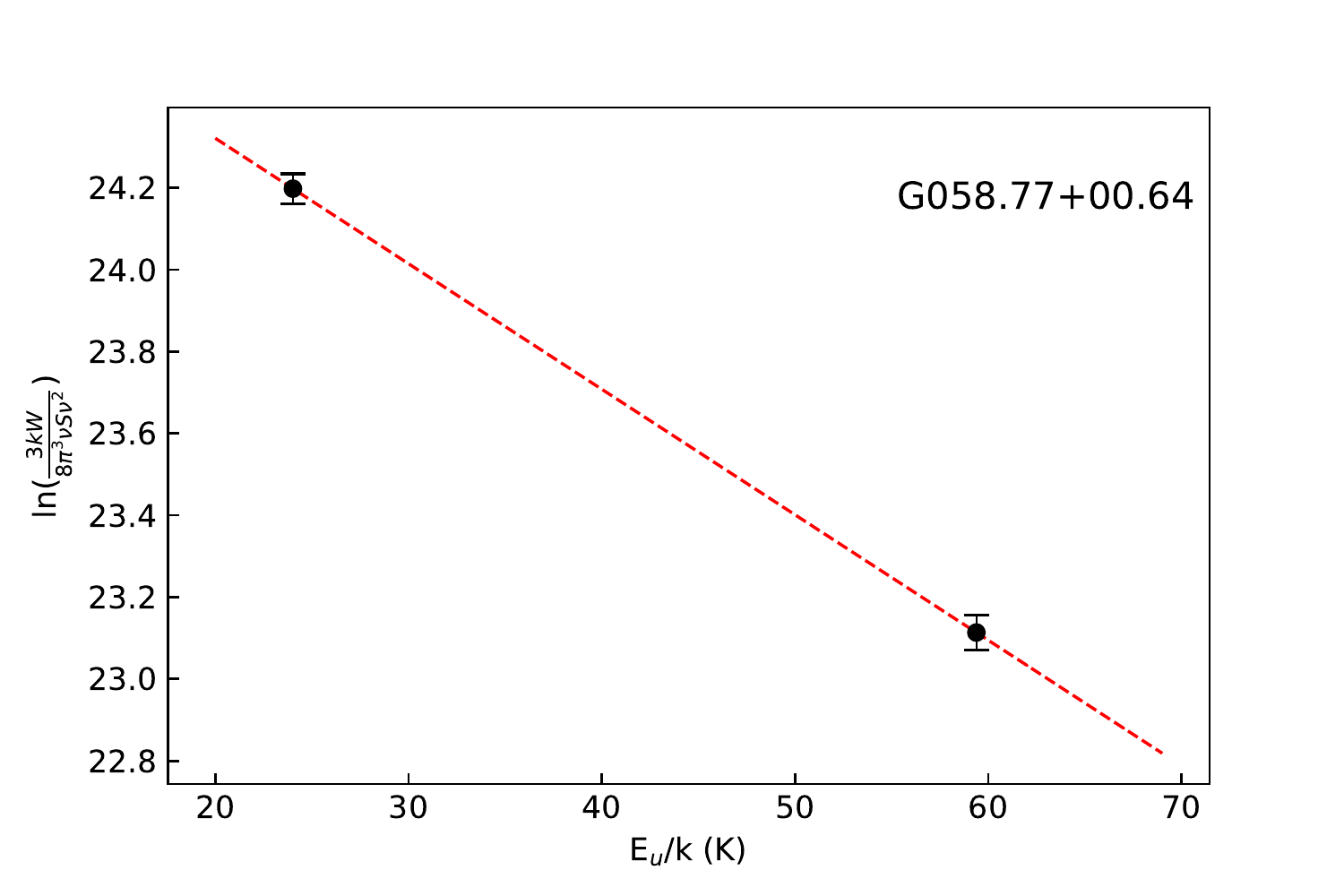} 
    \includegraphics[width=0.3\textwidth]{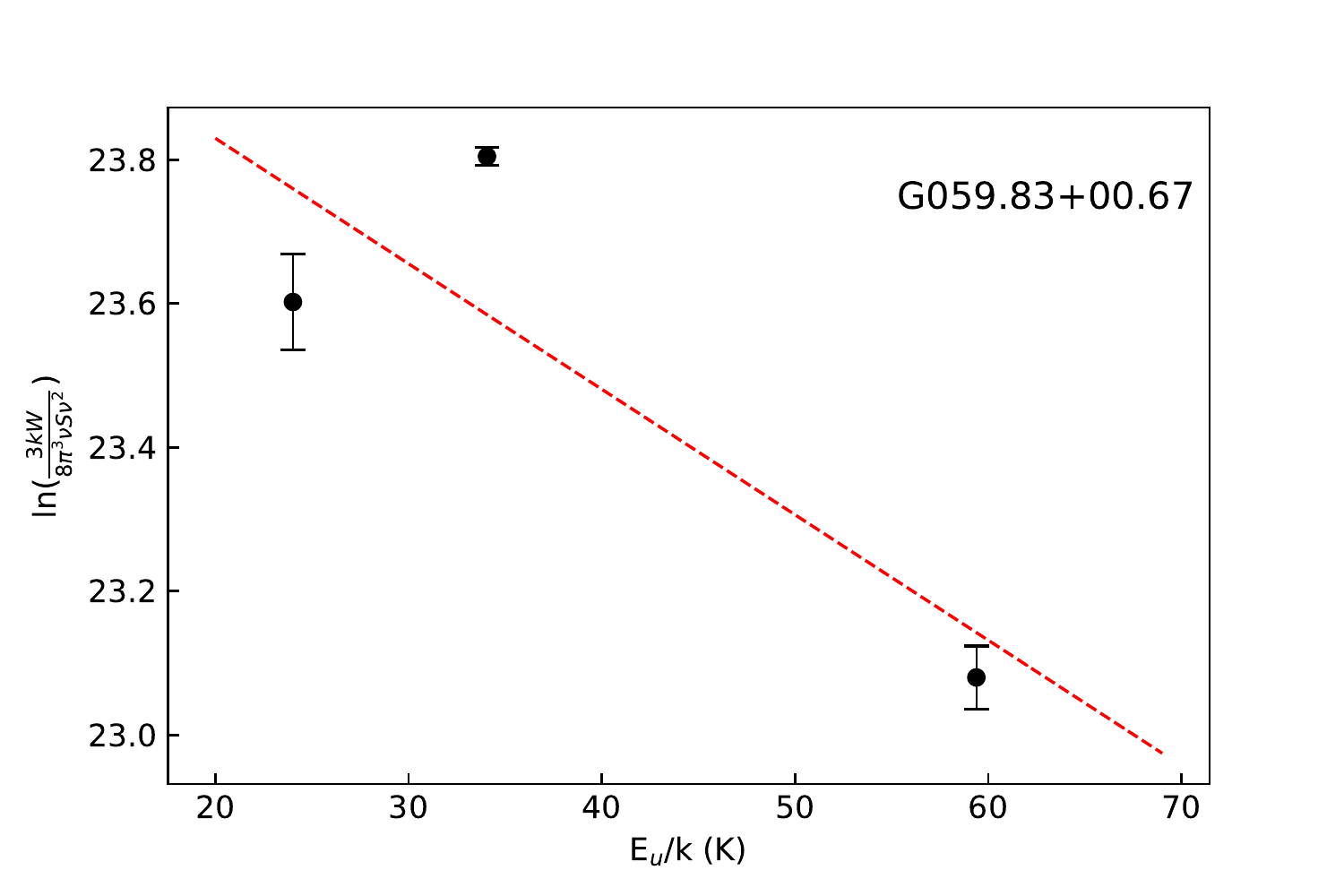} 
    \includegraphics[width=0.3\textwidth]{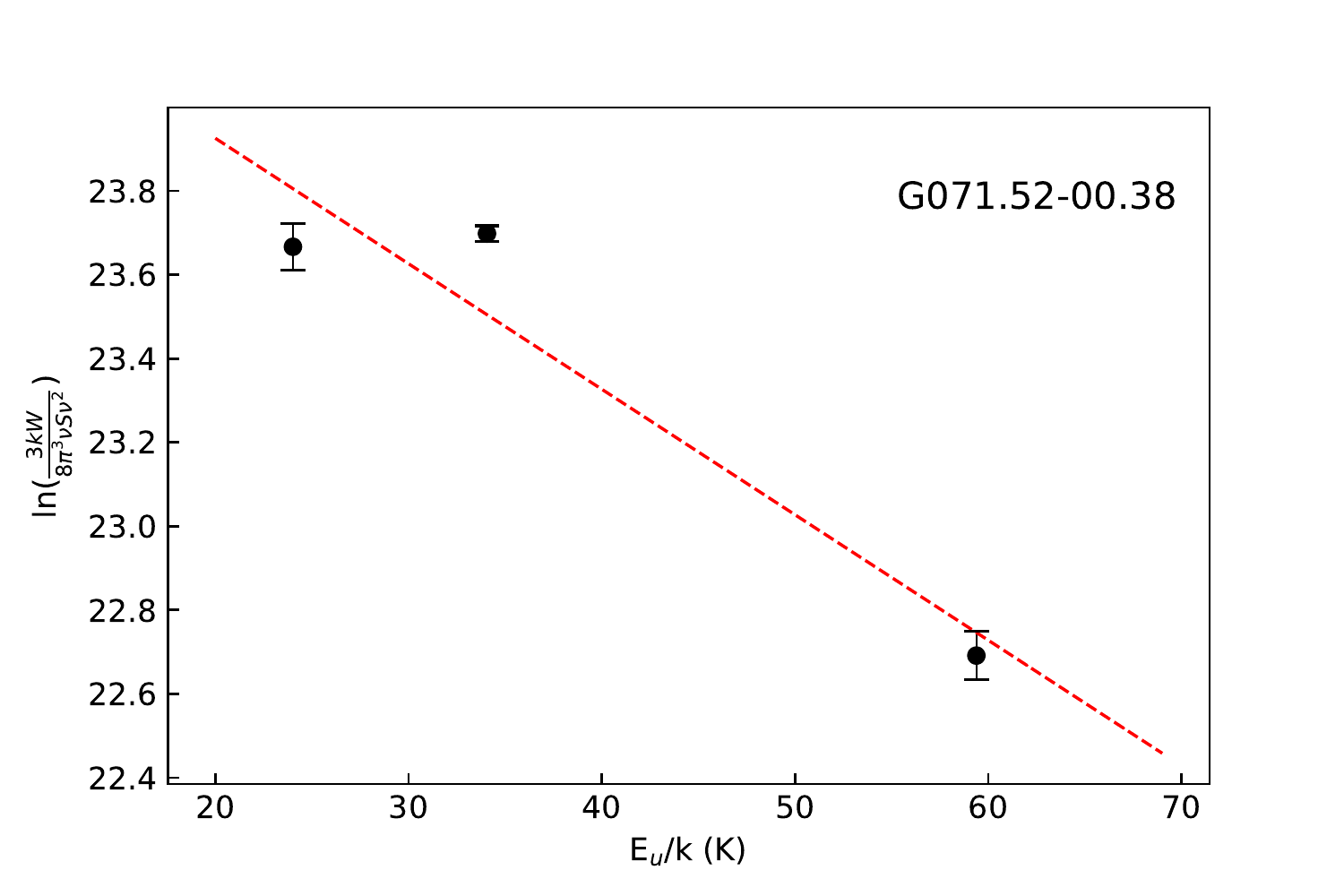}
    \includegraphics[width=0.3\textwidth]{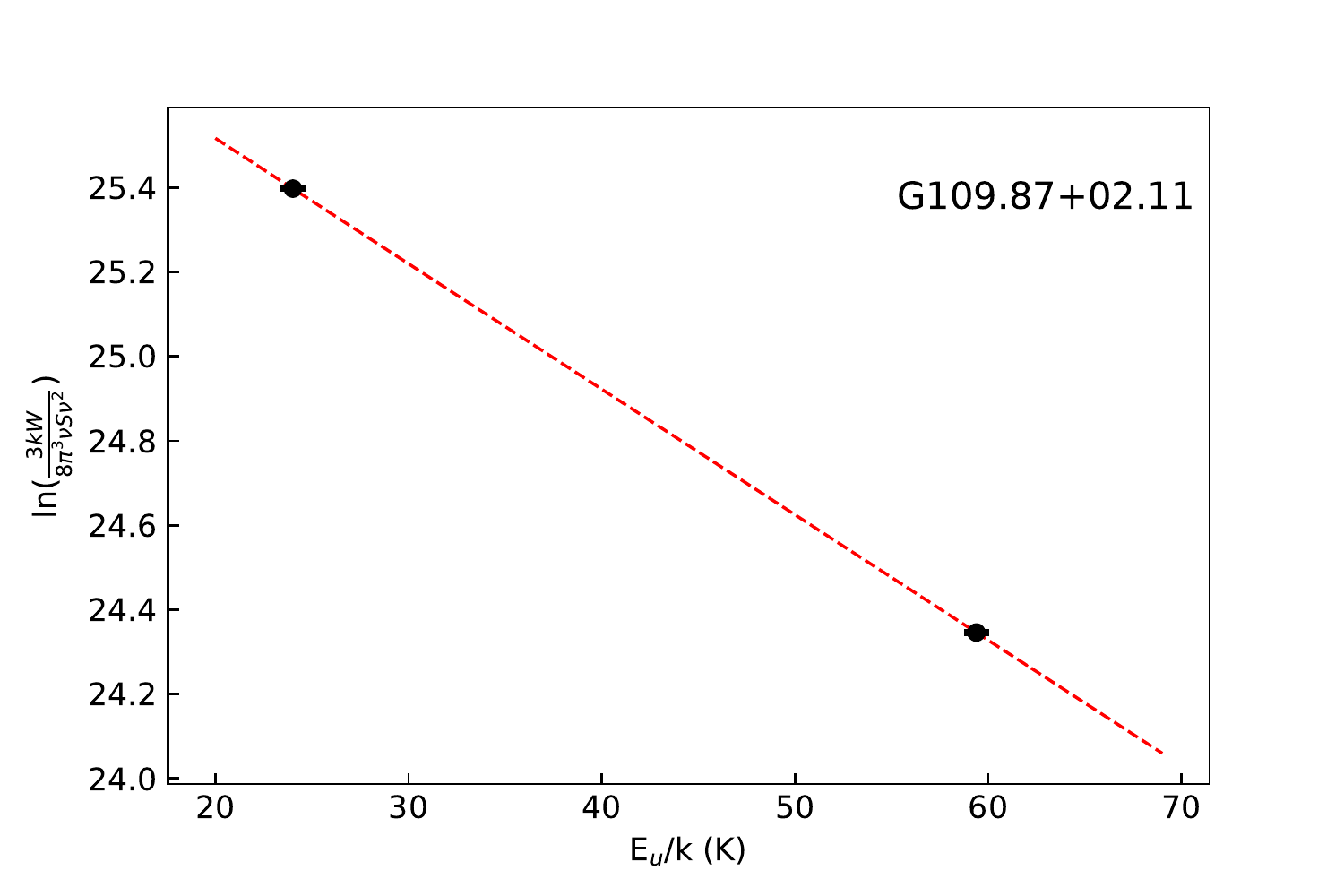} 
    \caption{Continued.}
    \label{fig5}
\end{figure*}

\begin{figure}
\centering
\includegraphics[width=\textwidth]{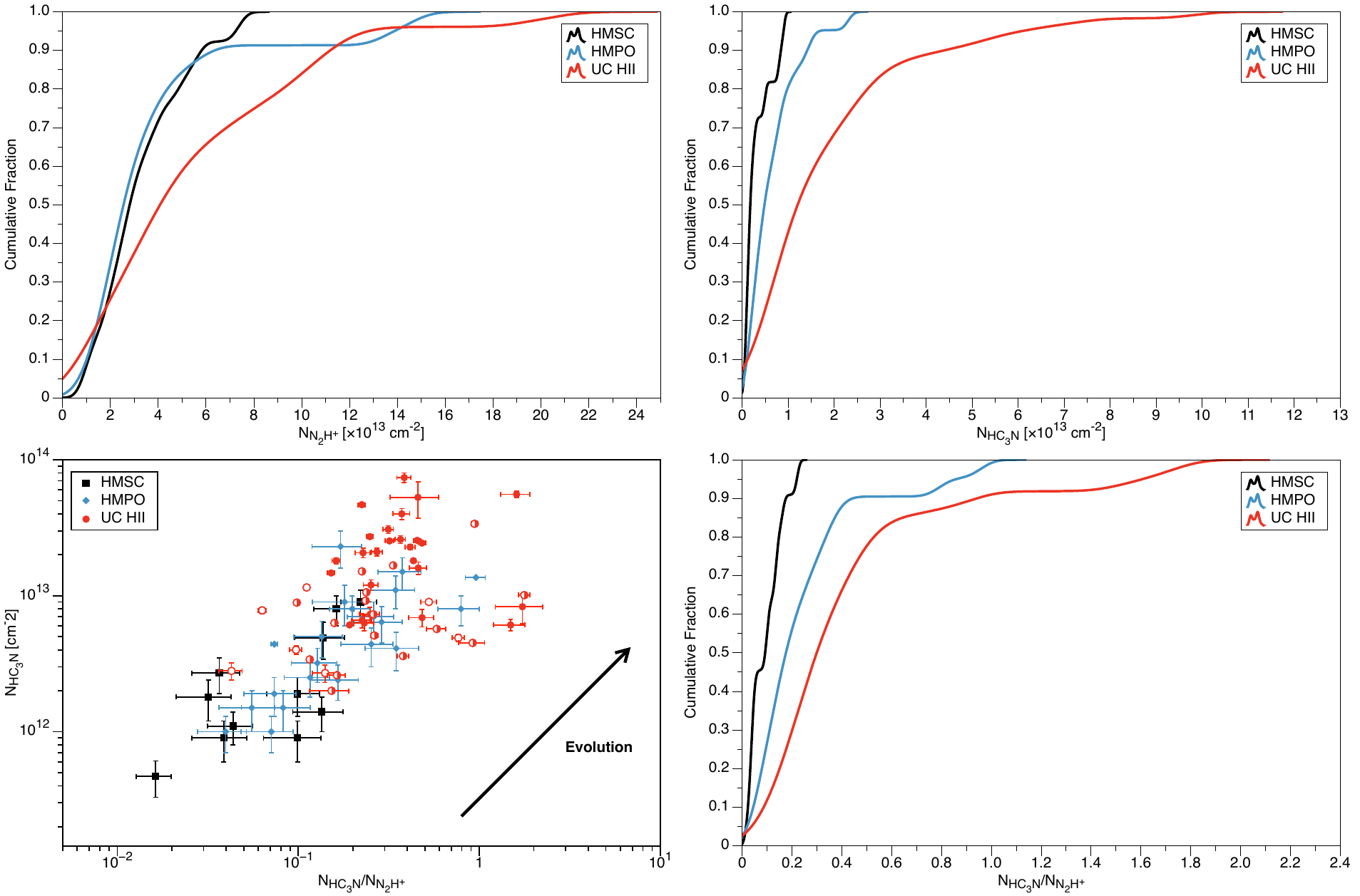}
\caption{The cumulative distributions for $N\rm (N_{2}H^{+})$ (top left panel), $N\rm (HC_{3}N)$ (top right) and their ratio values (bottom right) for HMSC, HMPO and UC H{\sc ii} region samples. The $N\rm (HC_{3}N)$/$N\rm (N_{2}H^{+})$ ratio is plotted against the $N\rm (HC_{3}N)$ (bottom left) and one clear evolution trend of the ratio can be found, i.e., from HMSC, HMPO to UC H{\sc ii} region stages (our data, red circles, half filling and empty ones for those sources where $N\rm (HC_{3}N)$ was derived using two HC$_{3}$N lines data and sources with only HC$_{3}$N J = 10$-$9 line, respectively). $N\rm (HC_{3}N)$ in HMSC and HMPO are taken from \citep{2019ApJ...872..154T}, $N\rm (N_{2}H^{+})$ in HMSC and HMPO are derived using the N$_{2}$H$^{+}$ (J = 1$-$0) data from \cite{2019ApJ...872..154T} and following the procedure of \cite{2009MNRAS.394..323P}.}
\label{fig6}
\end{figure}

\begin{figure}
\centering
\includegraphics[width=0.5\textwidth]{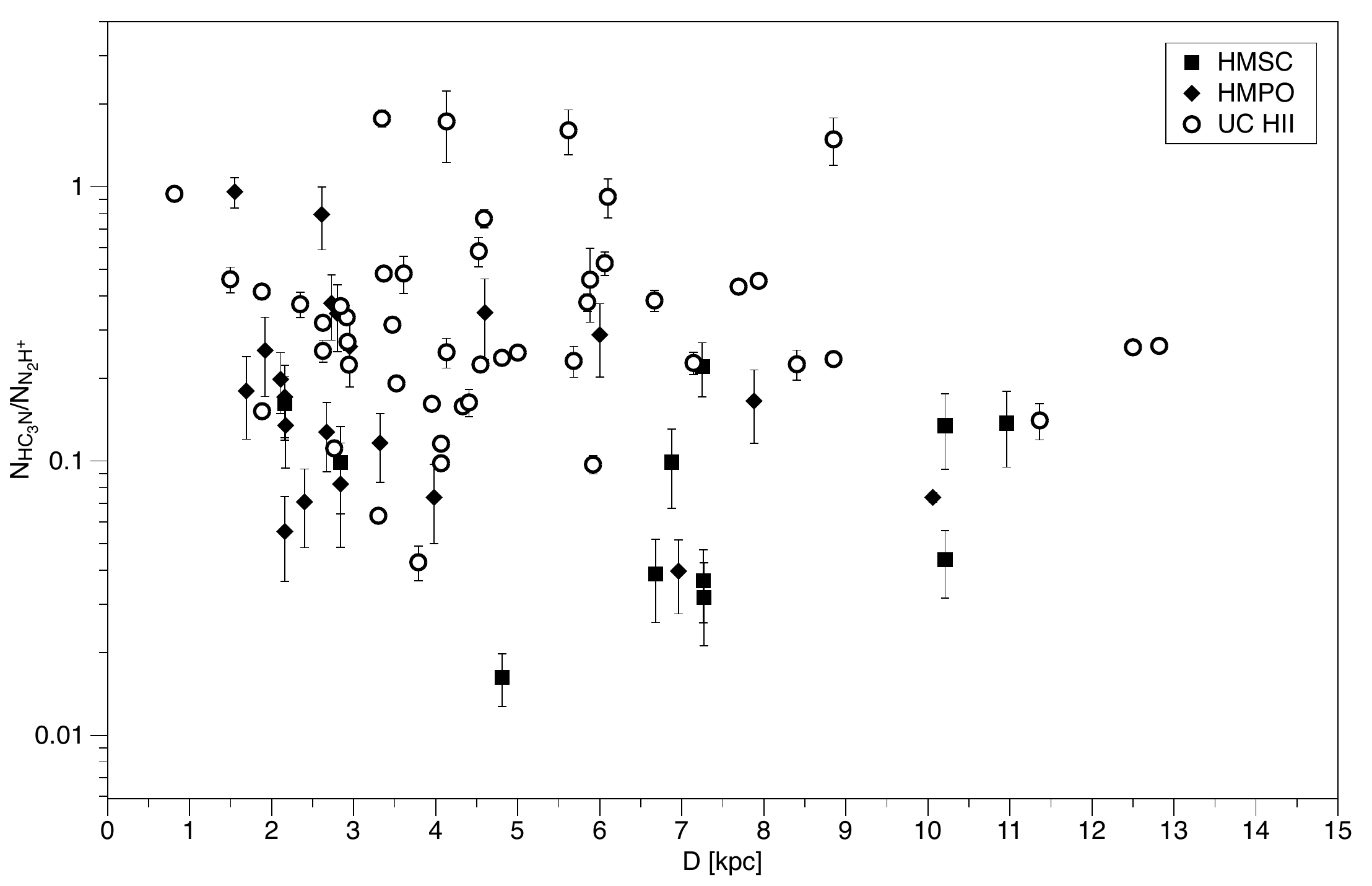}
    \caption{$N\rm (HC_{3}N)$/$N\rm (N_{2}H^{+})$ against the heliocentric distance and no significant variation can be found between them.}
    \label{fig7}
\end{figure}

\begin{figure}
\centering
\includegraphics[width=0.5\textwidth]{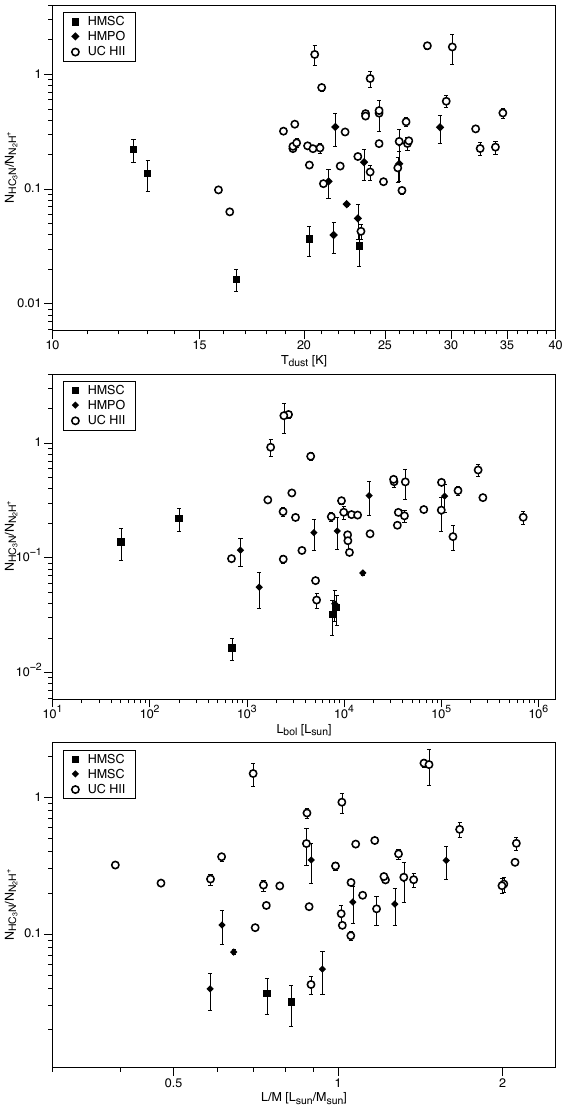}
    \caption{The correlations between $N\rm (HC_{3}N)$/$N\rm (N_{2}H^{+})$ and other evolutionary indicators, including dust temperature, bolometric luminosity, and luminosity-to-mass ratio.}
    \label{fig8}
\end{figure}

\end{CJK*}
\end{document}